%% file: paper.tex
\documentclass[12pt]{article}

\usepackage{amsthm,amsmath,amsfonts,graphicx,amssymb,mathtools}
\usepackage[colorlinks,citecolor=blue,urlcolor=blue]{hyperref}
\usepackage{natbib}
\usepackage{float}
\usepackage{bbm}
\usepackage{longtable}
\usepackage{caption}
\usepackage{subcaption}
\usepackage{enumitem}

\usepackage{adjustbox}

\usepackage{cleveref}

\theoremstyle{plain}
\newtheorem{theorem}{Theorem}

\newtheorem{proposition}{Proposition}
\newtheorem{corollary}{Corollary}
\newtheorem{assumption}{Assumption}
\theoremstyle{remark}
\newtheorem{remark}{Remark}

\usepackage[left=1in,right=1in,top=1in, bottom=1in]{geometry}

\usepackage{setspace}
\usepackage{pdfpages}
\begin{document}


	\title{Inference on model parameters with many L-moments}

\author{Luis A. F. Alvarez\thanks{Corresponding author. Department of Economics, University of São Paulo. \textit{E-mail address:} \href{mailto:luis.alvarez@usp.br}{luis.alvarez@usp.br}}
	\and Chang Chiann\thanks{Department of Statistics, University of S\~{a}o Paulo. \textit{E-mail address:} \href{mailto:chang@ime.usp.br}{chang@ime.usp.br}.} \and
	Pedro A. Morettin  \thanks{Department of Statistics, University of S\~{a}o Paulo. \textit{E-mail address:} \href{mailto:pam@ime.usp.br}{pam@ime.usp.br}.}}

	\date{November, 2025}
\maketitle
	\singlespacing
\begin{abstract}
			This paper studies parameter estimation using L-moments, an alternative to traditional moments with attractive statistical properties. The estimation of model parameters by matching sample L-moments is known to outperform maximum likelihood estimation (MLE) in small samples from popular distributions. The choice of the number of L-moments used in estimation remains \textit{ad-hoc}, though: researchers typically set the number of L-moments equal to the number of parameters, which is inefficient in larger samples. In this paper, we show that, by properly choosing the number of L-moments and weighting these accordingly, one is able to construct an estimator that outperforms MLE in finite samples, and yet retains asymptotic efficiency. We do so by introducing a generalised method of L-moments estimator and deriving its properties in an asymptotic framework where the number of L-moments varies with sample size. We then propose methods to automatically select the number of L-moments in a sample. Monte Carlo evidence shows our approach can provide mean-squared-error improvements over MLE in smaller samples, whilst working as well as it in larger samples. We consider extensions of our approach to the estimation of conditional models and a class semiparametric models. We apply the latter to study expenditure patterns in a ridesharing platform in Brazil.
			
				\vspace{1em}
			\noindent%
			\textit{Keywords:}  L-statistics; quantiles; generalised method of moments; tuning parameter selection methods; higher-order expansions.
			\vfill
\end{abstract}

\onehalfspacing
\section{Introduction}
\input{sections/introduction}
\input{sections/motivation}

\section{Asymptotic properties of the generalised method of L-moments estimator with many moments}
\label{properties}
\input{sections/properties}

\section{Choosing the number of L-moments in estimation}
\label{selection}
\input{sections/selection}

\section{Extensions}
\label{extensions}

\input{sections/extensions}

\section{Concluding remarks}
\label{conclusion}

\input{sections/conclusion}

\section*{Acknowledgments}
	Earlier working paper versions of this article circulated as ``Inference in parametric models with many L-moments''.  We thank Cristine Pinto, Marcelo Fernandes, Marcos Fernandes, Ricardo Masini, Silvia Ferrari and Victor Orestes and two anonymous referees for their useful comments and suggestions. We are also thankful to Ciro Biderman for sharing his dataset with us. All the remaining errors are ours. Luis Alvarez gratefully acknowledges financial support from Capes grant 88887.353415/2019-00 and Cnpq grant 141881/2020-8. Chang Chiann and Pedro Morettin gratefully acknowledge financial support from Fapesp grant 2018/04654-9 and 2023/02538-0.

{
	\singlespacing
\bibliographystyle{chicago} 
\bibliography{bibliography}       
}


\end{document}


\pagestyle{plain}
	
	\begin{abstract}
		
		This Supplemental Appendix presents the proofs of the main results in the paper, as well as details on the methods of selection of L-moments, the extensions to ``residual analysis'' and conditional models, and additional information on the Monte Carlo exercises. 
	\end{abstract}

	\maketitle

	\tableofcontents 
\newpage

	\appendix 

\section{Proof of main results in the text}

\subsection{Proof of \Cref{prop_consistency}}
\label{proof_consistency}
\begin{proof}
For $\theta \in \Theta$, define:
\begin{equation*}
    \begin{aligned}
    M(\theta) \coloneqq \left[\int_{\underline{p}}^{\bar{p}} \left(\hat{Q}_Y (u) - Q_Y(u|\theta) \right)  \mathbf{P}^R(u)' du \right] (  W^R) \left[\int_{\underline{p}}^{\bar{p}}\left(\hat{Q}_Y (u) - Q_Y(u|\theta) \right)  \mathbf{P}^R(u) du \right] \\
     M_0(\theta) \coloneqq \left[\int_{\underline{p}}^{\bar{p}} \left(Q_Y (u) - Q_Y(u|\theta) \right)  \mathbf{P}^R(u)' du \right] ( \Omega^R)  \left[\int_{\underline{p}}^{\bar{p}}\left(Q_Y (u) - Q_Y(u|\theta) \right)  \mathbf{P}^R(u) du \right] \\ 
    h^R(\theta) \coloneqq \int_{\underline{p}}^{\bar{p}} \left(\hat{Q}_Y (u) - Q_Y(u|\theta) \right)  \mathbf{P}^R(u) du  \\ 
       h^R_0(\theta) \coloneqq  \int_{\underline{p}}^{\bar{p}} \left(Q_Y (u) - Q_Y(u|\theta) \right)  \mathbf{P}^R(u) du \, .
    \end{aligned}
\end{equation*}

Then, proceeding similarly to Theorem 2.6 of \cite{Newey1994}, we have that, by application of the Cauchy-Schwarz inequality and the properties of the spectral norm:

\begin{equation*}
\begin{aligned}
    |M(\theta) -  M_0(\theta) | \leq |(h^R(\theta) - h^R_0(\theta))'W^R (h^R(\theta) - h^R_0(\theta))| + \\  |h^R_0(\theta)'(W^R + W^{R'}) (h^R(\theta) - h^R_0(\theta))|  +  |h^R_0(\theta)'(W^R - \Omega^R)  h^R_0(\theta)| \leq \\
    \leq \lVert W^R \rVert_2 \lVert h^R(\theta) - h^R_0(\theta) \rVert_2^2 + 2 \lVert W^R \rVert_2 \lVert h^R_0(\theta)  - h^R(\theta) \rVert_2 \lVert h^R_0(\theta) \rVert_2 +  \lVert W^R - \Omega^R \rVert_2 \lVert h^R_0(\theta) \rVert_2^2 \, .
\end{aligned}
\end{equation*}

We analyse the behaviour of each term separately. First, note that, by Bessel's inequality and \Cref{ass_consistency}:

\begin{equation*}
\begin{aligned}
    \lVert h^R(\theta) - h^R_0(\theta) \rVert_2^2 = \sum_{l=1}^R \left[ \int_{\underline{p}}^{\bar{p}} [\hat{Q}_Y(u) - {Q}_Y(u)]P_l(u) du \right]^2 \leq \lVert (\hat{Q}_Y(\cdot) - {Q}_Y(\cdot)) \mathbbm{1}_{[\underline{p},\bar{p}]} \rVert_{L^2[0,1]}^2 = o_{p^*}(1) \, ,
\end{aligned}
\end{equation*}
where the upper bound does not depend on $\theta$. Next, we have:

\begin{equation*}
    \lVert h^{R}_0(\theta)  \rVert_2^2 \leq \lVert (Q_Y(\cdot) - Q_Y(\cdot|\theta)) \mathbbm{1}_{[\underline{p},\bar{p}]} \rVert_{L^2[0,1]}^2 \leq 2 \sup_{\Delta \in \Theta} \lVert Q_{Y}(\cdot|\Delta) \mathbbm{1}_{[\underline{p},\bar{p}]}\rVert_{L^2[0,1]}  < \infty \, ,
\end{equation*}
where we use Bessel's inequality \citep[page 157]{Kreyszig1989} and the last part of \Cref{ass_identification}. Combining these facts with \Cref{ass_weights}, we obtain:

\begin{equation*}
    \sup_{\theta \in \Theta}| M(\theta) - M_0(\theta) | \overset{P^*}{\to} 0  \, .
\end{equation*}

Finally we verify the unique identifiability condition of \citet[Definition 3.1]{Potscher1997}. Since $M_0(\theta_0) = 0$, the condition subsumes to verifying that, for each $\epsilon > 0$:

\begin{equation*}
    \liminf_{T,R \to \infty} \inf_{\theta \in \Theta: \lVert \theta -\theta_0 \rVert_2\geq \epsilon} M_0(\theta) > 0 \, .
\end{equation*}

This condition is clearly implied by \Cref{ass_weights}. Applying Lemma 3.1. of \cite{Potscher1997}, we conclude that $\hat{\theta} \overset{P^*}{\to} \theta_0$, as desired.
\end{proof}

\subsection{Proof of Proposition \ref{prop_asymptotic_linear}}

\label{proof_asymptotic_linear}
\begin{proof}

 Following the usual argument in the Generalised Method of Moments literature \citep{Newey1994}, we first show that $\hat \theta_T$ satisfies a first order condition with high probability. Indeed, under \Cref{ass_mvt_weak}, straightforward application of the dominated convergence theorem reveals that $h^R(\theta) = \int_{\underline{p}}^{\bar{p}} \left(\hat{Q}_Y (u) - Q_Y(u|\theta) \right)  \mathbf{P}^R(u) du $ is differentiable on $\mathcal{O}$, with derivative given by differentiation under the integral sign. Moreover, since $\theta_0 \in \mathcal{O}$ and $\hat{\theta} \overset{p}{\to} \theta_0$, $\hat{\theta} \in \mathcal{O}$ with probability approaching one (wpa 1). It thus follows that
 
 \begin{equation}
 	\label{eq_foc}
 	\nabla_{\theta'} h^R(\hat{\theta}) 'W^R h^R(\hat{\theta}) = 0 \, ,
 \end{equation}
 holds wpa1, where $ \nabla_{\theta'} h^R(\tilde{\theta})$ denotes the Jacobian of $h^R$ with respect to $\theta$, evaluated at $\tilde{\theta}$. 
 
Next, since, for each $u \in [\underline{p},\bar{p}]$, $\theta \mapsto Q_Y(u|\theta)$ is continuously differentiable on $\mathcal{O}$, a mean-value-expansion yields that, with probability approaching one:

\begin{equation*}
	h^R(\hat{\theta}) = h^R(\theta_0) + \nabla_{\theta'} \widetilde{h^R(\theta)} (\hat{\theta} - \theta)\, ,
\end{equation*}
where $  \nabla_{\theta'} \widetilde{h^R(\theta)}$ is the $R \times d$ matrix where each line $l$ is equal to $ - \int_{\underline{p}}^{\overline{p}} \nabla_{\theta'} Q_Y(u|\tilde{\theta}(u)) P_l(u) du$, and $\tilde{\theta}(u)$ is a $u$-specific element in the line segment between $\hat{\theta}$ and $\theta_0$. Rearranging terms, adding and subtracting $\Omega^R$ yields:

\begin{equation*}
	\begin{aligned}
		\nabla_{\theta'} h^R(\hat{\theta})' \Omega^R h^R(\theta_0) + \nabla_{\theta'} h^R(\hat{\theta})' (W^R - \Omega^R) h^R(\theta_0)
		\\ =  - \nabla_{\theta'} h^R(\hat{\theta})' \Omega^R  \nabla_{\theta'} \widetilde{h^R(\theta)} (\hat{\theta} - \theta_0) -     \nabla_{\theta'} h^R(\hat{\theta})'(W^R - \Omega^R)  \nabla_{\theta'} \widetilde{h^R(\theta)} (\hat{\theta} - \theta_0)\, .
	\end{aligned}
\end{equation*}

The crucial step now is to work out asymptotic tightness of a normalization of $h^R(\theta_0)$. Observe that \Cref{ass_weak_convergence} entails that:

\begin{equation*}
	\left\lVert \sqrt{T} h^R(\theta_0) \right\rVert_2^2 \leq \sum_{l=1}^\infty \left| \int_{\underline{p}}^{\bar{p}} \sqrt{T}(\hat{Q}_Y(u) - Q_Y(u)) P_l(u) du\right|^2 \leq \lVert \sqrt{T} (\hat{Q}_Y(\cdot) - {Q}_Y(\cdot)) \mathbbm{1}_{[\underline{p},\bar{p}]} \rVert_{L^2[0,1]}^2 = O_{p^*}(1)\, .
\end{equation*}

The next step in the proof concerns the approximation of $\nabla_{\theta'} h^R(\hat{\theta})$ to $ \nabla_{\theta'} h^R(\theta_0)$ in the spectral norm. Notice that, by the properties of the spectral norm and Bessel's inequality:

\begin{equation*}
	\lVert \nabla_{\theta'} h^R(\hat{\theta}) -  \nabla_{\theta'} h^R(\theta_0) \rVert_2^2 \leq \sum_{s=1}^d \lVert  [\partial_{\theta_s} Q_{Y}(\cdot|\hat{\theta}) - \partial_{\theta_s} Q_{Y}(\cdot|\theta_0)]\mathbbm{1}_{[\underline{p}, \overline{p}]} \rVert_{L^2[0,1]}^2 \, .
\end{equation*}

We claim that, $\hat{\theta} \overset{P^*}{\to} \theta_0$, together with \Cref{ass_mvt_weak}, is sufficient to ensure the upper bound above is $o_{p^*}(1)$. Since $d$ is fixed, we may consider the argument for a fixed $s = 1,2,\ldots d$. Fix $\eta, \epsilon > 0$. Since, by assumption, $\partial_{s} Q_{Y}(u|\theta)$ is continuous at $\theta_0$, uniformly in $u$; there exists $\delta > 0$ such that:

\begin{equation*}
	\lVert \theta - \theta_0 \rVert_2\leq \delta  \implies |\partial_{s} Q_{Y}(u|\theta) - \partial_{s} Q_{Y}(u|\theta_0) | \leq \frac{\sqrt{\epsilon}}{{\bar{p}-\underline{p}}} \quad \forall u \in [\underline{p},\bar{p}] \, .
\end{equation*}

Now, since $\hat{\theta} \overset{P^*}{\to} \theta_0$, there exists $N \in \mathbb{N}$ such that, for all $T \geq N$:

\begin{equation*}
	P^*(\lVert \hat{\theta} - \theta_0 \rVert_2\leq \delta) \geq 1 - \eta \, ,
\end{equation*}
implying that, by monotonicity of the outer probability, for $T \geq N$:

\begin{equation*}
	P^*(\lVert [ \partial_{\theta_s} Q_{Y}(\cdot|\hat{\theta}) - \partial_{\theta_s} Q_{Y}(\cdot|\theta_0)]\mathbbm{1}_{[\underline{p}, \overline{p}]} \rVert_{L^2[0,1]}^2 \leq \epsilon  ) \geq 1 - \eta \, .
\end{equation*}

Since the choice of $\eta$ and $\epsilon$ is arbitrary, we obtain that:

\begin{equation*}
	\lVert  [\partial_{\theta_s} Q_{Y}(\cdot|\hat{\theta}) - \partial_{\theta_s} Q_{Y}(\cdot|\theta_0)]\mathbbm{1}_{[\underline{p}, \overline{p}]} \rVert_{L^2[0,1]}^2  = o_{p^*}(1) \, ,
\end{equation*}
and since $d$ is fixed, we conclude that:

\begin{equation*}
	\lVert \nabla_{\theta'} h^R(\hat{\theta}) -  \nabla_{\theta'} h^R(\theta_0) \rVert_2^2 = o_{p^*}(1) \, .
\end{equation*}

Next, we would like to similarly argue that  $\lVert \nabla_{\theta'} \widetilde{h^R(\theta)} - \nabla_{\theta'} h^R(\theta_0) \rVert_2^2 = o_{P^*}(1)$. The difficulty here is that each $u$ possesses its $u$-specific $\tilde{\theta}(u)$. Note, however, that by Bessel's inequality:

\begin{equation*}
	\lVert \nabla_{\theta'} \widetilde{h^R(\theta)} - \nabla_{\theta'} h^R(\theta_0) \rVert_2^2 \leq \sum_{s=1}^d \left[\int_{\underline{p}}^{\bar{p}} \left( \partial_{\theta_s} Q_Y(u|\tilde{\theta}(u)) - \partial_{\theta_s} Q_Y(u|\theta_0) \right) P_l(u) du\right]^2 \, .
\end{equation*}

Under Assumption \ref{ass_mvt_strong}, a mean-value expansion of the right-hand side above, followed by using H\"older's inequality, $\lVert P_l \rVert_{L^2[0,1]} = 1$, the Cauchy-Schwarz inequality, and that $\lVert \tilde{\theta} - \theta_0 \rVert_2\leq \lVert \hat{\theta} - \theta_0 \rVert_2$ for any $\tilde{\theta}$ in the line segment between $\hat{\theta}$ and $\theta_0$; yields

\begin{equation*}
	\begin{aligned}
		\left[\int_{\underline{p}}^{\bar{p}} \left( \partial_{\theta_s} Q_Y(u|\tilde{\theta}(u)) - \partial_{\theta_s} Q_Y(u|\theta_0) \right) P_l(u) du\right]^2 = \left[ \int_{\underline{p}}^{\bar{p}} \nabla_{\theta} \partial_{\theta_s}  Q_Y(u | \check{\theta}(u))' (\tilde{\theta}(u) - \theta_0) P_l(u) du\right]^2 \leq \\ 
		\leq \int_{\underline{p}}^{\bar{p}}    \left[ \nabla_{\theta} \partial_{\theta_s} Q_Y(u | \check{\theta}(u))' (\tilde{\theta}(u) - \theta_0) \right]^2 du \leq  \left(\int_{\underline{p}}^{\bar{p}}  \rVert \nabla_{\theta} \partial_{\theta_s} Q_Y(u | \check{\theta}(u)) \lVert_2^2 du \right)\cdot \lVert \hat{\theta} - \theta_0 \rVert_2^2 = o_p(1) \, ,
	\end{aligned}
\end{equation*}
as desired.

Next, using that $\lVert \nabla_{\theta'} h^R(\theta_0) \rVert_2^2 = O(1) $ (which follows from Bessel's inequality and the last part of \Cref{ass_mvt_weak}) and the previous results, we arrive at:

\begin{equation*}
	( \nabla_{\theta'} h^R(\theta_0)' \Omega^R \nabla_{\theta'} h^R(\theta_0) + r^{TR}) \sqrt{T}(\hat{\theta} - \theta_0) = - \nabla_{\theta'} h^R(\theta_0)' \Omega^R (\sqrt{T} h^R(\theta_0)) + o_{P^*}(1) \, ,
\end{equation*}
where the remainder $r^{TR}$ satisfies $\lVert r^{TR} \rVert_{2}^2 = o_{P^*}(1)$. To proceed, we need to ensure that the matrix $(\nabla_{\theta'} h^R(\theta_0)' \Omega^R \nabla_{\theta'} h^R(\theta_0) + r^{TR})$ is invertible with high probability. Notice that, under the condition in \Cref{ass_eigen}, we have, using the Bauer-Fike theorem \cite[Theorem VIII.3.1]{Bhatia1997}, that, wpa 1, 

\begin{equation*}
	\lambda_{\min}  \left(\nabla_{\theta'} h^R(\theta_0)' \Omega^R \nabla_{\theta'} h^R(\theta_0) + r^{TR} \right) > 0 \, ,
\end{equation*}
and further using that, for invertible matrices $A_0$ and $A$

\begin{equation*}
	\begin{aligned}
	\lVert  A^{-1} - A_0^{-1} \rVert_2 = \lVert A^{-1}(A_0 - A) A_0^{-1} \rVert_2\leq  \lVert A^{-1}\rVert_2 \lVert (A_0 - A) \rVert_2  \lVert A_0^{-1} \rVert_2  \, , \\
\lVert A^{-1} \rVert_2=	\lVert A^{-1}(A  - A_0) A_0^{-1} - A_0^{-1} \rVert_2 \leq \lVert  A_0^{-1}\rVert_2  +  \lVert A^{-1}\rVert_2 \lVert (A_0 - A) \rVert_2  \lVert A_0^{-1} \rVert_2 \, ,
	\end{aligned}
\end{equation*}

We obtain that:

\begin{equation*}
	\lVert  A^{-1} - A_0^{-1} \rVert_2 \leq \lVert A_0^{-1} \rVert_2 \frac{\lVert A_0 - A\rVert_2}{ \lVert A_0^{-1} \rVert^{-1}_2 - \lVert A - A_0\rVert_2 }  \, .
\end{equation*}

Taking $A_0 = \nabla_{\theta'}h^R(\theta_0)' \Omega^R \nabla_{\theta'} h^R(\theta_0)$ and $A = \nabla_{\theta'} h^R(\theta_0)' \Omega^R \nabla_{\theta'} h^R(\theta_0) + r^{TR}$, and using that \Cref{ass_eigen} implies $\lVert  \left(\nabla_{\theta'}h^R(\theta_0)' \Omega^R \nabla_{\theta'} h^R(\theta_0) \right)^{-1}\rVert_2$ is bounded above uniformly in $R$,\footnote{For a positive (semi)definite symmetric matrix, eigenvalues and singular values coincide. Consequently, $\lVert A_0^{-1} \rVert = \frac{1}{\lambda_{\min}(A_0)}$, which is bounded above  by  \Cref{ass_eigen}.} we conclude that:

\begin{equation*}
	\lVert  (\nabla_{\theta'}h^R(\theta_0)' \Omega^R \nabla_{\theta'} h^R(\theta_0) +r^{RT})^{-1} - (\nabla_{\theta'}h^R(\theta_0)' \Omega^R \nabla_{\theta'} h^R(\theta_0))^{-1} \rVert_2 = o_{P^*}(1) \, .
\end{equation*}

From which we conclude that, wpa 1:

\begin{equation*}
	\sqrt{T}(\hat{\theta} - \theta_0) = - ( \nabla_{\theta'} h^R(\theta_0)' \Omega^R \nabla_{\theta'} h^R(\theta_0))^{-1} \nabla_{\theta'} h^R(\theta_0)' \Omega^R (\sqrt{T} h^R(\theta_0)) + o_{P^*}(1) \, ,
\end{equation*}
proving the desired asymptotic linear representation.
\end{proof}

\section{Verification of the boundedness condition in Assumption \ref{ass_identification}  for the GEV and GPD distributions}
\subsection{Generalized Extreme Value} Consider a Generalized Extreme Value with location parameter $\xi \in \mathbb{R}$, scale parameter $\alpha > 0$ and shape parameter $k \in \mathbb{R}$. Denoting by $\theta = (\xi, \alpha, k)'$ the vector of parameters, we have that the quantile function is given by \citep[page 70]{hosking1986theory}:

$$Q(u|\theta) = \begin{cases}
 \xi + \alpha(1- (-\log(u))^k)/k & k \neq 0 \\
 \xi -\alpha \log(-\log(u))  & k = 0
 
 \end{cases}$$ 
 
 In this case, for $k > -1/2$, it follows that \citep[page 178]{singh1998entropy}:
 
 \begin{equation*}
 	\int_0^1 Q(u|\theta)^2 du  =\begin{cases}
\left(\xi + \frac{\alpha}{k}[1-\Gamma(1+k)]\right)^2 + \frac{\alpha^2}{k^2}\left(\Gamma(1+2k) - \Gamma(1+k)^2\right)\, , & k \neq 0 \\
\left(\xi -\alpha\Gamma'(1)\right)^2 +  \alpha^2\left(\Gamma''(1) - (\Gamma'(1))^2\right)\, , & k =0
 	\end{cases}\, ,
 \end{equation*}
 where $\Gamma$ denotes the Gamma function; and the integral under $k=0$ coincides with the limit when $k \to 0$ \citep[page 12]{kotz2000extreme}. By continuity of the Gamma function on $\mathbb{R}_{++}$, it follows that $\theta \mapsto 	\int_0^1 Q(u|\theta)^2 du  $ is continuous  on $\mathbb{R} \times \mathbb{R}_{++} \times (-1/2,\infty)$. Consequently, for any compact parameter space $\Theta \subseteq \mathbb{R} \times \mathbb{R}_{++} \times (-1/2,\infty)$, the uniform boundedness condition in Assumption \ref{ass_identification} will be satisfied with $0 = \underline{p} < \overline{p} = 1$. In addition, if we choose trimming constants $ 0 <  \underline{p}$ and $\overline{p}<1$, it is possible to consider a compact parameter space  $\Theta \subseteq \mathbb{R} \times \mathbb{R}_{++} \times\mathbb{R}$, since, in this case, for any $\theta \in \Theta$:
 
  \begin{equation*}
 	\int_{\underline{p}}^{\overline{p}} Q(u|\theta)^2 du \leq  \left(\overline{p} -\underline{p}\right) \left(Q(\underline{p}|\theta)^2\lor Q(\overline{p}|\theta)^2\right)\, ,
 \end{equation*}
 with both $Q(\overline{p}|\theta)$ and $Q(\overline{p}|\theta)$ continuous in $\theta$.
 
\subsection{Generalized Pareto}  Consider a Generalized Pareto Distribution with location parameter $\xi \in \mathbb{R}$, scale parameter $\alpha > 0$ and shape parameter $k \in \mathbb{R}$. Denoting by $\theta = (\xi, \alpha, k)'$ the vector of parameters, we have that the quantile function is given by \citep[page 67]{hosking1986theory}:

$$Q(u|\theta) = \begin{cases}
	\xi + \alpha(1- (1-u)^k)/k\, ,& k \neq 0 \\
	\xi -\alpha \log(1-u)\, ,  & k = 0
	
\end{cases}\, ,$$ 

If $k > -1/2$, we have that \citep{Hosking1987}:

 \begin{equation*}
	\int_{0}^{1} Q(u|\theta)^2 du  = 
		\left(\xi+\frac{\alpha}{(1+k)} \right)^2+ \frac{\alpha^2}{(1+k)^2(1+2k)}\, ,
	\end{equation*}
	from which it follows that the boundedness condition in Assumption \ref{ass_identification} is satisfied with $0=\underline{p}<\overline{p}=1$ for a parameter space $\Theta \subseteq \mathbb{R}\times \mathbb{R}_{++} \times (-1/2,\infty)$ such that $\Pi_{1,2}\Theta = \{(\theta_1,\theta_2):\theta \in \Theta\}$ is compact and $\inf\{\theta_3:\theta \in \Theta\} > -1/2$. Moreover, and similarly to the GEV case, if one considers a trimming constant $\overline{p} < 1$, then it is possible to consider parameter spaces $\Theta \subseteq \mathbb{R}\times \mathbb{R}_{++} \times \mathbb{R}$  such that $\{(\theta_1,\theta_2):\theta \in \Theta\}$ is compact and $\Pi_3\Theta = \{\theta_3:\theta \in \Theta\}$ is bounded below, since, in this case, $\int_{0}^{\overline{p}} Q(u|\theta)^2 du  \leq  \overline{p}\left(Q(\overline{p}|\theta)^2\lor Q(0|\theta)^2\right) = \overline{p} \left( Q(\overline{p}|\theta)^2 \lor \xi^2 \right) $, with $Q(\overline{p}|\theta)^2 \leq \left(\xi + \frac{\alpha}{k}\right)^2$ if $k > 0$.

\section{Relation between Assumptions and different notions of identification}
\subsection{Relation between the strong identifiability part of Assumption \ref{ass_identification} and the usual notion of identifiability in parametric families}
In what follows, consider the population version of the objective function stated in Assumption \ref{ass_identification} in the main text, with the choice $0=\underline{p}<\overline{p}=1$. 
$$V^*_R(\theta) \coloneqq \left[\int_{0}^{1} \left(Q_{Y}(u|\theta) - Q_{Y}(u|\theta_0)\right) \mathbf{P}^R(u)' du \right] \Omega^R \left[\int_{0}^{1} \left(Q_{Y}(u|\theta) - Q_{Y}(u|\theta_0)\right) \mathbf{P}^R(u) du \right]\, .$$
\begin{proposition}
	Suppose that $\Theta$ is compact, that $\theta \mapsto \int_0^1 Q(u|\theta)^2 du$ is bounded and $(\theta',\theta'') \mapsto \int_0^1 (Q(u|\theta')-Q(u|\theta''))^2 du$ is continuous, that the $\{P_l\}_l$ form an orthonormal basis in $L^2[0,1]$, and that the smallest eigenvalue of $\Omega_R$ is bounded away from zero, uniformly in $R$. Then the parametric family $\theta_0$ is identified in the usual sense (meaning $\theta \neq \theta_0 \implies F_\theta \neq F_{\theta_0}$) if, and only if, for every $\epsilon > 0$:
	$$\liminf_{R\to\infty}\inf_{\theta\in \Theta: \lVert\theta -\theta_0\rVert \geq \epsilon} V_R^*(\theta)>0$$
\end{proposition}
\begin{proof}
	Suppose that $\theta_0$ is not identified in the usual sense. Then there exists $\tilde{\theta} \in \Theta$ such that $\tilde{\theta} \neq \theta$ and $F_{\tilde{\theta}} = F_{\theta_0}$. Consequently, $Q(\cdot|\theta_0) = Q(\cdot|\tilde{\theta})$, and, taking $\epsilon^* = \lVert \tilde{\theta} - \theta_0\rVert > 0 $, we have:
	$$\inf_{\theta\in \Theta: \lVert\theta -\theta_0\rVert \geq \epsilon^*} V_R^*(\theta) = 0\,, \quad \forall R \in \mathbb{N}\, .$$
	
	In the other direction, suppose that $\theta_0$ is identified in the usual sense. Fix $\epsilon > 0$. Since $V^*(\theta)\coloneqq \int_0^1 (Q(u|\theta) -Q(u|\theta_0))^2du$ is continuous and bounded and $\{\theta \in \Theta: \lVert \theta-\theta_0\rVert \geq \epsilon \}$ is compact, identifiability in the usual sense implies that $\inf_{\theta \in \Theta: \lVert\theta-\theta_0 \rVert\geq\epsilon} V^*(\theta)>0$.\footnote{If not, there would exist $\tilde{\theta}\neq \theta_0$ such that $V^*(\theta)=0\implies F_{\tilde{\theta}}= F_{\theta_0}$.} Moreover, since $\Omega_R$ is symmetric and real, it admits an eigendecomposition  $D_R\Lambda_R D_R'$, where $\Lambda_R$ a diagonal matrix with the eigenvalues of $\Omega_R$, and  $D_R'D_R =D_RD_R' = \mathbb{I}_R$. This implies that, denoting by $D_{l,R}$ the $l$-th column of $D_R$ and defining $v_R(\theta) = \int_{0}^{1} \left(Q_{Y}(u|\theta) - Q_{Y}(u|\theta_0)\right) \mathbf{P}^R(u) du$:
	
	$$V_{R}^*(\theta) =\sum_{l=1}^{R} \lambda_{l,R} (D_{l,R}'v_R(\theta))^2 \geq \underline{\lambda}\sum_{l=1}^{R} (D_{l,R}'v_R(\theta))^2 = \underline{\lambda}v_R(\theta)'\left(\sum_{l=1}^{R} D_{l,R}D_{l,R}'\right) v_R(\theta) = \underline{\lambda}\lVert v_R(\theta) \rVert^2\, ,$$
	where $\underline{\lambda}>0$ is the uniform lower bound on the eigenvalues of the $\Omega_R$. Now, since the $\{P_l\}_{l \in \mathbb{N}}$ form an orthonormal basis in $L^2[0,1]$, it follows from Parseval identity \citep[page 170]{Kreyszig1989} that,  for every $\theta \in \Theta$, as $R \to \infty$, $\lVert v_R(\theta) \rVert^2  \uparrow V^*(\theta)$. Moreover, notice that, by the mean-value theorem, for every $\theta', \theta'' \in \Theta$:	
	\begin{equation}
		\label{eq_equicontinuity}
		\begin{aligned}
				| \lVert v_R(\theta') \rVert^2 - \lVert v_R(\theta'') \rVert^2|\leq 2 \tilde{C}_{\theta',\theta'' } | \lVert v_R(\theta') \rVert - \lVert v_R(\theta'')\rVert |\leq 2C^*\lVert v_R(\theta')- v_R(\theta'') \rVert   \leq \\ 2C^* \sqrt{\int_0^1 (Q(u|\theta') -Q(u|\theta''))^2du} \, ,
		\end{aligned}
	\end{equation}
	where $\tilde{C}_{\theta',\theta'' } \in [\lVert v_R(\theta') \rVert, \lVert v_R(\theta'') \rVert]$, with $\lVert v_R(\theta') \rVert 
	\lor \lVert v_R(\theta'') \rVert \leq \sup_{\theta \in \Theta}\sqrt{\int_0^1Q(u|\theta)^2 du} \eqqcolon C^*$ by Bessel inequality; and where the last inequality in \eqref{eq_equicontinuity} follows again by Bessel inequality. From \eqref{eq_equicontinuity}, we conclude that  the functions $\theta \mapsto \lVert v_R(\theta)\rVert^2$, $R \in \mathbb{N}$, are uniformly equicontinuous. Moreover, this sequence of functions is also uniformly bounded, as they converge pontwise monotonically to the bounded function $V^*$. Consequently, it follows by the Arzelà-Ascoli theorem that these functions converge uniformly to $V^*$, implying that: 
	$$\lim_{R\to \infty} \inf_{\theta \in \Theta: \lVert\theta-\theta_0 \rVert\geq\epsilon}  \lVert v_R(\theta)\rVert^2  =  \inf_{\theta \in \Theta: \lVert\theta-\theta_0 \rVert\geq\epsilon} V^*(\theta) > 0\, ,$$
thus yielding that $\liminf_{R\to \infty} \inf_{\theta\in \Theta: \lVert\theta -\theta_0\rVert \geq \epsilon} V_R^*(\theta) >0$.

\end{proof}
\subsection{Relation between eigenvalue assumption and identification}
\label{app_relation}
The goal of this section is to show how \Cref{ass_eigen} is related to identification. We consider a stronger version of \Cref{ass_identification} as follows:

\begin{asst}
	\label{ass_identification_relation}
	There exists $C > 0$ and $h: \mathbb{R}_+\mapsto \mathbb{R}_+$ such that, for every $R \in \mathbb{N}$ and $\epsilon > 0$:
	{\small
	\begin{equation*}
		\inf_{\theta \in \Theta: \lVert \theta - \theta_0 \rVert_2\geq \epsilon }  \left[\int_{\underline{p}}^{\bar{p}} \left(Q_{Y}(u|\theta) - Q_{Y}(u|\theta_0)\right) \mathbf{P}^R(u)' du \right] \Omega^R \left[\int_{\underline{p}}^{\bar{p}} \left(Q_{Y}(u|\theta) - Q_{Y}(u|\theta_0)\right) \mathbf{P}^R(u) du \right] \geq C h(\epsilon) \, ,
	\end{equation*}
}
	where $h(x) > 0$ for all $x > 0$ and $\lim_{x \to 0}\frac{h(x)}{x^2} = 1$.
\end{asst}

It is clear that \Cref{ass_identification_relation} implies \Cref{ass_identification}. Perhaps less obviously, \Cref{ass_identification_relation} implies \Cref{ass_eigen} under conditions that allow differentiability under the integral sign (\Cref{ass_mvt_weak}).

\begin{proposition}
	Suppose \Cref{ass_mvt_weak} holds. Then \Cref{ass_identification_relation} implies \Cref{ass_eigen}.
	\begin{proof}
		Suppose \Cref{ass_identification_relation} holds. Fix $\iota \in \mathbb{R}^d$, $\lVert \iota \rVert_2 = 1$. We then have that, by \Cref{ass_identification_relation}:
		
		{\small
		\begin{equation*}
			\left[\frac{1}{\epsilon}\int_{\underline{p}}^{\bar{p}} \left(Q_{Y}(u|\theta_0 + \epsilon \iota) - Q_{Y}(u|\theta_0)\right) \mathbf{P}^R(u)' du \right] \Omega^R \left[\frac{1}{\epsilon}\int_{\underline{p}}^{\bar{p}} \left(Q_{Y}(u|\theta_0 + \epsilon \iota) - Q_{Y}(u|\theta_0)\right) \mathbf{P}^R(u) du \right] > C \frac{h(\epsilon)}{\epsilon^2} \, .
		\end{equation*}
	}
		Taking limits yields that:
		
		\begin{equation*}
			\iota' \nabla_{\theta'} h^R(\theta_0)' \Omega^R \nabla_{\theta'} h^R(\theta_0) \iota \geq C \, .
		\end{equation*}
		
		Now, since $\nabla_{\theta'} h^R(\theta_0)' \Omega^R \nabla_{\theta'} h^R(\theta_0)$ is symmetric and real, it admits an eigendecomposition $P_R \Lambda_R P_R'$, where $P_RP_R' = P_R' P_R = \mathbb{I}_d$ and $\Lambda_R = \operatorname{diag}(\lambda_{1R}, \lambda_{2R} \ldots \lambda_{dR})$, with $\lambda_{1R} \leq \lambda_{2R} \ldots \leq \lambda_{dR}$ being the eigenvalues of $\nabla_{\theta'} h^R(\theta_0)' \Omega^R \nabla_{\theta'} h^R(\theta_0)$. This in turn implies that:
		
		\begin{equation*}
			\lambda_{1R} = \min_{x: \lVert x \rVert_2 = 1} x' \Lambda_R x = \min_{u: \lVert u \rVert_2 = 1} (P_R' u)' \Lambda_R (P_R' u) \geq C > 0 \, ,
		\end{equation*}
		which proves the desired result.
	\end{proof}
\end{proposition}

	\section{Comparison with series-IV estimator of \cite{Donald2003}}
	In this Appendix, we compare our proposed generalised L-moment estimator with the series-IV estimator introduced by \cite{Donald2003} in the context of inference based on conditional moment restrictions. Specifically, consider a scalar parameter $\beta_0 \in \mathbb{R}$ that is identified through a conditional moment restriction of the form:
	
	$$\mathbb{E}[Z(\beta_0)|X]  = 0\, ,$$
	and suppose one has access to a random sample $\{(Z_t(\cdot), X_t) \}_{t=1}^T$  from $(Z(\cdot),X)$, with $Z(b)$ having finite second moment for every $b \in \mathbb{R}$, and $\mathbb{V}[Z(b)|X] \leq C_b$ for some $C_b \in \mathbb{R}$ and every $b \in \mathbb{R}$. Let $\{p_l(\cdot)\}_{l \in \mathbb{N}}$ be a sequence of series transformations that is able to approximate $\mathbb{E}[Z(b)|X]$ in mean-squared error, for any $b \in \mathbb{R}$. For $R \in \mathbb{N}$, we denote by $P_R(X_i) = (p_1(X_i), p_2(X_i), \ldots, p_R(X_i))'$ and $\boldsymbol{P}_R = \begin{bmatrix}
	P_R(X_1)  & P_R(X_2) & \ldots & P_R(X_T)
	\end{bmatrix}'$. We assume that the basis functions are ``orthogonalised'', in the sense that $\boldsymbol{P}_R' \boldsymbol{P}_R = \mathbb{I}_R$. Do further define $\boldsymbol{z}(b) = (Z_1(b),\ldots Z_T(b))'$, and $\boldsymbol{X} = \begin{bmatrix}
	X_1 & X_2 & \ldots X_T
	\end{bmatrix}'$. \citeauthor{Donald2003}'s series-IV approach to estimating $\beta_0$ consists in minimizing the following criterion function:
	
	$$\hat{S}(b) =\sum_{l=1}^R \left( \sum_{t=1}^T Z_t(b) p_l(X_t) \right)^2 = \boldsymbol{z}(b)'\boldsymbol{P}_R \boldsymbol{P}_R' \boldsymbol{z}(b)\, . $$
	
	\subsection{Comparison between consistency arguments} The consistency argument in Theorem 5.1. of \citeauthor{Donald2003} relies on showing that, as $T,R\to \infty$:
	 $$\hat{s}(b)\coloneqq \frac{\hat{S}(b)}{T} \overset{p}{\to} \mathbb{E}[\left(\mathbb{E}[Z(b)|X]\right)^2] \eqqcolon s_0(b)\, .$$
	 
	 The error of estimating $s_0(b)$ by $\hat{s}(b)$ can be decomposed into three terms:
	 
	 \begin{equation*}
	 	\begin{aligned}
	 		 s_0(b) - \hat{s}(b)  =   \left( \mathbb{E}[\mathbb{E}[\boldsymbol{z}(b)|\boldsymbol{X}]^2]- \frac{1}{T}\mathbb{E}[\boldsymbol{z}(b)|\boldsymbol{X}]'\mathbb{E}[\boldsymbol{z}(b)|\boldsymbol{X}] \right) +\frac{1}{T}\mathbb{E}[\boldsymbol{z}(b)|\boldsymbol{X}]'(\mathbb{I}_T - \boldsymbol{P}_R\boldsymbol{P}_R')\mathbb{E}[\boldsymbol{z}(b)|\boldsymbol{X}] + \\	 		 
	 		 \frac{1}{T}\left(\mathbb{E}[\boldsymbol{z}(b)|\boldsymbol{X}]'\boldsymbol{P}_R\boldsymbol{P}_R'\mathbb{E}[\boldsymbol{z}(b)|\boldsymbol{X}] - \boldsymbol{z}(b)'\boldsymbol{P}_R\boldsymbol{P}_R'\boldsymbol{z}(b) \right) 
	 	\end{aligned} \, .
	 \end{equation*}
	 
	 The above error consists of three parts. The first term is $o_p(1)$ by the law of large numbers. The second term may be seen as an ``approximation bias'' component and is $o_p(1)$ by the series transformation approximation property.\footnote{Indeed, since, by assumption, there exists a sequence $\gamma_K \in \mathbb{R}^K$, $K \in \mathbb{N}$, such that $
	 \lim_{K \to \infty}\mathbb{E}[|Z(b) - \gamma_K'P_K(X) |^2] = 0$, and given that $\mathbb{I}_T  - \boldsymbol{P}_R\boldsymbol{P}_R'$ is a residual-maker matrix, it follows that:
	 $$\mathbb{E}\left[\left|\frac{1}{T}\mathbb{E}[\boldsymbol{z}(b)|\boldsymbol{X}]'(\mathbb{I}_T - \boldsymbol{P}_R\boldsymbol{P}_R')\mathbb{E}[\boldsymbol{z}(b)|\boldsymbol{X}] \right|\right] \leq \mathbb{E}[|Z(b) - \gamma_R'P_R(X) |^2] = o(1)$$
	 } The third term may be seen as a ``variance component'', whose proper control imposes restrictions on the rate of growth of $R$. Indeed, control of this term depends crucially on showing that:
	 
	 $$ \tilde{C} \coloneqq \frac{1}{T}\left(\left(\mathbb{E}[\boldsymbol{z}(b)|\boldsymbol{X}] -\boldsymbol{z}(b) \right) '\boldsymbol{P}_R\boldsymbol{P}_R'\left(\mathbb{E}[\boldsymbol{z}(b)|\boldsymbol{X}] -\boldsymbol{z}(b) \right) \right) $$
	 is $o_P(1)$. By the cyclic invariance of the trace operator, we have:
	
	\begin{equation*}
		\begin{aligned}
				\mathbb{E}\left[\tilde{C}|\boldsymbol{X}\right] =  \operatorname{tr}\left(\frac{1}{T}\left(\mathbb{E}\left[\left(\mathbb{E}[\boldsymbol{z}(b)|\boldsymbol{X}] -\boldsymbol{z}(b) \right) \left(\mathbb{E}[\boldsymbol{z}(b)|\boldsymbol{X}] -\boldsymbol{z}(b) \right)'\Big|\boldsymbol{X}\right] \boldsymbol{P}_R\boldsymbol{P}_R'\right) \right)=\\ \frac{1}{T}\sum_{t=1}^T \mathbb{V}[Z_t(b)|X_t] (\boldsymbol{P}_R \boldsymbol{P}_R')_{tt} \leq \\ \frac{C_b}{T} \sum_{t=1}^T  (\boldsymbol{P}_R \boldsymbol{P}_R')_{tt} = \frac{C_b K}{T}\, ,
		\end{aligned}
	\end{equation*}
	which shows that  the rate condition $\frac{R}{T} \to 0$ is sufficient to ensure that $\mathbb{E}\left[\tilde{C}\right] = o(1)$. Moreover, if $\mathbb{V}[Z(b)|X]$ is bounded \emph{below} by a constant $c_b > 0$, then the rate condition is also necessary for $\mathbb{E}\left[\tilde{C}\right] = o(1)$.  
	
	By Markov inequality, we conclude that $\frac{R}{T} \to 0$  is a sufficient condition to ensure that $\tilde{C} = o_p(1)$.  In addition, if $Z(b)$ has finite fourth moment and $\mathbb{V}[Z(b)|X]$ is bounded below by a constant $c_b > 0$, the rate requirement is also necessary for  $\tilde{C} = o_p(1)$, since, in this case, there exists a constant $\Lambda > 0$ not depending on $T$ or $R$ such that $\mathbb{E}[\tilde{C}^2] \leq \Lambda$. This implies that $\tilde{C}$ is uniformly integrable \citep[Theorem 4.6.2]{Durrett2019} and hence, if $\tilde{C} = o_p(1)$, one has that $\mathbb{E}[\tilde{C}] = o(1)$ \citep[Theorem 4.6.3]{Durrett2019}, thus implying that $\frac{R}{T} \to 0$ under the lower bound in the conditional variance.
	
	The previous discussion evidences that, in the context of the series-IV estimator of \cite{Donald2003}, the rate condition $R/T \to 0$ is essential for consistency. Why does our proposed generalised L-moment estimator not require such rate restriction? An inspection of the proof of Proposition \ref{prop_consistency} presented in Appendix \ref{proof_consistency} reveals that the ``analogous'' term to $\tilde{C}$ in this case is 
	
	$$\tilde{D} \coloneqq \left\lVert\int_{\underline{p}}^{\bar{p}}\left[ \left(\hat{Q}_Y(u)- Q_{Y}(u|\theta)\right) -  \left({Q}_Y(u)- Q_{Y}(u|\theta)\right)\right] \mathbf{P}^R(u)\ du\right\rVert_2^2 \, ,$$
	where we have assumed an identity weighting matrix for the sake of clarity and comparability with the series-IV estimator. Now, by Bessel's inequality, we have that:
$$\tilde{D} \leq  \lVert( \hat{Q}_Y(\cdot)- Q_{Y}(\cdot)) \mathbbm{1}_{[\underline{p},\bar{p}]} \rVert _{L^2[0,1]}^2 \, ,$$
	 where crucially the upper bound does not depend on $R$. Consequently, if $ \lVert( \hat{Q}_Y(\cdot)- Q_{Y}(\cdot)) \mathbbm{1}_{[\underline{p},\bar{p}]} \rVert _{L^2[0,1]}^2  \overset{p}{\to} 0$ (which is implied by uniform consistency of sample quantiles on $[\underline{p},\overline{p}]$), then $\tilde{D} = o_p(1)$. Therefore, it is the special structure of L-moments that enables us to dispense with rate requirements.
	 
	 \subsection{Comparison between linearisation arguments} The linearisation argument underlying the proof of Theorem 5.2 of \cite{Donald2003} assumes the rate restriction $R/T^2 \to 0$. Inspection of their argument reveals that this restriction is crucially used in order to ensure that term:
	 
	 $$\tilde{E} \coloneqq \frac{1}{\sqrt{T}} \nabla_b\boldsymbol{z}(\beta_0)' \boldsymbol{P}_R \boldsymbol{P}_R'\boldsymbol{z}(\beta_0)\, ,$$
	 satisfies:
	 
	 $$\tilde{E} =\frac{1}{\sqrt{T}}\mathbb{E}[\nabla_b\boldsymbol{z}(\beta_0)|\boldsymbol{X}]'\boldsymbol{P}_R \boldsymbol{P}_R'\boldsymbol{z}(\beta_0) + o_P(1)\, .$$
	 
	 In other words, the rate restriction $R/T^2 \to 0$ is used in order to ensure that the ``bias term''  $\frac{1}{T}\mathbb{E}\left[\nabla_b\boldsymbol{z}(\beta_0)' \boldsymbol{P}_R \boldsymbol{P}_R'\boldsymbol{z}(\beta_0)\right]  = \mathbb{E}[\nabla_b Z(\beta_0) \left( P_R(X)' P_R(X) \right) Z(\beta_0)] = O(R)$ is $o(T^{-1/2})$, and thus that $\tilde{E} = O_P(1)$ by application of a central limit theorem.\footnote{It has long been recognised in the literature \citep{Newey1990iv,Donald2001} that, in GMM estimation, controlling this form of ``own-observation-bias'' stemming from correlation between the ``individual gradient'' at the truth $\nabla_b Z_t(\beta_0)$ and the``individual moment'' at the truth $Z_t(\beta_0)$ typically requires, for asymptotic normality, stronger restrictions on the growth rate of $R$ than consistency does.} In contrast, inspection of the proof of Proposition \ref{prop_asymptotic_linear} provided in Appendix \ref{proof_asymptotic_linear} shows that the analogous term is (again assuming identity weights for clarity):
	 
	 	 $$\tilde{F} =  -\sqrt{T} \left(\int_{\underline{p}}^{\bar{p}}  \nabla_{\theta}Q_{Y}(u|\theta_0) \boldsymbol{P}_R(u)'du\right)  \int_{\underline{p}}^{\bar{p}} \left(\hat{Q}_Y(u)- Q_{Y}(u)\right) \boldsymbol{P}_R(u)du\, .$$
	 	 
	 	 Two important distinctions arise with respect to the term $\tilde{E}$. First, due to the additive separability between $\hat{Q}_Y(\cdot)$ and $Q_Y(\cdot|\theta)$ in the difference between theoretical and empirical  L-moments entering our estimator, the gradient present in $\tilde{F}$ is not affected by estimation error in $\hat{Q}_Y$. This contrasts with the term $\tilde{E}$, whose bias is precisely due to correlation between the gradient and the sample moments.\footnote{The absence of correlation between the Jacobian at the truth and the moments at the truth is a more general feature of minimum-distance-style estimators \citep{Newey2004}.}  Secondly, the special structure of L-moments enables us to straightforwardly apply Bessel's inequality to show that:
	 	 
	 	 	 	 $$\lVert\tilde{F}\rVert_2^2 \leq  \left(\sum_{j=1}^d \lVert \partial_{\theta_j} Q_Y(\cdot|\theta_0) \mathbbm{1}_{[\underline{p},\overline{p}]}\rVert_{L^2[0,1]}^2\right) \lVert \sqrt{T}(\hat{Q}_Y(\cdot) - Q_Y(\cdot)) \rVert_{L^2[0,1]}^2 \, ,$$
	 which ensures that $\tilde{F}$ is bounded in probability if $\lVert \sqrt{T}(\hat{Q}_Y(\cdot) - Q_Y(\cdot)) \rVert_{L^2[0,1]}^2$ is. Again, the special structure of L-moments allowed us to bound a crucial term without resort to rate requirements.
\section{Calculations for optimal weighting matrix in the iid case}
\label{app_computations}
Consider the optimal weighting matrix as in equation \eqref{eq_optimal_weights} of the main text. We focus on the case where $0=\underline{p} < \overline{p} = 1$ and the data is iid. Note that we may write: 

\begin{equation*}
	\Omega^R = \mathbb{E}\left[  \frac{B_T(U)}{f_{\theta_0}(Q_y(U))} \mathbf{P}^R(U) \frac{B_T(V)}{f_{\theta_0}(Q_y(V))} \mathbf{P}^R(V) \right]^{-} \, ,
\end{equation*}
where $U$ and $V$ are independent random variables, independent from the Brownian bridge $B_T$. By Foubini's theorem, we have:

\begin{equation*}
	\Omega^R = \mathbb{E}\left[ \frac{(U\land V - UV)}{f_{\theta_0}(Q_y(U))f_{\theta_0}(Q_y(V))} \mathbf{P}^R(U) \mathbf{P}^R(V) \right]^{-} \, . 
\end{equation*}

Since standard L-moments consist of a choice of weighting functions $\mathbf{P}^R$ where each entry is a linear combination of polynomials, it suffices, for the purposes of numerical computation, to analyse the formula for polynomials $U^r$ and $V^s$. In particular, we can estimate:

\begin{equation*}
	\begin{aligned}
		\Pi_{r,s} = \int_0^1 \int_0^1 \frac{(U\land V - UV)}{f_{\theta_0}(Q_y(U))f_{\theta_0}(Q_y(V))} U^r V^s dU dV \, ,
	\end{aligned}
\end{equation*}
using a first step consistent estimator $\tilde{\theta}$ of $\theta_0$, the empirical quantile function, and numerical integration as follows:

\begin{equation*}
	\hat{\Pi}_{r,s} = \frac{1}{H^2}\sum_{i=1}^H \sum_{j=1}^H \frac{\left[\left(\frac{i-0.5}{H}\right)\land \left(\frac{j-0.5}{H}\right)- \left(\frac{i-0.5}{H}\right))\left(\frac{j-0.5}{H}\right)\right]}{f_{\tilde{\theta}}\left(Q_Y\left(\left(\frac{i-0.5}{H}\right)\Big|\tilde{\theta}\right)\right) f_{\tilde{\theta}}\left(Q_Y\left(\left(\frac{j-0.5}{H}\right)\Big|\tilde{\theta}\right)\right)}\left(\frac{i-0.5}{H}\right)^r \left(\frac{j-0.5}{H}\right)^s \, ,
\end{equation*}
where $H$ is the number of grid points. Alternatively, we may use a nonparametric estimator for the quantile derivative $Q'_{Y}(u) = \frac{1}{f_Y(Q_Y(u))}$.

\section{Test statistic for overidentifying restrictions}
\label{app_test}
As noted in \Cref{test_over} in the main text, the strong approximation discussed in \Cref{as_distribution} motivates a test statistic for overidentifying restrictions. Suppose $R>d$. Denoting by $M(\cdot)$ the objective function of the estimator, we consider the test-statistic:

\begin{equation*}
	J \coloneqq T \cdot M(\hat{\theta}_T) \, .
\end{equation*}

Under the null that the model is correctly specified, i.e. that there exists $\theta \in \Theta$ such that $Q_Y(\cdot) = Q_Y(\cdot|\theta)$, the results in \Cref{as_distribution} of the main text reveal that:

\begin{equation*}
	\begin{aligned}
		J = T \cdot M(\hat{\theta}_T) =  \left[ \int_{\underline{p}}^{\bar{p}} \sqrt{T}\left[(\hat{Q}_Y(u) - Q_Y(u)) - \nabla_{\theta'}Q_Y(u|\theta_0)(\hat{\theta}-\theta_0)\right]\mathbf{P}^R(u) du \right] ' \Omega^R  \\ \left[ \int_{\underline{p}}^{\bar{p}} \sqrt{T}\left[(\hat{Q}_Y(u) - Q_Y(u)) - \nabla_{\theta'}Q_Y(u|\theta_0)(\hat{\theta}-\theta_0)\right]\mathbf{P}^R(u) du \right]  + o_{P^*}(1) = \\
		\lVert(\Omega^R)^{1/2}\left(\mathbb{I}_{R\times R} -  \nabla_{\theta'} h^R(\theta_0) ( \nabla_{\theta'} h^R(\theta_0)' \Omega^R \nabla_{\theta'} h^R(\theta_0))^{-1} \nabla_{\theta'} h^R(\theta_0)' \Omega^R \right) \sqrt{T}h_R(\theta_0) \rVert_{2} + o_p(1) \, .
	\end{aligned}
\end{equation*}

This approximation can be used, along with the Gaussian strong approximations discussed in this section, to approximate the distribution of the statistic under the null. Specifically, when the optimal weighting scheme \eqref{eq_optimal_weights} is used, it follows from the properties of idempotent matrices that the distribution of the test statistic may be approximated by a chi-squared distribution with $R-d$ degrees of freedom. To show that this approximation indeed conduces to valid inference, let $ \tilde{J}$ denote the distribution of the leading term in the representation above, with $\sqrt{T}h_R(\theta_0)$ replaced by the Gaussian approximating random variable. Let $e_{TR} = J - \tilde{J}$ be the approximation error. It follows by Lemma S.14	 of \cite{Fan2023} that, for any $\epsilon > 0$:

\begin{equation*}
	\sup_{c \in \mathbb{R}} |\mathbbm{P}[J \leq c] - \mathbbm{P}[\tilde{J} \leq c]| \leq  \sup_{c \in \mathbb{R}} \mathbbm{P}[c \leq \tilde{J} \leq c + \epsilon ] + \mathbbm{P}[|e_{TR}| > \epsilon ] \, .
\end{equation*}

In addition, by Theorem 2.7 in \cite{Gotze2019}, we have that, for some constant $C>0$:

$$\sup_{c \in \mathbb{R}} \mathbbm{P}[c \leq \tilde{J} \leq c + \epsilon ]\leq  \frac{C \epsilon}{\sqrt{(R-d)}}\, .$$

Consequently:

\begin{equation*}
	\sup_{c \in \mathbb{R}} |\mathbbm{P}[J \leq c] - \mathbbm{P}[\tilde{J} \leq c]| \leq   \frac{C \epsilon}{\sqrt{(R-d)}}  + \epsilon ] + \mathbbm{P}[|e_{TR}| > \epsilon ] \, ,
\end{equation*}

The approximation error $e_{TR}$ consists of two parts: the error due to linearisation, and the error due to approximating $\sqrt{T}h_R(\theta_0)$ by a Gaussian random variable. In light of our discussion in the main text, both errors are $o_{P^*}(1)$. As a consequence, for an arbitrary choice of $\epsilon > 0$, and as $T,R \to \infty$, we obtain that:

$$	\lim_{T,R\to \infty}\sup_{c \in \mathbb{R}} |\mathbbm{P}[J \leq c] - \mathbbm{P}[\tilde{J} \leq c]| = 0\, ,$$
justifying the validity of the chi-squared approximation when $R$ increases with the sample size. In contrast, if $R$ is held fixed, we have to explicitly take the rate of the error $e_{TR}$ into account. As we have discussed, this error can be decomposed into two parts: a linearization error, and a strong approximation error.
In \Cref{app_selection}, we provide conditions that ensure the linearization error is $O_P(T^{-1/2})$. The rate of the Gaussian approximation error depends on the dependence between observations and the assumptions on the distribution: Theorems \ref{thm_gaussian_iid} and \ref{thm_gaussian_mixing} in the main text provide rates in the iid and strongly mixing settings. Let $b_T$ denote the rate of the second type of error, and $c_T := T^{-1/2}\lor b_T $. In this case, by taking $\epsilon = (c_T)^\alpha$ for some $0 <\alpha < 1$, we are able to establish validity of the approximation with fixed $R$.

\section{Bootstrap-based inference}
\label{app_bootstrap}
In this Appendix, we show how one can leverage the Gaussian strong approximation result presented in the main text to perform bootstrap-based inference. We focus on the iid setting. Consider the asymptotic linear representation \eqref{eq_asymptotic linear_prop}. In the main text, we have shown that, under a Gaussian approximation, the term $\sqrt{T} h^R(\theta_0)$ can be approximated by the integral of a Brownian bridge. Consider, now, the alternative process:

\begin{equation*}
	A_T = -( \nabla_{\theta'} h^R(\theta_0)' \Omega^R \nabla_{\theta'} h^R(\theta_0))^{-1} \nabla_{\theta'} h^R(\theta_0)' \Omega^R \left[ \int_{\underline{p}}^{\bar{p}} \sqrt{T}(\check{Q}_Y(u) - \hat{Q}_Y(u))\mathbf{P}^R(u) du \right] \, ,
\end{equation*}
where $\check{Q}_Y(u)$ is the quantile function associated with distribution function $\check{F}_Y(y) = \sum_{t=1}^T \Delta_t \mathbbm{1} \{Y_t \leq y\}$, where $\Delta_t = \frac{Z_i}{\sum_{t=1}^T Z_t}$, and the $Z_t$ are iid random variables, independent from the data, with $\mathbbm{E}Z_t=1$, $\mathbbm{V}Z_t=1$, and a moment generating function (MGF) that exists on a neighborhood of zero. The distribution $\check{F}_Y(y)$ constructed in such way is known as a weighted bootstrap estimator of the empirical distribution $\hat{F}_Y$. The weighted bootstrap is quite general and encompasses, among others, the Bayesian bootstrap \citep{Rubin1981}.

If, in addition to the conditions in Theorem \ref{thm_gaussian_iid} of the main text, we assume $\sup_{y \in (a,b)} |f_Y'(y)| < \infty$  and $A = \lim_{y \downarrow a} f_Y(y) < \infty$, $B = \lim_{y \uparrow a} f_Y(y) < \infty$ with $\min\{A,B\} > 0$, then Theorem 7 in \cite{AlvarezAndrade2013} indicates that $ \left[ \int_{\underline{p}}^{\bar{p}} \sqrt{T}(\check{Q}_Y(u) - \hat{Q}_Y(u))\mathbf{P}^R(u) du \right]$ is strongly approximated by the integral of a Gaussian process that is \textbf{identically distributed} to the strong approximation of the term $\sqrt{T} h^R(\theta_0)$ obtained in the main text. Specifically, inspection of the proof in \cite{AlvarezAndrade2013} reveals that there exists a sequence of Brownian bridges $\{\tilde{B}_n\}_{n \in \mathbb{N}}$, where each $B_n$ is \emph{independent} from $Y_1,Y_2, \ldots, Y_n$, such that, as $T,R\to \infty$:

{\footnotesize$$\left\lVert \int_{\underline{p}}^{\bar{p}} \sqrt{T}(\check{Q}_Y(u) - \hat{Q}_Y(u))\mathbf{P}^R(u) du - \int_{\underline{p}}^{\bar{p}} \sqrt{T}(\check{Q}_Y(u) - \hat{Q}_Y(u))\mathbf{P}^R(u)\frac{1}{f_Y(Q_Y(u))} \tilde{B}_ndu  \right\rVert_2 \overset{\text{a.s.}}{=} O(\log n/\sqrt{n})\, .$$}

It thus follows by Markov inequality that:

{\footnotesize $$\mathbb{P}\left[\left\lVert \int_{\underline{p}}^{\bar{p}} \sqrt{T}(\check{Q}_Y(u) - \hat{Q}_Y(u))\mathbf{P}^R(u) du - \int_{\underline{p}}^{\bar{p}} \sqrt{T}(\check{Q}_Y(u) - \hat{Q}_Y(u))\mathbf{P}^R(u)\frac{1}{f_Y(Q_Y(u))} \tilde{B}_ndu  \right\rVert_2 > \epsilon \Bigg|Y_1,\ldots, Y_T\right]= o_{P}(1)\, ,$$}for every $\epsilon > 0$. Since the Brownian bridges are \textbf{independent} from the data, the conditional convergence justifies the use of the weighted bootstrap to approximate the distribution of $A_T$. Indeed, given consistent estimators of $\theta_0$ an $\Omega_R$, we can approximate the distribution of $A_T$ by generating a large number of simulations of the $\{Z_t\}_{t=1}^T$ and computing, for each simulation, the quantile function of the resulting weighted cdf. The distribution across simulations can then be used to approximate the distribution of the $A_T$.

\section{Inference based on Bahadur-Kiefer representation} \label{app_inference_bahadur} In this Appendix, we discuss how we can conduct inference by relying on a Bahadur-Kiefer representation. We first state the result of Kiefer, in the iid context, as extended by \cite{Csorgo1978}.

\begin{theorem}[Bahadur-Kiefer, \cite{Csorgo1978}]
	\label{thm_bahadur}
	Let $Y_1, Y_2 \ldots Y_T$ be an iid sequence of random variables with a continuous distribution function $F$ which is also twice differentiable on $(a,b)$, where $-\infty \leq a = \sup\{z: F(z) = 0 \}$ and $b=\inf\{z: F(z) = 1\}\leq \infty$. Suppose that $F'(z) = f(z) > 0$ for $z \in (a,b)$. Assume that, for $\gamma > 0$:
	
	\begin{equation*}
		\sup_{a < x < b} F(x)(1-F(x))\left|\frac{f'(x)}{f^2(x)}\right| \leq \gamma \ ,
	\end{equation*}
	where $f$ denotes the density of $F$. Moreover, assume  that $f$ is nondecreasing (nonincreasing) on an interval to the right of $a$ (to the left of $b$). We then have that
	
	\begin{equation}
		\label{eq_bahadur_kiefer}
		\begin{aligned}
			\sup_{0 < u < 1} |f(Q_Y(u)) \sqrt{T}(\hat{Q}_Y(u) -  Q_Y(u)) -  \sqrt{T}(\hat{F}_Y(Q_Y(u)) - F(Q_Y(u))| \overset{a.s.}{=} \\ \overset{a.s.}{=}  O(T^{-1/4} (\log T)^{1/2} (\log \log T)^{\frac{1}{4}}) \, .
		\end{aligned}
	\end{equation}
\end{theorem}

The result above could be used as the basis for an inferential procedure -- as well as for the computation of the optimal weights $\Omega^R$. Indeed, we note that, under the assumptions on the theorem above, $F(Q_Y(u)) = u$ and $\hat{F}_Y(Q_Y(u)) = \frac{1}{T} \sum_{t=1}^T \mathbbm{1}\{U_t \leq u\}$, where the $U_t \coloneqq F(Y_T)$ are iid uniform random variables. Suppose that $\int_{\underline{p}}^{\overline{p}}\frac{1}{f_Y(Q_Y(u))^2} du < \infty$.  Then, using the representation of the theorem above in the asymptotic linear representation \eqref{eq_asymptotic linear_prop} in the main text and applying Bessel's inequality, we get:

\begin{equation}
	\label{eq_asymptotic distr_kiefer}
	\begin{aligned}
		\sqrt{T}(\hat{\theta} - \theta_0) = \\ = -( \nabla_{\theta'} h^R(\theta_0)' \Omega^R \nabla_{\theta'} h^R(\theta_0))^{-1} \nabla_{\theta'} h^R(\theta_0)' \Omega^R \left[ \int_{\underline{p}}^{\bar{p}} \frac{\sqrt{T}(\hat{F}_Y(Q_Y(u)) - F_Y(Q_Y(u))}{f_Y(Q_Y(u))} \mathbf{P}^R(u) du \right]  + o_{P^*}(1) \, ,
	\end{aligned}
\end{equation}
where the distribution of the leading term is known (it could be simulated by drawing $T$ independent Uniform[0,1] random variables many times) up to $\theta_0$.

There is a sizeable literature on Bahadur-Kiefer representations in the context of dependent observations (see \cite{Kulik2007} and references therein). Nonetheless, in the context of dependent observations, it would be more difficult to use \eqref{eq_asymptotic distr_kiefer} as a basis for an inferential procedure, as in this case there would be dependence between the $U_t \coloneqq F_Y(Y_t)$ uniform random variables entering the empirical cdf. For that reason, our focus in this section is on the iid case.

Finally, to show the validity of our approach to inference based on drawing uniform random variables, we note that, under a Bahadur-Kiefer approximation, we have that:

\begin{equation*}
	V_{T,R}^{-1/2} \sqrt{T}(\hat{\theta}_T - \theta_0) = \frac{1}{\sqrt{T}}\sum_{t=1}^T \xi_{t}^{R,T} + o_p(1) \, ,
\end{equation*}
where $V_{T,R}$ is the variance of the leading term of the Bahadur-Kiefer representation, $\mathbbm{E}[\xi_{t}^{R,T}] = 0$ and $\mathbbm{V}\left[\frac{1}{\sqrt{T}}\sum_{t=1}^T \xi_{t}^{R,T}\right]=1$. In the iid context, it is immediate that the conditions of Lindeberg's CLT for triangular arrays \citep[Theorem 3.4.10]{Durrett2019} are satisfied, from which it follows that $V_{T,R}^{-1/2} \sqrt{T}(\hat{\theta}_T - \theta_0)  \overset{d}{\to} N(0,\mathbbm{I}_d)$.\footnote{In the dependent case, even though it is not feasible to leverage the Bahadur-Kiefer representation directly for inference, it is possible to adopt it to establish, under (possibly) additional assumptions, weak convergence of the estimator, by verifying the conditions of a CLT for triangular arrays under dependent data. For example, in the stationary mixing case, one could verify if the conditions of Theorem 4.4 in \cite{Rio2017} hold.} Observe that as a byproduct of such convergence, we obtain that the Kolmogorov distance between the distribution of $\sqrt{T}(\hat{\theta_T}-\theta_0)$ and that of the leading term of the representation  \eqref{eq_bahadur_kiefer} goes to zero, analogously to the result in equation \eqref{eq_kolmogorov_bound} in the main text. This result justifies our approach to inference.

We collect the discussion of this section in the next corollary.

\begin{corollary}
	\label{corollary_kiefer}
	Suppose Assumptions \ref{ass_consistency}-\ref{ass_eigen} hold. Moreover, suppose a Bahadur-Kiefer representation such as \eqref{eq_bahadur_kiefer} is valid; and that $\int_{\underline{p}}^{\overline{p}}\frac{1}{f_Y(Q_Y(u))^2} du < \infty$. Then, as $T,R\to\infty$, the approximation \eqref{eq_asymptotic distr_kiefer} holds. In addition, under the conditions of \Cref{thm_bahadur}, $F_Y(Q_Y(u)) = u$ and $\hat{F}_Y(Q_Y(u)) = \frac{1}{T} \sum_{t=1}^T \mathbbm{1}\{U_t\leq u \}$, where the $\{U_t\}_{t=1}^T$ are iid Uniform[0,1] random variables. Moreover, under the previous assumptions, and as $T,R\to\infty$, $V_{T,R}^{-1/2} \sqrt{T}(\hat{\theta}_T - \theta_0)  \overset{d}{\to} N(0,\mathbbm{I}_d)$, where $V_{T,R}$ is the variance of the leading term in \eqref{eq_asymptotic distr_kiefer}; and a bound analogous to equation \eqref{eq_kolmogorov_bound} in the main text holds.
\end{corollary}

\begin{remark}[Optimal choice of weighting matrix under Bahadur-Kiefer approximation] \label{same_optimal}
	It should be noted that the optimal choice of weighting matrix under the Bahadur-Kiefer representation \textbf{coincides} with \eqref{eq_optimal_weights} in the main text. This is due to the fact that both the Brownian bridge and empirical distribution process share the same covariance kernel. 
\end{remark}

\begin{remark}[Distribution of the overidentifying test statistic in \Cref{test_over}] Note that we could use the distributional results in this section to compute the distribution of the test statistic in \Cref{test_over} in the main text under the null.
\end{remark}

\section{Asymptotic efficiency}
\label{app_efficiency}
In this Appendix, we analyse whether our L-moment estimator is asymptotically efficient. We consider the case where $0=\underline{p}<\overline{p}=1$, since in this case all information on the curve is used; for simplicity, we also focus on the iid case. In this setting, we will say our L-moment estimator is \emph{asymptotically efficient} if its asymptotic variance coincides with the inverse of the Fisher information matrix of the parametric model. Unless stated otherwise, we work under Assumptions \ref{ass_consistency}-\ref{ass_eigen} and those of \Cref{corollary_kiefer}. To proceed with the analysis, we introduce the alternative estimator:

\begin{equation}
	\label{eq_alternative}
	\tilde{\theta}_T \in \operatorname{argmin}_{\theta \in \Theta} \sum_{i \in \mathcal{G}_T} \sum_{j \in \mathcal{G}_T} (\hat{Q}_Y(i)-Q_Y(i|\theta)) \kappa_{i,j} (\hat{Q}_Y(j)-Q_Y(j|\theta)) \, ,
\end{equation}
for a grid of $G_T$ points $\mathcal{G}_T =\{g_1,g_2,\ldots, g_{G_T}\} \subseteq (0,1)$ and weights $\kappa_{i,j}$, $i,j \in \mathcal{G}_T$. This is a weighted version of a ``percentile-based estimator'', which is used in contexts where it is difficult to maximise the likelihood \citep{Gupta2001}. It amounts to choosing $\theta$ so as to match a weighted combination of the order statistics in the sample. 

Under regularity conditions similar to the ones in the main text,\footnote{We omit these conditions for brevity, but we note that, using the notation in \eqref{eq_first_order_alt}, since we assume $\lVert\sqrt{T}Q_{G_T}\rVert_{\infty} = O_p(1)$ (implied by \Cref{ass_weak_convergence}), it is crucial that $\lVert \partial Q_{G_T}'\boldsymbol{\kappa} \rVert_{\infty} = O_p(1)$, where $\lVert \cdot\rVert_{\infty}$ is the operator norm induced by the vector norm. This condition can be shown to hold for the optimal choice of weights described below under some conditions. We also require a restriction on the growth rate of $G_T$ so as to control the error of a mean-value expansion of increasing dimension.} the estimator in \eqref{eq_alternative} admits the following asymptotic linear representation as $T \to \infty$ and $G_T \to \infty$ at a rate:

\begin{equation}
	\label{eq_first_order_alt}
	\sqrt{T}(\tilde{\theta}_T-\theta_0) = -(\partial Q_{G_T} '\boldsymbol{\kappa}_{G_T} \partial Q_{G_T})^{-1} \partial Q_{G_T}' \boldsymbol{\kappa}_{G_T} \sqrt{T}Q_{G_T} + o_p(1) \, ,
\end{equation}
where $Q_{G_T} = Q_{G_T}(\theta_0) = (\hat{Q}_Y(g_1) - Q_Y(g_1|\theta_0),\ldots, \hat{Q}_Y(g_T) - Q_Y(g_T|\theta_0))'$; $\partial Q_{G_T}$ is the Jacobian matrix of $Q_{G_T}(\theta)$ evaluated at $\theta_0$; and $\boldsymbol{\kappa}_{G_T}$ is the matrix containing the $\kappa_{i,j}$. Using the Bahadur-Kiefer representation (assumed by \Cref{corollary_kiefer}), we arrive at:

\begin{equation}
	\label{eq_alt_bahadur}
	\sqrt{T}(\tilde{\theta}_T-\theta_0) = -(\partial Q_{G_T} '\boldsymbol{\kappa}_{G_T} \partial Q_{G_T})^{-1} \partial Q_{G_T}' \boldsymbol{\kappa}_{G_T}\left[ \boldsymbol{f}^{-1} * \sqrt{T}F_{G_T}\right] + o_p(1) \, ,
\end{equation}
where we define $F_{G_T} = (\hat{F}_Y(Q_Y(g_1)) -{F}_Y(Q_Y(g_1)),\ldots, \hat{F}_Y(Q_Y(g_{G_T})) - {F}_Y(Q_Y(g_{G_T})))'$; $\boldsymbol{f}^{-1} = (1/{f}_Y(Q_Y(g_1)), \ldots, 1/{f}_Y(Q_Y(g_T)))'$; and $*$ denotes entry-by-entry multiplication.

For a given $\mathcal{G}_T$, representation \eqref{eq_alt_bahadur} yields the following choice of optimal weighting matrix, $\boldsymbol{\kappa}^* = \mathbbm{V}[\boldsymbol{f}^{-1} * \sqrt{T}F_{G_T}]^{-1}$; and this implies that the variance of the leading term of \eqref{eq_alt_bahadur} under such choice is $\mathbb{V}^* = (\partial Q_{G_T} '\boldsymbol{\kappa}_{G_T} \partial Q_{G_T})^{-1}$. But, if we take the grid $\mathcal{G}_T$ as $\left\{\frac{1}{G_T+1}, \frac{2}{G_T+1}, \ldots, \frac{G_T}{G_T+1}\right\}$, it follows from Lemma C.1. in \cite{Firpo2021} that:

\begin{equation*}
	\mathbb{V}^* = ((\partial Q_{G_T} * ( \mathbf{1}_{d}' \otimes \boldsymbol{f} ))' \Sigma_{G_T}^{-1} (\partial Q_{G_T} * ( \mathbf{1}_{d}' \otimes \boldsymbol{f} ))^{-1} \, ,
\end{equation*}
where
\begin{equation*}
	(\Sigma_{G_T}^{-1})_{g_i,g_j} = \mathbbm{1}_{\{g_i = g_j\}} 2(G_T+1) - (\mathbbm{1}_{\{g_{i} = g_{j+1}\}}  + \mathbbm{1}_{\{g_{i} = g_{j-1}\}}) (G_T+1) \, .
\end{equation*}

It then follows that, for $d_1,d_2 \in \{ 1,2\ldots, d\}$ :

\begin{equation}
	\label{eq_variance_alt}
	\begin{aligned}
		\left(\mathbbm{V}^{*-1}\right)_{d_1,d_2} =  \\  (G_T+1)\sum_{i=2}^{G_T} f_Y(Q_Y(g_i)) \partial_{d_1} Q_Y(g_i|\theta_0)\left[ f_Y(Q_Y(g_{i}) \partial_{d_2} Q_Y(g_{i}|\theta_0) - f_Y(Q_Y(g_{i-1}) \partial_{d_2} Q_Y(g_{i}|\theta_0) \right] \\
		 -(G_T+1)\sum_{i=1}^{G_T-1} f_Y(Q_Y(g_i)) \partial_{d_1} Q_Y(g_i|\theta_0)\left[ f_Y(Q_Y(g_{i+1}) \partial_{d_2} Q_Y(g_{i}|\theta_0) - f_Y(Q_Y(g_{i}) \partial_{d_2} Q_Y(g_{i}|\theta_0) \right] \\
		+ (G_T+1)(f_Y(Q_Y(g_1)))^2  \partial_{d_1} Q_Y(g_1|\theta_0) \partial_{d_2} Q_Y(g_1|\theta_0) \\ +   (G_T+1) (f_Y(Q_Y(g_{G_T})))^2  \partial_{d_1} Q_Y(g_{G_T}|\theta_0) \partial_{d_2} Q_Y(g_{G_T}|\theta_0) \, .
	\end{aligned}
\end{equation}

Assuming the tail condition:\footnote{A similar tail condition is considered in a working paper version of \cite{Firpo2021}.}
{\footnotesize
\begin{equation}
	\label{eq_tail}
	\lim_{u \to 0} \frac{(f_Y(Q_Y(u)))^2  \partial_{d_1} Q_Y(u|\theta_0) \partial_{d_2} Q_Y(u|\theta_0) +  (f_Y(Q_Y(1-u)))^2  \partial_{d_1} Q_Y(1-u|\theta_0) \partial_{d_2} Q_Y(1-u|\theta_0)}{u} = 0 \, ,
\end{equation}} leads to the last two terms of \eqref{eq_variance_alt} being asymptotically negligible as $T \to \infty$. If we further assume the $u \mapsto f_Y(Q(u)) \partial_{d_1}Q_Y(u|\theta_0)$ are differentiable uniformly on $(0,1)$, it follows from Riemann integration that:
\begin{equation*}
	\lim_{T \to \infty}\left(\mathbbm{V}^{*-1}\right)_{d_1,d_2} = \int_0^1 \frac{d \left[f_Y(Q_Y(v))\partial_{d_1} Q_Y(v|\theta_0)\right]}{dv} \Bigg|_{v=u}  \frac{d \left[f_Y(Q_Y(v))\partial_{d_2} Q_Y(v|\theta_0)\right]}{dv} \Bigg|_{v=u} du \ .
\end{equation*}

But then, from the relation:

\begin{equation*}
	F_Y(Q_Y(u|\theta)|\theta) = u \implies f_Y(Q_Y(u|\theta)) \partial_d Q_Y(u|\theta)   =  - \partial_d F_Y(Q_Y(u)|\theta) \ ,
\end{equation*}
it follows, by exchanging the order of differentiation:

\begin{equation*}
	\begin{aligned}
	\frac{d \left[f_Y(Q_Y(v))\partial_{d_1} Q_Y(v|\theta_0)\right]}{dv} \Bigg|_{v=u} =  -\partial_d \left[f_Y(Q_Y(u)|\theta) \cdot \frac{d Q_Y(v)}{dv}\Big|_{v=u} \right]\Bigg|_{\theta=\theta_0} = \\  -\partial_d f_Y(Q_Y(u)|\theta_0) \cdot \frac{1}{f_Y(Q_Y(u))} \ ,
	\end{aligned}
\end{equation*}
and, using the quantile representation of a random variable, we conclude that:

\begin{equation*}
	\lim_{T \to \infty}\left(\mathbbm{V}^{*-1}\right)_{d_1,d_2} = (I(\theta_0))_{d_1,d_2} \ ,
\end{equation*}
where $I(\theta) = \mathbbm{E}[\nabla_{\theta} \log(f(Y|\theta)) \nabla_{\theta'}\log(f(Y|\theta))]$ is the Fisher information matrix. We have thus shown that, under the proposed grid and optimal weights, the asymptotic variance of the leading term of the first order representation \eqref{eq_alt_bahadur} of estimator \eqref{eq_alternative} converges to the inverse Fisher information. We summarise this point in the lemma below.

\begin{lemma}
	Consider the estimator \eqref{eq_alternative}. Assume that  representation \eqref{eq_alt_bahadur} holds. If, in addition, the tail condition \eqref{eq_tail} and the uniform differentiability condition in the text holds; then the variance of the leading term of representation \eqref{eq_alt_bahadur} converges to the inverse Fisher information.
\end{lemma}

How does the previous estimator relate to our L-moment estimator? Notice that, if the $\{P_l\}_{l\in \mathbbm{N}}$ form an orthonormal \textbf{basis} on $L^2[0,1]$, then, for $X \in L^2[0,1]$:

\begin{equation*}
	X(u) = \sum_{l=1}^\infty \left(\int_0^1 X(s) P_l(s)ds\right) P_l(u) \ .
\end{equation*}

Therefore, since $\{Q_Y(\cdot|\theta): \theta \in \Theta_0\} \subseteq L^2[0,1]$,\footnote{This is implied by \Cref{ass_identification}.} we have:

\begin{equation*}
	\begin{aligned}
		\sum_{i \in \mathcal{G}_T} \sum_{j \in \mathcal{G}_T} (\hat{Q}_Y(i)-Q_Y(i|\theta)) \kappa^*_{i,j} (\hat{Q}_Y(j)-Q_Y(j|\theta))  = \\  \sum_{k=1}^\infty \sum_{l=1}^\infty \left[\int_0^1(\hat{Q}_Y(u) - Q_Y(u|\theta)) P_k(u) du\right] \tilde{\kappa}_{k,l} \left[\int_0^1(\hat{Q}_Y(u) - Q_Y(u|\theta)) P_l(u)  du\right] \eqqcolon \tilde{A}_T^\infty(\theta) \ ,
	\end{aligned}
\end{equation*}
which shows that the optimal estimator we described is a generalised L-moment estimator which uses infinitely many L-moments and suitable weights $\tilde{\kappa}_{k,l} = \sum_{i \in \mathcal{G}_T} \sum_{j \in \mathcal{G}_T} P_k(i) \kappa_{i,j}^* P_l(j) $. Consider an alternative L-moment estimator that uses only the first $R$ L-moments and weights $\tilde{\boldsymbol{\kappa}}_R = (\tilde{\kappa}_{i,j})_{i,j = 1,\ldots R}$. Denote the estimator by $\check{\theta}_T$, and its objective function by $A^R_T(\theta)$. It can be shown that, for an identifiable parametric family and $\lVert \tilde{\boldsymbol{\kappa}}\rVert_2 = O(1)$, $\nabla_{\theta} A^R_T(\check{\theta}_T) = \nabla_{\theta}A^\infty_T(\check{\theta}_T) + o_{p^*}(T^{-1/2})$. This shows that the estimator admits the same first order representation as \eqref{eq_alt_bahadur}; and from the previous lemma we know the variance of the leading term of this representation converges to $I(\theta_0)^{-1}$.	It thus follows that $\check{\theta}_T$ is asymptotically efficient.\footnote{Let $\Psi_T$ denote the variance of the leading term of representation \eqref{eq_asymptotic distr_kiefer} of the estimator $\check{\theta}$. By weak convergence (the last part of \Cref{corollary_kiefer})  and Fatou's lemma, it follows that $\liminf_{T \to \infty}\xi'(\Psi_T^{-1/2} M_T {\Psi}_T^{-1/2} - \mathbb{I}_d)\xi \geq 0$ for any $\xi \in \mathbbm{R}^d$, where $\lim_T {M_T} = I(\theta_0)^{-1}$. It then follows that  $\lim_{T \to \infty} \Psi_T =  I(\theta_0)^{-1}$, from which we conclude that $\check{\theta}$ is asymptotically efficient. } But then, since the optimal weights \eqref{eq_optimal_weights} minimise the variance of the leading term in \eqref{eq_asymptotic distr_kiefer} (recall \Cref{same_optimal}), we conclude that they too must, asymptotically, yield a variance equal to $I(\theta_0)^{-1}$. This shows that the generalised L-moment estimator is efficient, in the sense that its asymptotic variance coincides with $I(\theta_0)^{-1}$.

We collect the discussion of this section in the corollary below:

\begin{corollary}
	Suppose the conditions of the previous lemma hold. Suppose the $\{P_l\}_{l \in \mathbbm{N}}$ constitute an orthonormal \textbf{basis}. Consider the estimator $\check{\theta}_T$ defined in the main text. Suppose that Assumptions \ref{ass_consistency}-\ref{ass_eigen} and those of \Cref{corollary_kiefer} hold with $W^R = \Omega^R = \boldsymbol{\kappa}_R$. We then have that, for an identifiable parametric family:
	\begin{equation*}
		\lim_{T,R \to \infty } V_{T,R}^*= I(\theta_0)^{-1} \ ,
	\end{equation*}
	where $V_{T,R}^*$ is the variance of the leading term of  \eqref{eq_asymptotic distr_kiefer} under the optimal choice of weights, i.e. $V_{T,R}^* = (\nabla_{\theta'} h^R(\theta_0)' \Omega_R^* \nabla_{\theta'} h^R(\theta_0))^{-1}$, where $\Omega_R^*$ is given by \eqref{eq_optimal_weights}.
\end{corollary}

\begin{remark}[Related estimators]
	Similarly to \eqref{eq_alternative}, we can show that estimators based on minimising the objective functions:
	
	\begin{equation}
		\label{eq_alternative_est}
		\begin{aligned}
			W^1(\theta) \coloneqq \int_{0}^{1} w(u)(\hat{Q}_Y(u) - Q_Y(u|\theta))^2 du \ , \\
			W^2(\theta) \coloneqq \int_{0}^{1}\int_{0}^{1} (\hat{Q}_Y(u) - Q_Y(u|\theta))w(u,v)(\hat{Q}_Y(v) - Q_Y(v|\theta)) dvdu \ ,
		\end{aligned}
	\end{equation}
	are also L-moment-based estimators which use infinitely many L-moments. A similar argument as the one in this section then shows that our generalised method of L-moments estimator under optimal weights will be at least as efficient as estimators based on minimising \eqref{eq_alternative_est}. However, given that we are able to control the number of L-moments used in estimation in finite samples, it is expected that our method will lead to nonasymptotic performance gains.
\end{remark}

\section{Monte Carlo exercise: additional results}
\subsection{Results for linear combinations of parameters} In this Appendix, we revisit the data-generating processes of the Monte Carlo exercises in Section \ref{monte_carlo} of the main text, but now consider linear combinations $\delta '\theta_0$, $\delta \in \mathbb{R}^d$, as the target parameters. Specifically, for a given L-moment-based estimator $\hat{\theta}_R$ and linear combination $\delta \in \mathbb{R}^d$, the relative RMSE, vis-à-vis the MLE, is given by:

$$\operatorname{RRMSE}(\delta, \hat{\theta}_{R}) \coloneqq \sqrt{\frac{\mathbb{E}[(\delta'\hat{\theta}_R - \delta'{\theta}_0)^2]}{ \mathbb{E}[(\delta'\hat{\theta}_{\text{MLE}}- \delta'{\theta}_0)^2]}} =\sqrt{ \frac{\delta'(\mathbb{E}[(\hat{\theta}_R-\theta_0)\hat{\theta}_R-\theta_0)'] \delta}{\delta'\mathbb{E}[(\hat{\theta}_{\text{MLE}}-\theta_0)(\hat{\theta}_{\text{MLE}}-\theta_0)'])\delta}} \,,$$

Since we have no direct interest in any particular linear combination $\delta$, we consider the relative RMSE under the most and least favourable values (directions) of $\delta$. These are defined, respectively, as:

\begin{align*}
	\underline{\operatorname{RRMSE}(\hat{\theta}_{R})} \coloneqq \min_{\delta \in \mathbb{R}^d: \delta\neq0}\operatorname{RRMSE}(\delta, \hat{\theta}_{R})  =\\ \sqrt{\lambda_{\text{min}}(\mathbb{E}[(\hat{\theta}_{\text{MLE}}-\theta_0)(\hat{\theta}_{\text{MLE}}-\theta_0)']^{-1/2}\mathbb{E}[(\hat{\theta}_{{R}}-\theta_0)(\hat{\theta}_{{R}}-\theta_0)'] \mathbb{E}[(\hat{\theta}_{\text{MLE}}-\theta_0)(\hat{\theta}_{\text{MLE}}-\theta_0)']^{-1/2})} \\
		\overline{\operatorname{RRMSE}(\hat{\theta}_{R})} \coloneqq \max_{\delta \in \mathbb{R}^d: \delta \neq 0}\operatorname{RRMSE}(\delta, \hat{\theta}_{R})  =\\ \sqrt{\lambda_{\text{max}}(\mathbb{E}[(\hat{\theta}_{\text{MLE}}-\theta_0)(\hat{\theta}_{\text{MLE}}-\theta_0)']^{-1/2}\mathbb{E}[(\hat{\theta}_{{R}}-\theta_0)(\hat{\theta}_{{R}}-\theta_0)'] \mathbb{E}[(\hat{\theta}_{\text{MLE}}-\theta_0)(\hat{\theta}_{\text{MLE}}-\theta_0)']^{-1/2})} \\
\end{align*}

Tables \ref{gev_table_linear} and \ref{gpd_table_linear} report the relative RMSE of the four estimators considered in the main text, under the most and least favourable directions $\delta$, with the choice of $R$ that minimizes the most (least) favourable-RMSE (this choice is reported in parentheses), respectively in the GEV and GPD exercises. Values above $1$ indicate that the MLE outperforms the L-moment estimator, under the stated choice of $R$, in the corresponding (most or least favourable) direction.  In the GEV design, two-step L-moment estimators are able to offer improvements over the MLE in the most favourable directions in smaller samples sizes (with RMSE improvements of nearly 8\%), while severely mitigating losses of first-step estimators in the least favourable directions. Indeed, first-step estimators in the GEV exercise perform quite poorly in the least-favourable direction, especially in the largest sample size, with root-mean-squared errors over 25\% larger than the MLE. In contrast,the càglàd two-step estimators has RMSE only 1.3\% larger than the MLE in the least favourable direction and largest sample size.  

Similarly to the main text, in the GPD design, first-step and two-step  estimators perform well relatively to the MLE, in both the most and least favourable directions, even in the largest sample size. Gains of the L-moment approach can reach 16\% in the smallest sample size and most-favourable direction (4\% in the smallest sample size and least favourable direction). Finally, we observe that, consistent with our theoretical results, the MSE-minimising number of L-moments for two-step estimator increases with sample size when we consider the least favourable direction, in both designs.

\input{tables/gev/gev_table_linear.tex}
\input{tables/gpd/gpd_table_linear.tex}
\subsection{Comparison with trimming approaches} In this Appendix, we compare our L-moment-based approach with maximum likelihood estimators that attempt to control the influence of extreme observations. We consider two approaches. In one of the approaches, we first estimate the model parameters via MLE, here denoted by $\tilde{\theta}$. We then discard (trim) those observations $Y_t$ such that $Y_t > Q(1-\epsilon|\tilde{\theta})$, where $\epsilon$ is a trimming proportion parameter. We then reestimate the model via MLE in the restricted set. We label this approach ``trimmed MLE''.\footnote{Our trimmed MLE approach may be seen as a one-step approximation to more complex trimming approaches where the indices of the discarded observations and the model parameters $\theta$ are simultaneously estimated \citep[e.g.][]{Hadi1997,Awasthi2022}.} We also consider an alternative, ``tilted MLE'' approach \citep{Choi2000}, that seeks to find $\theta$ by maximizing the following quantity over $\Theta$:
$$\sup_{(p)_i \in \mathcal{P}_\epsilon}\sum_{i=1}^T p_i \log(f(Y_i|\theta))\, ,$$
where $\mathcal{P}_\epsilon$ is the subset of the simplex $\Delta^{T-1}$ such that $D_{\text{KL}}((p_i)_i || (1/n)_i) = - \log(1-\epsilon)$, with $D_{\text{KL}}((p_i)_i || (1/n)_i)$ denoting the  Kullback-Leibler divergence of the uniform distribution on the indices $\{1,2\ldots,T\}$ from the distribution $(p_i)_{i=1}^T$ over  $\{1,2\ldots,T\}$. The estimator amounts to running the MLE in a ``reweighted'' dataset, where the weights are chosen in order to minimize the KL divergence of the model from the (reweighted) data, subject to the constraint that weights are not far astray from the untilted empirical distribution. As argued by \cite{Choi2000}, in the context of a data contamination model, the parameter $\epsilon \in [0,1)$ amounts to the proportion of observations allowed to be corrupted, i.e. that do not follow the parameteric model $f_\theta$ of interest.

Tables \ref{gev_table_additional} and \ref{gpd_table_additional} replicate the results for the Càglàd TS estimator under the MSE-minimizing choice of $R$ presented in Tables \ref{gev_table_mle} and \ref{gpd_table_mle} in the main text, and compare it with the trimmed and tilted MLE approaches.  For the trimming/tilting proportion $\epsilon$, we consider values $\epsilon \in \{0.1,0.01, 0.001\}$. For the tilted MLE, we also report in parantheses the percentage of cases where the method for finding the estimator proposed in \cite{Choi2000} did \emph{not} converge. We observe that the trimming approach never compares fabourably to the Càglàd TS estimator under the MSE-minimizing choice of $R$. As for the tilted MLE, it is able to compete with the Càglàd TS estimator for some combinations of tail quantiles and sample sizes, under a suitable choice of $\epsilon$. However, it is important to note that the competitiviness and overall performance of this method is \textbf{very} dependent on the trimming fraction $\epsilon$. For example, in the GEV design with $\tau  = 0.999$ and $T=500$, the relative RMSE changes from 1.6\% with $\epsilon = 0.1\%$ to $76.9\%$ when $\epsilon = 1\%$.  Such sensitivity limits applicability of this approach in these designs, especially since, to the best of our knowledge, there does not exist any method to tune the tilting proportion $\epsilon$ with an aim to obtain RMSE gains over the MLE. \footnote{\cite{Choi2000} propose a heuristic to select $\epsilon$ which consists in computing the QQ-plot that compares the reweighted empirical quantiles with the parametric quantiles obtained from the trimmed MLE estimator. The tilting proportion $\epsilon$ should then be chosen so as to make the QQ-plot ``close'' to the 45 degrees line. Their heuristic is motivated by data corruption concerns, though, and not RMSE reductions. Indeed, notice that, inasmuch as parametric tail quantile estimators offer MSE improvements over nonparametric empirical quantiles, one would expect differences between these estimators.}
\input{tables/gev/gev_additional.tex}
\input{tables/gpd/gpd_additional.tex}
\subsection{Results on confidence interval coverage and length} In this section, we study the coverage and length properties of confidence based on the normal approximations derived in the main text. We work in the same setting of the Monte Carlo exercise of Section \ref{monte_carlo} in the main text, where the goal was quantile estimation. We focus on the behaviour of the Càglàd two-step estimator.

\subsubsection{GEV design} To begin understanding the quality of the normal approximations derived in the main text, we analyse the coverage and length properties of confidence intervals based on normal critical values and the \emph{true} sampling variance of the estimators. Specifically, we study confidence intervals of the form $Q(\tau|\hat{\theta})  \pm\sqrt{\mathbb{V}[Q(\tau|\hat{\theta})} q_Z(1-(1-\beta)/2)$, where $\beta$ is the nominal coverage level, $q_Z(u)$ is the $u$-quantile of a standard-normal distribution, and $\mathbb{V}[Q(\tau|\hat{\theta}) ]$ is the true sampling-variance of the plug-in quantile estimator, which we recover from the Monte Carlo draws. We consider the case $\beta = 0.95$.

Figure \ref{fig:gev_coverage_true}  reports, in blue, the coverage of the confidence intervals based on the Càglád two-step estimator, for different values of $R$. In red, we also report the coverage of MLE-based CIs that use the true sampling variance and normal critical values. The nominal level $\beta =0.95$ is presented as a black horizontal line. As one can observe, coverage of the L-moment-based CIs is close to the nominal level for every combination of sample size and target quantile, suggesting that the strong approximations derived in the main text offer a good approximation to the designs at had. Moreover, and consistent with our theoretical results that do not impose any restrictions on the growth rate of $R$ in the derivation of the normal approximations, coverage is constant across the values of $R$. The MLE-based CIs also have coverage very close to the nominal in all sample sizes and target quantiles.

Figure \ref{fig:gev_length_true} reports how the length of the CIs based on the true sampling variance changes with different values of $R$. Length is reported as a proportion of the length of the MLE-based CIs, meaning that values above one indicate that the L-moment-based CIs are larger than those based on the MLE. As expected from our efficiency results, the length-minimising choice of $R$ is always competitive with the MLE, offering substantial improvements in length for smaller sample sizes/more extreme quantiles, and working as well as the MLE in the largest sample size.

Next, we analyse the behaviour of feasible versions of the above CIs that rely on estimators of the asymptotic variance. For the MLE, we rely on a delta-method approximation combined with an estimator of the asymptotic variance of $\hat{\theta}_{\text{MLE}}$ based on the Hessian of the objective function. For the two-step Càglád estimator, we rely on the delta-method-type result in \ref{corollary_tail} to estimate the variance $\mathbb{V}[Q(\tau|\hat{\theta}_R)]$, as:

$$ \widehat{\mathbb{V}[Q(\tau|\hat{\theta}_R)]}= \frac{\nabla_{\theta}Q(\tau|\hat{\theta}_R)' \hat{V} \nabla_{\theta}Q(\tau|\hat{\theta}_R)}{T}\, ,$$
where $\hat{V}$ is an estimator of the asymptotic variance of the optimally-weighted estimator $\hat{\theta}_R$, which is given by:
$$\hat{V} = \widehat{\mathbb{V}[\hat{\theta}_R]} =\left(\left(\int_0^1 \nabla_\theta Q(u|\hat{\theta}_R)\boldsymbol{P}_R(u)'du\right)\hat{\Omega}_R \left(\int_0^1 \boldsymbol{P}_R(u)\nabla_{\theta} Q(u|\hat{\theta}_R)'du\right)\right)^{-1}\, ,$$
with $\hat{\Omega}_R$ the estimator of the optimal weighting matrix used in the minimization of $\hat{\theta}_R$.

Figure \ref{fig:gev_coverage_est} presents the coverage of the feasible CIs that rely on estimators of the asymptotic variance. We see that, in the largest sample size ($T=500$), coverage is close to the nominal level for all target quantiles. Coverage is also quite close to the nominal level in sample sizes $T=50$ and $T=100$ at the median ($\tau =0.5$). However, in samples $T=50$ and $T=100$, as we move further into the tails, coverage tends to deteriorate for both the MLE-based as well as the L-moment-based CIs. In these settings, the L-moment-based CIs undercover more than the MLE-based CIs, in some cases by a small margin, but with especially large differences at the two tailmost quantiles ($\tau=0.99$ and $\tau=0.999$) in the smallest sample size ($T=50$). Finally, we note that, for a given $T$ and $\tau$, coverage is insensitive to the choice of $R$.

In order to better understand the drivers of undercoverage at the tails in smaller sample sizes, we report, in Figure \ref{fig:gev_length_est}, the median-length of the feasible CIs. We normalize these lengths by the length of the \emph{unfeasible MLE-based CI that relies on the true sampling-variance}. By observing the red lines, we note that the feasible MLE-based CIs understate the true sampling variance at the tails in smaller sample sizes, as the red line in these cases can be substantially below one. Similarly, by comparing the blue lines in Figure \ref{fig:gev_length_est} with the corresponding blue lines in Figure \ref{fig:gev_length_true}, we see that the feasible CIs based on the Càglád estimator share a similar pattern, understating the true length of the unfeasible Càglàd-based CIs in the extreme quantiles of smaller sample sizes.

To understand how much the understatement of the correct sampling variance contributes to undercoverage of the feasible confidence intervals in smaller sample sizes and extreme quantiles, we present, in Figure \ref{fig:gev_coverage_rescale}, the coverage of \emph{unfeasible} confidence intervals that, while still relying on estimators for the variance, rescale these by the amount of underestimation verified in the previous discussion.\footnote{Such unfeasible rescaling may be seen as ``a best case'' scenario for feasible strategies that attempt to ``bias-correct'' the standard error estimator, e.g. Welch-style corrections \citep[e.g.][]{Welch1951,Belloni2012,Imbens2016}, or variance estimators based in higher-order expansions such as those presented in Appendix \ref{higher_order_selector} for the generalised L-moment estimator.} We focus on sample sizes $T<500$ and quantiles $\tau > 0.5$. We note that, except for the smallest sample size and more extreme quantile, rescaling the variance estimator so that, on average, it correctly assesses sampling uncertainty does not seem to substantially improve the coverage of confidence intervals. This suggests that other elements are at play in driving the undercoverage. One obvious candidate is correlation between the variance estimator and the estimator for the target quantile, which, in smaller samples and more extreme quantiles, may generate  a non-normal reference distribution for the test that is inverted to construct the confidence interval.

To remove the effect of this correlation and improve coverage of the Càglád-based CI, we suggest a simple procedure based on our strong approximations. First, we note that the feasible CI is based on ``inversion'' of:

$$\mathbf{t} = \frac{{Q}(\tau|\hat{\theta}_R) - \hat{Q}(\tau|\theta_0)}{\sqrt{\widehat{\mathbb{V}[Q(\tau|\hat{\theta}_R)]}}} = \sqrt{T}\frac{{Q}(\tau|\hat{\theta}_R) - \hat{Q}(\tau|\theta_0)}{\sqrt{\nabla_{\theta}Q(\tau|\hat{\theta}_R)' \hat{V} \nabla_{\theta}Q(\tau|\hat{\theta}_R)}}\, .$$

Now, performing a first-order Taylor-expansion on $\mathbf{t}$ in terms of $\hat{\theta}_R$ separately in the numerator and in the variance estimator that enters the denominator, and leveraging the strong approximation of $\sqrt{T}(\hat{\theta}_R-\theta_0)$ in the main text suggests the following distributional approximation:\footnote{In our expansion, we explicitly do not expand $\hat{V}$ in terms of $\hat{\theta}_R$, because estimation error in $\hat{V}$ does not seem to be at the source of undercoverage in smaller sample sizes, given that the feasible CIs perform well at the median even in the smallest sample size.}

$$\widetilde{\mathbf{t}} = \frac{\sqrt{T} \nabla_{\theta}Q(\tau|\theta_0) Z_T}{\sqrt{\nabla_{\theta}Q(\tau|\theta_0)' {V} \nabla_{\theta}Q(\tau|\theta_0) + 2 \nabla_{\theta}Q(\tau|\theta_0)'{V} \nabla_{\theta\theta'}Q(\tau|\theta_0)Z_T/T}}$$
 where $Z_T \sim N(0,V)$, with $V = \mathbb{V}[\hat{\theta}_R]$. Notice that this approximation captures correlation between the numerator and denominator, as $Z_T$ appears in both terms. By replacing $\theta_0$ and $V$ with estimators and simulating $Z_T$, one may estimate the quantiles of $\widetilde{\mathbf{t}}$, which can then be used to construct confidence intervals of the form:

$$\left[Q(\tau|\hat{\theta}_R)- \widehat{q_{\widetilde{\mathbf{t}}}(1-(1-\beta)/2)}\sqrt{\widehat{\mathbb{V}[Q(\tau|\hat{\theta}_R)]}}, Q(\tau|\hat{\theta}_R)- \widehat{q_{\widetilde{\mathbf{t}}}((1-\beta)/2)}\sqrt{\widehat{\mathbb{V}[Q(\tau|\hat{\theta}_R)]}}\right]\, .$$

Figure \ref{fig:gev_coverage_correct} reports the coverage of our feasible corrected L-moment confidence intervals in blue. We compare these with the coverage of the feasible \emph{uncorrected} MLE confidence interval (in red). We note that our correction substantially improves coverage in those settings wherein the uncorrected CIs would most undercover. Except at $\tau = 0.999$ and $T=50$, coverage of the resulting CI is quite close to the nominal level. For the case $\tau = 0.999$ and $T=50$, coverage of the corrected CI becomes quite close to the uncorrected MLE, whereas previously the uncorrected Càglád CI undercovered by a larger margin. The remaining undercoverage in this case can be removed by ``bias-correcting'' the Càglád-variance estimator -- e.g. by estimating the higher-order variance of our estimators by leveraging the higher-order approximations in Appendix \ref{app_selection} (implemented, for example, via Algorithm \ref{alg:rmse})--, as the remaining amount of undercoverage corresponds to the improvement of rescaling over the feasible uncorrected CIs reported in Figures \ref{fig:gev_coverage_rescale}.

To understand how much is the increase in length imparted by our correction, the blue lines in  Figure \ref{fig:gev_length_correct} report the median relative length (vis-à-vis the unfeasible MLE-based CI that relies on the true sampling variance) of our corrected CIs. We also report the relative length of the uncorrected MLE-based CIs (red solid lines) and, for comparison, the relative length of an unfeasibly corrected MLE-based CI that uses the true quantiles of the sampling distribution of the t-statistic inverted in the construction of the confidence intervals (red dotted lines). The latter may be seen as a best-case scenario for the length of feasible corrections to the MLE that attempt to estimate the approximation to the distribution of the $t$-statistic. The benefits of our approach are clear: in four out of the five settings where our corrected L-moment-based CIs display coverage close to the nominal level, length is always below the length of the unfeasibly corrected MLE. The exception is the case $T=100$ and $\tau=0.99$, where length is on average 5 relative percentage points above the unfeasibly corrected MLE. Note, however, that one would expect such differences to vanish, or even revert, once we consider feasible corrections for the MLE, since this would introduce estimation error into the quantiles.

In the case where coverage of our correction is close to the uncorrected MLE ($\tau = 0.999$ and $T=50$), length of our corrected L-moment-based CIs is below \textbf{both} the uncorrected and feasibly uncorrected MLEs. This is also true at $\tau=0.9$ for the length-minimising choice of $R$.

 \subsubsection{GPD design} Figures \ref{fig:gpd_coverage_true} to \ref{fig:gpd_length_correct} report the preceding analyses in the GPD design. Overall patterns are similar to the GEV design. There are two important differences, though. First, in those settings where the feasible (uncorrected) MLE and L-moment-based CIs undercover, they undercover by a similar margin, even in the smallest sample size and tailmost quantile (see Figure \ref{fig:gpd_coverage_est}). This is different from the GEV design, where the L-moment CI undercovers by more than the MLE, especially in the smallest sample size and tailmost quantile. Secondly, we note that, in Figure \ref{fig:gpd_length_correct}, our proposed feasible correction to the L-moment CIs never underperforms the unfeasibly corrected MLE, with the blue solid line always below the red dotted line. In contrast, this happens in one of the six cases considered in the GEV design, though, as we discussed in the preceding section, we expect these differences to disappear once we consider a feasible correction to the MLE CI.

\input{plots/gev/coverage_plots}

\input{plots/gpd/coverage_plots}

\section{Details on selection methods}
\label{app_selection}
\subsection{Higher order expansion of the generalised L-moment estimator}
\label{higher_order_selector}

In this section, we derive a higher order expression for the L-moment estimator in \Cref{properties} of the main text. Our goal is to derive a representation of the estimator as follows:

\begin{equation}
	\label{eq_higher_taylor}
	\sqrt{T}(\hat{\theta}_T - \theta_0) = \Theta_1^T + \frac{\Theta_2^T}{\sqrt{T}} + \frac{\Theta_3^T}{T} + O_{P}(T^{-3/2}) \, ,
\end{equation}
for tight sequences of random variables $\Theta_1^T$, $\Theta_2^T$, $\Theta_3^T$. Under uniform integrability conditions on $\Theta_1^T$, $\Theta_2^T$, $\Theta_3^T$ and the remainder, representation \eqref{eq_higher_taylor} allows us to write:

\begin{equation}
	\label{eq_higher_mse}
	\mathbb{E}[T (\hat{\theta}_T - \theta_0) (\hat{\theta}_T - \theta_0)'] = \mathbb{E}\left[\left(\Theta_1^T + \frac{\Theta_2^T}{\sqrt{T}} + \frac{\Theta_3^T}{T} \right)\left(\Theta_1^T + \frac{\Theta_2^T}{\sqrt{T}} + \frac{\Theta_3^T}{T} \right)'\right] + O(T^{-3/2}) \, ,
\end{equation}
which may be used as a basis for a method of selecting $R$, provided the expectation on the right-hand side is estimable.\footnote{Even if the uniform integrability conditions that allow us to write \eqref{eq_higher_mse} from \eqref{eq_higher_taylor} do not hold, we can posit that our goal is to minimise the MSE of the leading term in \eqref{eq_higher_taylor}. This is the \cite{Nagar1959} style approach of \cite{Rothenberg1984} and \cite{Donald2001}. We return to this point later on.} The idea would be to choose $R$ so as to minimise a linear combination of an estimator of the expectation on the right-hand side. Alternatively, if the goal is to estimate a scalar function of the true parameter, $g(\theta_0)$, one could use the higher-order expansion \eqref{eq_higher_taylor} to construct the higher-order MSE of the estimator $g(\hat{\theta}_T)$. We return to this point in a remark by the end of this section.

To derive representation \eqref{eq_higher_taylor} for the L-moment estimator, we assume, in addition to Assumptions \ref{ass_consistency}-\ref{ass_eigen} in the main text, the following conditions.

\begin{assumption}
	\label{ass_matrix_strong}
	As $T, R \to \infty$, $\mathbb{P}[W_R^{-1} \ \text{and} \ \Omega_R^{-1} \ \text{exist}] \to 1$. We also assume that, as $T, R \to \infty$, $W_R^{-1}  = \Omega_R^{-1} + O_p(T^{-1/2})$.
\end{assumption}

\begin{assumption}
	\label{ass_mvt_strongest}
	$Q_{Y}(u|\theta)$ is five times continuously differentiable on $\mathcal{O}$,for each $u \in [\underline{p},\bar{p}]$. The partial derivatives of $Q_Y(u|\theta)$ with respect to $\theta$, up to the fourth order, are square integrable on $[\underline{p},\overline{p}]$, for each $\theta \in \mathcal{O}$. For ach $i,j,k,l,m \in \{1,2,\ldots, p\}$, the  partial derivatives satisfy $\sup_{\theta \in \mathcal{O}} \sup_{u \in [\underline{p},\bar{p}]}|\frac{\partial^5 Q_{Y}(u|\theta)}{\partial \theta_i \partial \theta_j \partial \theta_k \partial \theta_l \partial \theta_m }| < \infty$.
\end{assumption}

In the next proposition, we use Assumptions  \ref{ass_consistency}-\ref{ass_eigen}, \ref{ass_matrix_strong} and \ref{ass_mvt_strongest} to provide a higher order expansion of the $L$-moment estimator. Our proof strategy mimics that used in \cite{Newey2004} to derive a higher order expression for a GMM estimator with a fixed number of moments, but with additional care to take into account that $R \to \infty$ and the L-moment structure in our setting.\footnote{\cite{Donald2009} consider the higher order expansion of a GMM-type estimator with an increasing number of moment conditions, but their results hold for a special type of moment conditions, which inhibits direct application of their results  to our L-moment setting.}
\begin{proposition}
	\label{PROP_HIGHER_ORDER}
	Suppose Assumptions \ref{ass_consistency}-\ref{ass_eigen}, \ref{ass_matrix_strong} and \ref{ass_mvt_strongest} are satisfied. Then \eqref{eq_higher_taylor} holds for $\hat{\beta} = \begin{pmatrix}\hat{\theta}' & - h^R(\hat{\theta})' W^R  \end{pmatrix}'$ with the objects as follows:
	
	\begin{equation*}
		\begin{aligned}
			\Theta_1^T = - M_0^{-1} \sqrt{T} m(\beta_0) \, , \\
			\Theta_2^T =  M_0^{-1}\sqrt{T}(M-M_0)M_0^{-1}\sqrt{T}m(\beta_0) - \frac{M_0^{-1}}{2}\sum_{j} \left(M_0^{-1}\sqrt{T} m(\beta_0) \right)_j  \partial_{j} M M_0^{-1} \sqrt{T} m(\beta_0) \, , \\
			\Theta_3^T = - M_0^{-1}\sqrt{T}(M-M_0)M_0^{-1}\Theta_2^T - \frac{M_0^{-1}}{2}\sum_{j} (\Theta^1_T)_j  \partial_jM  \Theta^2_T - \frac{M_0^{-1}}{2}\sum_{j} (\Theta^2_T)_j  \partial_jM \Theta^1_T+ \\+ \frac{1}{6} \sum_{i,j} (M_0^{-1} \sqrt{T}m(\beta_0))_i (M_0^{-1} \sqrt{T}m(\beta_0))_j { \partial_{i,j}M }(M_0^{-1} \sqrt{T} m(\beta_0)) \, .
		\end{aligned} 
	\end{equation*}
	where $M_0$ and $m$ are defined in the proof of the theorem.
	\begin{proof}
		See \Cref{proof_prop_higher_order}.
	\end{proof}
	
\end{proposition}

The previous proposition yields a higher-order expansion of the L-moment estimator. Nonetheless, this expansion depends on two quantities whose moments may not be immediately computed: (i) the estimation error of the inverse of the weighting matrix, $(W^R)^{-1} - (\Omega^R)^{-1}$; (ii) moments of the (recentered) L-moment vector, $\sqrt{T}h^R(\theta_0)$. We deal with each term separately.

With regards to the estimation error of the inverse, it is possible to derive an $O_{P^*}(T^{-1})$ expansion of $\sqrt{T}((W^R)^{-1} - (\Omega^R)^{-1})$, which can then be plugged onto \eqref{eq_higher_taylor} to obtain an $O_{P^*}(T^{-3/2})$ expansion in terms of quantities whose moments may be estimated. In particular, if $(\Omega^R)^{-1}$ may be written as a function $M^{R}(\theta_0)$ (e.g. the optimal weights under a Gaussian approximation) and  $(W^R)^{-1} = M^R(\tilde{\theta}_T)$ for a preliminary estimator with representation $\sqrt{T}(\tilde{\theta}_T - \theta_0) = \Pi^1_T + \frac{\Pi^2_T}{\sqrt{T}} + O_p(T^{-1})  $ (e.g. the L-moment estimator with identity weights or the MLE estimator), then the result can be obtained under uniform differentiability conditions on $M^R(\cdot)$. We state these below:

\begin{lemma}
	Suppose $(\Omega^R)^{-1} = M^R(\theta_0) $ and that we estimate it by $(W^R)^{-1} = M^R(\tilde{\theta}_T) $, where $\tilde{\theta}_T$ is a preliminary estimator with representation $\sqrt{T}(\tilde{\theta}_T - \theta_0) = \Pi^1_T + \frac{\Pi^2_T}{\sqrt{T}} + O_p(T^{-1})  $. Suppose that   the entries in $M^R$ are three times continuously differentiable on $\mathcal{O}$. Let $\partial_{s_1, \ldots, s_k} M^R(\theta)$ be the $R \times R$ matrix with entry $(i,j)$ corresponding to the partial derivative $ \partial_{s_1, \ldots s_k}(M^R(\theta))_{i,j}$. Suppose that $\sum_{i=1}^d\lVert \partial_i M^{R}(\theta_0) \rVert_2^2 = O(1)$, $\sum_{i=1}^d\sum_{j=1}^d\lVert \partial_{i,j} M^R(\theta_0) \rVert_2^2 = O(1)$ and $\sup_{\theta \in \mathcal{O}}\sum_{i=1}^d\sum_{j=1}^d\lVert \partial_{i,j} M^{R}(\theta) \rVert_2^2$ $= O(1)$. Then the estimator satisfies:
	
	\begin{equation*}
		\sqrt{T}((W^R)^{-1} - (\Omega^R)^{-1}) = \Xi^1_T + \frac{\Xi^2_T}{\sqrt{T}} + O_p(T^{-1}) \, ,
	\end{equation*}
	where
	
	\begin{equation*}
		\begin{aligned}
			\Xi^1_T  = \sum_{i=1}^d \partial_i M^R(\theta_0) (\Pi_1^T)_i  \, , \\
			\Xi^2_T  = \sum_{i=1}^d \partial_i M^R(\theta_0) (\Pi_2^T)_i + \sum_{i=1}^d \sum_{j=1}^d \partial_{ij} M^R(\theta_0) \left[(\Pi_1^T)_i(\Pi_1^T)_j  \right]\, .
		\end{aligned}
	\end{equation*}
	\begin{proof}
		The proof follows by performing a third order mean value expansion and using the assumptions to show the third derivative term is $O_P(T^{-1})$.
	\end{proof}
\end{lemma}

As for computing moments of $\sqrt{T}h^R(\theta_0)$, one may be tempted to use the strong approximations considered in \Cref{properties} to obtain estimates of these. Nonetheless, we argue this approximation may not be desirable: in particular, it would imply that there is no bias in the estimation of L-moments, whereas it is known that the latter constitutes a large part of the mean squared error of quantile estimators \citep{Wuthrich2021}. To better formalise this notion, we follow \cite{Donald2001}, \cite{Donald2009} and \cite{Okui2009} in defining a \cite{Nagar1959} style approximation to the MSE $M_T\coloneqq \mathbb{E}\left[\left(\Theta_1^T + \frac{\Theta_2^T}{\sqrt{T}} + \frac{\Theta_3^T}{T} \right)\left(\Theta_1^T + \frac{\Theta_2^T}{\sqrt{T}} + \frac{\Theta_3^T}{T} \right)'\right]$ as the sum $\hat{V}_T+ \hat{H}_T$, where $\hat{V}_T$ is the first-order variance of the estimator, $\hat{H}_T$ are higher-order terms, and the approximation errors $\hat{E}_T \coloneqq M_T -(\hat{V}_T + \hat{H}_T) $ and $F_T \coloneqq T (\hat{\theta}_T - \theta_0) (\hat{\theta} - \theta_0) - \left(\Theta_1^T + \frac{\Theta_2^T}{\sqrt{T}} + \frac{\Theta_3^T}{T} \right)\left(\Theta_1^T + \frac{\Theta_2^T}{\sqrt{T}} + \frac{\Theta_3^T}{T} \right)' $ satisfy

\begin{equation}
	\frac{\lVert\hat{E}_T + F_T \rVert_2}{\lVert\hat{H}_T\rVert_2}= o_p(1) \, .
\end{equation}	

Clearly, the Gaussian approximation to the moments of  $\sqrt{T}h^R(\theta_0)$ is not ``\citeauthor{Nagar1959}'', as quantile estimators are generally second-order biased.

In a fully parametric setting and when the data is iid, we may use a parametric bootstrap approach to directly estimate $M_T$. Indeed, given a preliminary estimator $\tilde{\theta}$ of $\theta_0$, we may draw random samples with $T$ observations from $F_{\tilde{\theta}}$ and use these to simulate $\sqrt{T}h^R(\theta_0)$, which can then be used to approximate $\Theta_1$, $\Theta_2$ and $\Theta_3$. One then averages the simulated quadratic form over simulations to estimate $M_T$. Differently from a Gaussian approximation, this approach immediately incorporates higher-order biases, as the simulated approximation to $\sqrt{T}h^R(\theta_0)$ will generally have non-zero mean. Importantly, however, such approach is limited to the iid setting (or, more generally, settings where the sampling mechanism is known). It is not immediately extended to semiparametric settings either, as in these cases the distribution of the data is not fully specified.\footnote{See \Cref{app_extension_res} and \cite{alvarez2024learning} for extensions of the L-moment approach to semiparametric settings.} Given these limitations, it is important to consider alternative approaches to approximating moments of $\sqrt{T}h^R(\theta_0)$.

To circumvent the limitations in the previous paragraph, one could follow the approach used in the weighting matrix and try to find an $O_p(T^{-3/2})$ expansion of $\sqrt{T}h^R(\theta_0)$ in terms of estimable terms, which could then be plugged onto \eqref{eq_higher_taylor} to obtain a ``feasible'' $O_p(T^{-3/2})$ expansion. Higher-order expansions of quantile estimators have long been hindered by the ``non-smoothness'' of the estimation procedure. Recently, \cite{Wuthrich2021} have been able to solve this problem by casting quantile estimation as a particular case of quantile regression \citep{Koenker1978} and relying on empirical process machinery. However, their approach entails a higher-order expansion only up to order $O_P(T^{-3/4})$. If one wishes to compute additional higher-order terms, one could alternatively use the results in \cite{Lee2017,Lee2018}, who rely on \cite{Phillips1991}'s heuristic -- whereby a nonsmooth estimator is assumed to satisfy a first-order condition in terms of a (nonsmooth) subgradient, which then allows a Taylor expansion in terms of the Dirac delta function --, to obtain higher-order expansions of quantile estimators. However, as pointed out by \cite{Wuthrich2021}, \citeauthor{Phillips1991}'s heuristic does not account for $O_P(T^{-1/2})$ terms stemming from the nonunicity of quantile estimators, and these terms can have nonnegligible effects on the higher-order bias of the estimator.

\begin{remark}[Higher-order expansions for other scalar quantities] \label{remark_higher_scalar} Suppose we want to estimate a scalar function $g_T(\theta_0)$ of the true parameter, where we allow the function to vary with sample size. Suppose each $g_t$, $t \in \mathbb{N}$, is four times continuously differentiable; and define the rate $ \sup_{\theta \in \Theta} \sum_{i=1}^p\sum_{j=1}^p\lVert \partial_{i,j}\nabla_{\theta \theta'}g_t(\theta)\rVert_2= O(\xi_T)$. A fourth order mean-value expansion then yields:
	\begin{equation}
		\label{eq_expansion_function}
		\begin{aligned}
			\sqrt{T} (g_T(\hat{\theta}_T) - g_T(\theta_0)) = \nabla_{\theta'}g_T(\theta_0) \sqrt{T}(\hat{\theta}_T - \theta_0) +  \sqrt{T} (\hat{\theta}_T - \theta_0)' \nabla_{\theta \theta'} g_T(\theta_0) (\hat{\theta}_T - \theta_0) + \\ \sum_{i=1}^p \sqrt{T}(\hat{\theta}_{i,T} - \theta_{i,0})(\hat{\theta}_T - \theta_0) \partial_i\nabla_{\theta \theta'} g_T(\theta_0) (\hat{\theta}_T - \theta_0) +\\ \sum_{i=1}^p \sum_{j=1}^p \sqrt{T}(\hat{\theta}_{i,T} - \theta_{i,0})(\hat{\theta}_{j,T} - \theta_{j,0})(\hat{\theta}_T - \theta_0) \partial_{ij}\nabla_{\theta \theta'} g_T(\tilde{\theta}_T) (\hat{\theta}_T - \theta_0) \, ,
		\end{aligned}
	\end{equation}
	where $\tilde{\theta}_T$ lies in the line segment between $\theta_0$ and $\hat{\theta}_T$. Under uniform integrability of $\sqrt{T}(\hat{\theta}_T - \theta_0)$, it follows that:
	
	\begin{equation*}
		\begin{aligned}
			\mathbb{E}[T (g_T(\hat{\theta}_T) - g_T(\theta_0))^2] =  \mathbb{E}\Big[\Big(\nabla_{\theta'}g_T(\theta_0) \sqrt{T}(\hat{\theta}_T - \theta_0) + \sqrt{T} (\hat{\theta}_T - \theta_0)' \nabla_{\theta \theta'} g_T(\theta_0) (\hat{\theta}_T - \theta_0) + \\ \sum_{i=1}^p \sqrt{T}(\hat{\theta}_{i,T} - \theta_{i,0})(\hat{\theta}_T - \theta_0) \partial_i\nabla_{\theta \theta'} g_T(\theta_0) (\hat{\theta}_T - \theta_0)\Big)^2\Big] + O\left((\psi_T)\cdot(\xi_TT^{-3/2})\right) \, ,
		\end{aligned}
	\end{equation*}
	where $\psi_T$ is the order of the sum of the first three terms in the mean-value expansion \eqref{eq_expansion_function}. We can then plug \eqref{eq_higher_mse} on the above to obtain an expansion in terms of estimable terms. Useful sequences of $g_T$ would be $g_T(\theta) = Q_Y(u_T|\theta)$ (in quantile estimation) or $g_T(\theta) = F_\theta(p_T)$ (in probability estimation).
\end{remark}
\subsection{A Lasso-based alternative}
\label{lasso_selector}
In this section, we briefly review the selection method proposed by \cite{Luo2016} in the GMM context. As discussed in the main text, our generalised $L$-moment estimator can be seen as combining the $R$ moments used in estimation into $d$ linear restrictions via the mapping:

\begin{equation}
	\hat{A}_R h_R(\hat{\theta}) = 0 \, ,
\end{equation}
where the combination matrix is estimated as:

\begin{equation}
	\label{est_combination}
	\hat{A}_R = \nabla_{\theta'} h^R(\hat{\theta})' W^R \, .
\end{equation}

The poor behaviour of the L-moment estimator with large $R$ may be partly attributed to the estimation of \eqref{est_combination}. Indeed, we note that the term $\Theta^2_T/\sqrt{T}$ in the higher order expansion of Proposition \ref{PROP_HIGHER_ORDER} is closely related to the estimation error of $\Omega^R$ and $\nabla_{\theta'} h^R(\theta_0)$; and correlation between the estimation error of these quantities with $h^R(\theta_0)$ affects the bias due to this term.\footnote{Notice that correlation between $\nabla_{\theta'} h^R(\hat{\theta})$ and $h^R(\theta_0)$ is due solely to estimation error of $\hat{\theta}$, since the Jacobian $\nabla_{\theta'} h^R(\theta_0)$ is nonstochastic in our setting. This is a more general feature of minimum-distance-style estimators, and stands in contrast with GMM estimators where the Jacobian of the empirical moment condition at the truth is random, which possess an additional bias term due to this additional source of correlation \citep{Newey2004}.} Suppose $\Xi = (W^R)^{-1}$ exists (as in an estimator of the optimal weighting matrix). Instead of estimating $A_R = \nabla_{\theta'} h^R(\theta_0)' \Omega^R$ by $\hat{A}_R$, \cite{Luo2016} propose to estimate the $j$-th row of $A_l$ as (adapting their program to our context):

\begin{equation}
	\label{eq_alg_luo}
	\tilde{\lambda}_j \in \operatorname{argmin}_{\lambda \in \mathbb{R}^R} \frac{1}{2}\lambda'\Xi \lambda - \lambda' \nabla_{\theta'} h^R(\tilde{\theta})e_j + \frac{k}{T}\sum_{l=1}^R \nu_l^j  \cdot |\lambda_l| \, ,
\end{equation}
for penalties $k \geq 0$,$\nu_l^j \geq 0$, $l=1,\ldots R$; and where $\tilde{\theta}$ is a preliminary estimator and $e_j$ is a $d\times 1$ vector with one in the $j$-th entry and zero elsewhere. Observe that, when the penalties are set to zero, the solution is $\hat{\lambda}_j = \Xi^{-1}\nabla_{\theta'} h^R(\tilde{\theta})e_j$, which coincides with the $j$-th row of $\hat{A}_R$. In general, however, the penalties will induce sparsity on the estimated rows, so only a few entries are selected. Importantly, \eqref{eq_alg_luo} can be efficiently estimated by quadratic programming algorithms.\footnote{For example, the \texttt{quadprog} package in \texttt{R} \citep{Berwin2011}.} The problem is also well-defined even if $\Xi$, but not $\Xi^{-1}$, exists.

Once the $d$ rows of $A_R$ are estimated, we can stack them onto $\tilde{A}_R = \begin{bmatrix}
	\hat{\lambda}_1 &
	\hat{\lambda}_2 & \ldots & \hat{\lambda}_d
\end{bmatrix}'
$ and estimate $\theta_0$ by solving:
\begin{equation}
	\label{eq_lasso_pos}
	\tilde{A}_R h^R(\check{\theta})  =0 \, .
\end{equation}

Alternatively, we may adopt a ``post-Lasso'' procedure, which is known to reduce regularisation bias \citep{Belloni2012}. In our setting, this amounts to running our two-step  L-moment estimator using as targets those moments selected by matrix $\tilde{A}_R$, i.e. we use the moments given by the indices $\mathcal{I}_S = \{l \in \{1,2,\ldots, R\}: \hat{\lambda}_{ij} = 0 \text{ for  some } j =1 \ldots d \}$. 

In their paper, \cite{Luo2016} provide theoretical guarantees that, in the GMM context with iid data, if $\Xi$ is the inverse of the optimal weighting matrix and the true combination matrix $A_R$ is \textit{approximately sparse} -- in the sense that it is well approximated by a sparse matrix at a rate --, then the estimator based on \eqref{eq_lasso_pos} is asymptotically efficient under some additional conditions and as $T,R\to \infty$. Their result can be adapted to our L-moment context -- an extension we pursue in Supplemental Appendix \Cref{lasso_lmoments}. In what follows, we contrast the Lasso approach with the higher-order MSE method in a Monte Carlo exercise.

\subsection{Monte Carlo Exercise}
\label{monte_carlo_select}

We return to the Monte Carlo exercise in \Cref{properties}. We consider the behaviour of four estimators: (i) the L-moment estimator with identity weights and $R = d$ (\textbf{FS}); (ii) the two-step generalised L-moment estimator with $R$ selected in order to minimise the approximate RMSE of the quantile one wishes to estimate (\textbf{TS RMSE}); (iii) the generalised L-moment estimator with optimal weights and Lasso selection (\textbf{TS Lasso}); and (iv) the post-Lasso estimator that runs the two-step generalised estimator using only the moments selected in the Lasso procedure (\textbf{TS Post-Lasso}). For conciseness, we only consider estimators based on the ``càglàd'' L-moment estimator \eqref{eq_est_lmoments_quantile}. As in \Cref{properties} of the main text, we compare the root-mean-squared error (RMSE) of each approach with that obtained from a MLE plug-in.

A few details with regards to the methods used are in order. First, in method (ii), the approximate RMSE of the target quantile is computed using a parametric bootstrap and the expansion in \Cref{remark_higher_scalar}, \textbf{up to second order}.  We do so by considering a third order mean-value expansion of the target quantile (i.e. we discard the third order terms in \eqref{eq_expansion_function} when estimating the RMSE), and by working with the expansion \eqref{eq_higher_taylor} of the L-moment estimator up to order $T^{-1}$. See Algorithm \ref{alg:rmse} for the pseudo-code of our resulting approach and the script \texttt{selection.R} in the accompanying Github reposition for a computational implementation in \texttt{R}. We discard third-order terms from the expansion for computational reasons -- doing so allows us to compute the required derivatives inexpensively using closed-form expressions for Jacobians and Hessians involved in these expressions.\footnote{For those distributions for which an applied user does not have the required Jacobians and Hessians in closed form, our computational implementation computes the required derivatives using differentiation routines available in the R package \texttt{autodiffr}, which serves as a wrapper to Julia routines that compute Jacobians and Hessians efficiently using automatic differentiation.}$^,$\footnote{While it is certainly possible to use symbolic differentiation (e.g. Mathematica routines) to obtain closed-form expressions for third-order derivatives for the GEV and GPD families, the evaluation  of such derivatives, as well as the computation of the cross-products involved in the higher-order term $\Theta^3_T$,  start to become computationally prohibitive as we consider larger candidate values of $R$.} As can be seen from the formulae in \Cref{remark_higher_scalar}, if $\hat{\theta}-\theta_0$ is (approximately) symmetrically distributed, then dropping third order terms should not affect the bias term that composes the MSE estimate, though it could change the estimated RMSE by changing higher-order variance-related terms. In such settings, one would thus expect the RMSE formula that ignores third order terms to skew selection towards lower bias-inducing choices of $R$. Given the nonlinear nature of the problem, one expects such choices to lead to smaller values of $R$ than including further higher-order terms would. In future research, it would be interesting to assess whether there are gains in including third-order terms in the estimated higher-order MSE. In our Monte Carlo simulations, we run our selection approach with $R$ ranging from $d$ to $T \land 100$.

As for the Lasso-based approaches, we follow the recommendations in \cite{Belloni2012} and \cite{Luo2016} when setting the penalty $k$ and the moment-specific loadings $v_l^j$. Specifically the penalty $k$ follows the rule in \cite{Luo2016}. In setting  the penalty-specific loading, we observe that \Cref{assumption_penalties} in \Cref{lasso_lmoments} requires that, for each program $j = 1 \ldots d$ and variable $l=1,\ldots R$, the penalty $v^j_l k/T$ dominate, with high probability, the derivative with respect to the $l$-th variable in the optimisation, evaluated at the target sparse approximation. Following \cite{Luo2016}, we ensure this by setting $v^j_l$ equal to an estimate of an upper bound to the standard error of the gradient at the sparse approximation. We estimate this bound by first using the Delta-method to compute an approximate variance to the entries of the matrices $\Xi$ and $\nabla_{\theta'}h^R(\tilde{\theta})$ that enter the quadratic program in equation \eqref{eq_alg_luo}.\footnote{Notice that, per the discussion in Section \ref{lasso_selector}, $\Xi$ is an estimator of the matrix whose (generalised) inverse corresponds to the optimal weighting scheme. In our application, we compute this estimator and the gradient $\nabla_{\theta'}h^R(\tilde{\theta})$ by taking $\tilde{\theta}$ as the FS estimator. } We then use these variance estimates in the binomial search Algorithm 1 of \cite{Luo2016} to find an upper bound to the standard error of the gradient at the sparse approximation. Our penalty is ``coarse'', in the sense that we do not refine the upper bound by using the results of a previous Lasso estimator and then iterating this formula. Given the findings in \cite{Belloni2012} and \cite{Luo2016}, one would expect that such refinements would lead to less stringent regularisation, though we leave the design of a proper refinement algorithm for future research. Importantly, for each program $j$, we modify the loadings $v^j_l$ to be equal to zero for $l=1,2,\ldots, d$, i.e. we do not regularise the first $d$ L-moments, so they are effectively ``always included'' in the second step estimation. For the Lasso estimators, we consider a maximum number of allowed L-moments (the number $R$ in \eqref{eq_alg_luo}) as $2 (T\land 100)$.

Tables \ref{gev_table_select} and \ref{gpd_table_select} replicate the results from Tables \ref{gev_table_select_summary} and \ref{gpd_table_select_summary} in the main text, but including the TS Lasso procedure. We note that the TS Lasso tends to perform poorly. This is due to the large regularisation bias imparted by the ``coarse'' penalty, which tends to dominate the RMSE. The Post-Lasso approach is able to attenuate such bias and produce estimators with desirable properties; still, in smaller samples, the relative performance of this approach vis-à-vis TS RMSE can be unfavourable.
\input{tables/rmse_algorithm}

	\input{tables/gev/gev_table_select}

	\input{tables/gpd/gpd_table_select}

\subsection{Proof of Proposition \ref{PROP_HIGHER_ORDER}}
\label{proof_prop_higher_order}

\begin{proof}
	On $\hat{\theta}_T \in \mathcal{O}$ and existence of $(W^R)^{-1}$ and $(\Omega^R)^{-1}$, the estimator satisfies the following first order condition:
	
	\begin{equation*}
		\nabla_{\theta'} h^R(\hat{\theta}) 'W^R h^R(\hat{\theta}) = 0 \, ,
	\end{equation*}
	which may be written as \citep{Newey2004}:
	
	\begin{equation*}
		\begin{pmatrix}
			-\nabla_{\theta'} h^R(\hat{\theta})'\hat{\lambda} \\ 
			- h^R(\hat{\theta})- (W^R)^{-1}\hat{\lambda}\end{pmatrix} = 0 \, ,
	\end{equation*}
	where, by \Cref{prop_asymptotic_linear}, $\hat{\theta} - \theta_0 = O_p(T^{-1/2})$ and:
	
	\begin{equation*}
		\lVert\hat{\lambda} \rVert_2 \leq \lVert W^R \rVert_2 \lVert (\hat{Q}_Y(\cdot) - Q_Y(\cdot|\hat{\theta})) \mathbbm{1}_{[\underline{p},\bar{p}]} \rVert_{L^2[0,1]} \, ,
	\end{equation*}
	implying that $\lVert\hat{\lambda} \rVert_2 = O_p(T^{-1/2})$.
	
	Put  $\beta \coloneqq (\theta',\lambda')'$. Let:
	\begin{equation*}
		m(\beta) \coloneqq \begin{pmatrix}	-\nabla_{\theta'} h^R(\theta)'{\lambda} \\ 
			- h^R(\theta)- (W^R)^{-1}\lambda\end{pmatrix} \, .
	\end{equation*}
	
	The estimator solves $m(\hat{\beta})=0$. Let $\lambda_0  \coloneqq 0_{R \times 1}$ and $\beta_0 \coloneqq (\theta_0',\lambda_0')'$. 	On $\hat{\theta}_T \in \mathcal{O}$ and existence of $(W^R)^{-1}$ and $(\Omega^R)^{-1}$,  a fourth order mean-value expansion of $\hat{\beta}$ around $\beta_0$ yields:
	
	\begin{equation*}
		\begin{aligned}
			0 = m(\hat{\beta}) = m(\beta_0) + M (\hat{\beta} - \beta_0) + \frac{1}{2}\sum_{j} (\hat{\beta}_j - \beta_{j0})  \partial_j M(\hat{\beta} - \beta_0) + \\
			+ \frac{1}{6} \sum_{i,j}(\hat{\beta}_i - \beta_{i0}) (\hat{\beta}_j - \beta_{j0}) { \partial_{i,j}M }(\hat{\beta} - \beta_0) + \frac{1}{24} \sum_{g,i,j}(\hat{\beta}_g - \beta_{g0}) (\hat{\beta}_i - \beta_{i0}) (\hat{\beta}_j - \beta_{j0})  {  {\widetilde{\partial_{g,i,j}M} } }(\hat{\beta} - \beta_0) \, ,
		\end{aligned}
	\end{equation*}
	where $M = \nabla_{\beta'} m(\beta_0)$ and $\partial_j M$ is the $ (d+R) \times (d+R)$ matrix with entry $(l,k)$ equal to $\frac{\partial m^l(\beta_0)}{\partial \beta_j \partial \beta_k}$. Similarly, ${\partial_{i,j}M}$ is a $ (d+R) \times (d+R)$  matrix with entry $(l,k)$ equal to $\frac{\partial m^l(\theta_0)}{\partial \beta_{i} \beta_{j} \partial \beta_k}$; and  $\widetilde{\partial_{g,i,j}M}$ is a $ (d+R) \times (d+R)$  matrix with the fourth order partial derivatives evaluated at $u$-specific $\tilde{\theta}(u)$ in the line segment between hat $\theta_0$ and $\hat{\theta}$.
	
	Next, we observe that:
	
	\begin{equation*}
		M = \begin{pmatrix}
			0 & - \nabla_{ \theta'} h^R(\theta_0)'\\
			- \nabla_{ \theta'} h^R(\theta_0) & - (W^R)^{-1}
		\end{pmatrix} \, .
	\end{equation*}
	
	Letting: 
	
	\begin{equation*}
		M_0 \coloneqq \begin{pmatrix}
			0 & - \nabla_{ \theta'} h^R(\theta_0)'\\
			- \nabla_{ \theta'} h^R(\theta_0) & - (\Omega^R)^{-1}
		\end{pmatrix} \, .
	\end{equation*}
	
	Then by \Cref{ass_matrix_strong}, $\lVert M - M_0 \rVert_2 = O_p(T^{-1/2})$. Moreover, note that:
	
	\begin{equation*}
		M_0^{-1} = \begin{pmatrix}
			\Omega^*_R & -\Omega_R^* \nabla_{ \theta'} h^R(\theta_0)' \Omega_R \\
			- \Omega_R  \nabla_{ \theta'} h^R(\theta_0)\Omega_R^*  & -\Omega_R +  \Omega_R  \nabla_{ \theta'} h^R(\theta_0) \Omega_R^* \nabla_{ \theta'} h^R(\theta_0)' \Omega_R	\end{pmatrix} \, ,
	\end{equation*}
	where $\Omega^*_R = (\nabla_{ \theta'} h^R(\theta_0)' \Omega_R  \nabla_{ \theta'} h^R(\theta_0))^{-1}$, which exists by \Cref{ass_eigen}. Rearranging, we have:
	
	\begin{equation}
		\begin{aligned}
			\label{eq_expansion_term}
			(\hat{\beta} - \beta_0) = - M_0^{-1}m(\beta_0) - M_0^{-1}(M - M_0)(\hat{\beta}-\beta_0) - \frac{M_0^{-1}}{2}\sum_{j} (\hat{\beta}_j - \beta_{j0})  \partial_j M(\hat{\beta} - \beta_0) - \\ - \frac{M_0^{-1}}{6} \sum_{i,j}(\hat{\beta}_i - \beta_{i0}) (\hat{\beta}_j - \beta_{j0}) {  {{\partial_{i,j}M} }}(\hat{\beta} - \beta_0) - \\ - \frac{M_0^{-1}}{24} \sum_{g,i,j}(\hat{\beta}_g - \beta_{g0}) (\hat{\beta}_i - \beta_{i0}) (\hat{\beta}_j - \beta_{j0}){  {{\partial_{g,i,j}M} } }(\hat{\beta} - \beta_0) - \\
			- \frac{M_0^{-1}}{24} \sum_{g,i,j}(\hat{\beta}_g - \beta_{g0}) (\hat{\beta}_i - \beta_{i0}) (\hat{\beta}_j - \beta_{j0})  \left[  \widetilde{\partial_{g,i,j}M}  - \partial_{g,i,j}M\right](\hat{\beta} - \beta_0) \, .
		\end{aligned}
	\end{equation}
	
	Our first goal is to show that, on $\hat{\theta}_T \in \mathcal{O}$ and existence of $(W^R)^{-1}$ and $(\Omega^R)^{-1}$,  $(\hat{\beta}- \beta_0) = -M_0^{-1} m(\beta_0) + O_p(T^{-1})$. We split the proof in several steps. First, note that $\lVert M_0^{-1} \rVert_2 = O(1)$. Indeed, for a block-matrix, it follows by the properties of the operator norm that:
	
	\begin{equation*}
		\left\lVert \begin{bmatrix}A& B \\ C & D\end{bmatrix}\right\rVert_2 \leq \lVert A \rVert_2  + \lVert B \rVert_2 + \lVert C \rVert_2 + \lVert D \rVert_2 \, .
	\end{equation*}
	
	Then, since $\lVert \Omega_R^*\rVert_2 = O(1)$ (\Cref{ass_eigen}), $\Omega_R = O(1)$  (\Cref{ass_weights}), $\lVert \nabla_{\theta'} h^R(\theta_0) \rVert_2^2 \leq \operatorname{tr}( \nabla_{\theta'}  h^R(\theta_0)' \nabla_{\theta'}  h^R(\theta_0) ) \leq \sum_{s=1}^p\int_{\underline{p}}^{\bar{p}} |\partial_{\theta_s} Q_Y(u|\theta_0)|^2 du < \infty$ (\Cref{ass_mvt_weak}), it follows that $\lVert M_0^{-1}\rVert_2 = O(1)$.
	
	Next, we claim that, except for the first term, all terms on the right-hand side of \eqref{eq_expansion_term} are $O_p(T^{-1})$. Clearly, $\lVert (M-M_0)(\hat{\beta}-\beta_0) \rVert = O_p(T^{-1})$   As for the third term, one needs to characterize $\partial_j M$. For $j \leq d$, we get:
	
	\begin{equation*}
		\partial_j M = \begin{pmatrix}
			0 & - \partial_j \nabla_{\theta'} h^R(\theta_0)' \\
			- \partial_j \nabla_{\theta'} h^R(\theta_0) & 0
		\end{pmatrix} \, ,
	\end{equation*}
	whereas, for $j \geq d+1$:
	
	\begin{equation*}
		\partial_j M = \begin{pmatrix}
			-\nabla_{\theta \theta'} h_{j-d}(\theta_0)  & 0 \\
			0 & 0
		\end{pmatrix} \, .
	\end{equation*}
	
	This implies that:
	
	\begin{equation*}
		\begin{aligned}
			\lVert\sum_{j} (\hat{\beta}_j - \beta_{j0})  \partial_j M(\hat{\beta} - \beta_0)\rVert_2 \leq \\ \leq \lVert \hat{\beta} - \beta_0 \rVert_2 \cdot \sum_{j=1}^{d+R} \lvert \hat{\beta}_j - \beta_{j0} \rvert \cdot \lVert\partial_j M_j \rVert_2  \leq \lVert \hat{\beta} - \beta_0 \rVert_2^2 \sqrt{\sum_{j=1}^{d+R}  \lVert\partial_j M_j \rVert_2^2} = O_p(T^{-1}) \, ,
		\end{aligned}
	\end{equation*}
	where we used that $ \sum_{j=1}^{p+R}  \lVert\partial_j M_j \rVert_2^2= O(1)$, which follows from Bessel's inequality and Assumption \Cref{ass_mvt_strongest}.
	
	Next, we characterize $\partial_{ij} M$. For $1 \leq i, j \leq d$:
	
	\begin{equation*}
		\partial_{ij} M = \begin{pmatrix}
			0 & - \partial_{ij} \nabla_{\theta'} h^R(\theta_0)' \\
			- \partial_{ij} \nabla_{\theta'} h^R(\theta_0) & 0
		\end{pmatrix} \, ,
	\end{equation*}
	whilst, for $i \leq d$ and $j \geq d+1$:
	\begin{equation*}
		\partial_{ij} M = \begin{pmatrix}
			- \partial_i \nabla_{\theta \theta'} h_{j-d}(\theta_0) & 0 \\
			0 & 0
		\end{pmatrix}
	\end{equation*}
	and, finally, for $i,j \geq d+1$:
	\begin{equation*}
		\partial_{ij} M = \begin{pmatrix}
			0 & 0 \\
			0 & 0
		\end{pmatrix} \, ,
	\end{equation*}
	which, when using Bessel's inequality and \Cref{ass_matrix_strong}, implies that:
	
	\begin{equation*}
		\lVert\sum_{i,j}(\hat{\beta}_i - \beta_{i0}) (\hat{\beta}_j - \beta_{j0}) {  {{\partial_{i,j}M} }}(\hat{\beta} - \beta_0) \rVert_2 \leq \lVert \hat{\beta}-\beta_0 \rVert_2^3 \cdot \sqrt{\sum_{i,j} \lVert \partial_{i,j}M\rVert_2^2} = O_p(T^{-3/2}) \, .
	\end{equation*}
	
	By a similar argument, we can show that the fifth term, which involves fourth order derivatives, is $O_p(T^{-2})$. Finally, we can use the last part of \Cref{ass_mvt_strongest} in a similar way as in the proof of \Cref{prop_asymptotic_linear} to show that the last term is $O_p(T^{-2})$.
	
	Next, using the $O_p(T^{-1})$ representation of $\hat{\beta}-\beta_0$, we get that, on $\hat{\theta}_T \in \mathcal{O}$ and existence of $(W^R)^{-1}$ and $(\Omega^R)^{-1}$:
	
	\begin{equation*}
		\hat{\beta} - \beta_0 = -M_0^{-1} m(\beta_0)  + M_0^{-1}(M-M_0)M_0^{-1}m(\beta_0) - \frac{M_0^{-1}}{2}\sum_{j} \left(M_0^{-1}m(\beta_0) \right)_j  \partial_{j} M M_0^{-1}m(\beta_0) +O_p(T^{-3/2}) \, .
	\end{equation*} 
	
	Plugging this expression back onto \eqref{eq_expansion_term} and disconsidering terms that are $O_p(T^{-2})$ allows us to define $\Theta_1^{T}$, $\Theta_2^{T}$ and $\Theta_3^{T}$ as in \eqref{eq_higher_taylor}. To conclude, we must show that focusing on the event that the inverse exists and $\hat{\theta}_T \in \mathcal{O}$ does not change the rates we have obtained. In particular, note that we have already shown that:
	
	\begin{equation*}
		\sqrt{T}(\hat{\theta}- \theta_0) =  \mathbbm{1}_{I_T}\left[ \Theta_1^T + \frac{\Theta_2^T}{\sqrt{T}} + \frac{\Theta_3^T}{T}  + O_{P^*}(T^{-3/2}) \right]+ \mathbbm{1}_{I_T^\complement} \sqrt{T}(\hat{\theta}_T - \theta_0) \, ,
	\end{equation*} 
	where $I_T$ is the event that  $\hat{\theta}_T \in \mathcal{O}$ and that $(W^R)^{-1}$ and $(\Omega^R)^{-1}$ exist. By \Cref{ass_matrix_strong} and $\theta_0 \in \mathcal{O}$, we know that $\mathbbm{1}_{I_T} \overset{p}{\to} 1 $ and $\mathbbm{1}_{I_T^\complement} \sqrt{T}(\hat{\theta}_T - \theta_0) = o_p(1)$.\footnote{In a similar vein, we have implicitly used that $\mathbbm{1}_{\hat{\theta}_T \in \mathcal{O}} \overset{p}{\to} 1 $ and $\mathbbm{1}_{\hat{\theta}_T \notin \mathcal{O}} \sqrt{T}(\hat{\theta}_T - \theta_0) = o_p(1)$ in the proof of the linear representation of \Cref{prop_asymptotic_linear}.} To show the rates derived from \eqref{eq_expansion_term} are not affected, we show that $\mathbbm{1}_{I_T^\complement} = o_p(T^{-3/2})$. Indeed, fix $\epsilon > 0$ and note that there exists $T^* \in \mathbb{N}$ such that, for $T \geq T^*$:
	
	\begin{equation*}
		T^{3/2}\mathbbm{1}_{I_T^\complement} > \epsilon \iff \mathbbm{1}_{I_T^\complement} = 1 \, ,
	\end{equation*}
	but $\mathbb{P}[I_T^\complement] \to 0$, which proves the desired result. Using that $\mathbbm{1}_{I_T^\complement} = o_p(T^{-3/2})$, we can write 
	
	\begin{equation*}
		\begin{aligned}
			\sqrt{T}(\hat{\theta}- \theta_0) =   \Theta_1^T + \frac{\Theta_2^T}{\sqrt{T}} + \frac{\Theta_3^T}{T}  + O_{P^*}(T^{-3/2})+ \\ \mathbbm{1}_{I_T^\complement} \sqrt{T}(\hat{\theta}_T - \theta_0) - \mathbbm{1}_{I_T^\complement}\left[ \Theta_1^T + \frac{\Theta_2^T}{\sqrt{T}} + \frac{\Theta_3^T}{T}  + O_{P^*}(T^{-3/2})\right] =   \Theta_1^T + \frac{\Theta_2^T}{\sqrt{T}} + \frac{\Theta_3^T}{T}  + O_{P^*}(T^{-3/2}) \, , 
		\end{aligned}
	\end{equation*} 
	which proves the result.
\end{proof}

\subsection{Conditions for validity of Lasso approach}
\label{lasso_lmoments}
This appendix presents sufficient conditions for the validity of the Lasso approach outlined in the main text and detailed in Appendix \ref{lasso_selector}. We adapt the conditions in \cite{Luo2016} to our setting. In addition to the assumptions in the main text, we require sparse eigenvalue conditions that enable invertibility (and bounded spectral norm) of ``small'' submatrices of $\Xi$; and approximate sparsity of the combination matrix $A_R$. We state these assumptions below. In what follows, define, for $v \in \mathbbm{R}^n$, $\lVert v \rVert_0  \coloneqq \# \{j: v_j \neq 0\}$.

\begin{assumption}[Approximate sparsity of combination matrix]
	\label{ass_app_sparsity}
	For $j=1,\ldots, d$, let $\lambda^*_j \coloneqq A_R' e_j$. We assume that, for each $j$, there exist constants $K^l_j$, $K^u_j$ and a sequence of vectors $\bar{\lambda}_d \in \mathbbm{R}^R$ such that, as $T,R\to \infty$:
	\begin{enumerate}
		\item  $\lVert \bar{\lambda}_j \rVert_0 = s_T $
		\item $\lVert \bar{A}_R - A_R\rVert_2 = o\left(1\right)$, where $\bar{A}_R = \begin{bmatrix}
			\bar{\lambda}_1 & \bar{\lambda}_2^{*} & \ldots & \bar{\lambda}_{d}^{*}
		\end{bmatrix}'$.
	\end{enumerate}
\end{assumption}

\begin{assumption}[Sparse eigenvalue and spectral norm condition]
	Let:
	\begin{equation*}
		\begin{aligned}
			\kappa(s,\Xi) = \min_{\delta \in \mathbb{R}^R: \lVert \delta \rVert_0 \leq s, \lVert \delta \rVert_2 = 1  } \delta' \Xi \delta \, , \\ 
			\phi(s,\Xi) = \max_{\delta \in \mathbb{R}^R: \lVert \delta \rVert_0 \leq s, \lVert \delta \rVert_2 = 1  } \delta' \Xi \delta \, .
		\end{aligned}
	\end{equation*}
	We assume there exist constants $0 < \kappa_1 \leq \kappa_2$ such that:
	\begin{equation*}
		\lim_{T,R \to \infty} \mathbb{P}[\kappa_1 \leq \kappa(s_T \log(T),\Xi) \leq \phi(s_T \log(T),\Xi) \leq \kappa_2] = 1 \, .
	\end{equation*}
\end{assumption}
The last assumption restricts the penalties $k$ and $\nu^j_i$, $i=1,\ldots R$, $j=1,\ldots, d$. In particular, we require these penalties to be sufficiently harsh so as to dominate the gradient $\hat{S}_j(\lambda) = \Xi \lambda - \nabla_{\theta'} h^R(\tilde{\theta}) e_j$ of the unpenalised objective function, evaluated at the sparse approximation $\bar{\lambda}_j$.
\begin{assumption}[Penalties]
	\label{assumption_penalties}
	The penalties satisfy:
	\begin{enumerate}
		\item For a sequence $\alpha_T$ converging to zero such that $\alpha_T R \to \infty$:
		\begin{equation*}
			\mathbbm{P}\left[\max_{j=1,\ldots, d} \max_{i=1,\ldots R} \left|(S_j(\bar{\lambda}_j))_i/\nu^j_i\right|\leq \frac{k}{T} \right] \geq 1 - \alpha_T \, ,
		\end{equation*} 
		where $k =(1+\epsilon)\sqrt{T\Phi^{-1}(1-\frac{\alpha_T}{4Rd})}$ for some $\epsilon > 0$, and where $\Phi$ denotes the cdf of a normal distribution.
		\item There exist constants $a>0$ and $b < \infty$, such that:
		\begin{equation*}
			\lim \mathbb{P}\left[a \leq \min_{j=1,\ldots d}\min_{i=1,\ldots, R} \nu^j_i \leq  \max_{j=1,\ldots d}\max_{i=1,\ldots, R} \nu^j_i \leq b\right] = 1 \, .
		\end{equation*}
	\end{enumerate}
\end{assumption}
Under Assumptions \ref{ass_app_sparsity}-\ref{assumption_penalties}, it follows, by application of Lemma 26 in \cite{Luo2016}, that there exists a constant $K_\lambda$ and a sequence of $\epsilon_T$ converging to zero, such that, with probability at least $1 - \epsilon_T$

\begin{equation}
	\label{eq_bound_sparse}
	\max_{j=1,\ldots, d}\lVert \bar{\lambda}_j - \hat{\lambda}_j \rVert_1 \leq K_\lambda \sqrt{\frac{s^2_T \log(\frac{R d}{\alpha_T})}{T}} \, ,
\end{equation}
where $\hat{\lambda}_j$ denotes the solution to program \eqref{eq_alg_luo}. 

The bound in \eqref{eq_bound_sparse}, together with Assumption 12, implies that:

\begin{equation*}
	\lVert \tilde{A}_R - A_R \rVert_2 \leq \lVert \tilde{A}_R - A_R \rVert_F  +\lVert \bar{A}_R - A_R \rVert_2 = O_p\left(\sqrt{\frac{s^2_T \log(\frac{R d}{\alpha_T})}{T}} \right) \, .
\end{equation*}

If we assume that $\frac{s_T^2 \log(R)}{T} \to 0$, then $\lVert \tilde{A}_R - A_R \rVert_2 = o_p(1)$, and the Lasso approach consistently estimates the combination matrix.  We can then derive the properties of the Lasso-based estimator in a similar vein as to Propositions \ref{prop_consistency} and \ref{prop_asymptotic_linear}  in \Cref{properties}. To see this, we observe that the estimator $\hat{\theta}^{\text{selected}}$ solves $S(\hat{\theta}^{\text{selected}}) = 0$, where:

\begin{equation*}
	S(\theta) = \tilde{A}_S h^R(\theta) \, .
\end{equation*}

If we define the population objective as $S_0(\theta)  = A_R [\int_{\underline{p}}^{\overline{p}} (\hat{Q}_Y(u) - Q_Y(u|\theta)\boldsymbol{P}^R(u)]du$, then we can proceed as in the consistency proof of Proposition \ref{prop_consistency}. Similarly, we can proceed as in the proof of Proposition \ref{prop_asymptotic_linear} to obtain an asymptotic linear representation of the estimator.

\section{Details on extensions}
\subsection{``Residual'' analysis of semi- and nonparametric models}
\label{app_extension_res}

Consider the model for a scalar outcome $Y$ outlined in the main text:

\begin{equation}
	\begin{aligned}
		Y = h(\epsilon, X; \gamma_0), \quad \gamma \in \Gamma \subseteq \mathcal{B}\, , \\
		\epsilon \sim F_{\theta_0}, \quad \theta_0 \in \Theta \subseteq \mathbb{R}^d \, ,
	\end{aligned}
\end{equation}
where we assume that the researcher has access to a first-step nonparametric estimator of $\gamma_0$; and that the scalar $\epsilon$ follows a continuous distribution function with true density $f_\epsilon$. We assume that $\Gamma$ is a subset of the Banach space $(\mathcal{B},\lVert \cdot \rVert_{\mathcal{B}})$. We consider the case where the data consists of independent copies of $(Y,X)$.

Following \cite{Ichimura2021}, let $\gamma(F)$ denote the probability limit of the first-step estimator when the distribution of $Z:=(X,Y)$ is $F$. Denote by $F_0$ the true distribution of $Z$, thus $\gamma_0 = \gamma(F_0)$.  We seek to find a function $\boldsymbol{b}(u;Z;\gamma(F), \psi(F))$ such that $\mathbb{E}_{F}[\boldsymbol{b}(u;Z;\gamma(F), \psi(F))] = 0$, $\mathbb{E}_{F}[(\boldsymbol{b}(u;Z;\gamma, \psi(F))^2] < \infty$, and for every distribution function $H$ (restricted except for regularity conditions) and $l \in \mathbb{N}$:

\begin{equation}
	\label{eq_def_influence}
	\frac{\partial}{\partial \tau}\int_{\underline{p}}^{\overline{p}}Q_{h^{-1}(Y,X; \gamma_\tau(H|F))}(u) P_l(u) du \Big|_{\tau = 0} = \int_{\underline{p}}^{\overline{p}} \int \boldsymbol{b}(u;z;\gamma(F), \psi(F)) P_l(u) H(dz) du \, ,
\end{equation}
where $\psi(F)$ denotes a set of nuisance parameters, and, for $\tau \in [0,1]$, $\gamma_\tau(H|F) = \gamma( F + \tau(H-F)))$. The function $\boldsymbol{b}$ is known as a \emph{first-step influence function}, quantifying the impact of estimating $\gamma(F)$ on the estimator of the L-moments. 

Following \cite{Chernozhukov2022}, we will focus on correcting L-moments by a sample average of an estimated version of the first-step influence function. To see why, take $F$ equal to the empirical distribution function of $Z$, and $H=F_0$. We may then consider the following \emph{distributional Taylor expansion} \citep{kennedy2023semiparametric} around $F$:
\begin{equation}
	\label{eq_taylor_semi}
	\begin{aligned}
		\sqrt{T}\left(\int_{\underline{p}}^{\overline{p}}Q_{h^{-1}(Y,X; \hat\gamma)}(u) P_l(u) du - \int_{\underline{p}}^{\overline{p}}Q_{h^{-1}(Y,X; \gamma_0)}(u) P_l(u) du \right)= \\
	-\sqrt{T}	\frac{\partial}{\partial \tau}\int_{\underline{p}}^{\overline{p}}Q_{h^{-1}(Y,X; \gamma_\tau(F_0|F))}(u) P_l(u) du \Big|_{\tau = 0}  +\sqrt{T}  \int_{\underline{p}}^{\overline{p}} E_\text{lin}(u) P_l(u)du = \\
- \int_{\underline{p}}^{\overline{p}} \int \boldsymbol{b}(u;z;\hat{\gamma}, \hat{\psi}) P_l(u) F_0(dz) du + \sqrt{T}\int_{\underline{p}}^{\overline{p}}	 E_\text{lin}(u) P_l(u)du , ,
	\end{aligned}
\end{equation}
where the second equality follows from \eqref{eq_def_influence}. By Bessel's inequality, $T \lVert	\int_{\underline{p}}^{\overline{p}} E_\text{lin}(u) \boldsymbol{P}^R(u)du  \rVert^2\leq T	\int_{\underline{p}}^{\overline{p}} E_\text{lin}(u)^2 du $ . Under smoothness conditions on the model, the linearisation error can be further shown to be bounded above by $T	\int_{\underline{p}}^{\overline{p}} E_\text{lin}(u)^2 du  \leq C T \lVert \hat{\gamma} - \gamma_0 \rVert_\mathcal{B}^4$ (see the discussion on Assumption 3.4 in \cite{Chernozhukov2018}; and the examples in Section 4.3 of \cite{kennedy2023semiparametric}). Therefore, in these settings, if $ \lVert \hat{\gamma} - \gamma_0 \rVert_\mathcal{B} = o_{F_0}(T^{-1/4})$, one would have that $T \lVert	\int_{\underline{p}}^{\overline{p}} E_\text{lin}(u) \boldsymbol{P}^R(u)du  \rVert^2 = o_{F_0}(1)$, and  \eqref{eq_taylor_semi} would imply that correcting by an average of the first-step influence function may be employed to remove the effect of first-step estimation.

In the following examples, we provide calculations for the function $$\boldsymbol{B}_l(Z;\gamma_0, \psi_0) \coloneqq \int_{\underline{p}}^{\overline{p}} \boldsymbol{b}(u;Z;\gamma_0;\psi_0) P_l(u) du$$
in three models. For that, we rely on the following observation:

\begin{equation}
	\label{eq_trick}
	\begin{aligned}
		F_{h^{-1}(Y,X;\gamma_\tau)}(Q_{h^{-1}(Y,X;\gamma_\tau)}(u)) = u \, \quad \forall u \in (0,1), \tau \in (0,1) \implies \\ \frac{\partial}{\partial \tau}Q_{h^{-1}(Y,X; \gamma_\tau)}(u) \Big|_{\tau = 0} = - \frac{1}{f_\epsilon(Q_\epsilon(u))} \frac{\partial }{\partial \tau} \mathbb{P}_{F_0} [Y\leq h(Q_\epsilon(u),X;\gamma_\tau)] \Big|_{\tau=0} \, .
	\end{aligned}
\end{equation}

Whenever the order of differentiation and integration can be exchanged in \eqref{eq_def_influence}, representation \eqref{eq_trick} will allow us to apply the results of \cite{Ichimura2021} in our context.

\begin{example}[Semiparametric conditional mean model]
	The model is given by:
		\begin{equation*}
		\begin{aligned}
			Y = \gamma_0(X) + \epsilon\, ,  \\
		\epsilon \sim F_{\theta_0}, \quad \theta_0 \in \Theta \subseteq \mathbb{R}^d \, \, .
		\end{aligned}
	\end{equation*}
	where for every distribution $H$, the probability limit of the preliminary consistent estimator of $\gamma_0$ is given by:
	
	$$\gamma(H) \in \operatorname{argmin}_{w\in \mathcal{W}} \mathbb{E}_H[(Y-w(X))^2]  \, ,$$
	with $\mathcal{W}$ ia closed linear subspace of $L_2(F_{0,X})$ that does not depend on $H$. Notice that $ \frac{\partial }{\partial \tau} \mathbb{P}_{F_0} [Y\leq Q_\epsilon(u) + \gamma_\tau(X)] \Big|_{\tau=0} = \frac{\partial }{\partial \tau} \mathbb{E}_{F_0}[\mathbb{P}_{F_0} [Y\leq Q_\epsilon(u) + \gamma_\tau(X)|X]] \Big|_{\tau=0} $. If the order of differentiation and integration can be exchanged, Proposition 1 of \cite{Ichimura2021} yields:
		
	$$\boldsymbol{B}_l(Z;\gamma, \psi) = - \int_{\underline{p}}^{\overline{p}}  \psi(u|X)(Y-\gamma(X)) P_l(u)du \, , $$
	where 
	$$\psi_0(u|X) \in \operatorname{arg min}_{w \in \mathcal{W}} \mathbb{E}_{F_0}[(\omega_{0}(u|X) - w(X))^2] \, , $$
	with  $\omega_0(u|X) =  \frac{1}{f_{\epsilon}(Q_\epsilon(u))} f_{\epsilon|X}(Q_\epsilon(u)|X) $.
\end{example}

\begin{example}[Semiparametric conditional quantile model]
	The model is given by
	
	\begin{equation*}
		\begin{aligned}
			Y = \gamma_0(\epsilon|X)\, , \\
		\epsilon|X \sim \operatorname{Uniform}[0,1]\, ,
		\end{aligned}
	\end{equation*}
	where, for each $\tau \in (0,1)$ and distribution $H$:
	$$\gamma(H)(\tau|X) \in \operatorname{argmin}_{w \in \mathcal{W}}\mathbb{E}_H[|Y-w(X)|(\tau \mathbf{1}\{Y-w(X) > 0\}  + (1-\tau )\mathbf{1}\{Y-w(X) < 0\}) ] \, ,$$
	with the class $\mathcal{W}$ defined as in the previous example. In this model, the L-moment approach may be used as a specification testing tool, since the model implies $\epsilon \sim \operatorname{Uniform}[0,1]$.
	
	In this setting, if $\mathcal{W}$ contains constant functions, Proposition 1 of \cite{Ichimura2021} yields:
	
	$$\boldsymbol{B}_l(Z;\gamma, \psi) = - \int_{\underline{p}}^{\overline{p}}(u - \mathbf{1}\{Y < \gamma(u|X)\}) P_l(u) du$$
\end{example}

\begin{example}[Linear-nonparametric instrumental variable model]
	\label{example_linear_iv}
	We consider the linear-nonparametric model of \cite{Athey2019}:

	\begin{equation*}
		\begin{aligned}
			Y =  \alpha_0(X) + \beta_0(X)W + \epsilon \\
					\epsilon \sim F_{\theta_0}, \quad \theta_0 \in \Theta \subseteq \mathbb{R}^d \, \, ,
		\end{aligned}
	\end{equation*}
	where $\mathbb{E}_{F_0}[\epsilon|X]=0$, but $W$ is believed to be endogenous. We assume that the researcher has access to a scalar instrumental variable $S$ satisfying $\mathbb{E}_{F_0}[\epsilon|X,S]=0$. Following \cite{Athey2019}, we consider consistent preliminary estimators of $\alpha$ and $\beta$ whose probability limits are characterized by, for each $H$:
		\begin{equation}
		\begin{aligned}
			\mathbb{E}_H[(Y - \alpha(H)(X) - \beta(H)(X) W )|X] = 0 \, , \\ 
					\mathbb{E}_H[S(Y - \alpha(H)(X) - \beta(H)(X) W )|X] = 0 \, .
					\end{aligned}
		\end{equation}
		
		In this case, if $\mathbb{E}_{F_0}[S|X,W] = a_0(X) + b_0(X)W$ with $\mathbb{P}_{F_0}[b_0(X)\neq0]=1$ and $\mathbb{E}_{F_0}[W|X,S] = c_0(X) + d_0(X)S$ with $\mathbb{P}_{F_0}[d_0(X)\neq0]=1$,  it follows by Proposition 3 of \cite{Ichimura2021} that:
		
		$$\boldsymbol{B}_l(Z;\gamma, \psi) = -\int_{\underline{p}}^{\overline{p}}  \left(g(u|X)\frac{(S-a(X))}{b(X)} + h(u|X)\right)(Y-\alpha(X)-\beta(X)W)P_l(u) du\, ,$$
		
		where:
		
		$$(g_0(u|X),h_0(u|X)) \in \operatorname{argmin}_{o,p \in L^2(F_{0,X})} \mathbb{E}_{F_0}\left[( \omega_0(u|X,W) - o(X) - p(X)W)^2\right]\, ,$$
		with  $\omega_0(u|X,S) = \frac{1}{f_{\epsilon}(Q_\epsilon(u))} f_{\epsilon|X,W}(Q_\epsilon(u)|X,W) $.
\end{example}

Once the form of correction is known, we follow the double machine learning literature \citep{Chernozhukov2018,Chernozhukov2022,kennedy2023semiparametric} and propose a sample-split bias-corrected version of the L-moment estimator. Fix $\kappa \in (0,1)$. Partition the sample into two blocks, with $T_1 = \lfloor T \kappa \rfloor$ and $T_2 = T-\lfloor T \kappa \rfloor$ observations. Let $\mathcal{I}_1$ and $\mathcal{I}_2$ denote the set of indices of each partition. We propose to estimate $\theta_0$ through the following steps:

\begin{enumerate}
	\item Using $\mathcal{I}_1$, estimate $\gamma_0$. Denote by $\hat{\gamma}$ this first-step estimator.
	
	\item Using $\mathcal{I}_2$, compute, for each $u \in (0,1)$, $\hat{Q}_{\hat{\epsilon}}(u)$ as the $u$-th empirical quantile of $\{h^{-1}(Y_i,X_i;\hat{\gamma}): i \in \mathcal{I}_2\}$. 
	\item Using $\mathcal{I}_1$, estimate $\psi_0$. Denote by  $\hat{\psi}$ this estimator.
	\item Compute the influence function adjustment:
	
	$$\widehat{\boldsymbol{A}} = -\frac{1}{ T_2}\sum_{i \in \mathcal{I}_2} \begin{pmatrix}
		\boldsymbol{B}_1(Z_i;\hat{\gamma},\hat{\psi}) \\ 
			\boldsymbol{B}_2(Z_i;\hat{\gamma},\hat{\psi}) \\
			\vdots \\
				\boldsymbol{B}_R(Z_i;\hat{\gamma},\hat{\psi}) 
	\end{pmatrix}\, .$$
	\item Estimate the model by:

	\begin{equation*}
		\begin{aligned}
			\hat{\theta}_{\mathcal{I}_2} \in \text{arg inf}_{\theta \in \Theta} \left[\int_{\underline{p}}^{\bar{p}} \left(\hat{Q}_{\hat{\epsilon}} (u) - Q_\epsilon(u|\theta)  \ \right)  \mathbf{P}^R(u)' du - \hat{\boldsymbol{A}} \right] W^R\left[\int_{\underline{p}}^{\bar{p}}\left(\hat{Q}_{\hat{\epsilon}} (u) - Q_Y(u|\theta)   \right)  \mathbf{P}^R(u) du - \hat{\boldsymbol{A}} \right]\, ,
		\end{aligned} 
	\end{equation*}
	where $W^R$ is a possibly estimated weighting matrix.
\end{enumerate} 
	
	For sample-splitting not to result in efficiency losses, one may adopt \textit{cross-fitting}, i.e. one swaps the roles of $\mathcal{I}_1$ and $\mathcal{I}_2$ in the above sequence, and compute the final estimator as $\hat{\theta} = \kappa\hat{\theta}_{\mathcal{I}_1} + (1-\kappa)\hat{\theta}_{\mathcal{I}_2}$ and .
	
Consistency and asymptotic linearity of $\hat{\theta}_{\mathcal{I}_2}$ follows by standard uniform differentiability conditions, with proofs similar to those of Propositions \ref{proof_consistency} and \ref{proof_asymptotic_linear}, if the crucial conditions:

$$T\left\lVert	\int_{\underline{p}}^{\overline{p}} E_\text{lin}(u) \boldsymbol{P}^R(u)du  \right\rVert^2 = o_{F_0}(1) \, ,$$
and
$$\sqrt{T}\left\lVert \hat{\boldsymbol{A}} - \boldsymbol{A} - \tilde{\boldsymbol{A}}   \right\rVert_2 = o_{F_0}(1) \, ,$$
hold. Here, $\boldsymbol{A}$ is given by:

	$${\boldsymbol{A}} = -\frac{1}{ T_2}\sum_{i \in \mathcal{I}_2} \begin{pmatrix}
	\boldsymbol{B}_1(Z_i;\gamma_0,\psi_0) \\ 
	\boldsymbol{B}_2(Z_i;\gamma_0,\psi_0) \\
	\vdots \\
	\boldsymbol{B}_R(Z_i;\gamma_0,\psi_0) 
\end{pmatrix}\, ,$$
and $\tilde{\boldsymbol{A}}$ is given by:
$$
\tilde{\boldsymbol{A}} = - \begin{pmatrix}
\int	\boldsymbol{B}_1(z;\hat{\gamma}, \hat{\psi}) F_0(dz)\\ 
	\vdots \\
\int	\boldsymbol{B}_R(z;\hat{\gamma}, \hat{\psi}) F_0(dz)
\end{pmatrix} $$

In our three examples, that $\sqrt{T}\lVert \hat{\boldsymbol{A}} - \boldsymbol{A} - \tilde{\boldsymbol{A}} \rVert_2 = o_{F_0}(1)$ holds follows by, first, applying Bessel's inequality to show that 

$$T\lVert \hat{\boldsymbol{A}} - \boldsymbol{A} - \tilde{\boldsymbol{A}}  \rVert_2^2 \leq \int_{\underline{p}}^{\overline{p}} T \left(\frac{1}{T_2}\sum_{i \in \mathcal{I}_2} \left[\boldsymbol{b}(u,Z_i;\gamma_0,\psi_0) -  \boldsymbol{b}(u,Z_i;\hat\gamma,\hat\psi) -\int   \boldsymbol{b}(u,z;\hat\gamma,\hat\psi) F_0(dz)\right]\right)^2 du  \, .$$

That the right-hand side converges in probability to zero can then be established under weak consistency requirements on $\hat{\gamma}$ and $\hat{\psi}$. Importantly, one does not have to resort to Donsker conditions, due to the special structure provided by sample-splitting \citep[see][Lemma 1 and the accompanying discussion]{kennedy2023semiparametric}.

Under the above conditions, the estimator admits the following asymptotic linear representation:

\begin{equation*}
	\footnotesize 
	\sqrt{T_2}(\hat{\theta}_{\mathcal{I}_2} - \theta_0) = - ( J^R(\theta_0)' \Omega^R  J^R(\theta_0))^{-1} J^R(\theta_0)' \Omega^R \left( \sqrt{T_2}  \int_{\underline{p}}^{\bar{p}} (\hat{Q}_{\hat{\epsilon}}(u) - Q_{h^{-1}(Y,X;\hat\gamma)}(u)) \mathbf{P}^R(u) du  - \sqrt{T_2}  \boldsymbol{A}  \right)+ o_{F_0}(1) \, ,
\end{equation*}
where $J^R(\theta) =- \int_{\underline{p}}^{\bar{p}} \nabla_{\theta'} Q_{\epsilon}(u|\theta) \mathbf{P}^R(u) du $, and, for each $\gamma \in \Gamma$, $Q_{h^{-1}(Y,X;\gamma)}(u)$ is the population quantile function of $h^{-1}(Y,X;\gamma)$.

If we assume that the conditions on the distribution of $h^{-1}(Y,X;\gamma)$  ensuring that a Bahadur-Kiefer approximation as discussed in Appendix \ref{app_inference_bahadur} hold uniformly over $\gamma$ in a neighbourhood of $\gamma_0$, then we may use these results as a tool for inference. For example, if the conditions in the statement of Theorem \ref{thm_bahadur} hold in such way, then we are able to show that:

\begin{equation*}
	\sqrt{T_2}(\hat{\theta}_{\mathcal{I}_2} - \theta_0) = - ( J^R(\theta_0)' \Omega^R  J^R(\theta_0))^{-1} J^R(\theta_0)' \Omega^R\left(\sqrt{T}_2(\boldsymbol{C}-\boldsymbol{A})\right)+ o_{F_0}(1) \,,\, ,
\end{equation*}
with 
where,

$$\boldsymbol{C} = \frac{1}{T_2}\sum_{i \in \mathcal{I}_2} \int_{\underline{p}}^{\overline{p}} \frac{1}{f_{h^{-1}(Y,X;\hat \gamma)}(Q_{h^{-1}(Y,X;\hat\gamma)}(u))}(\mathbf{1}\{h^{-1}(Y_i,X_i;\hat \gamma)\leq Q_{h^{-1}(Y,X;\hat\gamma)}(u) \} - u) \boldsymbol{P}_R(u) du \,.$$

Notice that, due to the nature of sample-splitting, conditional on $\hat{\gamma}$, both $\boldsymbol{C}$ and $\boldsymbol{A}$ are zero-mean random-variables whose variances and covariance can be estimated by sample-analogs. This result serves as a basis for inference in this setting. Moreover, the conditional variance of $\sqrt{T_2}\left[\boldsymbol{C}  - \boldsymbol{A}\right]$ can be used to compute the optimal weighting scheme to be used in step (5) of our estimation.  

\subsection{Details on prediction intervals}
\subsubsection{Asymptotic validity}
We consider a setting that nests our application in the main text as a particular case. Specifically, we consider that the researcher posits the following potential outcome model for the untreated outcome that is observed absent a policy intervention in a panel of $n$ units, indexed by $i$, over $T$ periods, indexed by $t$:

\begin{equation}
	\begin{aligned}
		Y_{it}(0) = h_{t}(\epsilon_{it}, X_{it}; \gamma_0), \quad \gamma \in \Gamma \subseteq \mathcal{B}\, , \\
		\epsilon_{i,t} \sim F_{\theta_0}, \quad \theta_0 \in \Theta \subseteq \mathbb{R}^d \, ,
	\end{aligned} 
\end{equation}
where $e\mapsto h_{t}(e, x;\gamma_0)$ is strictly increasing, for every $x$ in the support of $X_{it}$; and the distribution of $Y_{i,t}(0)$ has no point masses. The treatment intervention date $t^*$ takes values in a set $\mathcal{T}\subseteq \mathbb{N}\cup\{\infty\}$, which may include $t^*=\infty$ if, for some realisation of the random variables, the intervention would never take place. As in the main text, we assume that $t^*$ is independent of $\epsilon_{i,t^*}$. Fix a significance level $\alpha \in (0,1)$. Let $\hat{\gamma}_{n,\tau}$ and $\hat{\theta}_{n,\tau}$ denote the (generally unfeasible) estimators of $\gamma_0$ and $\theta_0$ computed using the \emph{untreated} potential outcomes of the $n$ individuals in the panel and the first $\tau$ periods. The \textbf{actual} estimators are given by $\hat{\gamma}_{n,t^*-1}$ and $\hat{\theta}_{n,t^*-1}$  We assume that, for every $\tau < \sup \mathcal{T} \land T$, $\operatorname{plim}_{n\to \infty} \sup_{x \in \operatorname{supp}X_{i,\tau+1}} |h_{\tau+1}(Q_{\hat{\theta}_{n,\tau}}(1-\alpha),x; \hat{\gamma}_{n,\tau}) -  h_{\tau+1}(Q_{\theta_0}(1-\alpha), x; \gamma_0)| = 0$. In the setting of the main text, this assumption requires that there always be, over alternative realisations of sampling uncertainty, sufficient pre-treatment periods to estimate the outcome model $Y_{it}(0)$ using lags as instruments.

Our proposed lower confidence region for a unit $i$ is given by:

$$\hat{\mathcal{I}}_{i,1-\alpha} = [Y_{i,t^*}-h_{t^*}(Q_{\hat{\theta}_{n,t^*-1}}(1-\alpha),X_{i,t^*}; \hat{\gamma}_{n,t^*}) ,\infty)\, ,$$
and, we may assume, without loss, that, in the event that $t^* > T$, the confidence region is estimated without uncertainty, as, in practice, nothing is reported in this case by the researcher and thus no ``mistakes'' are made.

The asymptotic validity of the confidence region can be shown by relying on a similar argument to the one in Appendix A of \cite{AlvarezFerman2024}. Specifically, we observe that, under the stated assumptions, $h_{t^*}(Q_{\hat{\theta}_{n,t^*-1}}(1-\alpha),X_{i,t^*}; \hat{\gamma}_{n,t^*-1}) \overset{p}{\to} h_{t^*}(Q_{\theta_0}(1-\alpha), X_{i,t^*}; \gamma_0)$, since, for any tolerance $\nu > 0$:

\begin{equation*}
	\begin{aligned}
	\lim_{n\to \infty}	\mathbb{P}[|h_{t^*}(Q_{\hat{\theta}_{n,t^*-1}}(1-\alpha),X_{i,t^*}; \hat{\gamma}_{n,t^*-1}) -h_{t^*}(Q_{\theta_0}(1-\alpha), X_{i,t^*}; \gamma_0)|>\nu] \leq \\
		\lim_{n\to \infty} \sum_{\tau \in  \mathcal{T}} \mathbb{P}\left[\left\{\sup_{x \in \operatorname{supp}X_{i,\tau}} |h_{\tau}(Q_{\hat{\theta}_{n,\tau-1}}(1-\alpha),x; \hat{\gamma}_{n,\tau-1}) -  h_{\tau}(Q_{\theta_0}(1-\alpha), x; \gamma_0)| > \nu \right\}\cap \{t^* = \tau\}\right] = \\ 
			\sum_{\tau \in  \mathcal{T}} \lim_{n\to \infty} \mathbb{P}\left[\left\{\sup_{x \in \operatorname{supp}X_{i,\tau}} |h_{\tau}(Q_{\hat{\theta}_{n,\tau-1}}(1-\alpha),x; \hat{\gamma}_{n,\tau-1}) -  h_{\tau}(Q_{\theta_0}(1-\alpha), x; \gamma_0)| > \nu \right\}\cap \{t^* = \tau\}\right] = 
		0 \, \,
	\end{aligned}
\end{equation*}
(we pass the limit under the sum because at most the $T < \infty$ first terms in the sum are different than zero). Consequently, given that the distribution of $Y_{i,t}(0)$ has no point masses, it follows by the continuous mapping theorem that:

\begin{equation*}
	\begin{aligned}
\mathbbm{1}\{Y_{it^*}(1)-Y_{it^*}(0) \in \mathcal{I}_{i,1-\alpha}\} = \mathbbm{1}\{Y_{it^*}(0) \in (-\infty,h_{t^*}(Q_{\hat{\theta}_{n,t^*-1}}(1-\alpha),X_{i,t^*}; \hat{\gamma}_{n,t^*-1})  ]\} \overset{p}{\to} \\
\mathbbm{1}\{Y_{it^*}(0) \in (-\infty,h_{t^*}(Q_{\theta_0}(1-\alpha),X_{i,t^*}; \gamma_0)  ]\} = \mathbbm{1}\{\epsilon_{it^*} \leq Q_{\theta_0}(1-\alpha)\}  \, ,
	\end{aligned}
\end{equation*}
where the last passage followed by the map $h_t$ being strictly increasing in the first entry. Asymptotic coverage then follows by the bounded convergence theorem and the independence between $t^*$ and $\epsilon_{t^*}$, by noting that:

\begin{equation*}
	\begin{aligned}
		\lim_{n \to \infty} \mathbb{P}[Y_{it^*}(1)-Y_{it^*}(0) \in \mathcal{I}_{i,1-\alpha} ] \ = \mathbb{P}[\epsilon_{it^*} \leq Q_{\theta_0}(1-\alpha)] = \
		\sum_{\tau \in \mathcal{T}} \mathbb{P}[t^*=\tau] \mathbb{P}[\epsilon_{i,\tau} \leq Q_{\theta_0}(1-\alpha)|t^*=\tau] = \\ \sum_{\tau \in \mathcal{T}} \mathbb{P}[t^*=\tau] \mathbb{P}[\epsilon_{i,\tau} \leq Q_{\theta_0}(1-\alpha)] = 1-\alpha
	\end{aligned}
\end{equation*}

\subsubsection{Bonferrroni correction}

The prediction interval presented in the previous section has an asymptotic justification. However, in finite samples, estimation error of $\theta_0$ and $\gamma_0$ may induce miscoverage. Following an idea similar to \cite{Cattaneo2021}, we consider two simple Bonferroni-style adjustments that  may improve upon coverage. The first adjustment relies on the assumption that $\epsilon_{i,t^*}$ is independent of the data used to estimate $\theta_0$ and $\gamma_0$. In this case, for $\beta \in (0,1)$ and $\phi \geq 0$, a simple calculation reveals that coverage probability of the modified interval  $ \hat{I}_{i,\beta} - \phi$ can be bounded below by:

\begin{equation*}
	\begin{aligned}
		\mathbb{P}[Y_{it^*}(1)-Y_{it^*}(0) \in \hat{I}_{i,\beta}-\phi] = \mathbb{P}[Y_{it^*}(0)\leq h_{t^*}(Q_{\hat{\theta}_{n,t^*-1}}(1-\beta),X_{i,t^*}; \hat{\gamma}_{n,t^*-1}) + \phi  ] \geq \\ \mathbb{P}[\{\epsilon_{it^*} \leq Q_{\theta_0}(1-\beta)\} \cap  \{h_{t^*}(Q_{\hat{\theta}_{n,t^*-1}}(1-\beta),X_{i,t^*}; \hat{\gamma}_{n,t^*-1})  -h_{t^*}(Q_{\theta_0}(1-\beta),X_{i,t^*}; \gamma_0)  \geq -\phi \}] =\\ 
		 (1-\beta) \mathbb{P}[\ \{h_{t^*}(Q_{\hat{\theta}_{n,t^*-1}}(1-\beta),X_{i,t^*}; \hat{\gamma}_{n,t^*-1})  -h_{t^*}(Q_{\theta_0}(1-\beta),X_{i,t^*}; \gamma_0)  \geq -\phi \}]\\
	\end{aligned}\,.
\end{equation*}
Now, if $h_{t^*}(Q_{\hat{\theta}_{n,t^*-1}}(1-\beta),X_{i,t^*}; \hat{\gamma}_{n,t^*-1})  -h_{t^*}(Q_{\theta_0}(1-\beta),X_{i,t^*}; \gamma_0)  $ is approximately normally distributed around zero in larger samples, which can be the case if a central limit theorem applies to this term, and if the variance of this term can be estimated,  then one can calibrate $\phi$ and $\beta$ based on the normal distribution as to ensure coverage at the $(1-\alpha)$ level. Notice that there is some flexibility in the choice of $\phi$ and $\beta$ in this case, which can be chosen so as to maximizie the low endpoint of the interval.

The previous correction relied on independence between $\epsilon_{i,t^*}$ and the data used in the estimation. If that is not the case, than a simpler Bonferroni bound may be obtained by applying the union bound to the probability of \emph{miscoverage} of $ \hat{I}_{i,\beta}-\phi$. In this case, one calibrates $\beta$ and $\phi$ such that $\beta + \mathbb{P}[\ \{h_{t^*}(Q_{\hat{\theta}_{n,t^*-1}}(1-\beta),X_{i,t^*}; \hat{\gamma}_{n,t^*-1})  -h_{t^*}(Q_{\theta_0}(1-\beta),X_{i,t^*}; \gamma_0)  \leq -\phi \}] \leq \alpha$. Notice that this adjustment is necessarily more conservative than the previous one, though the difference can be small.

\subsection{Conditional models}
\label{app_extension_cond}

Suppose the researcher postulates a conditional model $\{Q_{Y|X}(|X;\theta): \theta \in \Theta\}$, where $Y$ is a scalar outcome of interest, $X$ are a set of controls, and $\Theta \subseteq \mathbb{R}^d$. We consider the estimator $\hat{\theta}$ of the true parameter $\theta_0$:
\begin{equation}
	\label{eq_est_conditional}
	\hat \theta \in \operatorname{argmin}_{\theta \in \Theta}  \frac{1}{T}\sum_{t=1}^T \left\lVert \Omega(X_t)^{1/2}\left(\int_{\underline p}^{\overline p}\left( \hat Q_{Y|X}(u|X_t) - Q_{Y|X}(u|X_t;\theta) \right) \boldsymbol{P}_R(u)du\right) \right\rVert_2^2 \, ,
\end{equation}
where $\Omega$ is a $R\times R$ symmetric positive semidefinite weighting function of the controls $X$, and $\hat{Q}_{Y|X}$ is a preliminary nonparametric estimator of the conditional quantile process $(u,x) \mapsto Q_{Y|X}(u|x)$. The formulation may be seen as an extension of \cite{Ai2003}'s approach to models defined by conditional moments to a conditional L-moment setting.

Suppose that we rely on the nonparametric quantile series regression estimator of \cite{Belloni2019}, and that the conditions underlying their Comment 3 and Theorem 2 are satisfied. In this case, under identifiability and uniform differentiability conditions similar to those used in Propositions \ref{prop_consistency} and \ref{prop_asymptotic_linear}, it is possible to show that the estimator admits the asymptotic linear representation.
\begin{equation}
	\begin{aligned}
		\sqrt{T}(\hat\theta - \theta_0)  =  \\  -\left(\frac{1}{T} \sum_{t=1}^T \left(\int_{\underline{p}}^{\overline{p}} \partial_{\theta'} Q_{Y|X}(u|X_t;\theta_0) \boldsymbol{P}_R(u) du \right)' \Omega(X_t) \left( \int_{\underline{p}}^{\overline{p}} \partial_{\theta'} Q_{Y|X}(u|X_t;\theta_0) \boldsymbol{P}_R(u) du \right) \right)^{-1} \times \\  \left(\frac{1}{T}\sum_{t=1}^T \left(\int_{\underline{p}}^{\overline{p}} \partial_{\theta'} Q_{Y|X}(u|X_t;\theta_0) \boldsymbol{P}_R(u) du \right)' \Omega(X_t) \left( \int_{\underline{p}}^{\overline{p}}\sum_{s=1}^T\frac{1}{\sqrt T} Z_t'J^{-1}_T(u)Z_s (u - \mathbf{1}\{U_s \leq u\}) \boldsymbol{P}_R(u) du\right)\right) \\ 
		+o_{P}(1) \, ,
	\end{aligned}
\end{equation}
where $Z_t$ is the vector of transformations of $X_t$ used in the series estimator; $$ J_T(u) = \frac{1}{T}\sum_{t=1}^T f_{Y|X}(Q_{Y}(u|X_t)|X_t) Z_t Z_t',$$ and $\{U_t\}_{t=1}^T$ are independent uniform random variables, independent from $\{X_t\}_{t=1}^T$. It then follows that the optimal weighting scheme is given by:

$$\Omega(X_t)^* = \mathbb{V}\left[\left( \int_{\underline{p}}^{\overline{p}}\sum_{s=1}^T\frac{1}{\sqrt T} Z_t'J^{-1}_T(u)Z_s (u - \mathbf{1}\{U_s \leq u\}) \boldsymbol{P}_R(u) du\right)\Big|X_1,\ldots, X_T\right]^{-}\, .$$

This optimal weighting scheme can be estimated by relying on an estimator of $u \mapsto J_T(u)$ (\cite{Belloni2019} discuss nonparametric estimators of this function; a semiparametric version of this quantity that relies on a preliminary estimator of $\theta_0$ can also be used); and on simulation from uniform random variables.

Inference using normal critical values can be performed under the assumptions underlying Theorem 5 of \cite{Belloni2019}, which ensures a strong approximation of the series estimator to a Gaussian process. A weighted bootstrap approximation can also be used, if the assumptions underlying~ Theorem 6 of \cite{Belloni2019} hold.

Next, we note that, by considering $u$-specific orthogonalizations of the $Z_t$ when estimating a quantile function $x \mapsto Q_{Y|X}(u|x)$, it is without loss to assume that, for every $u$:

$$ \sqrt{f_{Y|X}(Q_{Y}(u|X_t)|X_t)f_{Y|X}(Q_{Y}(u|X_s)|X_s)} Z_t '  Z_s = \mathbf{1}\{t=s\}\, .$$

Using this fact, we are able to show that the variance of the leading term of the linear representation of the optimally weighted generalized L-moment estimator is:

\begin{equation*}
\tiny	\begin{aligned}
 \left(\left(\frac{1}{T} \sum_{t=1}^T\int_{\underline{p}}^{\overline{p}} \partial_{\theta'} Q_{Y|X}(u|X_t;\theta_0) \boldsymbol{P}_R(u) du \right)'  \left(\frac{1}{ T}\sum_{t=1}^T\mathbb{V}\left[\int_{\underline{p}}^{\overline{p}} \frac{(u - \mathbf{1}\{U_t \leq u\})}{ f_{Y|X}(u|X_t)} \boldsymbol{P}_R(u) du \Bigg|X_t\right]\right)^{-} \left(\frac{1}{T} \sum_{t=1}^T \int_{\underline{p}}^{\overline{p}} \partial_{\theta'} Q_{Y|X}(u|X_t;\theta_0) \boldsymbol{P}_R(u) du \right)\right)^{-1} \, ,	
\end{aligned}
\end{equation*} which closely resembles the variance of the optimally weighted L-moment estimator in the \emph{unconditional} case. Indeed, the above expression differs from the unconditional version in that it averages the relevant matrices across the sample support points of $X$. Proceeding by analogy to \Cref{app_efficiency}, we are then able to show that, when we rely on an orthonormal \emph{basis} $\{P_l\}_l$ and $0 \leq \underline{p} < \overline{p}\leq 1$, our proposed estimator is efficient, in the sense that its asymptotic variance coincides with the inverse of the expected conditional Fisher information matrix.

\section{Analytical expressions for the Generalized Extreme Value and Generalized Pareto Distributions}

This Appendix contains analytical expressions of the theoretical L-moments, as well as the gradients and Hessians used in computing the L-moment estimator and its higher-order expansion for both the GEV and GPD families of distributions. The file \texttt{gev\_npd.nb} provides a Mathematica notebook that analytically derives some of these expressions.

\subsection{Generalized Extreme Value distribution}

\subsubsection{Theoretical L-moments and its derivatives} Following \cite{hosking1986theory}, the $l$-th probability-weighted moment of a GEV distribution with location parameter $m$, scale parameter $r$ and shape parameter $k$ is given by:

$$\int_0^1 Q_\theta(u) u^l du =
\begin{cases}
\frac{m}{l+1} +	\frac{r \left(1-(l+1)^{-k} \Gamma (k+1)\right)}{k (l+1)} & \text{if } k \neq 0 \\
\frac{m}{l+1} +	\frac{r \log (l+1)+\gamma  r}{l+1} & \text{if } k = 0
\end{cases} \, ,
$$
where $\Gamma$ denotes the Gamma function and $\gamma$ the Euler-Mascheroni constant.

The gradient of the probability-weighted moment is given by:

$$\begin{bmatrix}
	\frac{1}{l+1}\\\frac{1-(l+1)^{-k} \Gamma (k+1)}{k (l+1)}\\\frac{r \left((l+1)^{-k} \Gamma (k+1) \log (l+1)-(l+1)^{-k} \Gamma (k+1) \psi ^{(0)}(k+1)\right)}{k (l+1)}-\frac{r \left(1-(l+1)^{-k} \Gamma (k+1)\right)}{k^2 (l+1)}
\end{bmatrix}\, ,
$$
at $k\neq 0$, with $\psi^{(j)}(x)$ denoting the $j$-th derivative of $x\mapsto \log(\Gamma(x))$ (the polygamma function of order $j$) and:

$$\begin{bmatrix}
	\frac{1}{l+1} \\
	\frac{\log (l+1)+\gamma }{l+1} \\
	-\frac{r \left(6 \log ^2(l+1)+12 \gamma  \log (l+1)+6 \gamma ^2+\pi ^2\right)}{12 (l+1)}
\end{bmatrix}$$
at $k=0$. Finally, the Hessian of  the probability-weighted moment is given by: 
\vspace{2em}

\hspace{-5em}\resizebox{1.2 \textwidth}{!}{
$
	\begin{bmatrix}
		0 & 0 & 0 \\
		0 & 0 & \frac{(l+1)^{-k} \Gamma (k+1) \log (l+1)-(l+1)^{-k} \Gamma (k+1) \psi ^{(0)}(k+1)}{k (l+1)}-\frac{1-(l+1)^{-k} \Gamma (k+1)}{k^2 (l+1)} \\
		0 & \frac{(l+1)^{-k} \Gamma (k+1) \log (l+1)-(l+1)^{-k} \Gamma (k+1) \psi ^{(0)}(k+1)}{k (l+1)}-\frac{1-(l+1)^{-k} \Gamma (k+1)}{k^2 (l+1)} & \frac{2 r \left(1-(l+1)^{-k} \Gamma (k+1)\right)}{k^3 (l+1)}-\frac{2 r \left((l+1)^{-k} \Gamma (k+1) \log (l+1)-(l+1)^{-k} \Gamma (k+1) \psi ^{(0)}(k+1)\right)}{k^2 (l+1)}+\frac{r \left(-(l+1)^{-k} \Gamma (k+1) \log ^2(l+1)+(l+1)^{-k} (-\Gamma (k+1)) \psi ^{(0)}(k+1)^2-(l+1)^{-k} \Gamma (k+1) \psi ^{(1)}(k+1)+2 (l+1)^{-k} \Gamma (k+1) \psi ^{(0)}(k+1) \log (l+1)\right)}{k (l+1)} \\
	\end{bmatrix}		$}

\vspace{2em}
at $k \neq 0$ and:

$$
\begin{bmatrix}
	0 & 0 & 0 \\
	0 & 0 & -\frac{6 \log ^2(l+1)+12 \gamma  \log (l+1)+6 \gamma ^2+\pi ^2}{12 (l+1)} \\
	0 & -\frac{6 \log ^2(l+1)+12 \gamma  \log (l+1)+6 \gamma ^2+\pi ^2}{12 (l+1)} & \frac{r \left(2 \log ^3(l+1)+6 \gamma  \log ^2(l+1)+\left(6 \gamma ^2+\pi ^2\right) \log (l+1)+2 \gamma ^3+\gamma  \pi ^2-2 \psi ^{(2)}(1)\right)}{6 (l+1)} \\
\end{bmatrix}
$$
at $k=0$.

\subsubsection{Quantile function and its derivatives} Again following \cite{hosking1986theory}, the $u$-th quantile of a GEV distribution with location parameter $m$, scale parameter $r$ and shape parameter $k$ is given by:

$$Q_\theta(u) = \begin{cases}
	m + \frac{r \left(1 - (-\log(u))^k\right)}{k} & \text{if } k \neq 0 \\
	m - r \log(-\log(u)) & \text{if } k = 0 
\end{cases}$$

The gradient function of the $u$-th quantile is given by:

\[
\begin{bmatrix}
	1 \\
	\frac{1 - (-\log(u))^k}{k} \\
	-\frac{r \left(1 - (-\log(u))^k\right)}{k^2} - \frac{r \log(-\log(u)) (-\log(u))^k}{k}
\end{bmatrix}
\]
at $k \neq 0$ and:

\[
\begin{bmatrix}
	1 \\
	-\log(-\log(u)) \\
	-\frac{1}{2} r \log^2(-\log(u))
\end{bmatrix}
\]
at $k = 0$. The Hessian is given by:

\[
\begin{bmatrix}
	0 & 0 & 0 \\
	0 & 0 & -\frac{1 - (-\log(u))^k}{k^2} - \frac{\log(-\log(u)) (-\log(u))^k}{k} \\
	0 & -\frac{1 - (-\log(u))^k}{k^2} - \frac{\log(-\log(u)) (-\log(u))^k}{k} & \frac{2r (1 - (-\log(u))^k)}{k^3} + \frac{2r \log(-\log(u)) (-\log(u))^k}{k^2} - \frac{r \log^2(-\log(u)) (-\log(u))^k}{k}
\end{bmatrix}
\]
at $k \neq 0$, and:

\[
\begin{bmatrix}
	0 & 0 & 0 \\
	0 & 0 & -\frac{1}{2} \log^2(-\log(u)) \\
	0 & -\frac{1}{2} \log^2(-\log(u)) & -\frac{1}{3} r \log^3(-\log(u))
\end{bmatrix}
\]
at $k=0$. Finally, for estimating the higher-order terms pertaining to estimation of the optimal weighting function, we require to compute the gradient, with respect to the model parameters, of the quantile density function $Q^\prime(u|\theta) =  \frac{1}{f(Q(u|\theta)|\theta))}$. The gradient is given by:
\[
\begin{array}{ccc}
	0 & \frac{(-\log (u))^{\text{shape}-1}}{u} & \frac{\text{scale} \log (-\log (u)) (-\log (u))^{\text{shape}-1}}{u} \\
\end{array}
\]
at $k \neq 0$ and:
\[
\begin{bmatrix}
	0 \\ \dfrac{(-\log(u))^{k - 1}}{u} \\ \dfrac{r \log(-\log(u)) (-\log(u))^{k - 1}}{u}
\end{bmatrix}
\]
and
\[
\begin{bmatrix}
	0 \\ -\dfrac{1}{u \log(u)} \\ -\dfrac{r \log(-\log(u))}{u \log(u)}
\end{bmatrix}
\]
at $k=0$.
\subsection{Generalized Pareto Distribution}
\subsubsection{Theoretical L-moments and its derivatives} Following \cite{hosking1986theory}, the $l$-th probability-weighted moment of a GPD distribution with location parameter $m$, scale parameter $r$ and shape parameter $k$ is given by:

$$\int_0^1 Q_\theta(u) u^l du = \begin{cases}
\frac{m}{l+1}  +\frac{r \left(1-\frac{\Gamma (k+1) \Gamma (l+2)}{\Gamma (k+l+2)}\right)}{k (l+1)}& \text{if } k \neq 0 \\
\frac{m}{l+1}+\frac{r \Gamma (l+1) (\psi ^{(0)}(l+2)+\gamma )}{\Gamma (l+2)} & \text{if } k = 0
\end{cases}\,. $$

The gradient of the $l$-th probability-weighted moment is given by:

\[
\begin{bmatrix}
	\frac{1}{l+1} \\
	\frac{1 - \dfrac{\Gamma(k+1) \Gamma(l+2)}{\Gamma(k+l+2)}}{k(l+1)} \\
	\dfrac{r \left( \dfrac{\Gamma(k+1)\Gamma(l+2) \psi^{(0)}(k+l+2)}{\Gamma(k+l+2)} - \dfrac{\Gamma(k+1)\psi^{(0)}(k+1)\Gamma(l+2)}{\Gamma(k+l+2)} \right)}{k(l+1)} - \dfrac{r \left( 1 - \dfrac{\Gamma(k+1)\Gamma(l+2)}{\Gamma(k+l+2)} \right)}{k^2(l+1)}
\end{bmatrix}
\]
at $k\neq 0$ and:

\[
\begin{bmatrix}
	\frac{1}{l+1} \\
	\frac{\psi^{(0)}(l+2) + \gamma}{l+1} \\
	-\dfrac{r \Gamma(l+1) \left(6 \psi^{(0)}(l+2)^2 + 12 \gamma \psi^{(0)}(l+2) - 6 \psi^{(1)}(l+2) + 6 \gamma^2 + \pi^2 \right)}{12 \Gamma(l+2)}
\end{bmatrix}
\]
at $k=0$. The Hessian is given by:
\vspace{1em}

\hspace{-5em}\resizebox{1.2 \textwidth}{!}{
$
\begin{bmatrix}
	0 & 0 & 0 \\
	0 & 0 & 
	\frac{ \dfrac{\Gamma(k+1) \Gamma(l+2) \psi^{(0)}(k+l+2)}{\Gamma(k+l+2)} - \dfrac{\Gamma(k+1) \psi^{(0)}(k+1) \Gamma(l+2)}{\Gamma(k+l+2)} }{k(l+1)} - 
	\frac{ 1 - \dfrac{\Gamma(k+1)\Gamma(l+2)}{\Gamma(k+l+2)} }{k^2(l+1)} \\
	0 & 
	\frac{ \dfrac{\Gamma(k+1) \Gamma(l+2) \psi^{(0)}(k+l+2)}{\Gamma(k+l+2)} - \dfrac{\Gamma(k+1) \psi^{(0)}(k+1) \Gamma(l+2)}{\Gamma(k+l+2)} }{k(l+1)} - 
	\frac{ 1 - \dfrac{\Gamma(k+1)\Gamma(l+2)}{\Gamma(k+l+2)} }{k^2(l+1)} &
	
	\frac{2r \left(1 - \dfrac{\Gamma(k+1)\Gamma(l+2)}{\Gamma(k+l+2)}\right)}{k^3(l+1)} - 
	\frac{2r \left( \dfrac{\Gamma(k+1)\Gamma(l+2) \psi^{(0)}(k+l+2)}{\Gamma(k+l+2)} - \dfrac{\Gamma(k+1) \psi^{(0)}(k+1) \Gamma(l+2)}{\Gamma(k+l+2)} \right)}{k^2(l+1)} \\
	& & 
	+ \frac{r}{k(l+1)} \left(
	-\dfrac{\Gamma(k+1) \psi^{(0)}(k+1)^2 \Gamma(l+2)}{\Gamma(k+l+2)} 
	+ \dfrac{2 \Gamma(k+1) \psi^{(0)}(k+1) \Gamma(l+2) \psi^{(0)}(k+l+2)}{\Gamma(k+l+2)} \right. \\
	& & \left.
	+ \dfrac{\Gamma(k+1) \Gamma(l+2) \psi^{(1)}(k+l+2)}{\Gamma(k+l+2)} 
	- \dfrac{\Gamma(k+1) \Gamma(l+2) \psi^{(0)}(k+l+2)^2}{\Gamma(k+l+2)} 
	- \dfrac{\Gamma(k+1) \psi^{(1)}(k+1) \Gamma(l+2)}{\Gamma(k+l+2)}
	\right)
\end{bmatrix}
$}
at $k \neq 0$, and:
\vspace{1em}

\hspace{-5em}\resizebox{1.2 \textwidth}{!}{
	$
\begin{bmatrix}
	0 & 0 & 0 \\
	0 & 0 & -\dfrac{ \Gamma(1 + l) \left(6 \gamma^2 + \pi^2 + 12 \gamma \psi^{(0)}(2 + l) + 6 \psi^{(0)}(2 + l)^2 - 6 \psi^{(1)}(2 + l)\right)}{12 \Gamma(2 + l)} \\
	0 & -\dfrac{ \Gamma(1 + l) \left(6 \gamma^2 + \pi^2 + 12 \gamma \psi^{(0)}(2 + l) + 6 \psi^{(0)}(2 + l)^2 - 6 \psi^{(1)}(2 + l)\right)}{12 \Gamma(2 + l)} & 
	\begin{aligned}
		&\dfrac{r}{6} \Bigg( \dfrac{1}{\Gamma(2 + l) \Gamma(1 + l)} \Big( 2 \gamma^3 + \gamma \pi^2 + 6 \gamma \psi^{(0)}(2 + l)^2 \\
		&\quad + \psi^{(0)}(2 + l)\left(6 \gamma^2 + \pi^2 - 12 \psi^{(1)}(2 + l)\right) - 6 \gamma \psi^{(1)}(2 + l) - 2 \psi^{(2)}(1) \Big) \\
		&\quad + \dfrac{2 \left( \psi^{(0)}(2 + l)^3 + 3 \psi^{(0)}(2 + l) \psi^{(1)}(2 + l) + \psi^{(2)}(2 + l) \right)}{1 + l} \Bigg)
	\end{aligned}
\end{bmatrix}
$}
at $k=0$.

\subsubsection{Quantile function and its derivatives} The quantile function is given by:

\begin{equation*}
	Q_\theta(u) =
	\begin{cases}
		m + \frac{r \left(1 - (1 - u)^k\right)}{k} & \text{if } k \neq 0 \\
		m + \frac{r \left(1 - (1 - u)^k\right)}{k} & \text{if } k = 0 
	\end{cases}\, .
\end{equation*}

The gradient with respect to the model parameters is given by:

\[
\begin{bmatrix}
	1 \\
	\dfrac{1 - (1 - u)^k}{k} \\
	-\dfrac{r (1 - (1 - u)^k)}{k^2} - \dfrac{r (1 - u)^k \log(1 - u)}{k}
\end{bmatrix}
\]
at $k \neq 0$ and:
\[
\begin{bmatrix}
	1 \\
	-\log(1 - u) \\
	-\dfrac{1}{2} r \log^2(1 - u)
\end{bmatrix}
\]
at $k =0$. The Hessian is given by:

\[
\begin{bmatrix}
	0 & 0 & 0 \\
	0 & 0 & -\dfrac{1 - (1 - u)^k}{k^2} - \dfrac{(1 - u)^k \log(1 - u)}{k} \\
	0 & -\dfrac{1 - (1 - u)^k}{k^2} - \dfrac{(1 - u)^k \log(1 - u)}{k} & \dfrac{2r (1 - (1 - u)^k)}{k^3} + \dfrac{2r (1 - u)^k \log(1 - u)}{k^2} - \dfrac{r (1 - u)^k \log^2(1 - u)}{k}
\end{bmatrix}
\]
at $k \neq 0$ and:
\[
\begin{bmatrix}
	0 & 0 & 0 \\
	0 & 0 & -\dfrac{1}{2} \log^2(1 - u) \\
	0 & -\dfrac{1}{2} \log^2(1 - u) & -\dfrac{1}{3} r \log^3(1 - u)
\end{bmatrix}
\]
at $k=0$. Finally, the gradient of $Q_{\theta}'(u)$ with respect to $\theta$ is:

\[
\begin{bmatrix}
	0 \\ (1 - u)^{k - 1} \\ r (1 - u)^{k - 1} \log(1 - u)
\end{bmatrix}
\]
at $k\neq0$ and:
\[
\begin{bmatrix}
	0 \\ \dfrac{1}{1 - u} \\ \dfrac{r \log(1 - u)}{1 - u}
\end{bmatrix}
\]
at $k=0$.

\section{Simple sufficient conditions for $L^2(0,1)$ consistency of empirical quantiles}
The following is a useful lemma for establishing $L^2(0,1)$ convergence of empirical quantiles.

\begin{lemma}
Let $Y_1,\ldots, Y_T$ be a random sample from a distribution with finite  $(2+\delta)$-moment that admits Lebesgue density $f$ such that $u \mapsto f(Q_Y(u))$ is continuous, where $Q_Y$ is the quantile function of this distribution. Then, denoting by $\hat{Q}_Y$ the empirical quantiles from $Y_1,\ldots, Y_T$ , we have that, as $T \to \infty$:

	$$\int_0^1(\hat{Q}_Y(u) - Q_Y(u))^2du \overset{P}{\to} 0\ , .$$
\begin{proof}
Note that, by Fubini theorem, we can always write:
$$\mathbb{E}\left[\int_0^1(\hat{Q}_Y(u) - Q_Y(u))^2 du\right]= \int_0^1\mathbb{E}[(\hat{Q}_Y(u) - Q_Y(u))^2]du = \int_0^1 g_T(u) du\, ,$$
where $g_T(u) = \mathbb{E}[(\hat{Q}_Y(u) - Q_Y(u))^2]$. Now, given that the distribution has a finite moment, and under the stated smoothness assumptions on $f\circ Q_Y$, $\lim_{T\to \infty} g_T(u)= 0$ for every $u \in (0,1)$ by Proposition 1 of \cite{Mason1984}. But then, we observe that, taking $\rho = \frac{2+\delta}{2} > 1$:

\begin{equation*}
	\begin{aligned}
		\int_0^1 g_T(u)^\rho du \leq \int_0^1\mathbb{E}[|\hat{Q}_Y(u) - Q_Y(u)|^{2+\delta}] du \leq\\  2^{2+\delta}\left(\mathbb{E}\left[\int_0^1|\hat{Q}_Y(v)|^{2+\delta} dv\right] + \mathbb{E}
	\left[\int_0^1|{Q}_Y(v)|^{2+\delta}dv\right] \right) = 	 \\
		2^{2+\delta}\left(\mathbb{E}\left[\frac{\sum_{t=1}^T|Y_t|^{2+\delta}}{T}\right] + \mathbb{E}[|Y_1|^{2+\delta}]\right) = 	2^{2+1+\delta} \mathbb{E}[|Y_1|^{2+\delta}]<\infty\, ,
	\end{aligned}
\end{equation*}
where the first inequality follows from Lyapunov inequality, and Fubini theorem is used in the second inequality. Since the quantity $2^{2+1+\delta} \mathbb{E}[|Y_1|^{2+\delta}]$ does not depend on $T$, it follows from Theorem 4.6.2 in Durrett that the sequence $\{g_t\}_t$ is  uniformly integrable. Consequently, by Theorem 4.6.3 of \cite{Durrett2019},  $\mathbb{E}\left[\int_0^1(\hat{Q}_Y(u) - Q_Y(u))^2 du\right] \to 0$, and a final application of Markov inequality yields that $\int_0^1(\hat{Q}_Y(u) - Q_Y(u))^2 du \overset{P}{\to} 0$, as desired.
\end{proof}
\end{lemma}

In the previous proof, we relied on a finite $(2+\delta)$-moment to establish uniform integrability of $g_T$. If one replaces this moment condition with the assumption that there exist real constants $C, k_1, k_2$ such that $f(Q_Y(u))^{-1} \leq C u^{k_1}(1-u)^{k_2}, \forall u \in (0,1)$, then \citet[pages 248-249]{Mason1984} shows that the distribution has a moment (and consequently, $g_T(u) \to 0$ for every $u \in (0,1)$ by his Proposition 1) and that the $\{g_t\}_t$ are uniformly integrable. Consequently, $\int_0^1g_T(u) du \to 0$, and consistency in $L^2(0,1)$ holds. We state this alternative result below:

\begin{lemma}
	Let $Y_1,\ldots, Y_T$ be a random sample from a distribution that admits Lebesgue density $f$ such that $u \mapsto f(Q_Y(u))$ is continuous, where $Q_Y$ is the quantile function of this distribution. If there exist real constants $C, k_1, k_2$ such that $f(Q_Y(u))^{-1} \leq C u^{k_1}(1-u)^{k_2}, \forall u \in (0,1)$, then, denoting by $\hat{Q}_Y$ the empirical quantiles from $Y_1,\ldots, Y_T$ , we have that, as $T \to \infty$:
		$$\int_0^1(\hat{Q}_Y(u) - Q_Y(u))^2du \overset{P}{\to} 0\ , .$$
\end{lemma}
\bibliographystyle{chicago}
\bibliography{bibliography}

%% file: sections/introduction.tex
\label{introduction}
L-moments, expected values of linear combinations of order statistics, were introduced by \cite{Hosking1990} and have been successfully applied in areas as diverse as computer science \citep{Hosking2007,Yang2021a}, hydrology \citep{Wang1997,Sankarasubramanian1999,Das2021,Boulange2021}, meteorology \citep{Wang2013,Simkova2017,Li2021}  and finance \citep{Gourieroux2008,Kerstens2011}. By appropriately combining order statistics, L-moments offer robust alternatives to traditional measures of dispersion, skewness and kurtosis. Models fit by matching sample L-moments (a procedure labeled ``method of L-moments'' by \cite{Hosking1990}) have been shown to outperform maximum likelihood estimators in small samples from flexible distributions such as generalised extreme value \citep{Hosking1985,Hosking1990}, generalised Pareto \citep{Hosking1987,Broniatowski2016}, generalised exponential \citep{Gupta2001} and Kumaraswamy \citep{Dey2018}.

Statistical analyses of L-moment-based parameter estimators rely on a framework where the number of moments is fixed \citep{Hosking1990,Broniatowski2016}. Practitioners often choose the number of L-moments equal to the number of parameters in the model, so as to achieve the order condition for identification. This approach is generally inefficient.\footnote{In the generalised extreme value distribution, there can be asymptotic root mean-squared error losses of 30\% with respect to the MLE when the target estimand are the distribution parameters \citep{Hosking1985,Hosking1990}. In our Monte Carlo exercise, we verify root mean squared error losses of over 10\% when the goal is tail quantile estimation.} It also raises the question of whether overidentifying restrictions, together with the optimal weighting of L-moment conditions, could improve the efficiency of ``method of L-moments'' estimators, as in the framework of generalised-method-of-moment (GMM) estimation \citep{Hansen1982}. Another natural question would be how to choose the number of L-moments in finite samples, as it is well-known from GMM theory that increasing the number of moments with a fixed sample size can lead to substantial biases \citep{Koenker1999,Newey2004}. In the end, one can only ask if, by correctly choosing the number of L-moments and under an appropriate weighting scheme, it may not be possible to construct an estimator that outperforms maximum likelihood estimation of some target quantities in small samples from popular distributions and yet achieves the Cramér-Rao bound asymptotically. Intuitively, the answer appears to be positive, especially if one takes into account that \cite{Hosking1990} shows L-moments characterise distributions with finite first moments.

The goal of this paper lies in answering the questions outlined in the previous paragraph. Specifically, we propose to study L-moment-based estimation in a context where: (i) the number of L-moments varies with sample size; and (ii) weighting is used in order to optimally account for overidentifying conditions. In this framework, we introduce a ``generalised'' method of L-moments estimator and analyse its properties. We provide sufficient conditions under which our estimator is consistent and asymptotically normal; we also derive the optimal choice of weights and introduce a test of overidentifying restrictions. We then show that, under independent and identically distributed (iid) data and the optimal weighting scheme, the proposed generalised L-moment estimator achieves the Cramér-Rao lower bound. We provide simulation evidence that our L-moment approach outperforms (in a mean-squared error sense) MLE estimation of some target quantities in smaller samples from popular distributions; while working as well as the MLE in larger sample sizes. We then construct methods to automatically select the number of L-moments used in estimation. For that, we rely on higher order expansions of the method-of-L-moment estimator, similarly to the procedure of \cite{Donald2001} and \cite{Donald2009} in the context of GMM. We use these expansions to find a rule for choosing the number of L-moments so as to minimise the estimated (higher-order) mean-squared incurred when targeting a given transformation of the model parameters of interest. We also consider an approach based on $\ell_1$-regularisation \citep{Luo2016}. We provide computational code to implement both methods,\footnote{The repository \url{https://github.com/luisfantozzialvarez/lmoments_redux} contains \texttt{R} script that implements our main methods, as well as replication code for our Monte Carlo exercise and empirical application.} and evaluate their performance through Monte Carlo simulations. With these tools, we aim to introduce a fully automated procedure for estimating parametric models that is able to improve upon maximum likelihood estimation in small samples from popular distributions, and yet has the \emph{guarantee} of not underperforming in larger datasets. As our examples and simulations throughout the paper indicate, our approach seems especially useful in tail quantile estimation of heavy-tailed distributions.\footnote{More generally, the question of whether our approach will generate significant small sample root mean squared error gains over the MLE for a given family of distributions and transformation of the model parameters of interest must be answered on a case-by-case basis. We provide tools to automatically select the number of L-moments used in estimation based on higher-order expansions of the mean-squared error that may be applied to any estimation target that is a smooth transformation of the model parameters. Moreover, our asymptotic efficiency result ensures that, asymptotically, there will be no losses in adopting this approach vis-à-vis the MLE.}

We also consider two extensions of our main approach. First, we show how the generalised method-of-L-moment approach introduced in this paper can be extended to the estimation of conditional models. Second, we show how our approach may be used in the analysis of the ``error term'' in semiparametric models, an important task in specification testing and the construction of prediction bands. We apply the latter extension to study the tail behaviour of expenditure patterns in a ridesharing platform in São Paulo, Brazil. We provide evidence that the heavy-tailedness in consumption patterns persists even after partialing out the effect of unobserved time-invariant heterogeneity and observable heterogeneity in consumption trends. We also show how our estimators can be used to construct prediction bands for individual treatment effects when one is interested in causal inference on individualised interventions. With these extensions, we hope more generally to illustrate how the generalised-method-of-L-moment approach to estimation may be a convenient tool in a variety of settings, e.g. when a model's quantile function is easier to evaluate than its likelihood. The latter feature has been explored in followup work by one of the authors in semi- and nonparametric settings \citep{alvarez2023quantile,alvarez2024learning}.

\paragraph{Related literature} This paper contributes to two main literatures. First, we relate to a couple of papers that, building on \citeauthor{Hosking1990}'s original approach, propose new L-moment-based estimators. \cite{Gourieroux2008} introduce a notion of L-moment for conditional moments, which is then used to construct a GMM estimator for a class of dynamic quantile models. As we argue in more detail in \Cref{extensions}, while conceptually attractive, their estimator is not asymptotically efficient (vis-à-vis the conditional MLE), as it focuses on a finite number of moment conditions and does not optimally explore the set of overidentifying restrictions available in the parametric model. In contrast, our proposed extension of the generalised method-of-L-moment estimator to conditional models is able to restore asymptotic efficiency.  In an unconditional setting, \cite{Broniatowski2016} propose estimating distribution functions by relying on a \emph{fixed} number of L-moments and a minimum divergence estimator that nests the empirical likelihood and generalized empirical likelihood estimators as particular cases. Even though these estimators are expected to perform better than (generalized) method-of-L-moment estimators in terms of higher-order properties \citep{Newey2004}, both would be \emph{first-order} inefficient (vis-à-vis the MLE) when the number of L-moments is held fixed. In this paper, we thus focus on improving L-moment-based estimation in terms of first-order asymptotic efficiency, by suitably increasing the number of L-moments with sample size and optimally weighting the overidentifying restrictions, while retaining its known good finite-sample behaviour. We do note, however, that one of our suggested approaches to select the number of L-moments aims at minimising an estimate of the higher-order mean-squared error, which may be useful in improving the higher-order behaviour of estimators even when a bounded (as a function of sample sizes) number of L-moments is used in estimation.

Secondly, we contribute to a literature that seeks to construct estimators that, while retaining asymptotic (first-order) unbiasedness and efficiency, improve upon maximum likelihood estimation in finite samples. The classical method to achieve finite-sample improvements over the MLE is through (higher-order) bias correction \citep{Pfanzagl1978}. However, analytical bias corrections may be difficult to implement in practice, which has led the literature to consider jackknife and bootstrap corrections \citep{hahn2002higher}. More recently, \cite{Ferrari2010} introduced a maximum $Lq$-likelihood estimator for parametric models that replaces the log-density in the objective function of the MLE with $\frac{f(x)^{1-q} -1}{1-q}$, where $q >0$ is a tuning parameter. They show that, by suitably choosing $q$ in finite samples, one is able to trade-off bias and variance, thus enabling MSE improvements over the MLE. Moreover, if $q \to 1$ asymptotically at a rate, the estimator is asymptotically unbiased and achieves the Crámer-Rao lower bound. There are some important differences between our approach and maximum $Lq$-likelihood estimation, though. First, we note that the theoretical justification for our construction is distinct from their method. Indeed, for a fixed number of L-moments, our proposed estimator is first-order asymptotically unbiased, whereas  the maximum $Lq$-likelihood estimator is inconsistent in an asymptotic regime with $q$ fixed and consistent but first-order biased if $q\to 1$ slowly enough. Therefore, whereas the choice of the tuning parameter $q$ is justified as capturing a tradeoff between first-order bias and variance; the MSE-optimal choice of L-moments in our setting concerns a tradeoff between the first-order variance of the estimator and its higher-order terms. This is precisely what we capture in our proposal to select the number of L-moments by minimising an estimator of the higher-oder MSE; whereas presently no general rule for choosing the tuning parameter $q > 0$ in maximum $Lq$-likelihood estimation exists \citep{Yang2021b}.

\paragraph{Structure of paper} The remainder of this paper is organised as follows. In the next section, we briefly review L-moments and parameter estimation based on these quantities. \Cref{properties} works out the asymptotic properties of our proposed estimator. In \Cref{monte_carlo} we conduct a small Monte Carlo exercise which showcases the gains associated with our approach. \Cref{selection} proposes methods to select the number of L-moments and assesses their properties in the context of the Monte Carlo exercise of \Cref{monte_carlo}. \Cref{extensions} presents the extensions of our main approach, as well as the empirical application. \Cref{conclusion} concludes. The Supplemental Appendix presents the proofs of the main results in the paper, as well as additional details on the methods of selection of L-moments, and the extensions to “residual analysis” and conditional models.

%% file: sections/motivation.tex
\section{L-moments: definition and estimation}
\label{motivation}

Consider a scalar random variable $Y$ with distribution function $F$ and finite first moment. For $r \in \mathbb{N}$, \cite{Hosking1990} defines the $r$-th L-moment as:
\begin{equation}
\label{eq_lmoments_quantile}
    \lambda_r \coloneqq \int_{0}^1 Q_Y(u) P^*_{r-1}(u) d u \, ,
\end{equation}
where $Q_Y(u) \coloneqq \inf \{y \in \mathbb{R}: F(y) \geq u\}$ is the quantile function of $Y$, and the functions $P^*_{r}(u) = \sum_{k=0}^r (-1)^{r-k} \binom{r}{k} \binom{r+k}{k} u^k$, $r \in \{0\}\cup \mathbb{N}$, are shifted Legendre polynomials.\footnote{Legendre polynomials are defined by applying the Gramm-Schmidt orthogonalisation process to the polynomials $1, x, x^2 ,x^3 \ldots $ defined on $[-1,1]$ \citep[p. 176-180]{Kreyszig1989}. If $P_r$ denotes the $r$-th Legendre polynomial, shifted Legendre polynomials are related to the standard ones through the affine transformation $P_r^*(u) =  P_r(2u-1)$ \citep{Hosking1990}.} Expanding the polynomials and using the quantile representation of a random variable \citep[Theorem 14.1]{Billingsley2012}, we arrive at the equivalent expression:
\begin{equation}
\label{eq_lmoments_linear}
    \lambda_r = r^{-1} \sum_{k=0}^{r-1} (-1)^k \binom{r-1}{k} \mathbb{E}[\tilde{Y}_{(r-k):r}] \, ,
\end{equation}
where, $\tilde{Y}_{j:l}$ is the $j$-th order statistic of a random sample from $F$ with $l$ observations. Equation \eqref{eq_lmoments_linear} motivates our description of L-moments as the expected value of linear combinations of order statistics. Notice that the first L-moment corresponds to the expected value of $Y$.

To see how L-moments may offer ``robust'' alternatives to conventional moments, it is instructive to consider, as in \cite{Hosking1990}, the second L-moment. In this case, we have:
\begin{equation*}
    \lambda_2 = \frac{1}{2}\mathbb{E}[\tilde{Y}_{2:2} - \tilde{Y}_{1:2}]= \frac{1}{2} \int \int \left( \max\{y_1, y_2 \} - \min\{y_1, y_2 \} \right) F(d y_1) F(d y_2) = \frac{1}{2} \mathbb{E}|\tilde{Y}_1 - \tilde{Y}_2| \, ,
\end{equation*}
where $\tilde{Y}_1$ and $\tilde{Y}_2$ are independent copies of $Y$. This is a measure of dispersion. Indeed, comparing it with the variance, we have:
\begin{equation*}
    \mathbb{V}[Y] = \mathbb{E}[(Y-\mathbb{E}[Y])^2] = \mathbb{E}[Y^2] - \mathbb{E}[Y]^2 = \frac{1}{2}\mathbb{E}[(\tilde{Y}_1 - \tilde{Y}_2)^2] \, ,
\end{equation*}
from which we note that the variance puts more weight to larger differences.

Next, we discuss sample estimators of L-moments. Let $Y_1, Y_2 \ldots Y_T$ be an identically distributed sample of $T$ observations, where each $Y_t$, $t=1,\ldots, T$, is distributed according to $F$. A natural estimator of the $r$-th L-moment is the sample analog of \eqref{eq_lmoments_quantile}, i.e.
\begin{equation}
\label{eq_est_lmoments_quantile}
    \hat{\lambda}_r = \int_{0}^1 \hat{Q}_Y(u)P_{r-1}^*(u) du \, ,
\end{equation}
where $\hat{Q}_Y$ is the left-continuous (\textit{càglàd}) empirical quantile process:
\begin{equation}
\label{empirical_qt}
    \hat{Q}_Y(u) = Y_{i:T}, \quad \text{if} \ \ \  \frac{i-1}{T} < u \leq \frac{i}{T} \, ,
\end{equation}
with $Y_{i:T}$ being the $i$-th sample order statistic. The estimator given by \eqref{eq_est_lmoments_quantile} is generally biased \citep{Hosking1990,Broniatowski2016}. When observations $Y_1, Y_2, \ldots Y_T$ may be assumed to be independent, researchers thus often resort to an unbiased estimator of $\lambda_r$, which is given by an empirical analog of \eqref{eq_lmoments_linear}:
\begin{equation}
\label{eq_est_lmoments_linear}
    \tilde{\lambda}_r = r^{-1} \sum_{k=0}^{r-1} (-1)^k \binom{r-1}{k} \binom{T}{r}^{-1} \sum_{1\leq i_1, i_2 \leq \ldots \leq i_r \leq T}Y_{i_{r-k}:T} \, .
\end{equation}

In practice, it is not necessary to iterate over all size $r$ subsamples of $Y_1, \ldots Y_T$ to compute the sample $r$-th L-moment through \eqref{eq_est_lmoments_linear}. \cite{Hosking1990} provides a direct formula that avoids such computation.

We are now ready to discuss the estimation of parametric models based on matching L-moments. Suppose that $F$ belongs to a parametric family of distribution functions $\{F_\theta: \theta \in \Theta\}$, where $\Theta \subseteq \mathbb{R}^d$ and $F = F_{\theta_0}$ for some $\theta_0 \in \Theta$. Let $l_r(\theta) \coloneqq \int_0^1 P^*_{r-1}(u) Q(u|\theta) du$ denote the theoretical $r$-th L-moment, where $Q(\cdot|\theta)$ is the quantile function associated with $F_\theta$. Let $H^R(\theta) \coloneqq (\lambda_1(\theta), \lambda_2(\theta), \ldots, \lambda_R(\theta))' $, and $\hat{H}^R$ be the vector stacking estimators for the first $R$ L-moments (e.g. \eqref{eq_est_lmoments_quantile} or \eqref{eq_est_lmoments_linear}). Researchers then  usually estimate $\theta_0$ by solving:
\begin{equation*}
    H^d(\theta) - \hat{H}^d = 0 \, .
\end{equation*}

As discussed in \Cref{introduction}, this procedure has been shown to lead to efficiency gains over maximum likelihood estimation in small samples from several distributions. Nonetheless, the choice of L-moments $R = d$ appears rather \textit{ad-hoc}, as it is based on an order condition for identification. One may then wonder whether increasing the number of L-moments used in estimation -- and weighting these properly --, might lead to a more efficient estimator in finite samples. Moreover, if one correctly varies the number of L-moments with sample size, it may be possible to construct an estimator that does not underperform MLE even asymptotically. The latter appears especially plausible if one considers the result in \cite{Hosking1990}, who shows that L-moments characterise a distribution with finite first moment. 

In light of the preceding discussion, we propose to analyse the behaviour of the ``generalised'' method of L-moments estimator:
\begin{equation}
\label{eq_many_l_moments}
    \hat{\theta} \in \text{arg inf}_{\theta \in \Theta} (H^R(\theta) - \hat{H}^R)'W^R (H^R(\theta) - \hat{H}^R)\, ,
\end{equation}
where $R$ may vary with sample size; and $W^R$ is a (possibly estimated) weighting matrix. In \Cref{properties}, we work out the asymptotic properties of this estimator in a framework where both $T$ and $R$ diverge, i.e. we index $R$ by the sample size and consider the asymptotic behaviour of the estimator in the large-sample limit, for sequences $(R_T)_{T \in \mathbb{N}}$ where $\lim_{T\to \infty}R_T=\infty$.\footnote{We maintain the indexing of $R$ by $T$ implicit to keep the notation concise. In this setting, the phrase ``as $T,R\to \infty$'' should be read as meaning that a property holds in the limit $T \to \infty$, for any sequence $(R_T)_{T \in \mathbb{N}}$ with $\lim_{T\to \infty}R_T = \infty$. The phrase ``as $T,R\to \infty$ with $\phi(R,T)\to c''$, where $\phi: \mathbb{N}^2 \mapsto \mathbb{R}$ and $c \in \mathbb{R}$, should be read as meaning that a property holds in the limit $T \to \infty$, for any sequence $(R_T)_{T \in \mathbb{N}}$ with $\lim_{T\to \infty}R_T = \infty$ and $\lim_{T \to \infty}\phi(R_T,T)=c$. }$^,$\footnote{As it will become clear in \Cref{properties}, our framework nests the setting with fixed $R$ as a special case by properly filling the weighting matrix $W^R$ with zeros.} We derive sufficient conditions for an asymptotic linear representation of the estimator to hold. We also show that the estimator is asymptotically efficient, in the sense that, under iid data and when optimal weights are used, its asymptotic variance coincides with the inverse of the Fisher information matrix. In \Cref{monte_carlo}, we conduct a small Monte Carlo exercise which showcases the gains associated with our approach. Specifically, we show that, for the task of tail quantile estimation, our L-moment approach entails mean-squared error gains over MLE in smaller samples, and performs as well as it in larger samples.\footnote{In the Supplemental Appendix, we consider an alternative setting where the estimation target consists of linear combinations of the model parameters, and also verify gains in adopting our approach.} In light of these results, in \Cref{selection} we propose to construct a semiautomatic method of selection of the number of L-moments by working with higher-order expansions of the mean-squared error of the estimator -- in a similar fashion to what has already been done in the GMM literature \citep{Donald2001, Donald2009, Okui2009, Abadie2019}. We also consider an approach based on $\ell_1$-regularisation borrowed from the GMM literature \citep{Luo2016}. We then return to the Monte Carlo setting of \Cref{monte_carlo} in order to assess the properties of the proposed selection methods. In Section \ref{extensions}, we consider extensions of our main approach to the estimation of conditional models and a class of semiparametric models.

In this paper, we will focus on the case where estimated L-moments are given by \eqref{eq_est_lmoments_quantile}. As shown in \cite{Hosking1990}, under random sampling and finite second moments, for each  $r \in \mathbb{N}$,  $\hat{\lambda}_r - \tilde{\lambda}_r = O_p(T^{-1})$, which implies that the estimator \eqref{eq_many_l_moments} using either \eqref{eq_est_lmoments_quantile} or \eqref{eq_est_lmoments_linear} as $\hat{H}^R$ are first-order asymptotically equivalent \emph{when $R$ is fixed}. However, in an asymptotic framework where $R$ increases with the sample size, this need not be the case. Indeed, note that, for $r > T$, $\tilde{\lambda}_r$ is not even defined, whereas $\hat \lambda_r$ is. Relatedly, the simulations in \Cref{monte_carlo} show that, for values of $R$ close to $T$, the generalised-method-of L-moment estimator \eqref{eq_many_l_moments}  based on $\tilde{\lambda}_r$ breaks down, whereas the estimator based on $\hat\lambda_r$ does not.\footnote{The latter phenomenon is corroborated by our theoretical results in \Cref{properties} for the estimator based on $\hat \lambda_r$, which allow $R$ to be much larger that $T$.} We thus focus on the properties of the estimator that relies on $\hat \lambda_r$, as it is especially well-suited for settings where one may wish to make $R$ large.

\begin{remark}[On computation of integrals]
	\label{remark_computation}
	Parameter estimation through L-moments hinges crucially on computation of integrals $\int_{0}^1 Q(u) u^r du$, for a given quantile function $Q$ and $r \in \{0,1,\ldots,R-1\}$. These quantities, known in the literature as probability-weighted moments \citep{Landwehr1979}, are directly available in closed form when $Q$ is stepwise-constant (as when $Q$ are the empirical quantiles), though this need not be the case for the theoretical moments  $\int_0^1 Q(u|\theta) u^r du$ from parametric families of distributions. \cite{hosking1986theory} provides closed-form formulae for the probability-weighted moments of several families of distributions, including those from the Generalised Pareto (GPD) and Generalised Extreme Value (GEV) families that are popular in the modelling of extreme events, and which we consider in the Monte Carlo exercise of Section \ref{monte_carlo}.\footnote{For completeness, we reproduce the closed-form expressions of these integrals in the GEV and GPD families in Supplemental Appendix M.} For those distributions where the integral does not admit a closed-form expression, the R package \texttt{lmom} \citep{Hosking2024} provides access to Fortran routines that compute probability-weighted moments using numerical integration.
\end{remark}

\begin{remark}[L-moment estimation as a computationally attractive alternative to the MLE]
	There are instances where L-moment-based parameter estimates are easily computable, whereas maximum likelihood estimation can be computationally complicated. One example is that given by \emph{quantile mixtures}, e.g. when the quantile function of the distribution of interest is given by $\theta_1 Q_1(u) + \theta_2 Q_2(u)$, with $\theta_1$ and $\theta_2$ unknown constants and $Q_1$ and $Q_2$ known quantile functions whose L-moments are easily computable. In this case, the conventional method-of L-moment estimator collapses to solving a linear system, and finding our generalised method-of-L-moment  estimator amounts to solving a quadratic program. In contrast, estimation of quantile mixtures through maximum likelihood can be much more complicated computationally, as it involves, for each candidate parameter value, differentiation of the inverse of the quantile function, which is generally unavailable in closed-form. The estimation of quantile mixtures through the method-of-L-moments has been studied by \cite{karvanen2006estimation} and \cite{Gourieroux2008} in the context of modelling asset returns, while \cite{alvarez2023quantile}, building on our proposed generalised-method-of-L-moment estimator, study quantile mixture models as a general tool for approximating a distribution of interest, with a particular focus in causal inference on distributional outcomes in observational settings.
\end{remark}

%% file: sections/properties.tex
\subsection{Setup} As in the previous section, we consider a setting where we have a sample with $T$ identically distributed observations, $Y_1, Y_2 \ldots Y_T$, $Y_t \sim F$ for $t = 1 , 2 \ldots T$, where $F$ belongs to a parametric family $\{F_\theta : \theta \in \Theta\}$, $\Theta \subseteq \mathbb{R}^d$; and $F = F_{\theta_0}$ for some $\theta_0 \in \Theta$. We will analyse the behaviour of the estimator:
\begin{equation}
	\small
\label{eq_objective_function}
    \hat{\theta} \in \underset{\theta \in \Theta}{\text{arg inf}} \, \, \sum_{k =1}^R \sum_{l=1}^R\left(\int_{\underline{p}}^{\bar{p}} \left[ \hat{Q}_Y (u)  - Q_Y(u|\theta)\right]  P_k(u) d u \right)w_{k,l}^R\left(\int_{\underline{p}}^{\bar{p}} \left[ \hat{Q}_Y (u)  - Q_Y(u|\theta)\right] P_l(u) d u \right) \, ,
\end{equation}
where $\hat{Q}_Y(\cdot)$ is the empirical quantile process given by \eqref{empirical_qt}; $Q_{Y}(\cdot|\theta)$ is the quantile function associated with $F_{\theta}$;  $\{w_{k,l}^R\}_{1\leq k, l \leq R}$ are a set of (possibly estimated) weights; $\{P_{k}\}_{1\leq k \leq R}$ are a set of quantile ``weighting'' functions with $\int_{0}^1 P_k(u)^2 du = 1$; and $0 \leq \underline{p} < \bar{p} \leq 1$. This setting encompasses the generalised-method-of-L-moment estimator discussed in the previous section. Indeed, by choosing $P_k(u) = \sqrt{2(2k-1)}\cdot P^*_{k-1}(u)$, where $P^*_{k}(u)$ are the shifted Legendre polynomials on $[0,1]$, and $0 = \underline{p} < \bar{p} = 1$, we have the generalised L-moment-based estimator in \eqref{eq_many_l_moments} using \eqref{eq_est_lmoments_quantile} as an estimator for the L-moments.\footnote{The rescaling by $\sqrt{2(2k-1)}$ is adopted so the polynomials have unit $L^2[0,1]$-norm.} We leave $\underline{p} < \bar{p}$ fixed throughout.\footnote{In \Cref{rmk_trimming} later on, we briefly discuss an extension to sample-size-dependent trimming.} All limits are taken \textbf{jointly} with respect to $T$ \textbf{and} $R$.

To facilitate analysis, we let $\mathbf{P}^R(u) \coloneqq (P_1(u), P_2(u) \ldots P_R(u))' $; and write $W^R$ for the $R \times R$ matrix with entry $W_{i,j}^R = w_{i,j}^R$. We may then rewrite our estimator in matrix form as:
\begin{equation*}
	\small
   \begin{aligned}
    \hat{\theta} \in \text{arg inf}_{\theta \in \Theta} \left[\int_{\underline{p}}^{\bar{p}} \left(\hat{Q}_Y (u) - Q_Y(u|\theta) \right)  \mathbf{P}^R(u)' du \right] W^R\left[\int_{\underline{p}}^{\bar{p}}\left(\hat{Q}_Y (u) - Q_Y(u|\theta) \right)  \mathbf{P}^R(u) du \right]\, .
\end{aligned} 
\end{equation*}

\subsection{Consistency}
In this section, we present conditions under which our estimator is consistent. 

We impose the following assumptions on our environment. In what follows, we write $Q_Y(\cdot) = Q_Y(\cdot|\theta_0)$.
\begin{assumption}[Consistency of empirical quantile process] \label{ass_consistency} The empirical quantile process is uniformly consistent on $(\underline{p},\bar{p})$, i.e.
\begin{equation}
\label{eq_uniform_cons_quantile}
 \sup_{u \in (\underline{p},\bar{p})}\lvert \hat{Q}_{Y}(u) - Q_Y(u) \rvert \overset{P}{\to} 0 \, .
\end{equation}
\end{assumption}

Assumption \ref{ass_consistency} is satisfied in a variety of settings. For example, if $Y_1$, $Y_2$ \ldots $Y_T$ are iid and the family $\{F_{\theta}:\theta \in \Theta\}$ is continuous with a (common) compact support; then \eqref{eq_uniform_cons_quantile} follows with $\underline{p} = 0$ and $\overline{p} = 1$ \citep[Proposition 2.1.]{Coutrix2016}. \cite{Yoshihara1995} and \cite{Portnoy1991} provide sufficient conditions for uniform consistency \eqref{eq_uniform_cons_quantile} to hold when observations are dependent. We also note that, for the result in this section, it would have been sufficient to assume convergence in probability in the $L^2(\underline{p},\bar{p})$ norm.\footnote{Under random sampling from a distribution with Lebesgue density $f$ such that $u\mapsto f(Q_Y(u))$ is continuous on $(0,1)$, empirical quantiles are consistent in $L^2(0,1)$ if one of the two conditions hold: the distribution has finite $(2+\delta)$-moment, or there exist  real constants $C, k_1, k_2$ such that $f(Q_Y(u))^{-1} \leq C u^{k_1}(1-u)^{k_2}, \forall u \in (0,1)$. See Supplemental Appendix N for a proof.   } We only state results in the $\sup$-norm because convergence statements regarding the empirical quantile process available in the literature are usually proved in $L^\infty (\underline{p},\bar{p})$. 

\begin{assumption}[Quantile weighting functions]
\label{ass_ortho}
The functions $\{P_l : l \in \mathbb{N}\}$ constitute an orthonormal sequence on $L^2[0,1]$.
\end{assumption}

Assumption \ref{ass_ortho} is satisfied by (rescaled) shifted Legendre polynomials, shifted Jacobi polynomials and other weighting functions.

Next, we impose restrictions on the estimated weights. In what follows, we write, for a $c\times d$ matrix $A$, $\lVert A \rVert_2 = \sqrt{\lambda_{\max}(A'A)}$.

\begin{assumption}[Estimated weights]
\label{ass_weights}
    There exists a sequence of nonstochastic symmetric positive semidefinite matrices $\Omega^R$ such that, as $T,R\to \infty$, $\lVert W^R - \Omega^R  \rVert_2 = o_{P^*}(1)$;\footnote{The notation $o_{P*}(1)$ expresses convergence in outer probability to zero. We state our main assumptions and results in outer probability in order to abstract from measurability concerns. We note these results are equivalent to convergence in probability when the appropriate measurability assumptions hold.} $\lVert \Omega^R \rVert_{2} = O(1)$. 
\end{assumption}

Assumption \ref{ass_weights} restricts the range of admissible weight matrices. Notice that $W^R = \Omega^R = \mathbb{I}_R$ trivially satisfies these assumptions. By the triangle inequality, Assumption \ref{ass_weights} implies that $\lVert W^R \rVert_2 = O_{P^*}(1)$.

Finally, we introduce our identifiability assumption. For some $X \in L^2[0,1]$, let $\lVert X \rVert_{L^2[0,1]} = \left(\int_{0}^1 X(u)^2 du \right)^{\frac{1}{2}}$:

\begin{assumption}[Strong identifiability and suprema of $L^2$ norm of parametric quantiles]
\label{ass_identification}
 For each $\epsilon > 0$:
   {\footnotesize \begin{equation*}
        \liminf_{R \to \infty} \inf_{\theta \in \Theta: \lVert \theta - \theta_0 \rVert_2\geq \epsilon }  \left[\int_{\underline{p}}^{\bar{p}} \left(Q_{Y}(u|\theta) - Q_{Y}(u|\theta_0)\right) \mathbf{P}^R(u)' du \right] \Omega^R\left[\int_{\underline{p}}^{\bar{p}} \left(Q_{Y}(u|\theta) - Q_{Y}(u|\theta_0)\right) \mathbf{P}^R(u) du \right] > 0 \, .
    \end{equation*}}
    Moreover, we require that  $\sup_{\theta \in \Theta} \lVert Q_{Y}(\cdot|\theta) \mathbbm{1}_{[\underline{p},\bar{p}]} \rVert _{L^2[0,1]} < \infty$.
\end{assumption}

The first part of this assumption is closely related to the usual notion of identifiability in parametric distribution models. Indeed, if $0=\underline{p}<\overline{p}=1$, $\Theta$ is compact, $\theta \mapsto \lVert Q(\cdot|\theta) \rVert_{L^2[0,1]}$ is bounded and $(\theta',\theta'') \mapsto \lVert Q(\cdot|\theta') - Q(\cdot|\theta'') \rVert_{L^2[0,1]}$ is continuous,  the $\{P_l\}_l$ constitute an orthonormal \emph{basis} in $L^2[0,1]$ (this is the case for rescaled shifted Legendre polynomials), and if the smallest eigenvalue of $\Omega_R$ is bounded away from zero uniformly in $R$ (for example, if we take  $W^R = \mathbb{I}_R$), then the first part is equivalent to identifiability of the parametric family $\{F_\theta\}_\theta$ (see Supplemental Appendix C.1 for a proof).

As for the second part of the assumption,  we note that boundedness of the $L^2$ norm of parametric quantiles uniformly in $\theta$ is satisfied in several settings. If the parametric family $\{F_\theta : \theta \in \Theta\}$ has common compact support, then the assumption is trivially satisfied. Alternatively, if we assume $\Theta$ is compact and $Q_{Y}(u|\theta)$ is jointly continuous and bounded on $ [\underline{p},\bar{p}] \times \Theta$, then the condition follows from Weierstrass' theorem, as in this case: $\sup_{\theta \in \Theta} \lVert Q_{Y}(\cdot|\theta) \mathbbm{1}_{[\underline{p},\bar{p}]} \rVert _{L^2[0,1]} \leq  \sqrt{\bar{p} - \underline{p}} \cdot  \sup_{(\theta,u) \in  \Theta \times [\underline{p},\bar{p}]} |Q_{Y}(u|\theta)| < \infty$. More generally, if the support of the family under consideration is unbounded, then we may ensure that the assumption is satisfied by considering compact parameter spaces, or by a proper choice of trimming constants $\underline{p}$ and $\overline{p}$. For example, if we assume that $\Theta$ is compact, and that $\theta \mapsto \int_{\underline{p}}^{\overline{p}}Q_{Y}(u|\theta)^2 du$ is continuous and bounded on $\Theta$, then the  condition is satisfied. Supplemental Appendix B shows that, for the GEV and GPD families of distributions mentioned in Remark \ref{remark_computation} and considered in the Monte Carlo Exercise, the uniform boundedness assumption is satisfied with $0=\underline{p} < \overline{p}=1$ and compact parameter spaces in the region where the distributions have finite variance. Moreover, we show that, by taking $0<\underline{p}$ and $\overline{p}<1$ in the GEV family and $\overline{p}<1$ in the GPD family, it is possible to extend these parameter spaces to the region where the distributions have infinite variance.\footnote{As we show in the Appendix, it is actually possible to extend the parameter space to regions where even the first moment does not exist, since, in this case, even though untrimmed L-moments are not defined, trimmed L-moments are. We discuss a data-driven method to select the trimming constants in Remark \ref{rmk_trimming}.}

Under the previous assumptions, the estimator is consistent. 

\begin{proposition}
\label{prop_consistency}
Suppose Assumptions \ref{ass_consistency} to \ref{ass_identification} hold. Then $\hat{\theta} \overset{P^*}{\to} \theta_0$ as $R,T \to \infty$.
\end{proposition} 
\begin{proof}
	See Supplemental Appendix A.1.
\end{proof}

\begin{remark}
	\label{rmk_rate_cons}
Note that Proposition \ref{prop_consistency} does not impose any restrictions on the rate of growth of L-moments. This stands in contrast with consistency results in the literature exploring the behaviour of GMM in asymptotic sequences with an increasing number of moments. For example, when estimating a finite-dimensional parameter identified by a conditional moment restriction through many unconditional moments that span the available restrictions, the series-IV estimator proposed by \cite{Donald2003} is consistent in an asymptotic regime where the number of moments $R$ satisfies $R/T\to 0$. Intuitively, one needs to impose this growth restriction in order to control the variance of an increasing number of moments, even in this particular case where moments are derived from series regressors. In contrast, the special structure of L-moments in our setting, being written as the projection coefficients of the \emph{same} quantile function on an orthonormal sequence in $L^2[0,1]$, enables us to properly control the variance even when $R$ is arbitrarily large, for Bessel's inequality \citep[page 157]{Kreyszig1989} implies that, for every $R$, $\left\lVert\int_{\underline{p}}^{\bar{p}} \left(\hat{Q}_Y(u)- Q_{Y}(u)\right) \mathbf{P}^R(u) du \right\rVert_2 \leq \lVert( \hat{Q}_Y(\cdot)- Q_{Y}(\cdot)) \mathbbm{1}_{[\underline{p},\bar{p}]} \rVert _{L^2[0,1]}$, with the upper bound crucially not depending on $R$. See Supplemental Appendix D for a detailed  comparison between the consistency arguments underlying our Proposition \ref{prop_consistency} and Theorem 5.1 of \cite{Donald2003}.
\end{remark}

\subsection{Asymptotic linear representation} In this section, we provide conditions under which the estimator admits an asymptotic linear representation. In what, follows, define $h^R(\theta) := \int_{\underline{p}}^{\bar{p}} \left(\hat{Q}_Y (u) - Q_Y(u|\theta) \right)  \mathbf{P}^R(u) du $; and write $ \nabla_{\theta'} h^R(\tilde{\theta})$ for the Jacobian of $h^R$ with respect to $\theta$, evaluated at $\tilde{\theta}$.  We assume that:

\begin{assumption}
	\label{ass_mvt_weak}
	There exists an open ball $\mathcal{O}$ in $\mathbb{R}^d$ containing $\theta_0$ such that $\mathcal{O} \subseteq \Theta$ and $Q_{Y}(u|\theta)$ is differentiable on $\mathcal{O}$, uniformly in $u \in [\underline{p},\bar{p}]$. Moreover, $\theta \mapsto Q_{Y}(u|\theta)$ is \textbf{continuously} differentiable on $\mathcal{O}$ for each $u$; and, for each $\theta \in \mathcal{O}$, $\nabla_{\theta'} Q_Y(\cdot|\theta)$ is square integrable  on $[\underline{p},\bar{p}]$.
\end{assumption}

\begin{assumption}
	\label{ass_weak_convergence}
	$\sqrt{T}( \hat{Q}_Y(\cdot) - Q_Y(\cdot))$ converges weakly in $L^\infty (\underline{p},\bar{p})$ to a zero-mean Gaussian process $B$ with continuous sample paths and covariance kernel $\Gamma$.
\end{assumption}

\begin{assumption}
	\label{ass_mvt_strong}
	$Q_{Y}(u|\theta)$ is \textbf{twice} continuously differentiable on $\mathcal{O}$, for each $u \in [\underline{p},\overline{p}]$. Moreover,  $ \sup_{\theta \in \mathcal{O}}\sup_{u \in [\underline{p},\bar{p}]}\lVert \nabla_{\theta \theta'}Q_{Y}(u|\theta)\rVert_2 < \infty$.
\end{assumption}

\begin{assumption}
\label{ass_eigen}
The smallest eigenvalue of $\nabla_{\theta'} h^R(\theta_0)' \Omega^R \nabla_{\theta'} h^R(\theta_0)$ is bounded away from $0$, uniformly in $R$.
\end{assumption}

Assumption \ref{ass_mvt_weak} requires $\theta_0$ to be an interior point of $\Theta$. It also implies the objective function is countinuously differentiable on a neighbourhood of $\theta_0$, which enables us to linearise the first order condition satisfied with high probability by $\hat \theta$.

Weak convergence of the empirical quantile process (Assumption \ref{ass_weak_convergence}) has been derived in a variety of settings, ranging from iid data \citep[Corollary 21.5]{Vaart1998} to nonstationary and weakly dependent observations \citep{Portnoy1991}. In the iid setting, if the family $\{F_{\theta}:\theta \in \Theta\}$ is continuously differentiable with strictly positive density $f_\theta$ over a (common) compact support; then weak-convergence holds with $\underline{p} = 0$ and $\overline{p} = 1$. In this case, the covariance kernel is $\Gamma(i,j) = \frac{(i\land j - ij)}{f_Y(Q_{Y}(i))f_Y(Q_{Y}(j)) }$. Similarly to the discussion of Assumption \ref{ass_consistency}, Assumption \ref{ass_weak_convergence} is stronger than necessary: it would have been sufficient to assume $\lVert \sqrt{T} (Q_Y(\cdot) - \hat{Q}_Y(\cdot)) \mathbbm{1}_{[\underline{p},\bar{p}]} \rVert_{L^2[0,1]}^2 = O_{P^*}(1)$, which is implied by weak convergence in $L^2(\underline{p},\overline{p})$ (see \cite{Mason1984} and \cite{Barrio2005} for results in this direction). 

Assumption \ref{ass_mvt_strong} is a technical condition which enables us to provide an upper bound to the linearisation error of the first order condition satisfied by $\hat \theta$.

Assumption \ref{ass_eigen} is similar to the rank condition used in the proof of asymptotic normality of M-estimators \citep{Newey1994}, which is known to be equivalent to a local identification condition under rank-regularity assumptions \citep{Rothenberg1971}. In our setting, where $R$ varies with sample size, we show in Supplemental Appendix C.2 that a stronger version of Assumption \ref{ass_identification} implies Assumption \ref{ass_eigen}.

Under Assumptions \ref{ass_consistency}-\ref{ass_eigen}, we have that:

\begin{proposition}
\label{prop_asymptotic_linear}
Suppose Assumptions \ref{ass_consistency}-\ref{ass_eigen} hold. Then, as $T,R\to \infty$, the estimator admits the asymptotic linear representation:
\begin{equation}
\label{eq_asymptotic linear_prop}
      \sqrt{T}(\hat{\theta} - \theta_0) = - ( \nabla_{\theta'} h^R(\theta_0)' \Omega^R \nabla_{\theta'} h^R(\theta_0))^{-1} \nabla_{\theta'} h^R(\theta_0)' \Omega^R (\sqrt{T} h^R(\theta_0)) + o_{P^*}(1) \, .
\end{equation}
\end{proposition}
\begin{proof}
	See Supplemental Appendix A.2.
\end{proof}

In the next subsection, we work out an asymptotic approximation to the distribution of the leading term  in \eqref{eq_asymptotic linear_prop}. 

\begin{remark}
	\label{remark_rate_linear}
	Note that our linearisation result in Proposition \ref{prop_asymptotic_linear} does not impose any restrictions on the rate of growth of L-moments, which again contrasts with existing results in the GMM literature \citep{Koenker1999,Donald2003,Han2006}, where rate restrictions are typically required in order to establish an asymptotic linear representation. This difference may again be attributed to the special structure of L-moments in our setting. Indeed, whereas the asymptotic normality result on the series-IV estimator discussed in \cite{Donald2003} assumes the rate restriction $R/T^2\to 0$ in order to control a crucial bias term in the asymptotic linear representation stemming from correlation between the gradient of the empirical moment condition at the true parameter and sample moments at the truth, the fact that, in our setting, the gradient of the difference between empirical and theoretical L-moments at the truth is not affected by estimation error of the empirical quantiles $\hat{Q}_Y$, coupled with Bessel's inequality, enables us to control the linearisation error without such restriction. See Supplemental Appendix D for further discussion.
\end{remark}

\subsection{Asymptotic distribution} \label{as_distribution} Finally, to work out the asymptotic distribution of the proposed estimator, we rely on \emph{a strong approximation concept}. The idea is to construct, in the \emph{same} underlying probability space, a sequence of Brownian bridges that approximates, in the supremum norm, the empirical quantile process. This can then be used to conduct inference based on a Gaussian distribution. In Supplemental Appendix H, we alternatively show how a \emph{Bahadur-Kiefer representation} of the quantile process can be used to conduct inference in the iid case. In this alternative, one approximates the distribution of the leading term of \eqref{eq_asymptotic linear_prop} by a transformation of independent Bernoulli random variables.

We first consider a strong approximation to a Gaussian process in the iid setting. We state below a classical result, due to \cite{Csorgo1978}:

\begin{theorem}[\cite{Csorgo1978}]
\label{thm_gaussian_iid}
Let $Y_1, Y_2 \ldots Y_T$ be an iid sequence of random variables with a continuous distribution function $F$ which is also twice differentiable on $(a,b)$, where $-\infty \leq a = \sup\{z: F(z) = 0 \}$ and $b=\inf\{z: F(z) = 1\}\leq \infty$. Suppose that $F'(z) = f(z) > 0$ for $z \in (a,b)$. Assume that, for $\gamma > 0$:
\begin{equation*}
    \sup_{a < x < b} F(x)(1-F(x))\left|\frac{f'(x)}{f^2(x)}\right| \leq \gamma \, ,
\end{equation*}
where $f$ denotes the density of $F$.  Moreover, assume  that $f$ is nondecreasing (nonincreasing) on an interval to the right of $a$ (to the left of $b$). Then, if the underlying probability space is rich enough, one can define, for each $t \in \mathbb{N}$, a Brownian bridge $\{B_t(u) : u \in [0,1]\}$ such that, if $\gamma < 2$:
\begin{equation}
\label{eq_strong_approx}
    \sup_{0 < u < 1} |\sqrt{T}f(Q_Y(u))(\hat{Q}_{Y}(u) - Q_{Y}(u)) - B_T(u)| \overset{a.s.}{=} O(T^{-1/2} \log(T)) \, ,
\end{equation}
and, if $\gamma \geq 2$
\begin{equation}
	\small
\label{eq_strong_approx_gamma}
\sup_{0 < u < 1} |\sqrt{T}f(Q_Y(u))(\hat{Q}_{Y}(u) - Q_{Y}(u))  - B_T(u)| \overset{a.s.}{=} O(T^{-1/2} (\log \log T)^{\gamma} (\log T)^{\frac{(1+\epsilon)}{(\gamma -1)}}) \, ,
\end{equation}
for arbitrary $\epsilon > 0$.
\end{theorem}

The above theorem is stronger than the weak convergence of Assumption \ref{ass_weak_convergence}. Indeed, Assumption \ref{thm_gaussian_iid} requires variables to be defined in the same probability space and yields explicit bounds in the sup norm; whereas weak convergence is solely a statement on the convergence of integrals \citep{Vaart1996}. Suppose the approximation \eqref{eq_strong_approx}/\eqref{eq_strong_approx_gamma} holds in our context. Let $B_T$ be as in the statement of the theorem, and assume in addition that $\int_{\underline{p}}^{\overline{p}}\frac{1}{f_Y(Q_Y(u))^2} du < \infty$. A simple application of Bessel's inequality then shows that:
\begin{equation}
	\small
\label{eq_asymptotic distr}
      \sqrt{T}(\hat{\theta} - \theta_0) = - ( \nabla_{\theta'} h^R(\theta_0)' \Omega^R \nabla_{\theta'} h^R(\theta_0))^{-1} \nabla_{\theta'} h^R(\theta_0)' \Omega^R \left[ \int_{\underline{p}}^{\bar{p}} \frac{B_T(u)}{f_Y(Q_Y(u))} \mathbf{P}^R(u) du \right]  + o_{P^*}(1) \, .
\end{equation}

Note that the distribution of the leading term in the right-hand side is known (by Riemann integration, it is Gaussian) up to $\theta_0$. This representation could thus be used as a basis for inference. The validity of such approach can be justified by verifying that the Kolmogorov distance between the distribution of $\sqrt{T}(\hat{\theta} - \theta_0)$ and that of the leading term of the representation goes to zero as $T$ and $R$ increase. We show that this indeed is true later on, where convergence in the Kolmogorov distance is obtained as a byproduct of weak convergence.

Next, we reproduce a strong approximation result in the context of dependent observations. The result is due to \cite{Fotopoulos1994} and \cite{Yu1996}.

\begin{theorem}[\cite{Fotopoulos1994,Yu1996}] \label{thm_gaussian_mixing}
Let $Y_1, Y_2 \ldots Y_T$ be a strictly stationary, $\alpha$-mixing sequence of random variables, with mixing coefficient satisfying $\alpha(t) = O(t^{-8})$. Let $F$ denote the distribution function of $Y_1$. Suppose the following \textbf{Csorgo and Revesz conditions} hold:

\begin{enumerate}
    \item[a.] $F$ is twice differentiable on $(a,b)$, where $-\infty \leq a = \sup\{z: F(z) = 0 \}$ and $b=\inf\{z: F(z) = 1\}\leq \infty$;
    \item[b.] $\sup_{0<s<1} |f'(Q_{Y}(s))| < \infty$; 
\end{enumerate}
as well as the condition:

\begin{itemize}
    \item[c.] $\inf_{0<s< 1} f(Q_Y(s)) > 0 $.
\end{itemize}

Let $\Gamma(s,t) \coloneqq \mathbb{E}[g_1(s) g_1(t)] + \sum_{n=2}^\infty \{ \mathbb{E}[g_1(s)g_n(t)] + \mathbb{E}[g_1(t)g_n(s)] \}$, where $g_n(u) \coloneqq \mathbbm{1}\{U_n \leq u\} - u$ and $U_n \coloneqq F(Y_n)$. Then, if the probability space is rich enough, there exists a sequence of Brownian bridges $\{\tilde{B}_n: n \in \mathbb{N}\}$ with covariance kernel $\Gamma$ and a positive constant $\lambda > 0$ such that:
\begin{equation}
\label{eq_strong_approx_mixing}
    \sup_{0 < u < 1} |\sqrt{T}(\hat{Q}_{Y}(u) - Q_{Y}(u))  - f(Q_Y(u))^{-1}\tilde{B}_T(u)| \overset{a.s.}{=} O((\log T)^{-\lambda}) \, .
\end{equation}
\end{theorem}

A similar argument as the previous one then shows that, under the conditions of the theorem above:
\begin{equation}
	\small
\label{eq_asymptotic distr_mixing}
      \sqrt{T}(\hat{\theta} - \theta_0) = -( \nabla_{\theta'} h^R(\theta_0)' \Omega^R \nabla_{\theta'} h^R(\theta_0))^{-1} \nabla_{\theta'} h^R(\theta_0)' \Omega^R \left[ \int_{\underline{p}}^{\bar{p}} \frac{\tilde{B}_T(u)}{f_Y(Q_Y(u))} \mathbf{P}^R(u) du \right]  + o_{P^*}(1) \, .
\end{equation}

Differently from the iid case, the distribution of the leading term on the right-hand side is now known up to $\theta_0$ \textbf{and the covariance kernel} $\Gamma$. The latter could be estimated with a \cite{Newey1987} style estimator.

To conclude the discussion, we note that the strong representation \eqref{eq_asymptotic distr} (resp. \eqref{eq_asymptotic distr_mixing}) allows us to establish asymptotic normality of our estimator. Indeed, let $L_T$ be the leading term of the representation on the right-hand side of \eqref{eq_asymptotic distr} (resp. \eqref{eq_asymptotic distr_mixing}), and $V_{T,R}$ be its variance. Observe that ${V}_{T,R}^{-1/2} L_T$ is distributed according to a multivariate standard normal. It then follows by Slutsky's theorem that $V_{T,R}^{-1/2}\sqrt{T}(\hat{\theta}-\theta_0) \overset{d}{\to} N(0, \mathbb{I}_d)$. Since pointwise convergence of cumulative distribution functions to a continuous distribution function implies uniform convergence \citep[page 438]{Parzen1960}, and given that  $V_{T,R}^{-1/2}$ is positive definite, we obtain that:
\begin{equation}
\label{eq_kolmogorov_bound} 
    \lim_{T,R \to \infty} \sup_{c \in \mathbbm{R}^d}|P[\sqrt{T}(\hat{\theta}-\theta_0) \leq c] - P[ L_T \leq c]| = 0 \, ,
\end{equation}
which justifies our approach to inference based on the distribution of the leading term on the right-hand side of \eqref{eq_asymptotic distr} (resp. \eqref{eq_asymptotic distr_mixing}).
	
We collect the main results in this subsection under the corollary below. 

\begin{corollary}
	\label{corollary_asymptotic}
Suppose Assumptions \ref{ass_consistency}-\ref{ass_eigen} hold. Moreover, suppose a strong approximation condition such as \eqref{eq_strong_approx}/\eqref{eq_strong_approx_gamma} or \eqref{eq_strong_approx_mixing} is valid; and, in addition, that $\int_{\underline{p}}^{\overline{p}}\frac{1}{f_Y(Q_Y(u))^2} du < \infty$. Then, as $T,R\to\infty$, the approximation \eqref{eq_asymptotic distr} (resp. \eqref{eq_asymptotic distr_mixing}) holds. Moreover, we have that, as $T,R\to\infty$, $V_{T,R}^{-1/2}\sqrt{T}(\hat{\theta}-\theta_0) \overset{d}{\to} N(0, \mathbb{I}_d)$ and that \eqref{eq_kolmogorov_bound} holds.
\end{corollary}

\begin{remark}[Optimal choice of weighting matrix under Gaussian approximation] Under \eqref{eq_asymptotic distr}, the optimal choice of weights that minimises the variance of the leading term is:
\begin{equation}
\label{eq_optimal_weights}
    \Omega_R^* = \mathbb{E} \left[ \left(\int_{\underline{p}}^{\bar{p}} \frac{B_T(u)}{f_Y(Q_y(u))} \mathbf{P}^R(u) du\right)  \left(\int_{\underline{p}}^{\bar{p}} \frac{B_T(u)}{f_Y(Q_y(U))} \mathbf{P}^R(u) du\right)' \right]^{-} \, ,
\end{equation}
where $A^-$ denotes the generalised inverse of a matrix $A$. This weight can be estimated using a preliminary estimator for $\theta_0$. An analogous result holds under \eqref{eq_asymptotic distr_mixing}, though in this case one also needs an estimator for the covariance kernel $\Gamma$. In Supplemental Appendix E, we provide an estimator for $\Omega_R$ in the iid case when the $\{P_l\}$ are shifted Legendre Polynomials. 
\end{remark}

\begin{remark}[A test statistic for overidentifying restrictions] \label{test_over} The strong approximation discussed in this subsection motivates a test statistic for overidentifying restrictions. Suppose $R>d$. Denoting by $M(\cdot)$ the objective function of \eqref{eq_objective_function}, we consider the test-statistic:
	\begin{equation*}
	J \coloneqq T \cdot M(\hat{\theta}_T) \, .
	\end{equation*}
	
An analogous statistic exists in the overidentified GMM setting \citep{Newey1994,Wooldridge2010}. Under the null that the model is correctly specified
(i.e. that there exists $\theta \in \Theta$ such that $Q_Y(\cdot) = Q_Y(\cdot|\theta)$), we can use the results in this section to compute the distribution of this test statistic. Specifically, if the optimal weighting scheme is adopted, the distribution of the test statistic under the null may be approximated by a chi-squared distribution with $R-d$ degrees of freedom. To establish this fact, we rely on an anticoncentration inequality due to \cite{Gotze2019}. See Supplemental Appendix F for details.
\end{remark}

\begin{remark}[Sample-size-dependent trimming]
	\label{rmk_rates_trimming}
	It is possible to adapt our assumptions and results to the case where the trimming constants $\underline{p}$, $\overline{p}$ are functions of the sample size. In particular, Theorem 6 of \cite{Csorgo1978} provides uniform strong approximation results for sample quantiles ranging from $ [1-\delta_T, \delta_T]$, where $\delta_T = 25 T^{-1}\log \log T$. This result imposes fewer restrictions on the distribution, and could be used as the basis for inference on a variable-trimming estimator. 
\end{remark}

\begin{remark}[Data-driven method to select trimming proportions]
	\label{rmk_trimming}
	If one wishes to adopt trimming, then, for a given $R$, a data-driven method for selecting $\underline{p}$ and $\overline{p}$ can be obtained by choosing these constants so as to minimise an estimate of the variance of the leading term in \eqref{eq_asymptotic linear_prop}. See \cite{Athey2021} for a discussion of this approach in estimating the mean of a symmetric distribution; and \cite{Crump2009} for a related approach when choosing trimming constants for the estimated propensity score in observational studies.
\end{remark}

\begin{remark}[Inference based on the weighted bootstrap]
In Supplemental Appendix G, we show how one can leverage the strong approximations discussed in this section to conduct inference on the model parameters using the weighted bootstrap.
\end{remark}

Finally, we observe that, in some settings, we are not interested in conducting inference on $\theta_0$, but rather on a sequence of scalar functions $g_T(\theta_0)$. Typical examples include the estimation of tail probabilities and quantiles. The following result, which is an immediate consequence of Corollary \ref{corollary_asymptotic}, provides conditions for inference based on the Delta Method to be valid in this setting.

\begin{corollary}
	\label{corollary_tail}
	Let $g_n: \mathbb{R}^d \mapsto \mathbb{R}$, $n \in \mathbb{N}$, be a sequence of functions such that there exists an open ball $\mathcal{B} \subseteq \mathbb{R}^d$ containing $\theta_0$ with: (1) each $g_n$ is continuously differentiable on $\mathcal{B}$, and (ii) the gradient functions $\{\nabla g_n:  n \in \mathbb{N}\}$ are equicontinuous on $\mathcal{B}$, with $\nabla g_n(\theta_0)\neq 0$ for every $n \in \mathbb{N}$. If the conditions of Corollary \ref{corollary_asymptotic}  are satisfied, then, as $T,R\to \infty$, $\frac{\sqrt{T}(g_T(\hat{\theta}_T) - g_T(\theta_0))}{\sqrt{\nabla g_T(\theta_0)'V_{T,R}\nabla g_T(\theta_0)}} \overset{d}{\to} N(0,1)$.
	
	\begin{proof}
		The conditions in the statement of the corollary imply that, as $T,R\to \infty$:
		$\sqrt{T}(g_T(\hat{\theta}_T) - g_T(\theta_0))= \nabla g_T(\theta_0)' \sqrt{T}(\hat{\theta}_T-\theta_0)+ o_{P^*}(1)$. The conclusion then follows from Corollary \ref{corollary_asymptotic}.
	\end{proof}
	
\end{corollary}

\subsection{Asymptotic efficiency} 
\label{efficiency}
Supplemental Appendix I discusses efficiency of our proposed L-moment estimator. Specifically, we show that, when no trimming is adopted ($0=\underline{p}<\overline{p}=1$), the optimal weighting scheme \eqref{eq_optimal_weights} is used, the $\{P_l\}_l$ constitute an orthonormal \emph{basis} in $L^2[0,1]$ (recall this is satisfied by shifted Legendre polynomials), and the data is iid, the generalised method of L-moments estimator is asymptotically efficient, in the sense that its asymptotic variance coincides  with the inverse of the Fisher information matrix of the parametric model.\footnote{In the dependent case, ``efficiency'' should be defined as achieving the efficiency bound of the semiparametric model that parametrises the marginal distribution of the $Y_t$, but leaves the time series dependence unrestricted up to regularity conditions \citep{Newey1990,Komunjer2010}. Indeed, in general, our L-moment estimator will be inefficient with respect to the MLE that models the dependency structure between observations. See \cite{Carrasco2014} for further discussion.} We leave details to the Supplemental Appendix, though we briefly outline the argument here. The idea is to consider the alternative estimator:
\begin{equation}
	\label{eq_alternative_text}
	\tilde{\theta}_T \in \operatorname{argmin}_{\theta \in \Theta} \sum_{i \in \mathcal{G}_T} \sum_{j \in \mathcal{G}_T} (\hat{Q}_Y(i)-Q_Y(i|\theta)) \kappa_{i,j} (\hat{Q}_Y(j)-Q_Y(j|\theta)) \, ,
\end{equation}
for a grid of $G_T$ points $\mathcal{G}_T =\{g_1,g_2,\ldots, g_{G_T}\} \subseteq (0,1)$ and weights $\kappa_{i,j}$, $i,j \in \mathcal{G}_T$. This is a weighted version of a ``percentile-based estimator'', which is used in contexts where it is difficult to maximise the likelihood \citep{Gupta2001}. It amounts to choosing $\theta$ so as to match a weighted combination of the order statistics in the sample. In the Supplemental Appendix, we show that, under a suitable sequence of gridpoints and optimal weights, this estimator is asymptotically efficient. We then show, by using the fact that the $\{P_l\}_l$  form an orthonormal basis, that estimator \eqref{eq_alternative_text} can be seen as a generalised L-moment estimator that uses infinitely many L-moments. The final step of the argument then consists in observing that a generalised L-moment estimator that uses a finite but increasing number of L-moments is asymptotically equivalent to estimator \eqref{eq_alternative_text}, which implies that a generalised L-moment estimator under optimal weights is no less efficient than the (efficient) percentile estimator.

\section{Monte Carlo exercise}
\label{monte_carlo}
In our experiments, we draw random samples $Y_1,Y_2,\ldots, Y_T$ from a distribution function $F = F_{\theta_0}$ belonging to a parametric family $\{F_{\theta}: \theta \in \Theta \}$. Following \cite{Hosking1990}, we consider the goal of the researcher to be estimating quantiles $Q_Y(\tau)$ of the distribution $F_{\theta_0}$ by using a plug-in approach: first, the researcher estimates $\theta_0$; then she estimates $Q_Y(\tau)$ by setting $\widehat{Q_Y(\tau)} = Q_Y(\tau|\hat{\theta})$. As in \cite{Hosking1990}, we consider $\tau \in \{0.9,0.99,0.999\}$. In order to compare the behaviour of alternative procedures in estimating more central quantiles, we also consider the median $\tau=0.5$. We analyse sample sizes $T \in \{50,100,500\}$. The number of Monte Carlo draws is set to $5,000$.

We compare the root mean squared error of four types of generalised method of L-moment estimators under varying choices of $R$ with the root mean squared error obtained were $\theta_0$ to be estimated via MLE. We consider the following estimators: (i) the generalised method of L-moments estimator that uses the càglàd L-moment estimates \eqref{eq_est_lmoments_quantile} and identity weights (\textbf{Càglàd FS});\footnote{To be precise, our choice of weights does not coincide with actual identity weights. Given that the coefficients of Legendre polynomials rapidly scale with $R$ -- and that this increase generates convergence problems in the numerical optimisation -- we work directly with the underlying estimators of the probability-weighted moments $\int_0^1 Q_Y(u)u^r du$ \citep{Landwehr1979}, of which L-moments are linear combinations. When (estimated) optimal weights are used, such approach is without loss, since the optimal weights for L-moments constitute a mere rotation of the optimal weights for probability-weighted moments, in such a way that the optimally-weighted objective function for L-moments and probability-weighted moments coincide. In other cases, however, this is not the case: a choice of identity weights when probability-weighted moments are directly targeted coincides with using $\boldsymbol{D}^{-1'}\boldsymbol{D}^{-1}$ as a weighting matrix for L-moments, where $\boldsymbol{D}$ is a matrix which translates the first $R$ probability-weighted moments onto the first $R$ L-moments. For small $R$, we have experimented with using the ``true'' L-moment estimator with identity weights, and have obtained the same patterns presented in the text.} (ii) a two step-estimator which first estimates (i) and then uses this preliminary estimator\footnote{This preliminary estimator is computed with $R=d$.} to estimate the optimal weighting matrix \eqref{eq_optimal_weights}, which is then used to reestimate $\theta_0$ (\textbf{Càglàd TS}); (iii) the generalised method of L-moments estimator that uses the unbiased L-moment estimates \eqref{eq_est_lmoments_linear} and identity weights (\textbf{Unbiased FS}); and (iv) the two-step estimator that uses the unbiased L-moment estimator in the first and second steps (\textbf{Unbiased TS}). The estimator of the optimal-weighting matrix we use is given in Supplemental Appendix E.

\subsection{Generalised Extreme value distribution (GEV)}

Following \cite{Hosking1985} and \cite{Hosking1990}, we consider the family of distributions
\begin{equation*}
    F_{\theta}(z) = \begin{cases}
    \exp\{- [1- \theta_2(x-\theta_1)/\theta_3]^{1/\theta_3}\}, & \theta_3 \neq 0 \\
    \exp\{-\exp(-(x-\theta_1)/\theta_2)\}, & \theta_3= 0
    \end{cases},
\end{equation*}
and $\theta_0 = (0,1,-0.2)'$.

\Cref{gev_table_mle} reports the RMSE of each procedure, divided by the RMSE of the MLE, under the choice of $R$ that achieves the smallest RMSE. Values above 1 indicate the MLE outperforms the estimator in consideration; and values below 1 indicate the estimator outperforms MLE. The value of $R$ that minimises the RMSE is presented under parentheses. Some patterns are worth highlighting. Firstly, the L-moment estimator, under a proper choice of $R$ and (estimated) optimal weights (two-step estimators) is able to outperform MLE in most settings, especially at the tail of the distribution function. Reductions in these settings can be as large as $31.9\%$. At the median, two-step L-moment estimators behave similarly to the MLE. The performance of two-step càglàd and unbiased estimators is also quite similar.  Secondly, the power of overidentifying restrictions is evident: except in three out of twenty-four cases, two-step L-moment estimators never achieve a minimum RMSE at $R=3$, the number of parameters. Two of these three exceptions are found at the smallest sample size ($T=50$), where the benefit of overidentifying restrictions may be outweighed by noisy estimation of the weighting matrix.\footnote{Indeed, as we discuss in Supplemental Appendix K, a higher-order expansion of our proposed estimator shows that correlation of the estimator of the optimal weighting matrix with sample L-moments plays a key role in the higher-order bias and variance of the two-step estimator.}  Thirdly, the relationship between $T$ and the MSE-minimising choice of $R$  in the two-step Càglàd estimator is monotonic when we move from the smallest ($T=50$) to the largest ($T=500$) sample size.\footnote{The optimal choice of $R$ also increases in three out of four quantiles when we move from $T=50$ to $T=100$. The exception occurs at the median, where the optimal choice decreases slightly from $R=12$ to $R=11$.  At $T=100$, the difference between $R=11$ and $R=12$ is negligible, though: choosing $R=11$ leads to a relative RMSE of $1.002787$, whereas the choice $R=12$ leads to a relative RMSE of $1.002800$.} This is consistent with our theoretical results: given $\sqrt{T}$-consistency of the estimators, as $T$ increases, one expects the contribution of the bias component in the RMSE to decrease, and, given asymptotic efficiency of the two-step estimator as $R$ diverges, a larger choice of $R$ may lead to variance reduction. Finally, the role of optimal weights is clear: first-step estimators tend to underperform the MLE as the sample size increases. In larger samples, and when optimal weights are not used, the best choice tends to be setting $R$ close to or equal to $3$, which reinforces the importance of weighting when overidentifying restrictions are included.

\begin{figure}[h]
	\centering

	\begin{subfigure}[H]{\textwidth}

		\centering
		\includegraphics[width=0.24\textwidth]{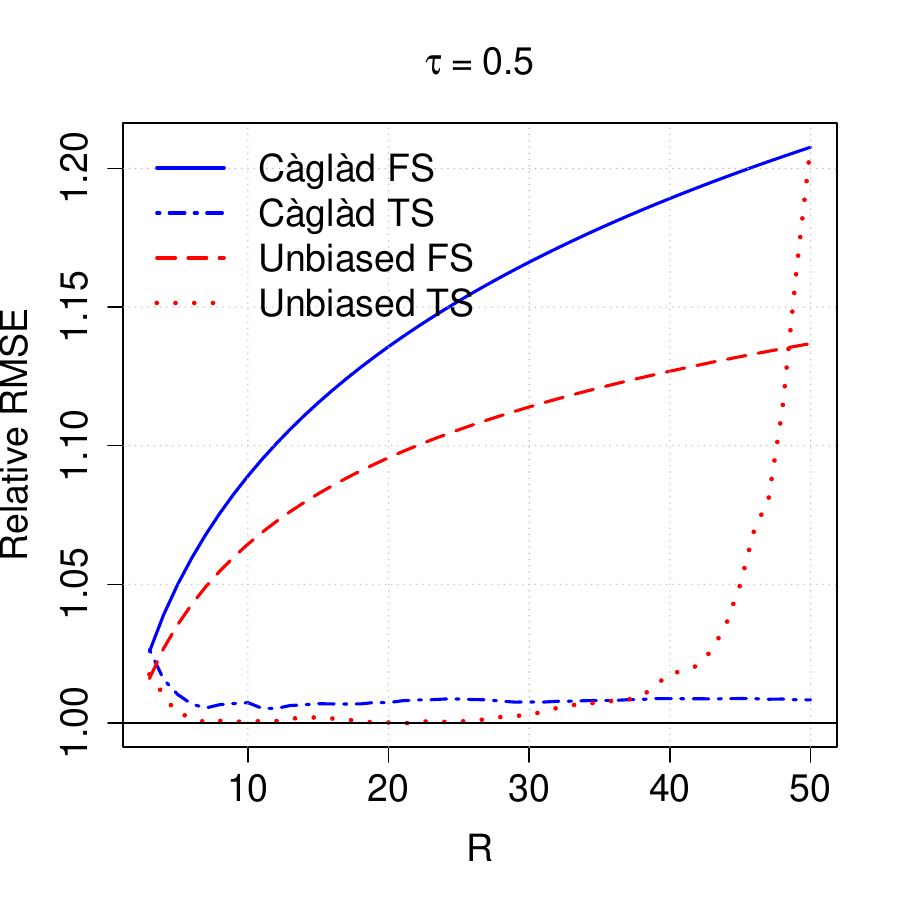}   \includegraphics[width=0.24\textwidth]{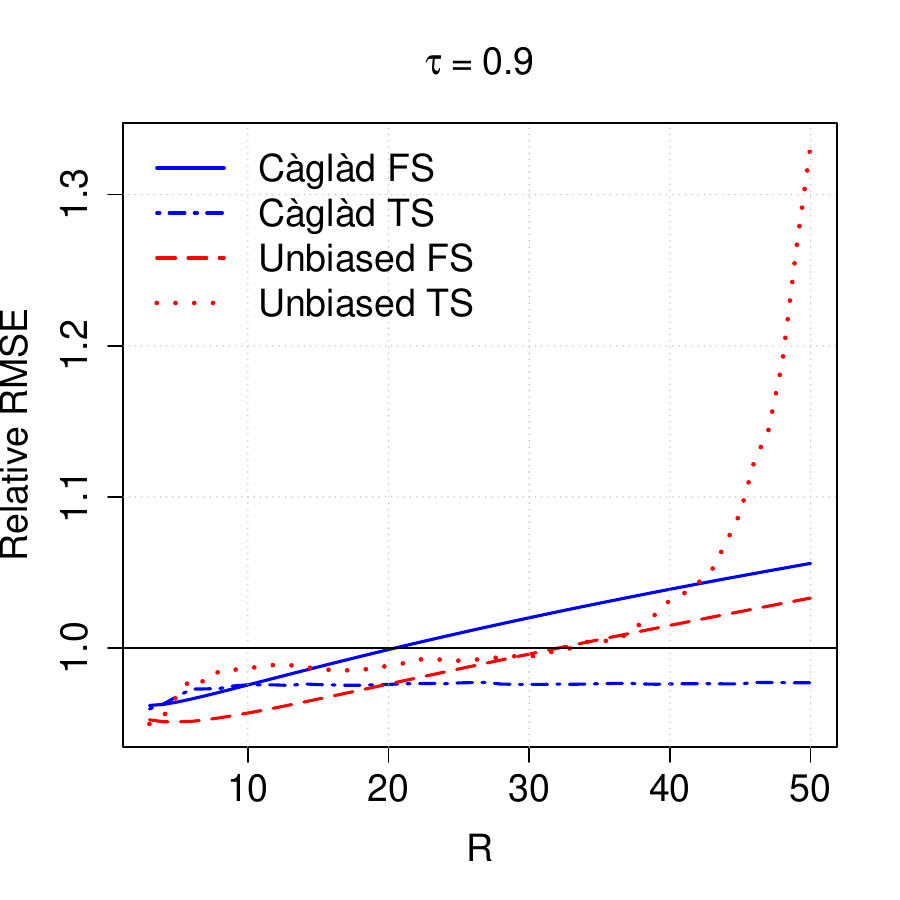}
		\includegraphics[width=0.24\textwidth]{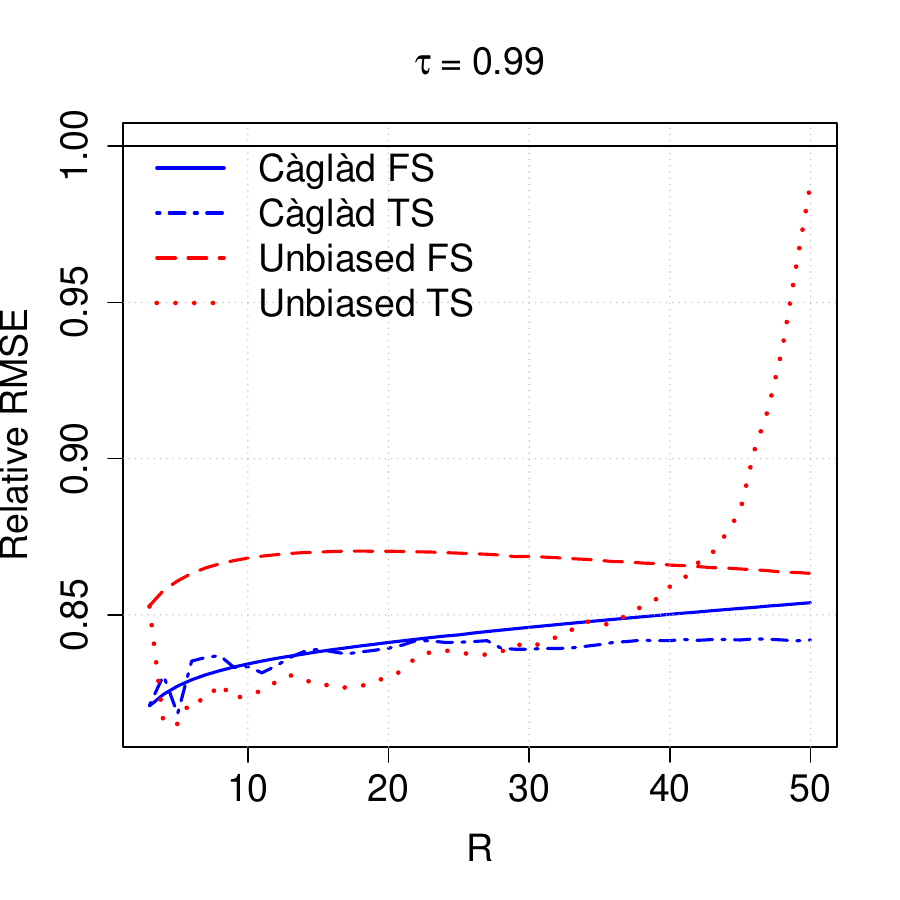}   \includegraphics[width=0.24\textwidth]{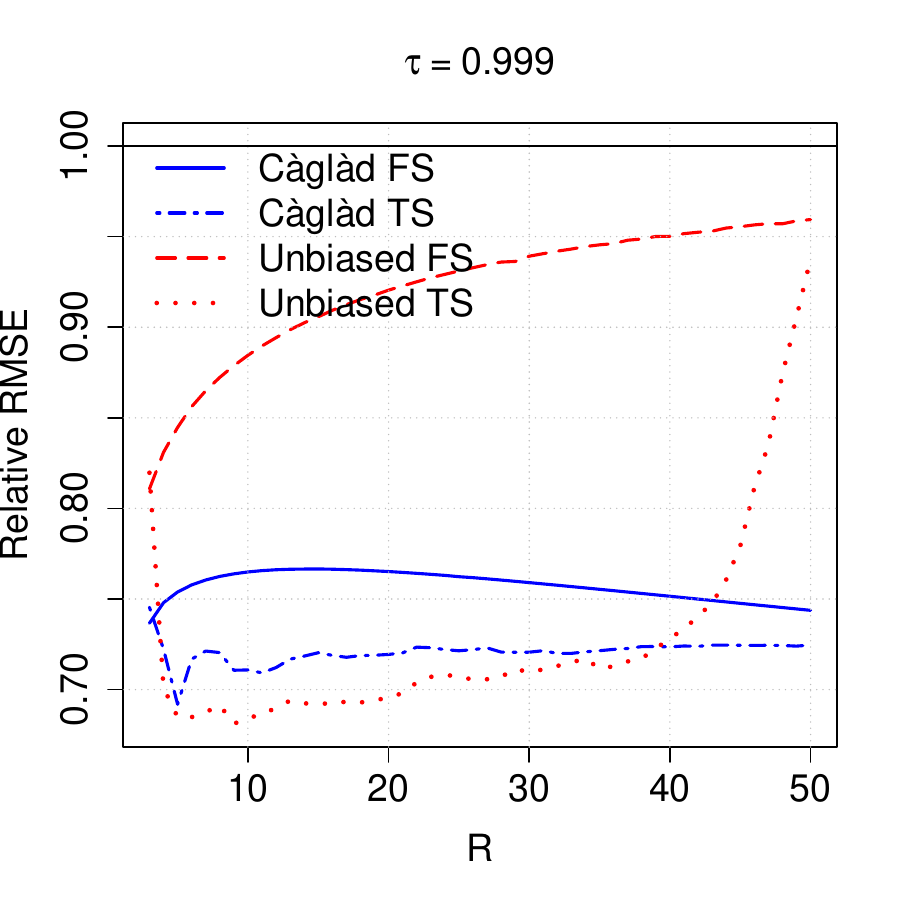}
		\caption{$T=50$}
	\end{subfigure}

	\begin{subfigure}[H]{\textwidth}

		\centering
		\includegraphics[width=0.24\textwidth]{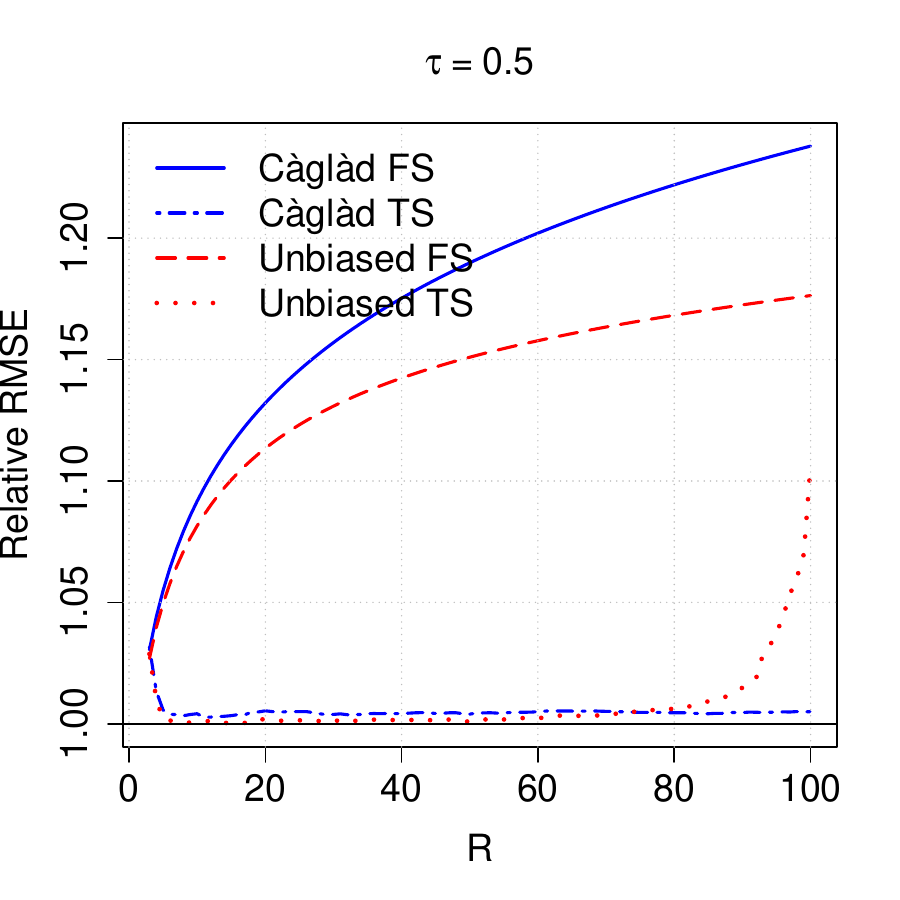}   \includegraphics[width=0.24\textwidth]{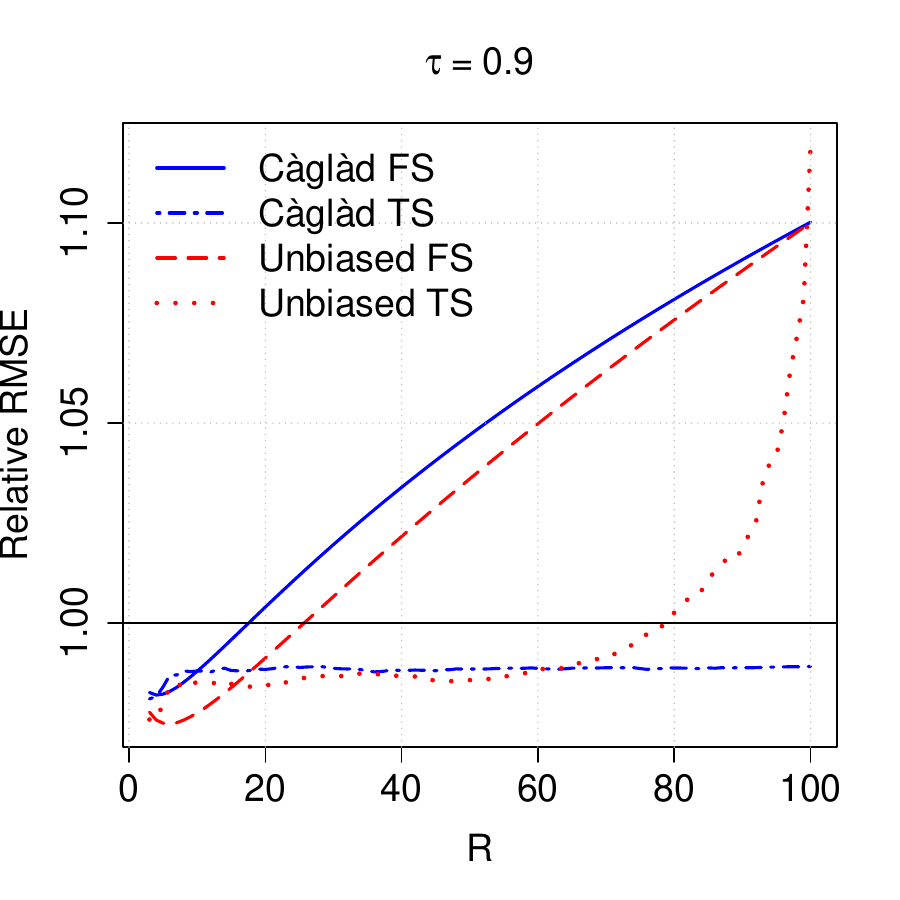}
		\includegraphics[width=0.24\textwidth]{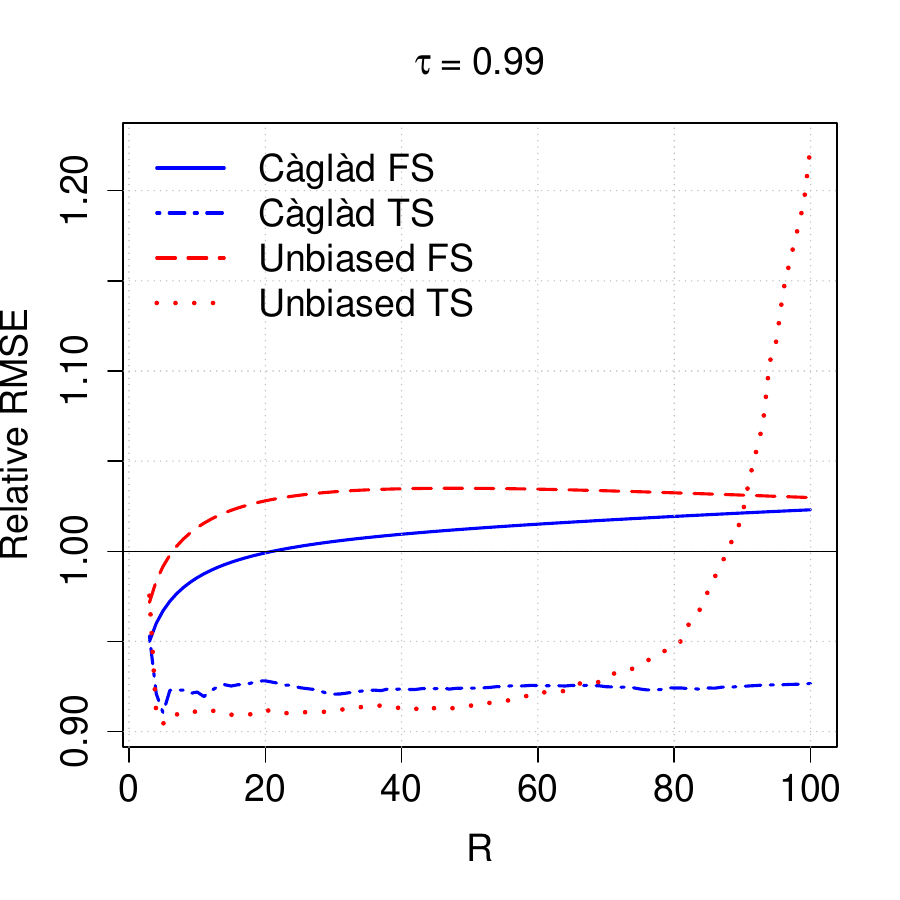}   \includegraphics[width=0.24\textwidth]{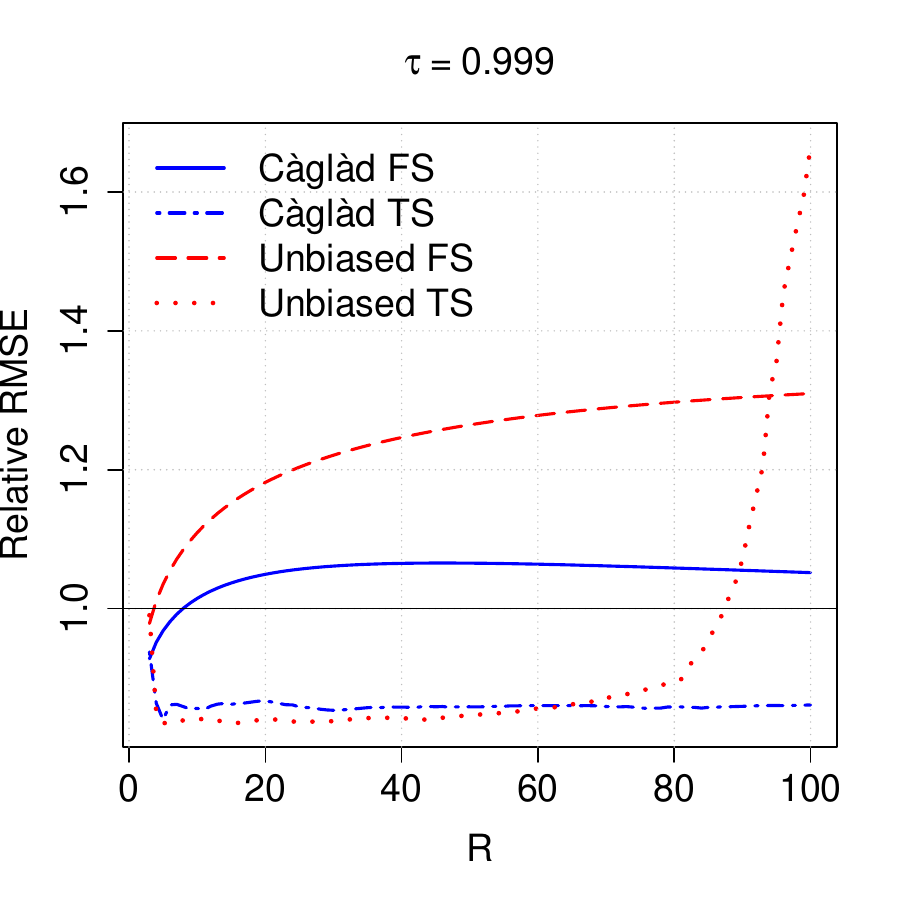}
		\caption{$T=100$}
	\end{subfigure}

	\begin{subfigure}[H]{\textwidth}

		\centering
		\includegraphics[width=0.24\textwidth]{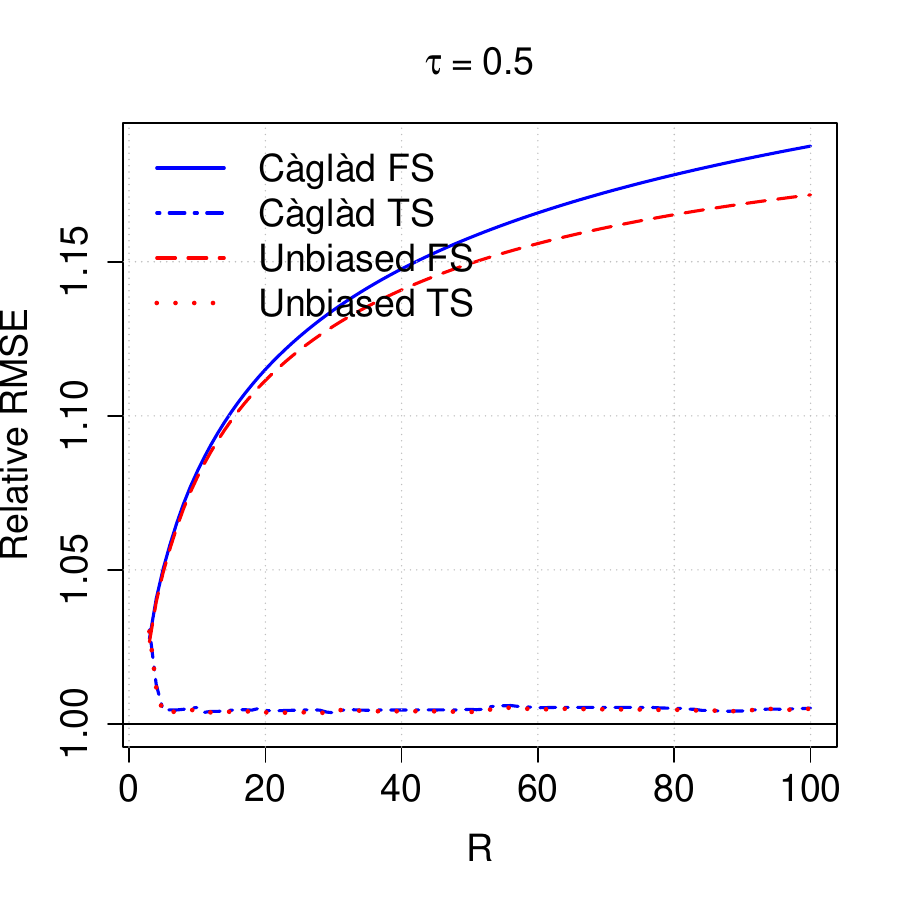}   \includegraphics[width=0.24\textwidth]{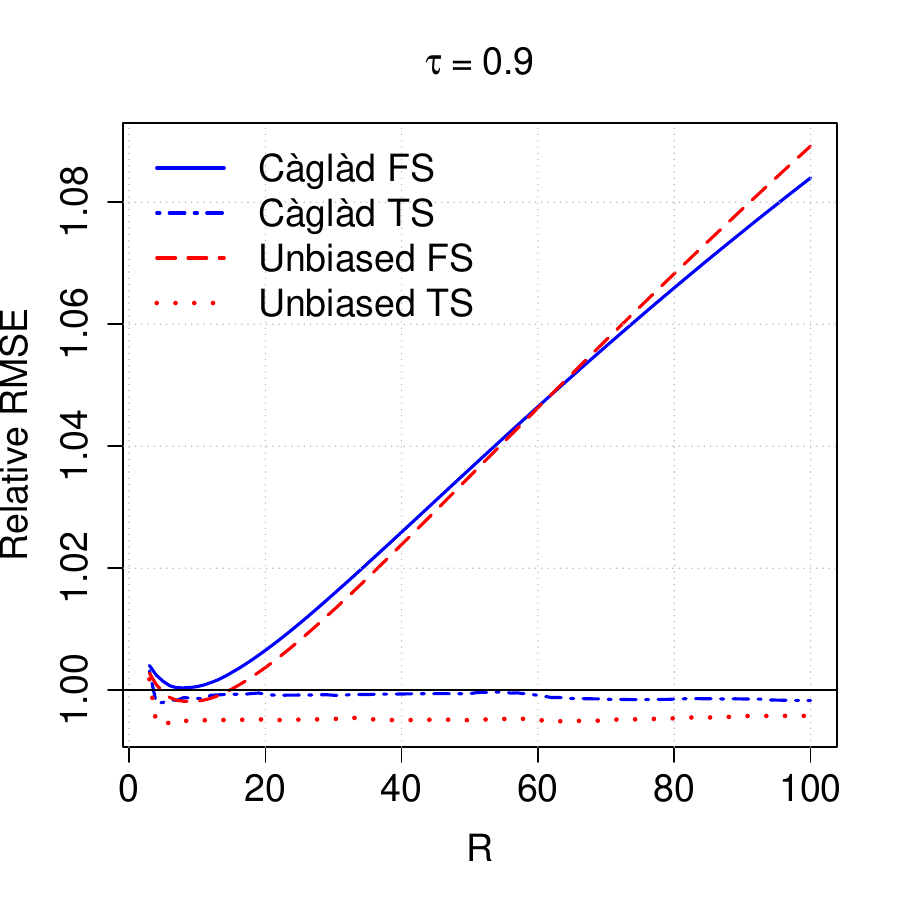}
		\includegraphics[width=0.24\textwidth]{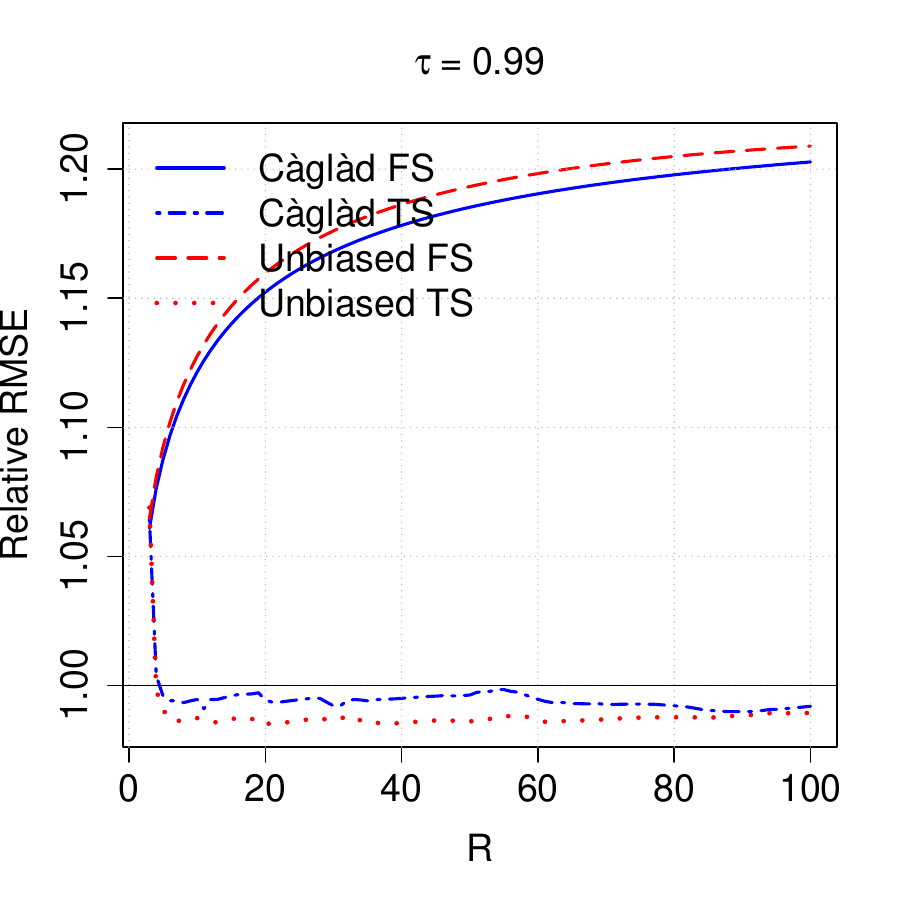}   \includegraphics[width=0.24\textwidth]{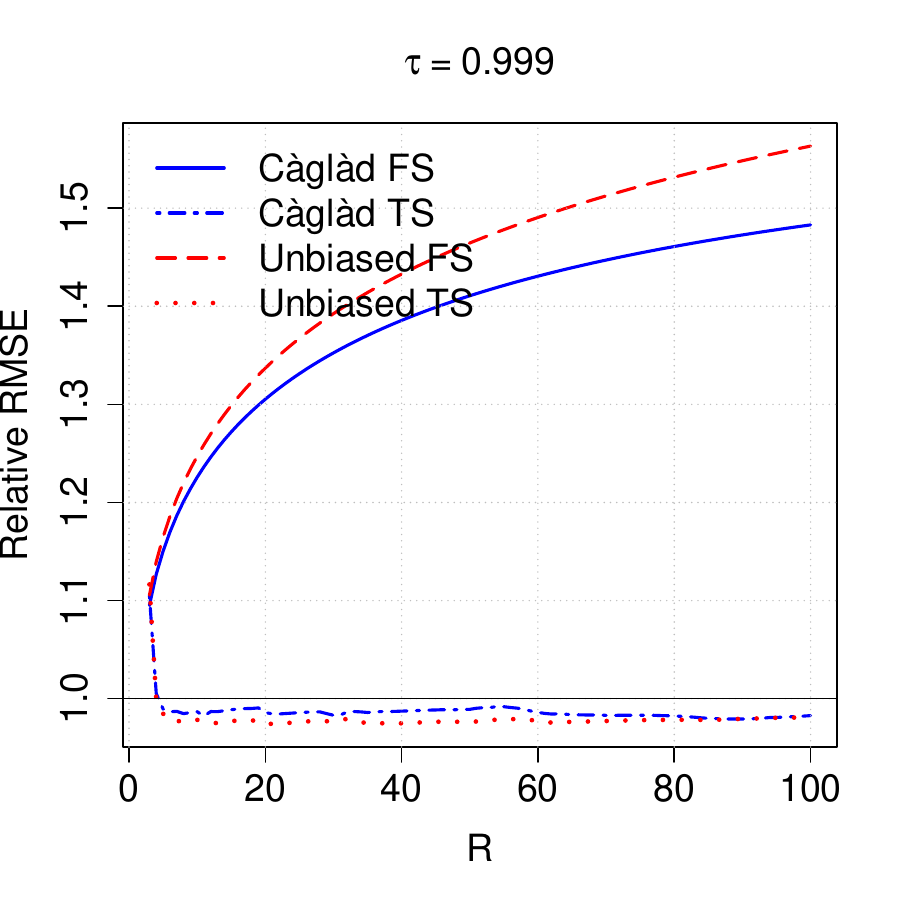}
		\caption{$T=500$}
	\end{subfigure}
	
	\caption{GEV: relative RMSE for different choices of $R$.}
	\label{fig:gev_mle}
\end{figure}

To better understand the patterns in the table, we report in \Cref{fig:gev_mle}, the relative RMSE curve for different sample sizes and choices of $R$. The role of optimal weights is especially striking: first-step estimators usually exhibit an increasing RMSE, as a function of $R$. In contrast, two-step estimators are able to better control the RMSE across $R$. It is also interesting to note that the two-step unbiased L-moment estimator behaves poorly when $R$ is close to $T$. This suggests that, in settings where one may wish to make $R$ large, the càglàd estimator is preferable.\footnote{This is also in accordance with our theoretical results for the càglàd-based estimator, which essentially place no restriction on the growth rate of $R$.} Finally, we note that, for two-step estimators, the RMSE curve is relatively flat over several regions of $R$. This implies that, if $R$ is chosen in these regions, then the RMSE of the resulting estimator is robust to (local) perturbations on the number of L-moments used in estimation. As we discuss in Section \ref{selection}, this flatness will be convenient when designing methods to automatically select $R$, since any method that sets this tuning parameter to be in the correct \emph{region} where RMSE is small should perform well. In contrast, if the RMSE curve were locally very sensitive to the choice of $R$, it could be unfeasible to obtain a sufficiently accurate assessment of the RMSE in finite samples that were to result in a good choice of $R$.

\input{tables/gev/gev_table_mle}

\subsection{Generalised Pareto distribution (GPD)}

Following \cite{Hosking1987}, we consider the family of distributions:
\begin{equation*}
	F_{\theta}(z) = \begin{cases}
		1 - (1-\theta_2x/\theta_1)^{-1/\theta_2}, & \theta_2 \neq 0 \\
		1 - \exp(-x/\theta_1), & \theta_2 = 0
	\end{cases},
\end{equation*}
and $\theta_0 = (1,-0.2)'$.

\begin{figure}[h]
	    \centering

     \begin{subfigure}[H]{\textwidth}

         \centering
         \includegraphics[width=0.24\textwidth]{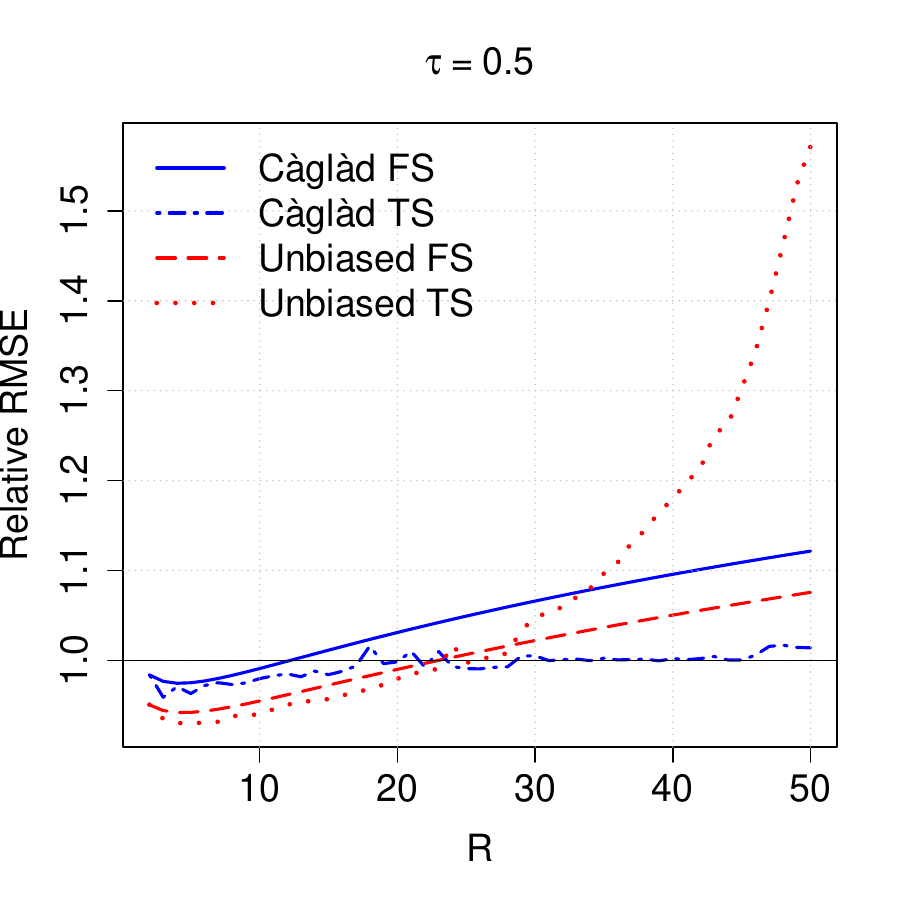}   \includegraphics[width=0.24\textwidth]{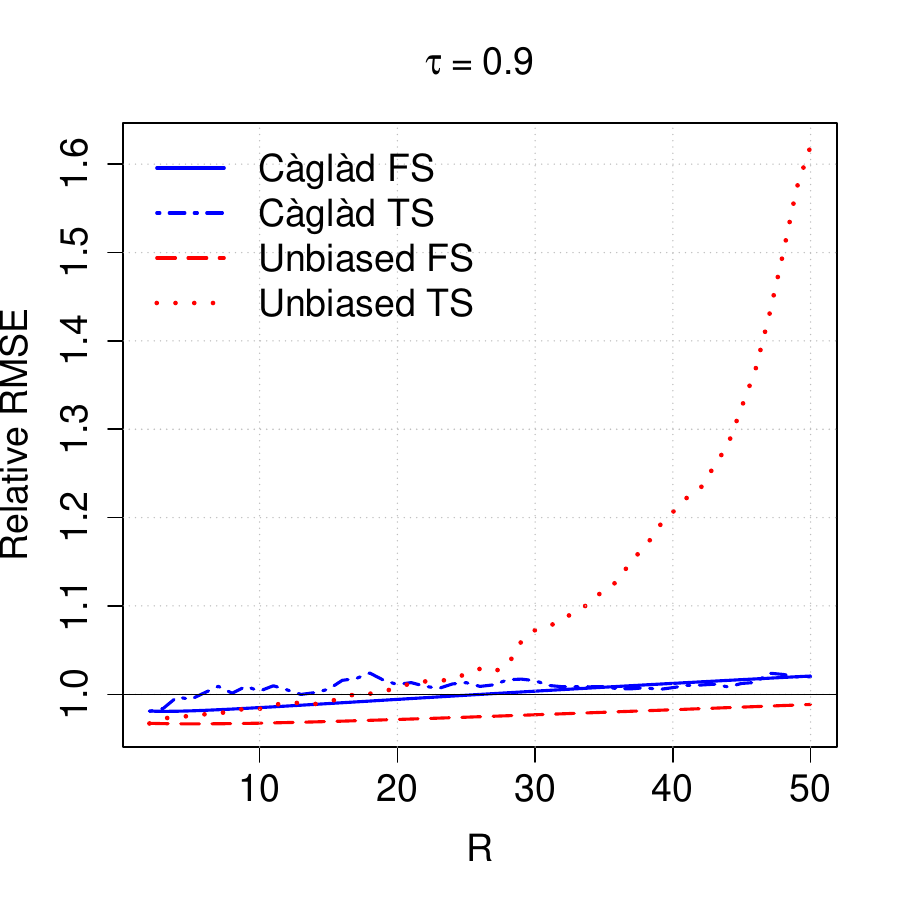}
         \includegraphics[width=0.24\textwidth]{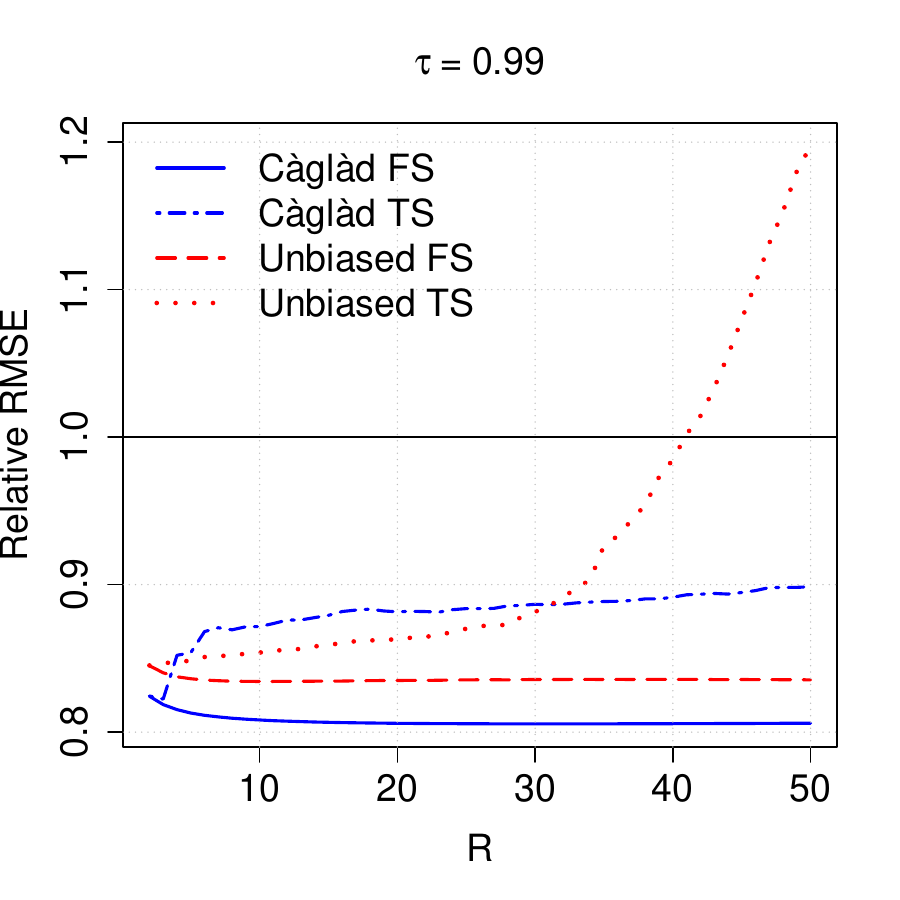}   \includegraphics[width=0.24\textwidth]{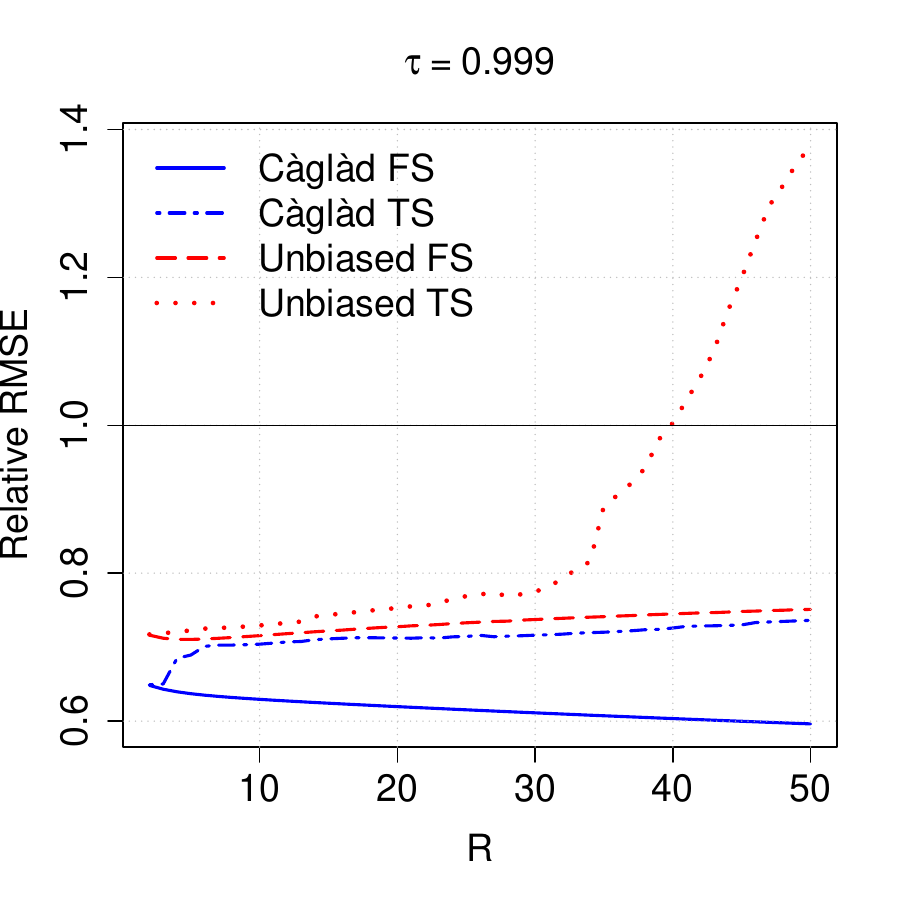}
                      \caption{$T=50$}
     \end{subfigure}

     \begin{subfigure}[H]{\textwidth}

         \centering
         \includegraphics[width=0.24\textwidth]{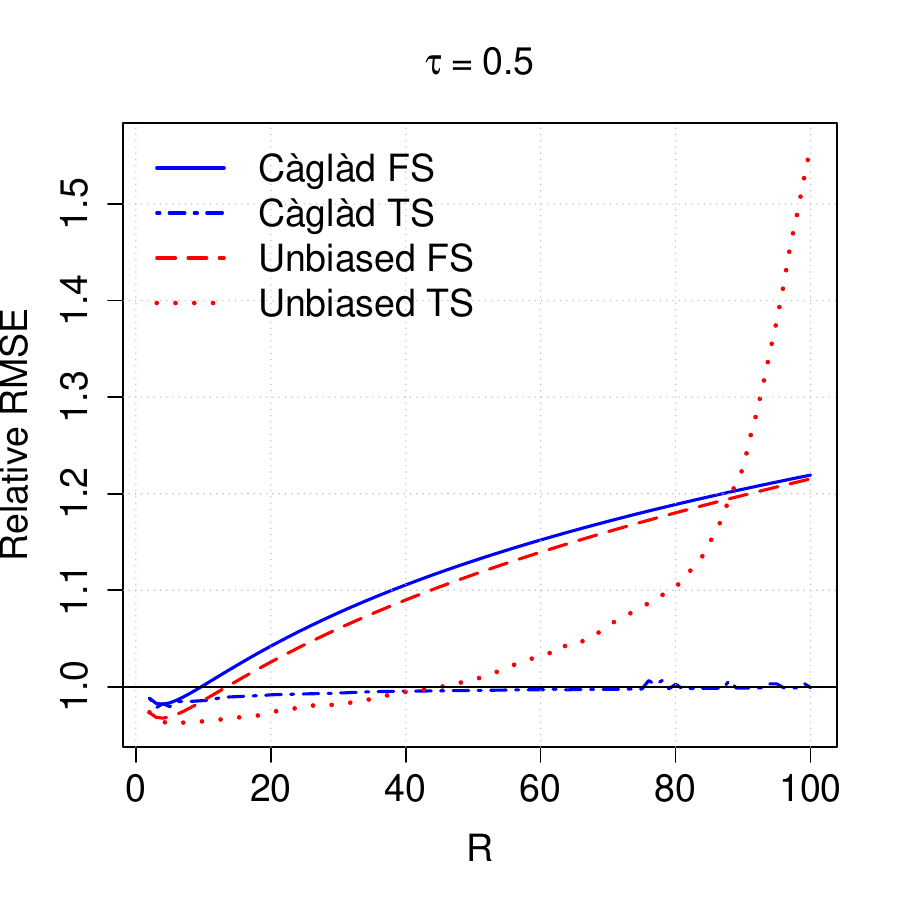}   \includegraphics[width=0.24\textwidth]{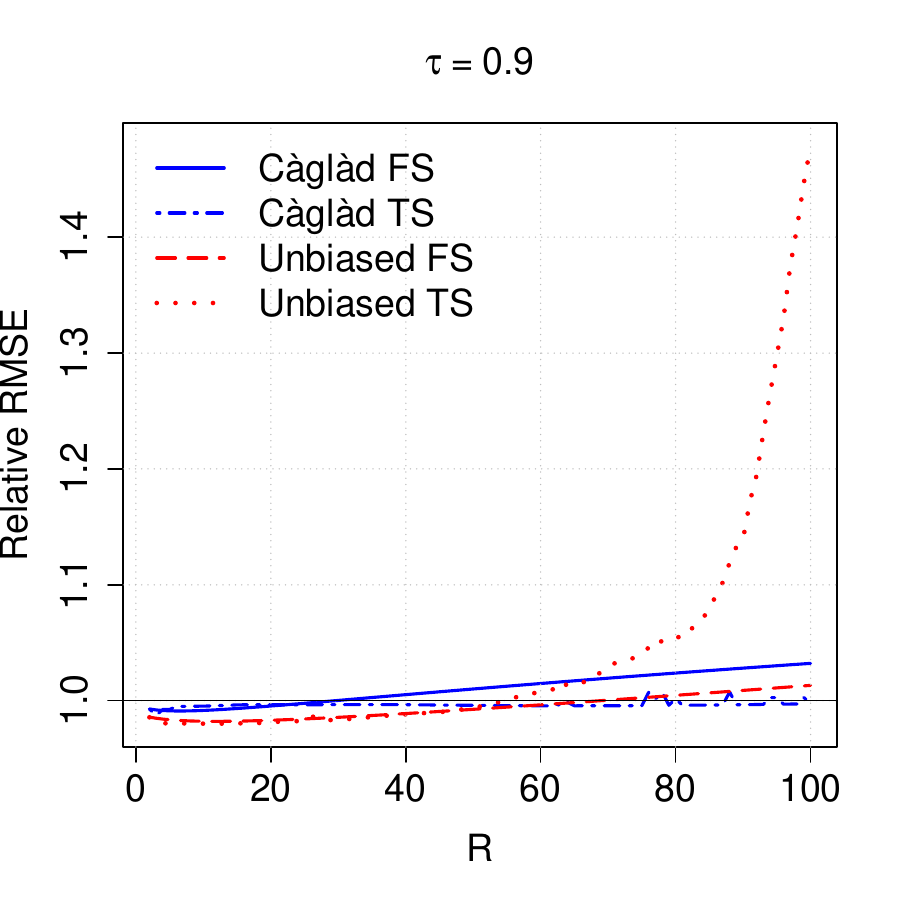}
         \includegraphics[width=0.24\textwidth]{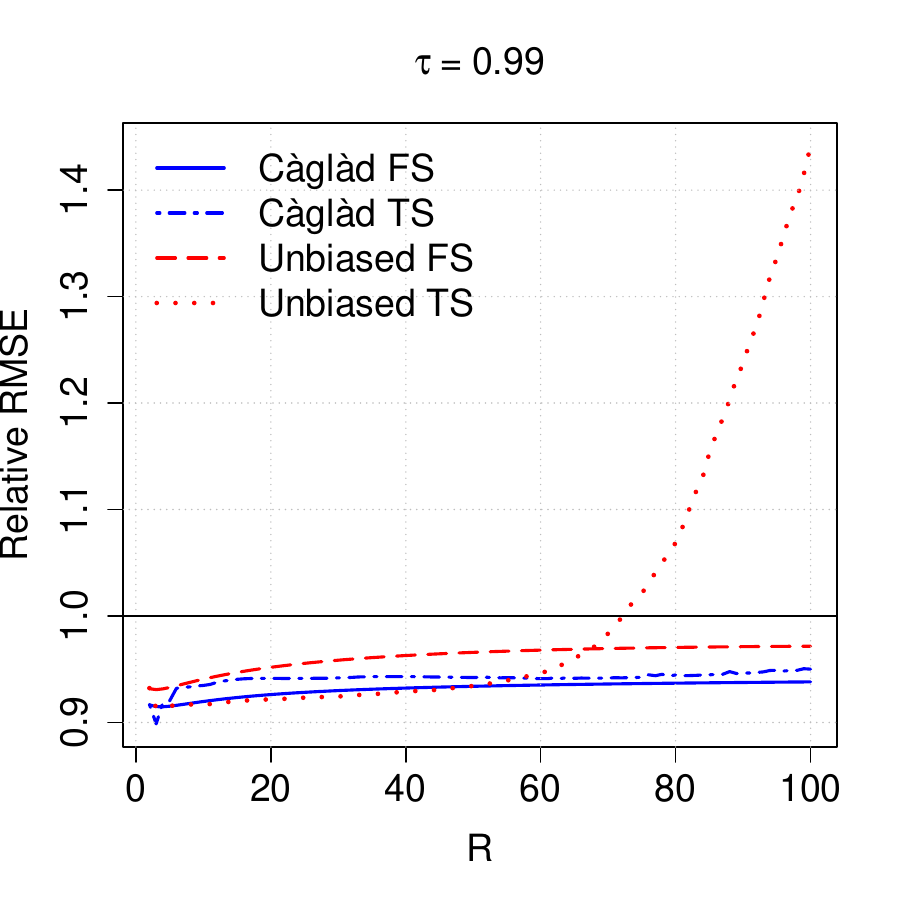}   \includegraphics[width=0.24\textwidth]{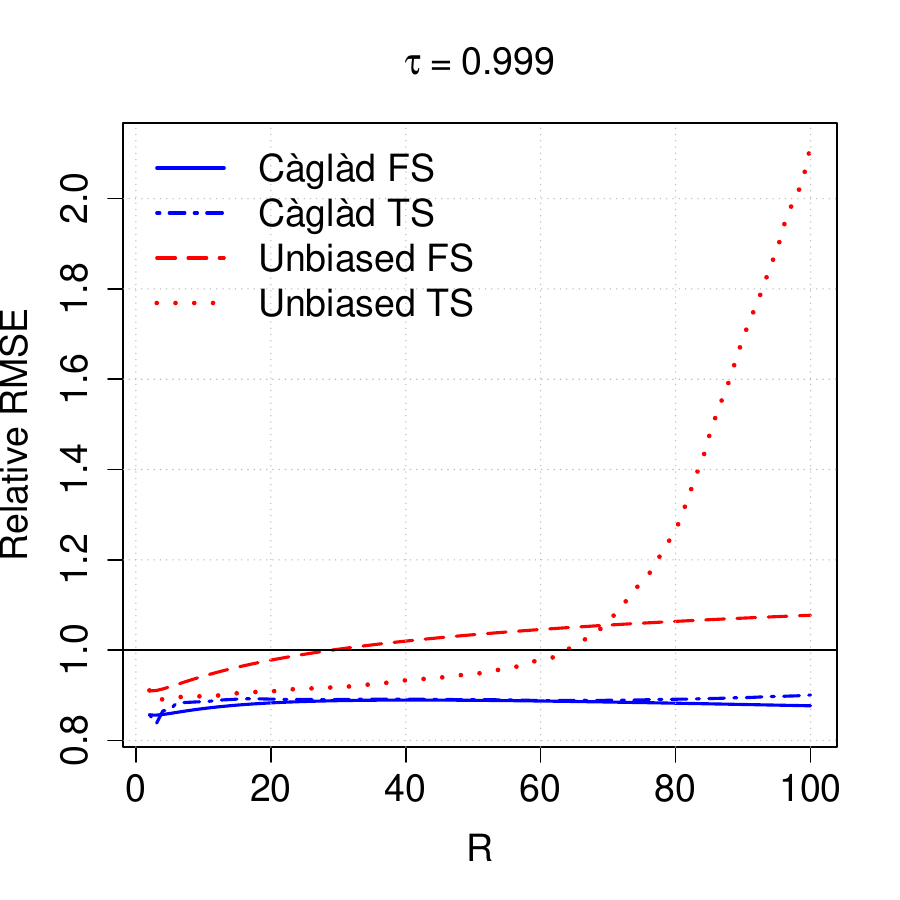}
                      \caption{$T=100$}
     \end{subfigure}

     \begin{subfigure}[H]{\textwidth}

         \centering
         \includegraphics[width=0.24\textwidth]{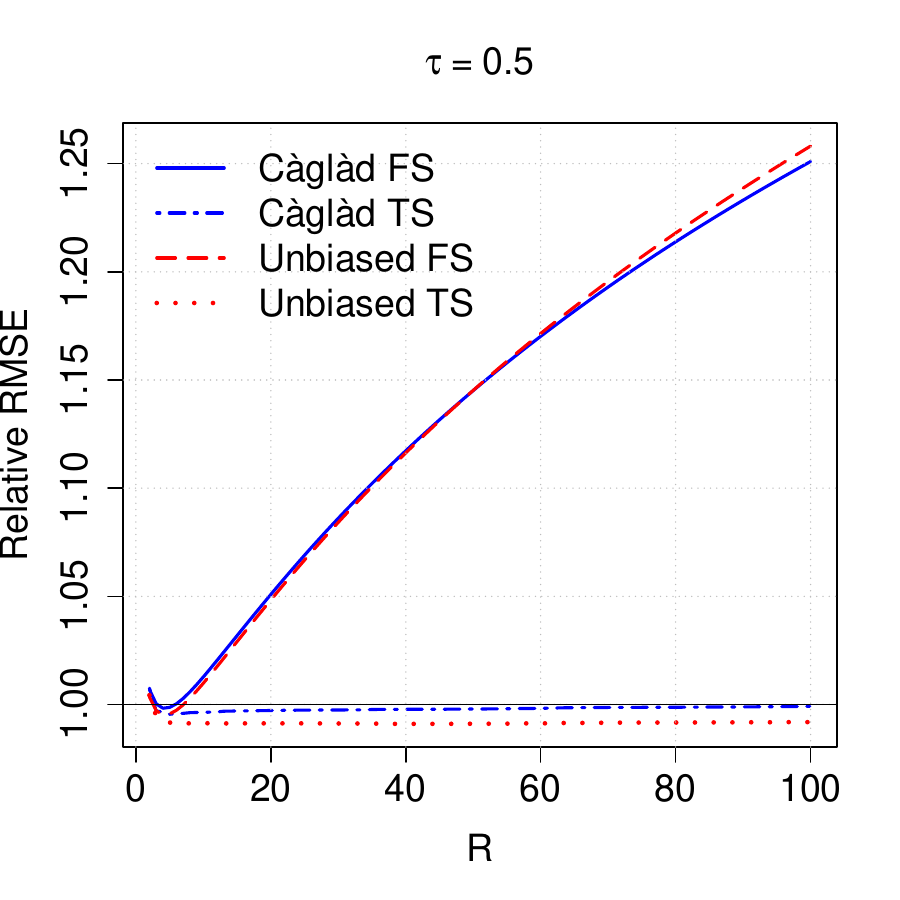}   \includegraphics[width=0.24\textwidth]{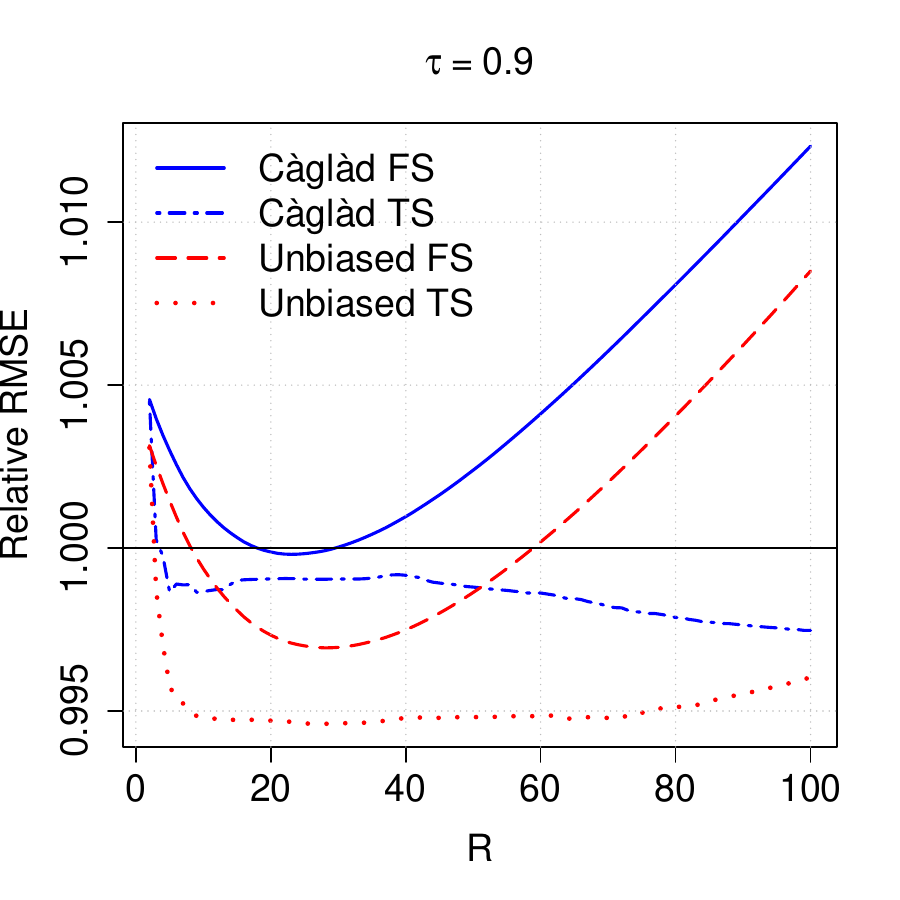}
         \includegraphics[width=0.24\textwidth]{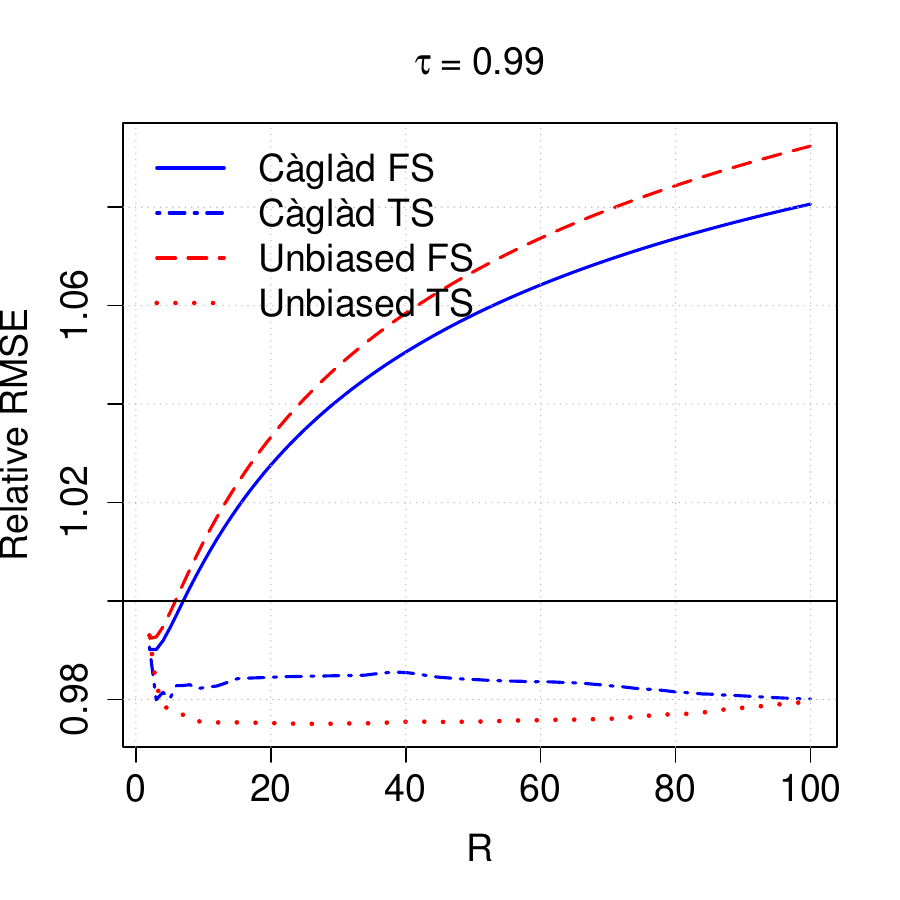}   \includegraphics[width=0.24\textwidth]{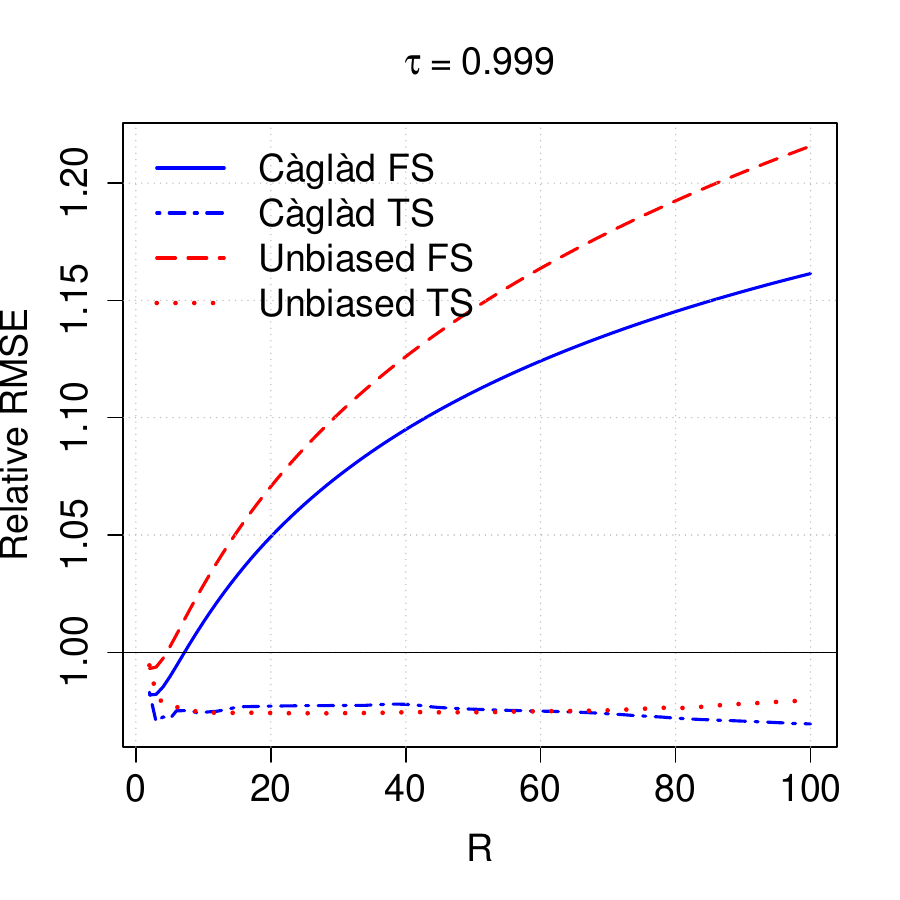}
                      \caption{$T=500$}
     \end{subfigure}

	         \caption{GPD: relative RMSE for different choices of $R$.}
\label{fig:gpd_mle}

\end{figure}

\Cref{gpd_table_mle} and \Cref{fig:gpd_mle} summarise the results of our simulation. Overall patterns are similar to the ones obtained in the GEV simulations. Importantly, though, estimation of the optimal weighting matrix impacts two-step estimators quite negatively in this setup. As a consequence, we verify that the choice of $R=2$ (i.e. a just-identified estimator that effectively does not rely on the weights) is optimal for TS estimators at five out of the eight cases in sample size $T=50$. This behaviour also leads to FS estimators, which do not use estimated weights, outperforming TS estimators at the $0.99$ and $0.999$ quantiles when $T=50$, and underperforming TS estimators in larger sample sizes by much smaller margins than in the GEV design.  Finally, we note that, in all cases, L-moment estimators compare favourably to the MLE.

\input{tables/gpd/gpd_table_mle}

\begin{remark}[Other target parameters]
	\label{other_target}
In this section, we have focused in a setting where the goal is quantile estimation. In Supplemental Appendix J.1, we consider instead a situation where the targets are linear combinations $\delta'\theta_0$ of the model parameters. Since we do not have any particular linear combination in mind, we consider choices of $\delta$ (directions) that lead to the most and least favourable relative RMSE vis-à-vis the MLE. In the GEV design, two-step estimators, under the optimal choice of $R$, are able to offer RMSE improvements of around 8\% in the most favourable direction and smaller sample sizes, while strongly mitigating the underperformance of first-step estimators in the least-favourable directions and larger sample sizes. Indeed, in the latter scenario, first-step estimators, even under an optimal choice of $R$, incur in RMSE losses relatively to the MLE of over 25\%; in contrast, this underperformance shrinks to only 1.3\% when càglàd two-step estimators are adopted. In the GPD design, both first-step and two-step estimators perform well relatively to the MLE, even when considering the least favourable directions and largest sample sizes, with gains reaching over 16\% in the smallest sample size and most favourable direction (and 4\% in the smallest sample size and least favourable direction).
\end{remark}

\begin{remark}[Comparison with trimming and tilting approaches]
	In both of our Monte Carlo exercises, the distributions exhibit heavy tails. In these settings, a natural approach would be to consider maximum likelihood estimators that take additional steps to limit the influence of extreme observations. We compare the behaviour of these estimators with our L-moment-based approach in Supplemental Appendix J.2. Specifically, we contrast our L-moment-based estimators with a trimming approach that discards extreme observations and computes MLE estimates in a restricted dataset, and also with a ``tilted'' MLE  that computes estimates in a reweighted dataset. In both the GEV and GDP designs, the Càglàd TS estimator under the RMSE-minimising choice of $R$ consistently outperforms the trimming approach. As for the tilted MLE, it is able to compete with the L-moment estimator at some combinations of tail quantiles and sample sizes, under a suitable amount of tilting. However, the competitiveness and overall performance of the tilted MLE estimator is extremely sensitive to the amount of tilting to which the data is subjected to, and, as we discuss in the Appendix, to the best of our knowledge, there currently do not exist methods to select the tilting proportion with an aim at MSE reductions.
\end{remark}

\begin{remark}[Size and coverage of confidence intervals based on the Gaussian approximation]
	Supplemental Appendix J.3 assesses the coverage and length of confidence intervals (CIs) derived from the normal approximation in Corollaries \ref{corollary_asymptotic} and \ref{corollary_tail}. We focus on the càglàd two-step estimator and provide two sets of results. First, to assess the quality of the normal approximations, we compute the coverage and length of confidence intervals for different target quantiles based on normal critical values and the true sampling variance of the estimators. We observe that these confidence intervals have coverage close to their nominal level even in the smallest sample size and more extreme quantiles, and that, consistent with our theoretical results that do not impose any growth restrictions on the number of L-moments in the derivation of the asymptotic normal approximation, coverage is approximately constant across choices of $R$. Confidence intervals based on the MLE estimator, normal critical values and the true sampling variance also exhibit correct coverage; however, their length is no less than the length of confidence intervals based on the generalized L-moment estimator under the length-minimising choice of $R$. Length reductions provided by the L-moment-based CIs can be substantial, especially at tail quantiles. We then compare the coverage and length of feasible versions of these confidence intervals that rely on estimators of the asymptotic variance. Both the MLE and L-moment-based confidence intervals display correct coverage at central quantiles in small sample sizes and at the tail when we consider larger samples. However, both methods display undercoverage at more extreme quantiles in smaller sample sizes, with L-moment CIs in some cases undercovering more than the MLE in the GEV design (differences in undercoverage are insubstantial in the GPD design). As we argue in the Appendix, this is driven partly by correlation between the asymptotic variance estimator and the target quantile estimator, which generates distortions in the sampling distribution of the $t$-test that is inverted to construct the confidence interval. Motivated by our strong approximation results, we provide a simple correction to the critical values used in the L-moment confidence interval that improves coverage in smaller sample sizes and more extreme quantiles, while still preserving reduced length over (coverage-corrected) MLE-based CIs.
\end{remark}

\begin{remark}
	We note that computation runtime of our L-moment estimators is quite fast in the GEV and GPD families. For example, the estimation of the parameters of a GEV distribution with the two-step càglàd L-moment estimator and a random sample with $T=500$ observations takes around half a second in a 2017 i7 Macbook Pro with 16GB RAM when $R=100$; around three seconds when $R=500$; and around ten seconds when $R=1,000$.
\end{remark}

%% file: tables/gev/gev_table_mle.tex
\begin{table}[h]
\centering
\caption{GEV : relative RMSE under MSE-minimising choice of $R$} 
\label{gev_table_mle}
\adjustbox{width=\textwidth}{
\begin{tabular}{|l|cccc|cccc|cccc|}
  \hline
  & \multicolumn{ 4 } {c|}{$T = 50 $}&\multicolumn{ 4 } {c|}{$T = 100 $}&\multicolumn{ 4 } {c|}{$T = 500 $} \\ & $\tau = 0.5 $ & $\tau = 0.9 $ & $\tau = 0.99 $ & $\tau = 0.999 $ & $\tau = 0.5 $ & $\tau = 0.9 $ & $\tau = 0.99 $ & $\tau = 0.999 $ & $\tau = 0.5 $ & $\tau = 0.9 $ & $\tau = 0.99 $ & $\tau = 0.999 $ \\ 
   \hline
Càglàd FS & 1.026 & 0.962 & 0.821 & 0.737 & 1.031 & 0.982 & 0.950 & 0.928 & 1.028 & 1.000 & 1.061 & 1.095 \\ 
   &  (3) &  (3) &  (3) &  (3) &  (3) &  (4) &  (3) &  (3) &  (3) &  (8) &  (3) &  (3) \\ 
   \hline
Càglàd TS & 1.005 & 0.960 & 0.818 & 0.692 & 1.003 & 0.981 & 0.910 & 0.840 & 1.004 & 0.998 & 0.990 & 0.979 \\ 
   &  (12) &  (3) &  (5) &  (5) &  (11) &  (3) &  (5) &  (5) &  (30) &  (4) &  (90) &  (90) \\ 
   \hline
Unbiased FS & 1.016 & 0.951 & 0.853 & 0.811 & 1.027 & 0.975 & 0.972 & 0.979 & 1.027 & 0.998 & 1.065 & 1.106 \\ 
   &  (3) &  (5) &  (3) &  (3) &  (3) &  (6) &  (3) &  (3) &  (3) &  (9) &  (3) &  (3) \\ 
   \hline
Unbiased TS & 1.000 & 0.950 & 0.815 & 0.681 & 1.000 & 0.976 & 0.904 & 0.834 & 1.003 & 0.994 & 0.985 & 0.974 \\ 
   &  (21) &  (3) &  (5) &  (9) &  (8) &  (4) &  (5) &  (5) &  (29) &  (7) &  (21) &  (21) \\ 
   \hline
\end{tabular}
}
\end{table}

%% file: tables/gpd/gpd_table_mle.tex
\begin{table}[h]
\centering
\caption{GPD : relative RMSE under MSE-minimising choice of $R$} 
\label{gpd_table_mle}
\adjustbox{width=\textwidth}{
\begin{tabular}{|l|cccc|cccc|cccc|}
  \hline
  & \multicolumn{ 4 } {c|}{$T = 50 $}&\multicolumn{ 4 } {c|}{$T = 100 $}&\multicolumn{ 4 } {c|}{$T = 500 $} \\ & $\tau = 0.5 $ & $\tau = 0.9 $ & $\tau = 0.99 $ & $\tau = 0.999 $ & $\tau = 0.5 $ & $\tau = 0.9 $ & $\tau = 0.99 $ & $\tau = 0.999 $ & $\tau = 0.5 $ & $\tau = 0.9 $ & $\tau = 0.99 $ & $\tau = 0.999 $ \\ 
   \hline
Càglàd FS & 0.975 & 0.981 & 0.806 & 0.596 & 0.982 & 0.991 & 0.915 & 0.856 & 0.998 & 1.000 & 0.990 & 0.982 \\ 
   &  (4) &  (3) &  (34) &  (50) &  (4) &  (7) &  (4) &  (3) &  (4) &  (23) &  (3) &  (2) \\ 
   \hline
Càglàd TS & 0.959 & 0.981 & 0.822 & 0.649 & 0.978 & 0.987 & 0.899 & 0.837 & 0.995 & 0.997 & 0.980 & 0.969 \\ 
   &  (3) &  (2) &  (3) &  (2) &  (3) &  (3) &  (3) &  (3) &  (5) &  (100) &  (3) &  (100) \\ 
   \hline
Unbiased FS & 0.942 & 0.967 & 0.834 & 0.711 & 0.967 & 0.982 & 0.931 & 0.910 & 0.995 & 0.997 & 0.992 & 0.993 \\ 
   &  (4) &  (5) &  (10) &  (4) &  (4) &  (12) &  (3) &  (2) &  (4) &  (28) &  (2) &  (2) \\ 
   \hline
Unbiased TS & 0.929 & 0.967 & 0.845 & 0.717 & 0.962 & 0.980 & 0.914 & 0.887 & 0.991 & 0.995 & 0.975 & 0.974 \\ 
   &  (5) &  (2) &  (2) &  (2) &  (5) &  (3) &  (3) &  (3) &  (39) &  (27) &  (27) &  (27) \\ 
   \hline
\end{tabular}
}
\end{table}

%% file: sections/selection.tex
The simulation exercise in the previous section evidences that the number of L-moments $R$ plays an important role in determining the relative behaviour of the generalised L-moment estimator. Indeed, Figures \ref{fig:gev_mle} and \ref{fig:gpd_mle} suggest that the RMSE of two-step estimators can be sensitive to the of number L-moments. For example, in the GEV design, at $T=500$ and $\tau = 0.999$, the RMSE of the two-step unbiased L-moment estimator is around 11\% \emph{larger} than the MLE when $R=3$, and around 2.3\% \emph{smaller} than the MLE when $R=6$. This indicates that designing a proper method to select $R$ is essential for competitiveness of the L-moment approach.  Moreover, the pattern of the curves in Figures \ref{fig:gev_mle} and \ref{fig:gpd_mle} suggests that there is great hope that such methods will perform well in practice. Indeed, given that the RMSE curve is flat over several regions of $R$, one should expect any method that sets $R$ to be in an appropriate \emph{region} where RMSE is small to perform well.\footnote{We thank two anonymous referees for pointing this out.} In contrast, if the RMSE curve were locally very sensitive to the choice of the number of L-moments, then one would require a rather sharp assessment of the RMSE to select $R$, which could be unfeasible in smaller sample sizes.

In light of these points, in this section we introduce (semi)automatic methods to select $R$. We briefly outline two approaches, with the details being left to Supplemental Appendix K. We then contrast these approaches in the context of the Monte Carlo exercise of \Cref{monte_carlo}.

In Supplemental Appendix K.1, we derive a higher-order expansion of the ``generalised'' L-moment estimator \eqref{eq_objective_function}. We then propose to choose $R$ by minimising the resulting higher-order mean-squared error of a suitable linear combination of the parameters. Similar approaches were considered in the GMM literature by \cite{Donald2001} -- where the goal is to choose the number of instruments in linear instrumental variable models --, and \cite{Donald2009} -- where one wishes to choose moment conditions in models defined by conditional moment restrictions (in which case infinitely many restrictions are available). Relatedly, \cite{Okui2009} considers the choice of moments in dynamic panel data models; and, more recently, \cite{Abadie2019} use higher order expansions to develop a method of choosing subsamples in linear instrumental variables models with first-stage heterogeneity. Importantly, our higher-order expansions can be used to provide higher-order mean-squared error estimates of target estimands $g_T(\theta_0)$, where $g_T$ is a function indexed by sample size. This can be useful when the parameter $\theta_0$ is not of direct interest. So, for example, if our goal is quantile estimation, we can choose $R$ so as to minimise the higher-order mean-squared error of estimating the target quantile.

In Supplemental Appendix K.2, we consider an alternative approach to selecting L-moments by employing $\ell_1$-regularisation. Following \cite{Luo2016}, we note that the first order condition of the estimator \eqref{eq_objective_function} may be written as: 
\begin{equation*}
	A_{R}h^R(\hat{\theta}) = 0 \, ,
\end{equation*} 
for a $d \times R$ matrix $A_R$ which combines the  L-moments linearly into $d$ restrictions. The idea is to estimate $A_R$ using a Lasso penalty. This approach implicitly performs moment selection, as the method yields exact zeros for several entries of $A_R$. In a GMM context, \cite{Luo2016} introduces an easy-to-implement quadratic program for estimating $A_R$ with the Lasso regularization. In the Supplemental Appendix, we show how this algorithm may be extended to our L-moment setting and provide conditions for its validity.

To conclude, we return to the Monte Carlo exercise of Section \ref{monte_carlo}. We contrast the RMSE (relatively to the MLE) of the original L-moment estimator due to \cite{Hosking1990} that sets $R=d$ (\textbf{FS}) with a two-step L-moment estimator where $R$ is chosen so as to minimise a higher-order MSE of the target quantile (\textbf{TS RMSE}), and a ``post-lasso'' estimator that estimates $\theta_0$ using only those L-moments selected by regularised estimation of $A_R$ (\textbf{TS Post-Lasso}). For brevity, we focus on estimators based on the \textit{càglàd} L-moments \eqref{eq_est_lmoments_quantile}. Additional details on the implementation of each method can be found in Supplemental Appendix K.3.

Tables \ref{gev_table_select_summary} and \ref{gpd_table_select_summary} present the results of the different methods in the GEV and GPD exercises. We report in parentheses the average number of L-moments used by each estimator. Overall, the TS RMSE estimator compares favourably to both \citeauthor{Hosking1990}'s original estimator and the MLE. In the GEV exercise, the TS RMSE estimator improves upon both the FS estimator and the MLE when $T<500$; and behaves similarly to the MLE and better than FS in the largest sample size. For example, at $T=500$, \citeauthor{Hosking1990}'s estimator has a 6.1\% (9.5\%) larger root-mean-squared error than then MLE at the 0.99 (0.999) quantile, whereas the relative performance of TS RMSE with respect to the MLE is 0.6\% (0.9\%). As for the GPD exercise, recall that this is a setting where estimation of the optimal weighting matrix impacts two-step estimators more negatively. Consequently, gains of TS RMSE over FS are more limited in this setting. Indeed, the average gain of TS RMSE over FS in the GPD exercise is 1.0 (relative) percentage points (pp), with the largest outperformance being 2.8 pp and an underperformance in $T=50$ and $\tau=0.9$ of 1.3 pp. In contrast, in the GEV exercise, the average gain of TS RMSE over FS is 3.2 pp, the largest gain is 8.6 pp and TS RMSE underperforms FS by only 0.4 pp at $T=100$ and $\tau=0.9$. More importantly, in both settings, our TS RMSE approach is able to \emph{simultaneously} generate gains over MLE in smaller samples and mitigate inefficiencies of \citeauthor{Hosking1990}'s original method in larger sample sizes. This phenomenon is especially pronounced at the tails of the distributions. 

With regards to the Post-Lasso method, we note that it behaves similarly to TS RMSE in larger sample sizes,\footnote{In the largest sample size of the GEV distribution, the Post-Lasso performs especially well, incurring in a gain of 10.2pp over the FS estimator at the $\tau=0.999$ quantile.} though it can perform somewhat unfavourably vis-à-vis the other L-moment alternatives in the smallest sample size (the method still improves upon MLE at tail quantiles in this scenario). As we discuss in Supplemental Appendix K.3, this issue can be partly attributed to a ``harsh'' regularisation penalty being used in the selection step. There is room for improving this step by relying on an iterative procedure to select a less harsh penalty (see \cite{Belloni2012} and \cite{Luo2016} for examples). We also discuss in the Supplemental Appendix that it could be possible to improve the TS RMSE procedure by including additional higher-order terms in the estimated approximate RMSE. We leave exploration of these improvements as future topics of research.

\begin{remark}
We remark that comparisons between the RMSE and post-Lasso approaches should also take computational concerns into consideration. Indeed, as summarised in the pseudo-code in the Supplemental Appendix (Algorithm K.1), our numerical implementation of the RMSE approach requires evaluation of cross-products, for different test values of $R$, between the gradient and Hessian of the theoretical L-moment functions at different choices of $R$ (which measure the sensitivity of estimates to the sample L-moments); the partial derivatives, with respect to $\theta$, of the quantile density function $Q'(u|\theta)$ at different values of $u$ (measuring higher-order terms pertaining to estimation of the optimal weighting matrix); and the gradient and Hessian, with respect to $\theta$, of the quantile function $Q(\tau|\theta)$ at the quantiles $\tau$ of interest, when the goal is quantile estimation (pertaining to the expansion of the RMSE of the target quantile). Even though these derivatives are available in closed form for the GEV and GPD families (see Supplemental Appendix M), and while we do provide \texttt{R} code that leverages fast automatic differentiation tools to evaluate the derivatives when these expressions are not available in closed form, computation of the RMSE approach is generally slower than the Post-Lasso method. Indeed, while the post-Lasso also requires computation of derivatives to estimate the Lasso penalty (see Supplemental Appendix K.3 for details), it does not hinge on the evaluation of cross-products  of these terms for different choices of $R$, which speeds up implementation considerably.\footnote{We remark that evaluation of the cross-products in the RMSE approach can be parallelized across different test values of $R$, an approach we adopt in our computational implementation. See the R script \texttt{selection.R} in the accompanying online repository for a generic implementation of our selection methods to any class of parametric distributions.} For comparison, in the GEV Monte Carlo exercise, with 500 observations, computing the higher-order RMSE estimate for the four target quantiles across test values $R \in \{3,\ldots, 100\}$ takes around 42 seconds in a 2017 i7 Macbook Pro with 16GB RAM (and around 10 seconds for test values  $R \in \{3,\ldots, 50\}$). In contrast, the Post-Lasso selection approach with a maximum allowed choice of $R_{\text{max}}=200$ takes around 6 seconds (and around 2 seconds  with $R_{\text{max}}=100$). Given that the Post-Lasso approach compares favourably to TS RMSE in sample sizes $T>50$, it may thus be preferable in these settings on computational grounds. One further advantage of this approach is its simplicity, as it delivers a single choice of $R$ irrespective of the target parameter.
\end{remark}

\input{tables/gev/gev_table_select_summary.tex}
\input{tables/gpd/gpd_table_select_summary.tex}

%% file: tables/gev/gev_table_select_summary.tex
\begin{table}[h]
\centering
\caption{GEV : relative RMSE under different selection procedures} 
\label{gev_table_select_summary}
\adjustbox{width=\textwidth}{
\begin{tabular}{|l|cccc|cccc|cccc|}
  \hline
  & \multicolumn{ 4 } {c|}{$T = 50 $}&\multicolumn{ 4 } {c|}{$T = 100 $}&\multicolumn{ 4 } {c|}{$T = 500 $} \\ & $\tau = 0.5 $ & $\tau = 0.9 $ & $\tau = 0.99 $ & $\tau = 0.999 $ & $\tau = 0.5 $ & $\tau = 0.9 $ & $\tau = 0.99 $ & $\tau = 0.999 $ & $\tau = 0.5 $ & $\tau = 0.9 $ & $\tau = 0.99 $ & $\tau = 0.999 $ \\ 
   \hline
FS &  1.026 &  0.962 &  0.821 &  0.737 &  1.031 &  0.983 &  0.950 &  0.928 &  1.028 &  1.004 &  1.061 &  1.095 \\ 
   &  (3) &  (3) &  (3) &  (3) &  (3) &  (3) &  (3) &  (3) &  (3) &  (3) &  (3) &  (3) \\ 
   \hline
TS RMSE &  1.008 &  0.964 &  0.794 &  0.674 &  1.004 &  0.987 &  0.923 &  0.865 &  1.005 &  0.999 &  1.006 &  1.009 \\ 
   &  (16.81) &  (3.66) &  (3.3) &  (3.41) &  (33.02) &  (4.17) &  (4.32) &  (4.57) &  (20.29) &  (35.51) &  (40.91) &  (43.17) \\ 
   \hline
TS Post-Lasso &  1.017 &  0.975 &  0.857 &  0.781 &  1.010 &  0.988 &  0.928 &  0.866 &  1.006 &  0.999 &  0.999 &  0.993 \\ 
   &  (7.99) &  (7.99) &  (7.99) &  (7.99) &  (9.24) &  (9.24) &  (9.24) &  (9.24) &  (9.95) &  (9.95) &  (9.95) &  (9.95) \\ 
   \hline
\end{tabular}
}
\end{table}

%% file: tables/gpd/gpd_table_select_summary.tex
\begin{table}[h]
\centering
\caption{GPD : relative RMSE under different selection procedures} 
\label{gpd_table_select_summary}
\adjustbox{width=\textwidth}{
\begin{tabular}{|l|cccc|cccc|cccc|}
  \hline
  & \multicolumn{ 4 } {c|}{$T = 50 $}&\multicolumn{ 4 } {c|}{$T = 100 $}&\multicolumn{ 4 } {c|}{$T = 500 $} \\ & $\tau = 0.5 $ & $\tau = 0.9 $ & $\tau = 0.99 $ & $\tau = 0.999 $ & $\tau = 0.5 $ & $\tau = 0.9 $ & $\tau = 0.99 $ & $\tau = 0.999 $ & $\tau = 0.5 $ & $\tau = 0.9 $ & $\tau = 0.99 $ & $\tau = 0.999 $ \\ 
   \hline
FS &                                                   0.984 &                                                   0.981 &                                                   0.824 &                                                   0.648 &                                                    0.988 &                                                    0.993 &                                                    0.917 &                                                    0.856 &                                                  1.007 &                                                  1.005 &                                                  0.990 &                                                  0.982 \\ 
   &  (2) &  (2) &  (2) &  (2) &  (2) &  (2) &  (2) &  (2) &  (2) &  (2) &  (2) &  (2) \\ 
   \hline
TS RMSE &                                                   0.964 &                                                   0.994 &                                                   0.817 &                                                   0.640 &                                                    0.980 &                                                    0.990 &                                                    0.896 &                                                    0.828 &                                                  0.997 &                                                  0.999 &                                                  0.978 &                                                  0.970 \\ 
   &  (2.86) &  (4.43) &  (2.61) &  (2.96) &  (3.59) &  (4.31) &  (3.02) &  (3.09) &  (5.34) &  (48.42) &  (31.11) &  (29.71) \\ 
   \hline
TS Post-Lasso &                                                   0.995 &                                                   0.999 &                                                   0.891 &                                                   0.741 &                                                    0.992 &                                                    0.999 &                                                    0.950 &                                                    0.905 &                                                  0.998 &                                                  1.000 &                                                  0.985 &                                                  0.977 \\ 
   &  (3.52) &  (3.52) &  (3.52) &  (3.52) &  (3.7) &  (3.7) &  (3.7) &  (3.7) &  (3.78) &  (3.78) &  (3.78) &  (3.78) \\ 
   \hline
\end{tabular}
}
\end{table}

%% file: sections/extensions.tex
\subsection{``Residual'' analysis in semi- and nonparametric models}

In this subsection, we consider a setting where a researcher has postulated a model for a scalar real-valued outcome $Y$:
\begin{equation}	
	\label{eq_fs_model}
	Y = h(\epsilon, X; \gamma_0), \quad \gamma \in \Gamma \subseteq \mathcal{B}\, ,
\end{equation}
where $h$ is a known mapping, $X$ is a vector of observable attributes taking values in $\mathcal{X}$, $\epsilon$ is an unobservable \emph{scalar} real-valued disturbance, and $\gamma_0$ is a nuisance parameter that is known to belong to a subset $\Gamma$ of a Banach space $(\mathcal{B},\lVert \cdot \rVert_{\mathcal{B}})$. We assume that, for each possible value $
(x, \gamma) \in \mathcal{X} \times \Gamma$, the map $e \mapsto h(e, x; \gamma)$ is invertible, and we denote its pointwise inverse by $h^{-1}(\cdot, x; \gamma)$.

We further assume that the researcher has access to an estimator of $\gamma_0$, and that her goal is to estimate a parametric model for the distribution of $\epsilon$, i.e. she considers the model:
\begin{equation}
	\label{eq_np_error}
	\epsilon \sim F_{\theta_0}, \quad \theta_0 \in \Theta \subseteq \mathbb{R}^d \, .
\end{equation}

Interest in \eqref{eq_np_error} nests different types of ``residual'' analyses, where one may wish to estimate \eqref{eq_np_error} with an aim to (indirectly) assess the appropriateness of \eqref{eq_fs_model} -- whenever theory imposes restrictions on the distribution of $\epsilon$ --, or as a means to construct unconditional prediction intervals for $Y$.

In Supplemental Appendix L.1, we show how our generalised L-moment approach may be adapted to estimate \eqref{eq_np_error}, while remaining agnostic about the first-step estimator of $\gamma_0$. We do so by borrowing insights from the double-machine learning literature \citep{Chernozhukov2018,Chernozhukov2022,kennedy2023semiparametric}. Specifically, we employ sample-splitting and debiasing to construct the generalised method-of-L-moment estimator:
 \begin{equation}
 	\label{eq_residual_ts}
	\begin{aligned}
		\hat{\theta} \in \text{arg inf}_{\theta \in \Theta} \left[\int_{\underline{p}}^{\bar{p}} \left(\hat{Q}_{\hat{\epsilon}} (u) - Q_\epsilon(u|\theta)  \ \right)  \mathbf{P}^R(u)' du - \hat{\boldsymbol{A}} \right] W^R\left[\int_{\underline{p}}^{\bar{p}}\left(\hat{Q}_{\hat{\epsilon}} (u) - Q_Y(u|\theta)   \right)  \mathbf{P}^R(u) du - \hat{\boldsymbol{A}} \right]\, ,
	\end{aligned} 
\end{equation}
where $\hat{Q}_{\hat{\epsilon}}(u)$ is the empirical quantile function of $\{h^{-1}(Y_i, X_i;\hat{\gamma}): i =1,\ldots, T\}$, with $\hat{\gamma}$ a first-step estimator of $\gamma_0$ computed from a sample independently from $\{(X_i,Y_i) : i=1,\ldots, T\}$. The adjustment term $\hat{\boldsymbol{A}}$ is an estimator of the first-step influence function \citep{Ichimura2021}, which reflects the impact of estimating $\hat{\gamma}$ on $\gamma \mapsto \hat{Q}_{h^{-1}(Y,X;\gamma)}$. The Supplemental Appendix provides the form of this correction in three examples. We also show that the asymptotic distribution of $\hat{\theta}$ may be computed as in the previous sections, provided that we adjust it to account for first-step estimation error. This correction can also be used to compute the optimal-weighting scheme of L-moments.

To illustrate this approach, we rely on expenditure data in a ridesharing platform collected by \cite{Biderman2018}. We observe weekly expenditures (in Brazilian reais) in the platform during eight weeks between August and Semptember 2018, for a subset of 3,961 users of the service in the municipality of São Paulo, Brazil. For these users, we also have access to survey data on sociodemographic traits and commuting motives. We denote by $Y_{it}$ the amount spent by user $i$ in week $t$, whereas $X_i$ collects their survey information.

\begin{figure}[h]

	\begin{subfigure}{0.5\textwidth}
		\includegraphics[scale=0.65]{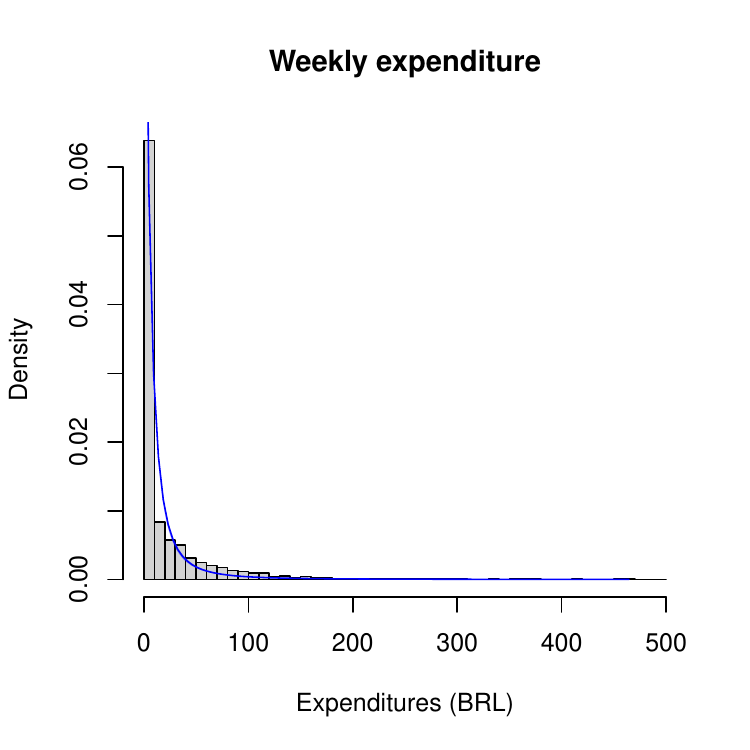}
				\caption{Levels}
		\label{app_level}
	\end{subfigure}
		\begin{subfigure}{0.5\textwidth}
		\includegraphics[scale=0.65]{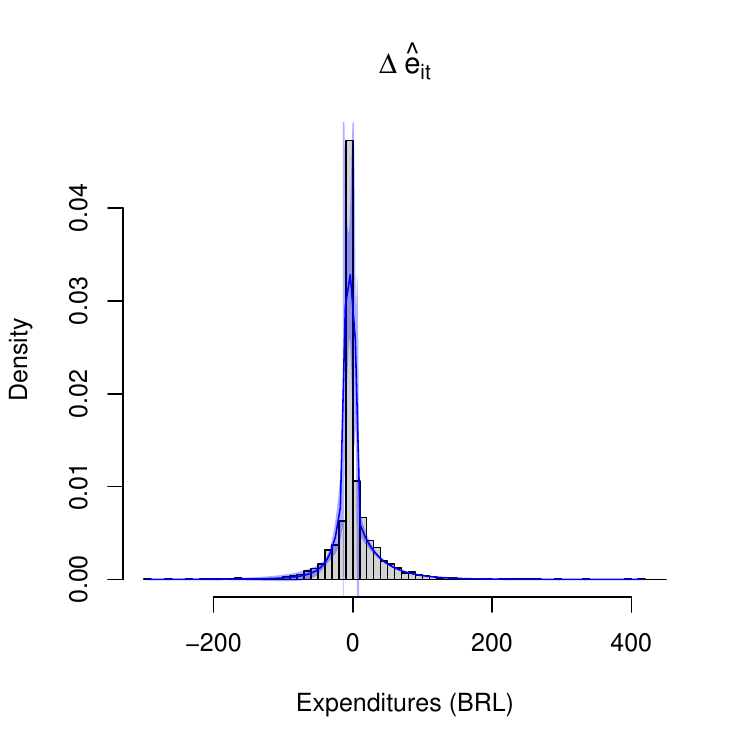}
			\caption{Error term}
		\label{app_err}
	\end{subfigure}
		\caption{Empirical application: expenditure patterns in mid-to-late September}
\label{fig_app}
\end{figure}

Panel \ref{app_level} plots the histogram for the distribution of expenditures in the penultimate week of our sample (mid-to-late September). The data clearly exhibits heavy tails: the maximum observed expenditure is 464.34 Brazilian reais, whereas average expenditure amounts to 20.81 reais.\footnote{For completeness, in late September 2018, 1USD=4 Brazilian reais.} Moreover, 58\% of individuals do not spend any money in rides during this period. The solid line presents the density of a GPD distribution fit to this data. The parameters of the distribution are estimated via the two-step generalized method-of L-moment estimator discussed in Section \ref{monte_carlo}, with $R=65$, which corresponds to the optimal choice for estimating several quantiles across the distribution, according to the RMSE criterion discussed in Section \ref{selection}. Even though the overidentifying restrictions test clearly rejects the null (p-value $\approx$ 0), we take the plotted density as further confirmation of heavy-tailedness of expenditure patterns, since the estimated GPD density understates mass at larger support points.

We seek to understand whether individual time-invariant heterogeneity, along with persistence in expenditure patterns, is able to explain the observed heavy-tailedness. To accomplish this, we posit the following model for the evolution of expenditures:
$$Y_{it} = \alpha_i + a(X_i) t + b(X_i) Y_{i,t-1} + \epsilon_{i,t} \, ,$$
where $\alpha_i$ is unobserved time-invariant heterogeneity (here treated as a fixed effect), and $\epsilon_{it}$ is time-varying idiosyncratic heterogeneity that is assumed to be, conditionally on $X_i$, independent across time. The coefficients $a(X_i)$ and $b(X_i)$ measure respectively deterministic trends and persistence in individual consumption patterns. We allow these coefficients to be nonparametric functions of survey information. Finally, we also assume that $\mathbb{E}[\epsilon_{it}|X_i]=0$, meaning that $(a(X_i), b(X_i))$ correctly capture mean heterogeneity in consumption trends attributable to $X_i$.

To estimate the above model, we take first-differences to remove the fixed effect, i.e. we consider:
\begin{equation}
	\label{eq_diff_model}
\Delta Y_{i,t} = a(X_i) + b(X_i) \Delta Y_{i,t-1} + \Delta \epsilon_{i,t}\,.
\end{equation}

Under the assumption that the $\epsilon_{i,t}$ are (conditionally on $X_i$) independent across time, we may then use $Y_{i,t-2}$ as a valid  instrument for the endogenous variable $\Delta Y_{i,t-1}$ \citep{Anderson1982,Arellano1991}. We estimate \eqref{eq_diff_model} using the instrumental forest estimator of \cite{Athey2019}, which assumes $a$ and $b$ to be in the closure of the linear span of regression trees.

Panel \ref{app_err} reports the histogram of the residuals $ \Delta \widehat \epsilon_{it}$ in mid-to-late September, where we adopt sample-splitting and estimate the functions $a(\cdot)$ and $b(\cdot)$ using data from the weeks prior to the penultimate week in the sample. The distribution is two-sided, with a large mass just below zero, suggesting that model \eqref{eq_diff_model} somewhat overpredicts expenditure variation in mid-to-late September. To assess whether the data exhibits heavy-tails, we estimate a GEV mixture model for the distribution of $\Delta \epsilon_{it}$, assuming that $\mathbb{P}[\Delta \epsilon_{it} \leq x] = \omega ( (1-\delta_1)/2 + \delta_1 F_{\theta_1}(\delta_1x)) + (1-\omega)( (1-\delta_2)/2 + \delta_2 F_{\theta_1}(\delta_2x))$, with $\omega \in [0,1]$, $\delta_1,\delta_2 \in \{-1,1\}$, and $F_{\theta_1}$ and $F_{\theta_2}$ belonging to the GEV family described in Section \ref{monte_carlo}. Our formulation allows for the left- and right-tails to exhibit different decay, e.g. if $\delta_1 = -1$ and $\delta_2 = -1$ and the shape parameter of $F_{\theta_1}$ is negative while $F_{\theta_1}$  is positive, then the left-tail behaves as a Fréchet, while the right-tail behaves as a Weibull distribution. Moreover, if the distributions being mixed by the weights $\omega$ exhibit disjoint supports, then the quantile function of the mixture admits a simple closed-form solution \citep{castellacci2012formula}, which enables us to rely on closed-form expressions for the L-moments of the GEV family to compute theoretical L-moments. We estimate the parameters $(\omega, \delta_1, \delta_2, \theta_1,\theta_2)$ by relying on the adjusted estimator \eqref{eq_residual_ts}, with two-step optimal weights and the form of correction $\hat{\boldsymbol{A}}$ derived in Example  3 in Supplemental Appendix L.1.

The solid line in Panel \ref{app_err} reports the fitted density of the GEV mixture, while the shaded area plots a 95\% uniform confidence band for the density function over the support of $\Delta \hat \epsilon_{it}$. The band is computed using the delta-method and sup-t critical values \citep{Freyberger2018}. Overall, the fit appears adequate, as evidenced by the overidentifying restrictions test not rejecting the null at the usual significance levels ($\text{p-value}\approx 1$). We then use our parameter estimates to test the null hypothesis that the left (right) tail exhibits exponentially light decay, against the alternative that it is heavy-tailed. This corresponds to testing whether the left-tail (right-tail) shape parameter is in the set $[0,1]$, against the alternative that it is not.\footnote{When the shape parameter equals zero, the GEV distribution collapses to a Gumbel distribution, which has exponentially light tails \citep{chernozhukov2011inference}. When the shape parameter is strictly greater than zero but smaller than one, the distribution collapses to a Weibull distribution with  shape parameter greater than one, a region for which the Weibull is known to have exponentially light tails \citep{foss2011introduction}. The region $(-\infty, 0)$ corresponds to a Fréchet distribution, whereas the region $(1, \infty)$ corresponds to a Weibull with shape parameter strictly greater than zero and strictly less than unity -- both cases corresponding to heavy-tailed distributions.} Upon computation of a 95\% confidence interval for the right- (left-) tail shape parameters, we verify that, while the confidence region for the right-tail shape parameter is entirely contained in the $(1,\infty)$ region, the confidence region for the left-tail shape parameter is $[-0.934, 0.124]$, which intersects with $[0,1]$. Therefore, at the 5\% significance level, we reject the null of exponentially light decay for the right-tail of the distribution, though we fail to do so for the left-tail. These results provide evidence that, even after accounting for heterogeneous persistence and trends, as well as time-invariant unobserved heterogeneity, weekly expenses still exhibit very positive idiosyncratic realizations. Such pattern is consistent with, in any given week, some individuals having to take very long trips (e.g. taking a ride to the airport), the demand for which may be hard to anticipate on the basis of observable traits.\footnote{We thank a referee for providing this interpretation of the conclusion of the test.}

As a final application of our approach, we show how our estimator of the distribution of $\Delta \epsilon_{it}$ may be used to construct prediction intervals for individual treatment effects \citep{Cattaneo2021,CWZ2021,CWZ2021b}, a useful tool in assessing the impacts of personalized interventions \citep{Kivaranovic2020,Lei2021}. Specifically, suppose that, in some week, the ridesharing company implements a personalized policy in the platform, e.g. a change in a parameter of the pricing algorithm that may result in disparate fees being charged across users. In this causal inference setting, the model \eqref{eq_diff_model} may be seen as a model for the untreated potential outcome, $\Delta Y_{i,t}(0)$, that is observed in the periods $t$ prior to the intervention at $t^*$, where $t^*$ is a random variable denoting the date when the intervention starts. Under the assumption that $t^*$ is independent of $\Delta \epsilon_{i,t^*}$,\footnote{In our example, this assumption allows the decision of when to implement the policy to depend on the values of $\alpha_i$, $X_i$ and the $Y_{i,t}$ prior to the treatment, but essentially excludes the possibility of the ridesharing platform, in its decision of when to change the pricing parameter, to rely on better predictions of post-treatment values of $\Delta Y_{it}(0)$ than those obtained from the predictable part of \eqref{eq_diff_model}, as this would introduce dependence between $t^*$ and the post-treatment variation in idiosyncratic components driving demand absent the intervention ($\Delta \epsilon_{i,t^*}$). See \cite{Ferman2021} and \cite{AlvarezFerman2024} for a discussion on the interpretation of similar assumptions in synthetic control designs.} the interval:

$$\hat{I}_{i,1-\alpha} =  [\Delta Y_{i,t^*} - \hat{a}(X_i) - \hat{b}(X_i)\Delta Y_{i,t^*-1} - Q_{\hat{\theta}}(1-\alpha) ,\infty)\, ,$$ 
is an asymptotically (in a regime where the number of users diverges) valid $(1-\alpha)$ prediction region for the individual treatment effect $Y_{it^*}(1)-Y_{it^*}(0)$, where $\hat{a}$, $\hat{b}$ and $\hat{\theta}$ are estimators computed using the pre-intervention sample, and asymptotic validity is meant as, when the number of users diverges:
$$\mathbb{P}[Y_{it^*}(1)- Y_{it^*}(0) \in \hat{I}_{i,1-\alpha}] \to 1-\alpha\, ,$$
(see Supplemental Appendix L.2 for details). The lower-bound of our one-sided prediction region has a Value-at-Risk type interpretation, representing the largest loss $c$ with the policy that cannot be rejected, in a test of the null $H_0: Y_{it^*}(1)- Y_{it^*}(0)\leq c$ against the alternative $H_1: Y_{it^*}(1)- Y_{it^*}(0)> c$, at the $\alpha$ significance level. Two-sided intervals can also be considered, as well as Bonferroni-style corrections to $\hat{I}_{i,1-\alpha}$ in order to account for estimation error of $\hat{a}$,$\hat{b}$ and $\hat{\theta}$ (see \cite{Cattaneo2021} and Supplemental Appendix L.2 for details).

As an illustration of our approach to constructing prediction intervals, we implement the intervals $\hat{I}_{i,0.95}$ in our data, assuming that the last week of September is the post-treatment period. Given that there was no known intervention in this period (i.e. $Y_{it^*}(1)-Y_{it^*}(0)$ is known to be zero for every unit), one would expect  that these regions would contain $0$ in approximately $95\%$ of the cases. This is indeed what we observe in the data: out of the 3,961 users in our sample, the corresponding individual prediction intervals do not contain zero in only 175 (4.4\%) of the cases.

\subsection{Conditional quantile models}

Let $Q_{Y|X}(\cdot|X)$ be the quantile  function of a conditional distribution function $F_{Y|X}(\cdot|X)$, where $Y$ is a scalar outcome and $X$ is a set of controls. Following \cite{Gourieroux2008}, we define the $r$-th conditional L-moment as:
\begin{equation}
	\label{eq_conditional_lmoment}
	\lambda_r(X) \coloneqq \int_{\underline{p}}^{\overline{p}} Q_{Y|X}(u|X)P_r(u)du \, .
\end{equation}

When $0 = \underline{p} \leq \overline{p} =1$, \cite{Gourieroux2008} note that $\lambda_r(X) = \mathbb{E}[Y P_l(F_{Y|X}(Y|X))|X]$. They suggest estimating $F_{Y|X}$ nonparametrically, and, for a fixed number $R$ of L-moments, to exploit the following $R K$ unconditional moments in the estimation of conditional parametric models $\{Q_{Y|X}(\cdot|\cdot;\theta): \theta \in \Theta\}$:
\begin{equation}
	\label{eq_uncond_moments}
\mathbb{E}\left[w(X) \otimes \left(Y \boldsymbol{P}^R(F_{Y|X}(Y|X)) - \int_{0}^{1} Q_{Y|X}(u|X;\theta_0)\boldsymbol{P}^R(u)du\right) \right] = 0\,, 
\end{equation}
where $w(X)$ is a $ R \times 1$ vector of transformations of $X$. 

In spite of its conceptual attractiveness -- L-moment estimation is cast as method-of-moment estimation --, formulation \eqref{eq_uncond_moments} does not directly extend to settings with trimming.\footnote{To implement trimming in this formulation would require nonparametric estimation of both the conditional distribution and conditional quantile functions, whereas our suggested approach solely relies on the latter.} Moreover, by working with fixed $R$ and $K$, it does not fully exploit the identifying information in the parametric model. In light of these points, Supplemental Appendix L.3 proposes an alternative method-of-L-moment estimator for conditional models. We propose to estimate \eqref{eq_conditional_lmoment} by directly plugging into the representation a nonparametric conditional quantile estimator. Following \cite{Ai2003}, we then optimally optimally exploit the conditional L-moment restrictions by weighting these using $R\times R$ weighting functionals $\Omega(X)$. In the Supplemental Appendix, we consider the case where we rely on the quantile series regression estimator of \cite{Belloni2019} for preliminary nonparametric estimation, though in principle any nonparametric estimator of conditional quantile functions for which an approximation theory is available could be used in this first step. Examples of such estimators include local polynomial quantile regression \citep{Yu1998,Guerre2012} and quantile regression forests \citep{Meinshausen2006,Athey2019}. Under regularity conditions, our estimator admits an asymptotic linear representation that can be used as a basis for inference, and for finding the optimal choice of functional $\Omega(X)$. Moreover, by suitably taking $R \to \infty$, we expect our optimally-weighted estimator to achieve good finite-sample performance, whilst retaining asymptotic efficiency -- indeed, as we argue in the Supplemental Appendix, the optimally-weighted estimator with no trimming and an orthonormal \emph{basis} choice of $\{P_l\}_l$ is asymptotically efficient as $T,R\to\infty$.

%% file: sections/conclusion.tex
This paper considered the estimation of parametric models using a ``generalised'' method of L-moments procedure, which extends the approach introduced in \cite{Hosking1990} whereby a $d$-dimensional parametric model for a distribution function is fit by matching the first $d$ L-moments. We have shown that, by appropriately choosing the number of L-moments and under an appropriate weighting scheme, we are able to construct an estimator that is able to outperform maximum likelihood estimation in small samples from popular distributions, and yet does not suffer from efficiency losses in larger samples. We have developed tools to automatically select the number of L-moments used in estimation, and have shown the usefulness of such approach in Monte Carlo simulations. We have also extended our L-moment approach to the estimation of conditional models, and to the ``residual analysis'' of semiparametric models. We then applied the latter to study expenditure patterns in a ridesharing platform in São Paulo, Brazil.

The extension of the generalised L-moment approach to other semi- and nonparametric settings appears to be an interesting venue of future research. The L-moment approach appears especially well-suited to problems where semi- and nonparametric maximum likelihood estimation is computationally complicated, but evaluation of integrals of quantiles is not. In followup work, \cite{alvarez2023quantile} propose using the generalised method-of-L-moment approach to estimate nonparametric quantile mixture models, while \cite{alvarez2024learning} introduce an efficient generalised L-moment estimator for the semiparametric models of treatment effects of \cite{Athey2021}. The study of such extensions in more generality is a topic for future research.

%% file: tables/gev/gev_table_linear.tex
\begin{table}[H]
\centering
\caption{GEV : maximal and minimal relative RMSE for linear combintations of parameters (MSE-minimising choice of $R$)} 
\label{gev_table_linear}
\begin{tabular}{|l|ccc|ccc|}
  \hline
  & \multicolumn{ 3 } {c|}{ Most favourable $\delta$ }&\multicolumn{ 3 } {c|}{ Least favourable $\delta$ } \\ & T= 50 & T= 100 & T= 500 & T= 50 & T= 100 & T= 500 \\ 
   \hline
Càglàd FS & 0.957 & 0.993 & 1.000 & 1.077 & 1.142 & 1.253 \\ 
   &  (3) &  (3) &  (5) &  (3) &  (3) &  (3) \\ 
  Càglàd TS & 0.922 & 0.966 & 0.999 & 1.032 & 1.016 & 1.013 \\ 
   &  (11) &  (11) &  (3) &  (4) &  (5) &  (30) \\ 
  Unbiased FS & 0.929 & 0.970 & 0.995 & 1.113 & 1.172 & 1.259 \\ 
   &  (3) &  (5) &  (5) &  (3) &  (3) &  (3) \\ 
  Unbiased TS & 0.928 & 0.969 & 0.994 & 1.037 & 1.013 & 1.018 \\ 
   &  (3) &  (3) &  (5) &  (5) &  (7) &  (20) \\ \hline
  \end{tabular}
\end{table}

%% file: tables/gpd/gpd_table_linear.tex
\begin{table}[h]
\centering
\caption{GPD : maximal and minimal relative RMSE for linear combintations of parameters (MSE-minimising choice of $R$)} 
\label{gpd_table_linear}
\begin{tabular}{|l|ccc|ccc|}
  \hline
  & \multicolumn{ 3 } {c|}{ Most favourable $\delta$ }&\multicolumn{ 3 } {c|}{ Least favourable $\delta$ } \\ & T= 50 & T= 100 & T= 500 & T= 50 & T= 100 & T= 500 \\ 
   \hline
Càglàd FS & 0.895 & 0.931 & 0.988 & 1.004 & 1.004 & 1.004 \\ 
   &  (6) &  (4) &  (4) &  (3) &  (3) &  (6) \\ 
  Càglàd TS & 0.874 & 0.912 & 0.977 & 0.996 & 1.001 & 1.002 \\ 
   &  (3) &  (3) &  (3) &  (3) &  (3) &  (7) \\ 
  Unbiased FS & 0.838 & 0.902 & 0.980 & 0.960 & 0.982 & 1.000 \\ 
   &  (6) &  (4) &  (3) &  (3) &  (3) &  (7) \\ 
  Unbiased TS & 0.838 & 0.878 & 0.962 & 0.960 & 0.977 & 0.997 \\ 
   &  (3) &  (5) &  (34) &  (2) &  (10) &  (38) \\ \hline
  \end{tabular}
\end{table}

%% file: tables/gev/gev_additional.tex
\begin{table}[H]
\centering
\caption{GEV : comparison with Trimmed and Tilted MLE methods} 
\label{gev_table_additional}
\adjustbox{width=\textwidth}{
\begin{tabular}{|l|cccc|cccc|cccc|}
  \hline
  & \multicolumn{ 4 } {c|}{$T = 50 $}&\multicolumn{ 4 } {c|}{$T = 100 $}&\multicolumn{ 4 } {c|}{$T = 500 $} \\ & $\tau = 0.5 $ & $\tau = 0.9 $ & $\tau = 0.99 $ & $\tau = 0.999 $ & $\tau = 0.5 $ & $\tau = 0.9 $ & $\tau = 0.99 $ & $\tau = 0.999 $ & $\tau = 0.5 $ & $\tau = 0.9 $ & $\tau = 0.99 $ & $\tau = 0.999 $ \\ 
   \hline
Càglàd TS & 1.005 & 0.960 & 0.818 & 0.692 & 1.003 & 0.981 & 0.910 & 0.840 & 1.004 & 0.998 & 0.990 & 0.979 \\ 
   &  (12) &  (3) &  (5) &  (5) &  (11) &  (3) &  (5) &  (5) &  (30) &  (4) &  (90) &  (90) \\ 
   \hline
Trimming 10\% & 1.193 & 2.177 & 1.505 & 0.886 & 1.374 & 3.033 & 2.367 & 1.619 & 2.266 & 6.677 & 5.847 & 4.548 \\ 
   &  &  &  &  &  &  &  &  &  &  &  &  \\ 
  Trimming 1\% & 1.005 & 1.093 & 1.020 & 0.950 & 1.006 & 1.169 & 1.075 & 0.931 & 1.013 & 1.723 & 1.813 & 1.645 \\ 
   &  &  &  &  &  &  &  &  &  &  &  &  \\ 
  Trimming 0.1\% & 1.000 & 1.006 & 1.003 & 1.000 & 1.001 & 1.019 & 1.015 & 1.004 & 1.002 & 1.053 & 1.063 & 1.037 \\ 
   &  &  &  &  &  &  &  &  &  &  &  &  \\ 
   \hline
Tilting 10\% & 1.846 & 2.258 & 1.223 & 0.736 & 2.428 & 3.296 & 2.000 & 1.273 & 4.886 & 7.573 & 5.224 & 3.799 \\ 
   & (38\%) & (38\%) & (38\%) & (38\%) & (31\%) & (31\%) & (31\%) & (31\%) & (28\%) & (28\%) & (28\%) & (28\%) \\ 
  Tilting 1\% & 1.110 & 1.136 & 0.818 & 0.687 & 1.242 & 1.466 & 1.036 & 0.792 & 1.922 & 2.936 & 2.231 & 1.769 \\ 
   & (5\%) & (5\%) & (5\%) & (5\%) & (1\%) & (1\%) & (1\%) & (1\%) & (0\%) & (0\%) & (0\%) & (0\%) \\ 
  Tilting 0.1\% & 1.005 & 0.961 & 0.880 & 0.844 & 1.026 & 1.018 & 0.905 & 0.852 & 1.133 & 1.335 & 1.137 & 1.016 \\ 
   & (4\%) & (4\%) & (4\%) & (4\%) & (1\%) & (1\%) & (1\%) & (1\%) & (0\%) & (0\%) & (0\%) & (0\%) \\ 
   \hline
\end{tabular}
}
\end{table}

%% file: tables/gpd/gpd_additional.tex
\begin{table}[H]
\centering
\caption{GPD : comparison with Trimmed and Tilted MLE methods} 
\label{gpd_table_additional}
\adjustbox{width=\textwidth}{
\begin{tabular}{|l|cccc|cccc|cccc|}
  \hline
  & \multicolumn{ 4 } {c|}{$T = 50 $}&\multicolumn{ 4 } {c|}{$T = 100 $}&\multicolumn{ 4 } {c|}{$T = 500 $} \\ & $\tau = 0.5 $ & $\tau = 0.9 $ & $\tau = 0.99 $ & $\tau = 0.999 $ & $\tau = 0.5 $ & $\tau = 0.9 $ & $\tau = 0.99 $ & $\tau = 0.999 $ & $\tau = 0.5 $ & $\tau = 0.9 $ & $\tau = 0.99 $ & $\tau = 0.999 $ \\ 
   \hline
Càglàd TS & 0.959 & 0.981 & 0.822 & 0.649 & 0.978 & 0.987 & 0.899 & 0.837 & 0.995 & 0.997 & 0.980 & 0.969 \\ 
   &  (3) &  (2) &  (3) &  (2) &  (3) &  (3) &  (3) &  (3) &  (5) &  (100) &  (3) &  (100) \\ 
   \hline
Trimming 10\% & 1.081 & 2.162 & 1.558 & 0.767 & 1.172 & 2.972 & 2.364 & 1.475 & 1.651 & 6.471 & 5.614 & 3.984 \\ 
   &  &  &  &  &  &  &  &  &  &  &  &  \\ 
  Trimming 1\% & 1.031 & 1.092 & 1.053 & 0.949 & 1.029 & 1.167 & 1.138 & 0.929 & 1.047 & 1.716 & 2.002 & 1.751 \\ 
   &  &  &  &  &  &  &  &  &  &  &  &  \\ 
  Trimming 0.1\% & 1.000 & 1.001 & 0.999 & 0.998 & 1.008 & 1.010 & 1.016 & 1.000 & 1.018 & 1.048 & 1.101 & 1.063 \\ 
   &  &  &  &  &  &  &  &  &  &  &  &  \\ 
   \hline
Tilting 10\% & 2.171 & 2.497 & 1.215 & 1.054 & 3.088 & 3.517 & 1.841 & 1.037 & 6.902 & 7.843 & 4.497 & 2.851 \\ 
   & (12\%) & (12\%) & (12\%) & (12\%) & (6\%) & (6\%) & (6\%) & (6\%) & (2\%) & (2\%) & (2\%) & (2\%) \\ 
  Tilting 1\% & 1.054 & 1.204 & 0.835 & 0.725 & 1.333 & 1.505 & 0.981 & 0.762 & 2.554 & 2.985 & 1.875 & 1.323 \\ 
   & (1\%) & (1\%) & (1\%) & (1\%) & (0\%) & (0\%) & (0\%) & (0\%) & (0\%) & (0\%) & (0\%) & (0\%) \\ 
  Tilting 0.1\% & 0.907 & 0.984 & 0.910 & 0.910 & 0.988 & 1.029 & 0.917 & 0.875 & 1.226 & 1.352 & 1.076 & 0.954 \\ 
   & (3\%) & (3\%) & (3\%) & (3\%) & (0\%) & (0\%) & (0\%) & (0\%) & (0\%) & (0\%) & (0\%) & (0\%) \\ 
   \hline
\end{tabular}
}
\end{table}

%% file: plots/gev/coverage_plots.tex
\begin{figure}[H]
	\centering

	\begin{subfigure}[H]{\textwidth}

		\centering
		\includegraphics[width=0.24\textwidth]{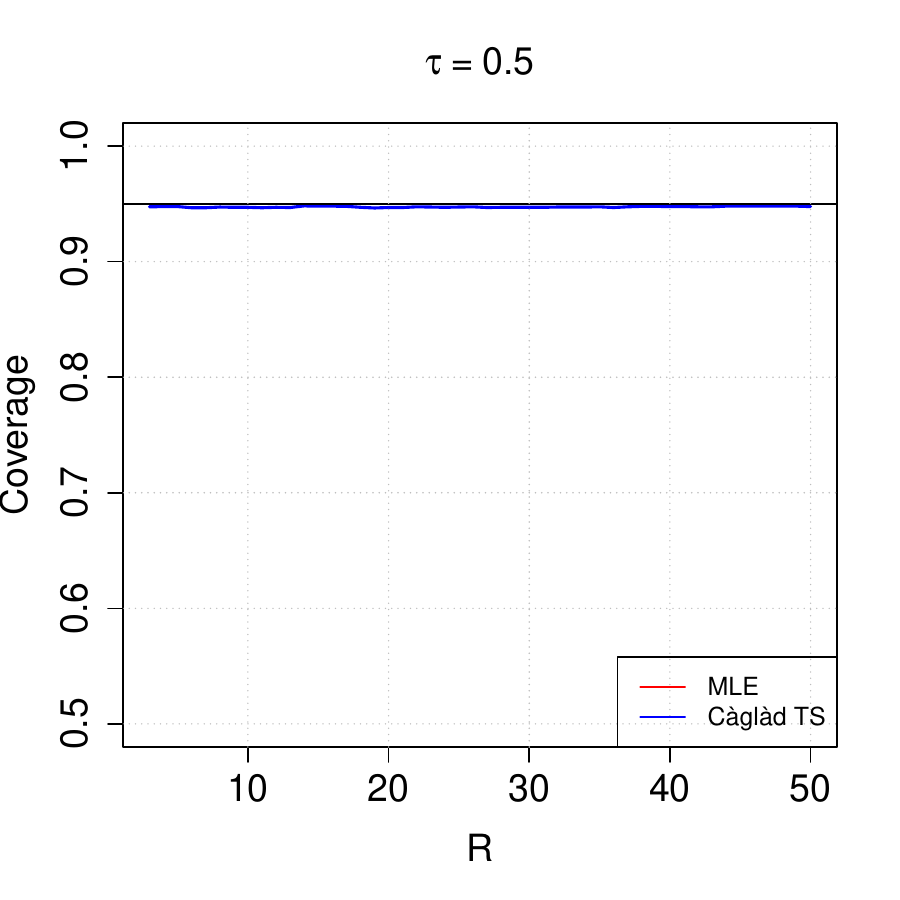}   \includegraphics[width=0.24\textwidth]{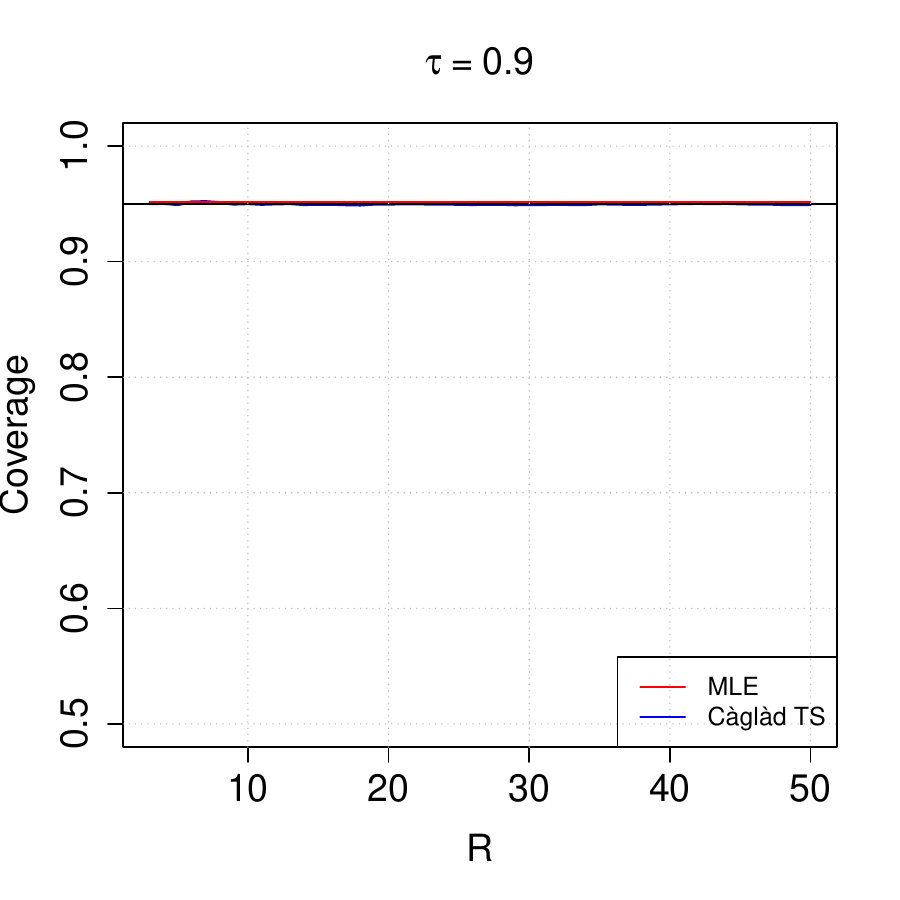}
		\includegraphics[width=0.24\textwidth]{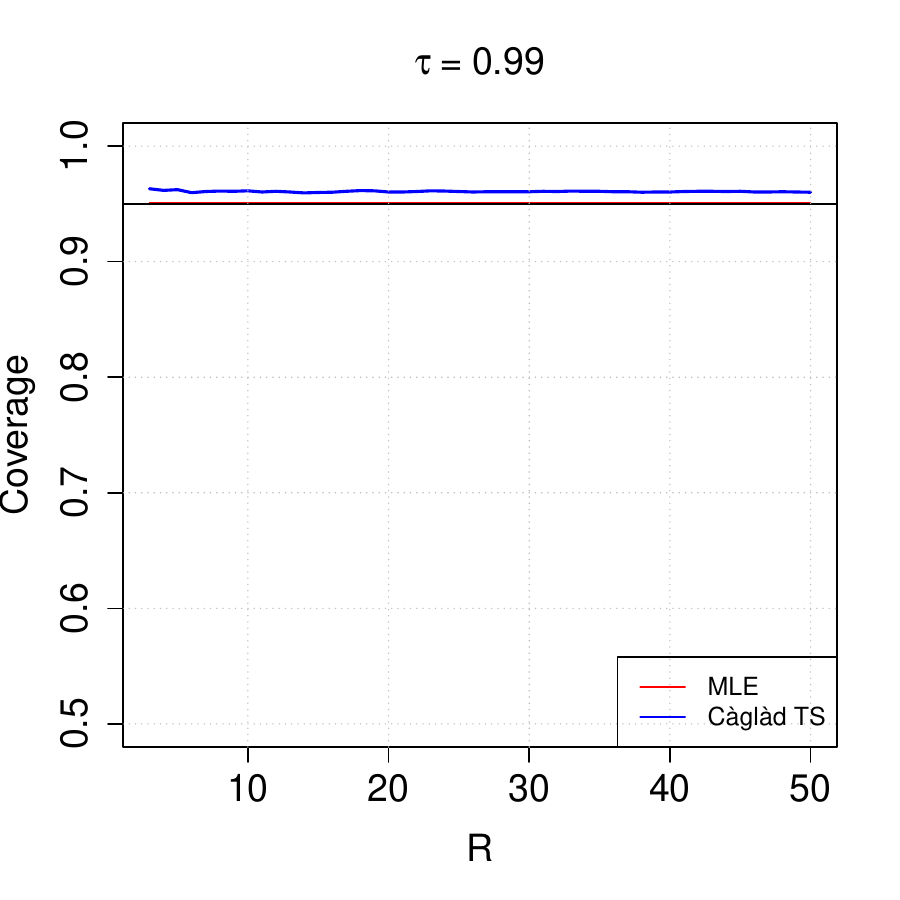}   \includegraphics[width=0.24\textwidth]{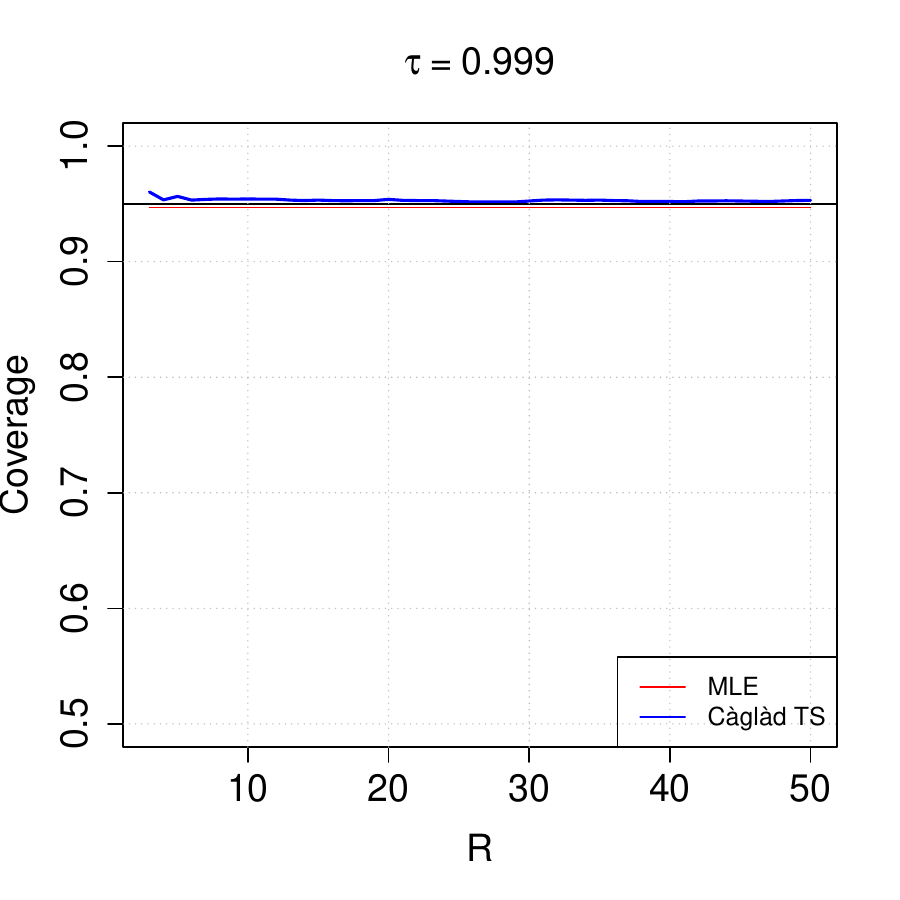}
		\caption{$T=50$}
	\end{subfigure}

	\begin{subfigure}[H]{\textwidth}

		\centering
		\includegraphics[width=0.24\textwidth]{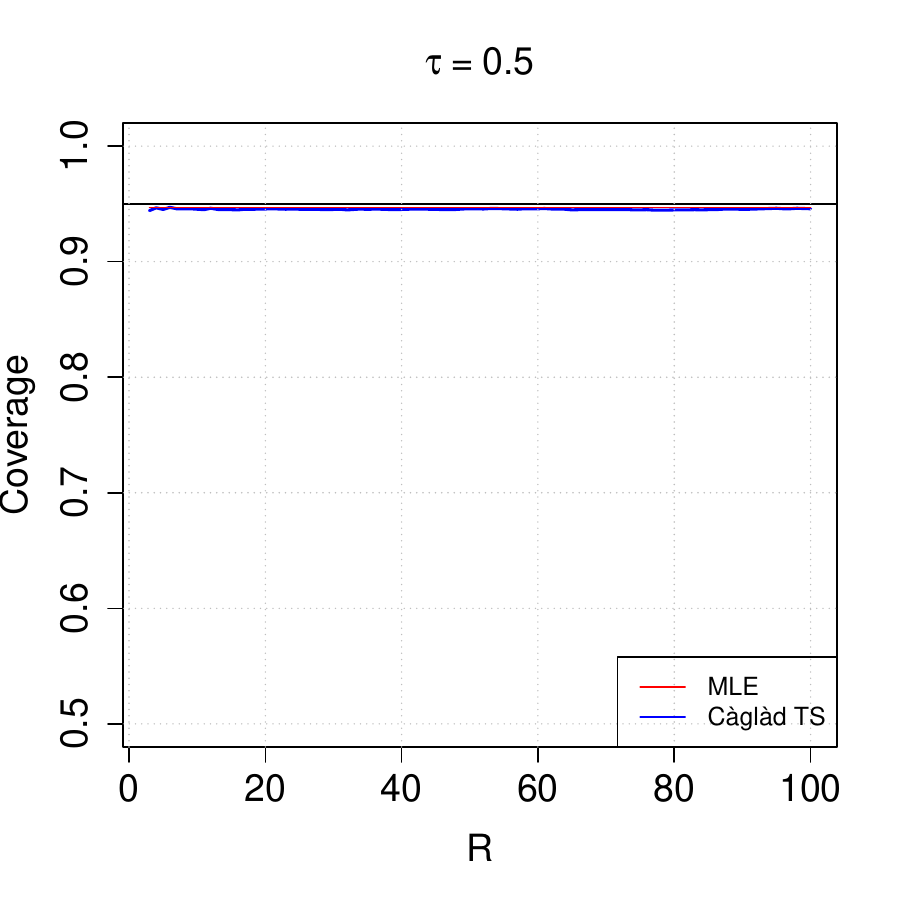}   \includegraphics[width=0.24\textwidth]{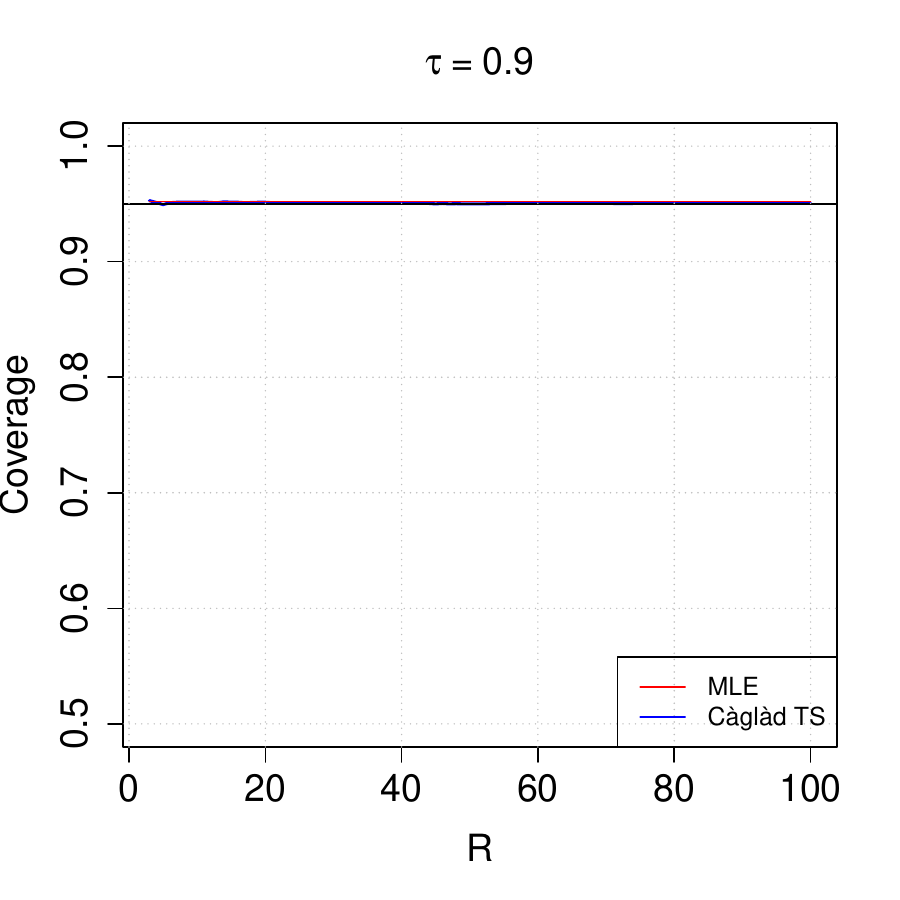}
		\includegraphics[width=0.24\textwidth]{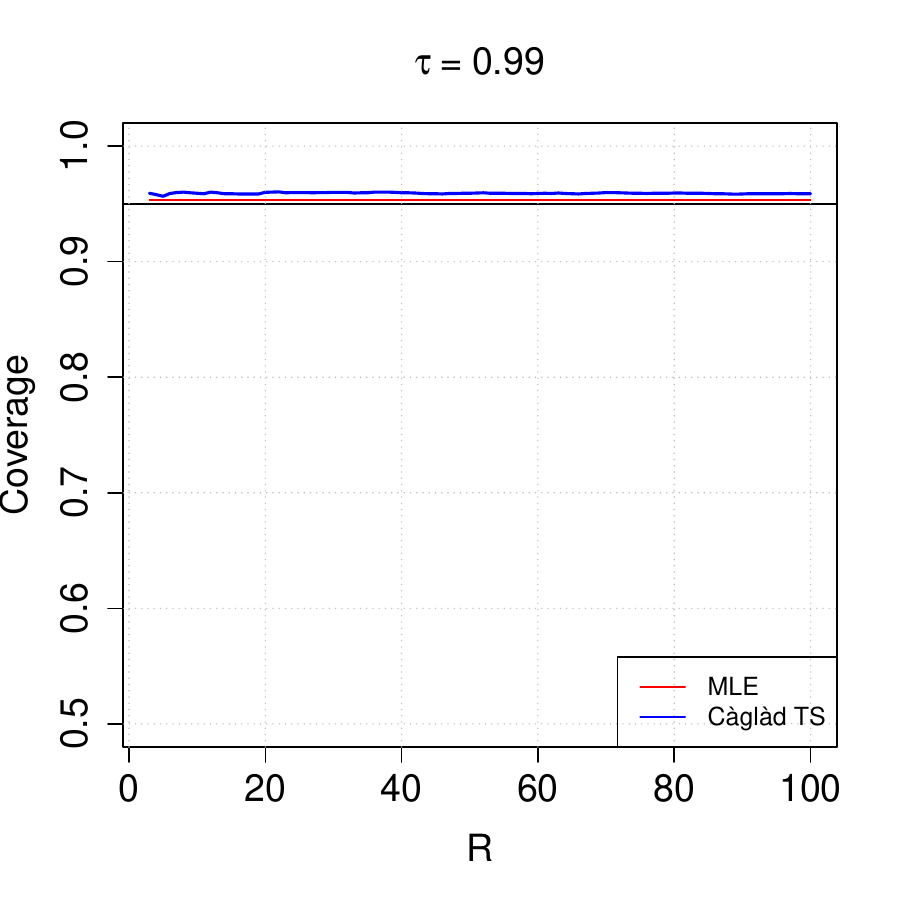}   \includegraphics[width=0.24\textwidth]{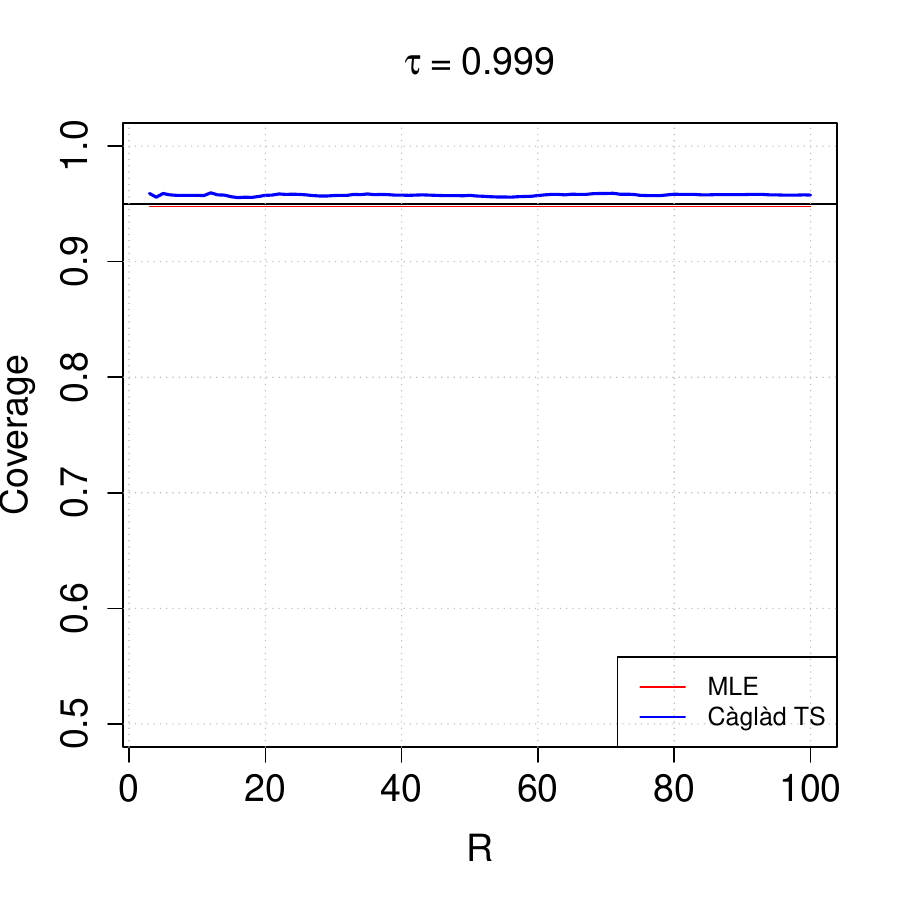}
		\caption{$T=100$}
	\end{subfigure}

	\begin{subfigure}[H]{\textwidth}

		\centering
		\includegraphics[width=0.24\textwidth]{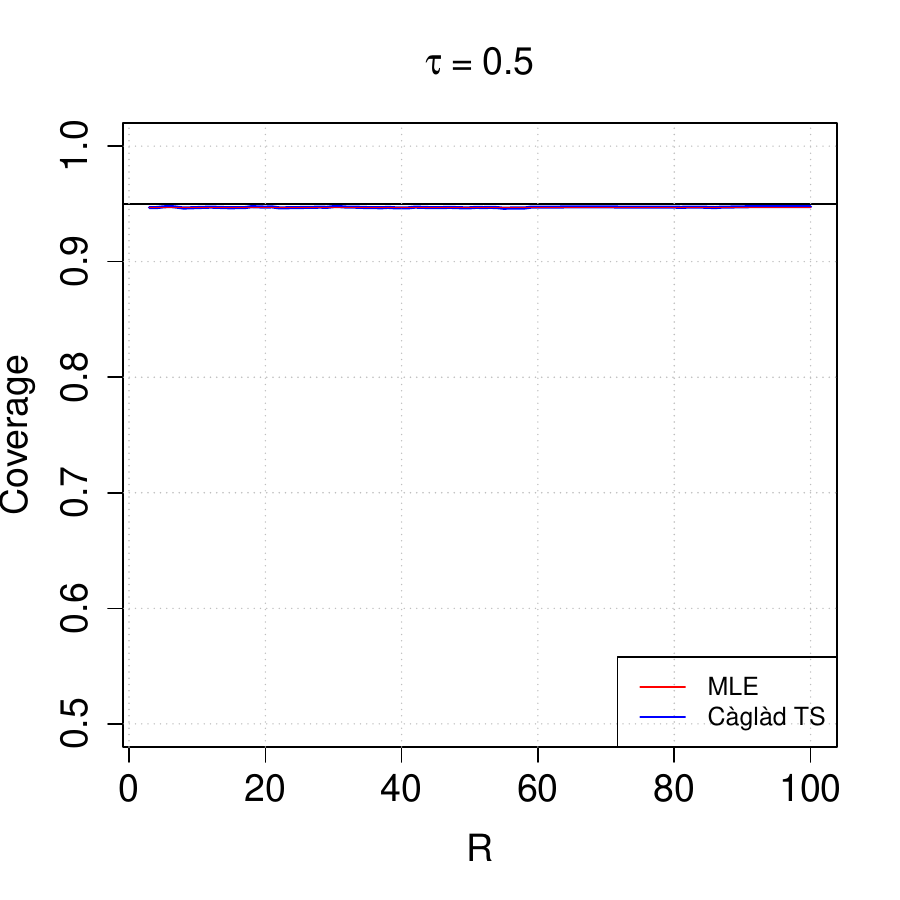}   \includegraphics[width=0.24\textwidth]{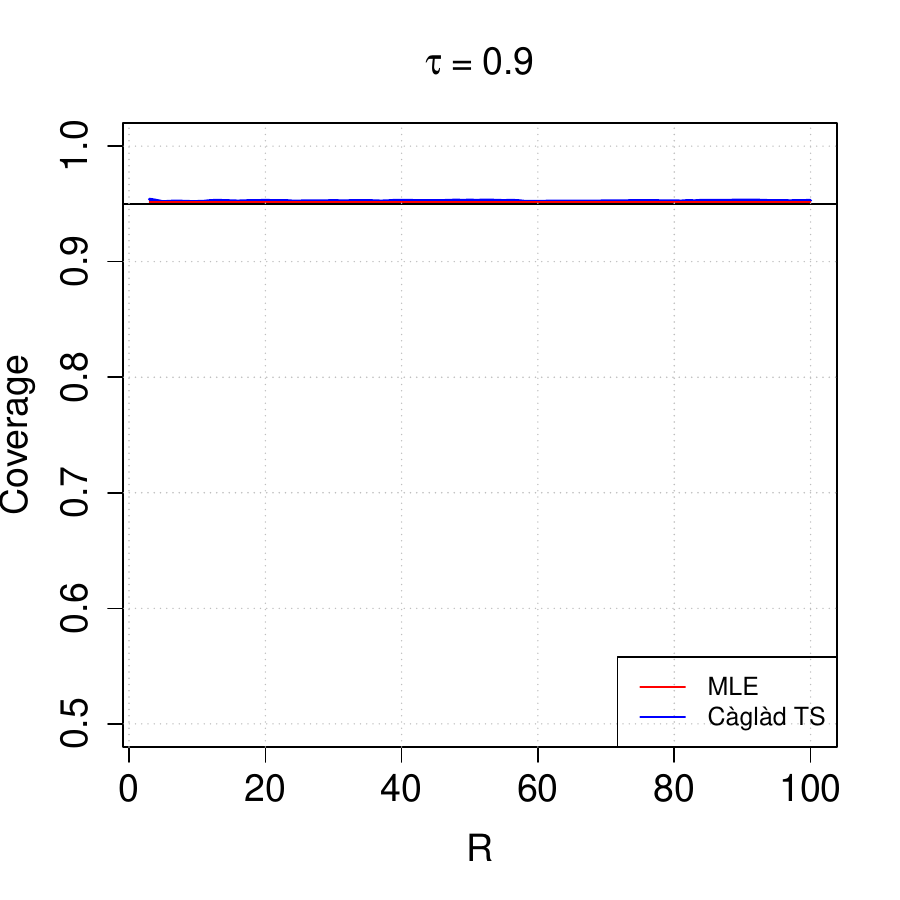}
		\includegraphics[width=0.24\textwidth]{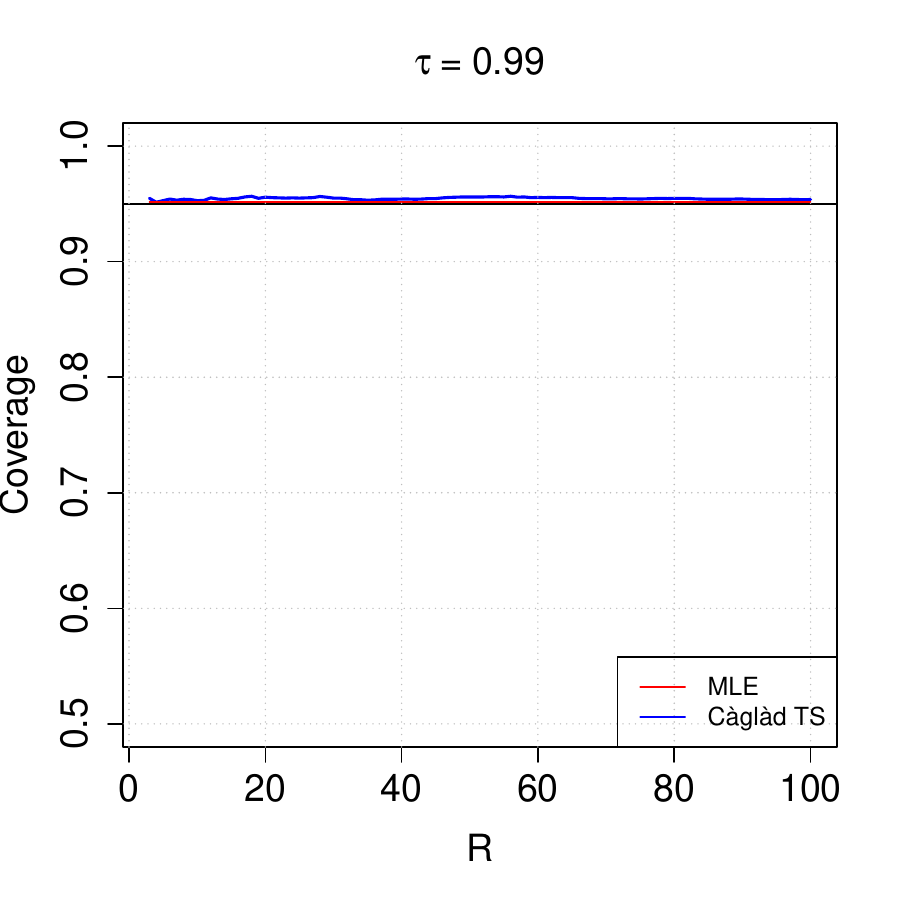}   \includegraphics[width=0.24\textwidth]{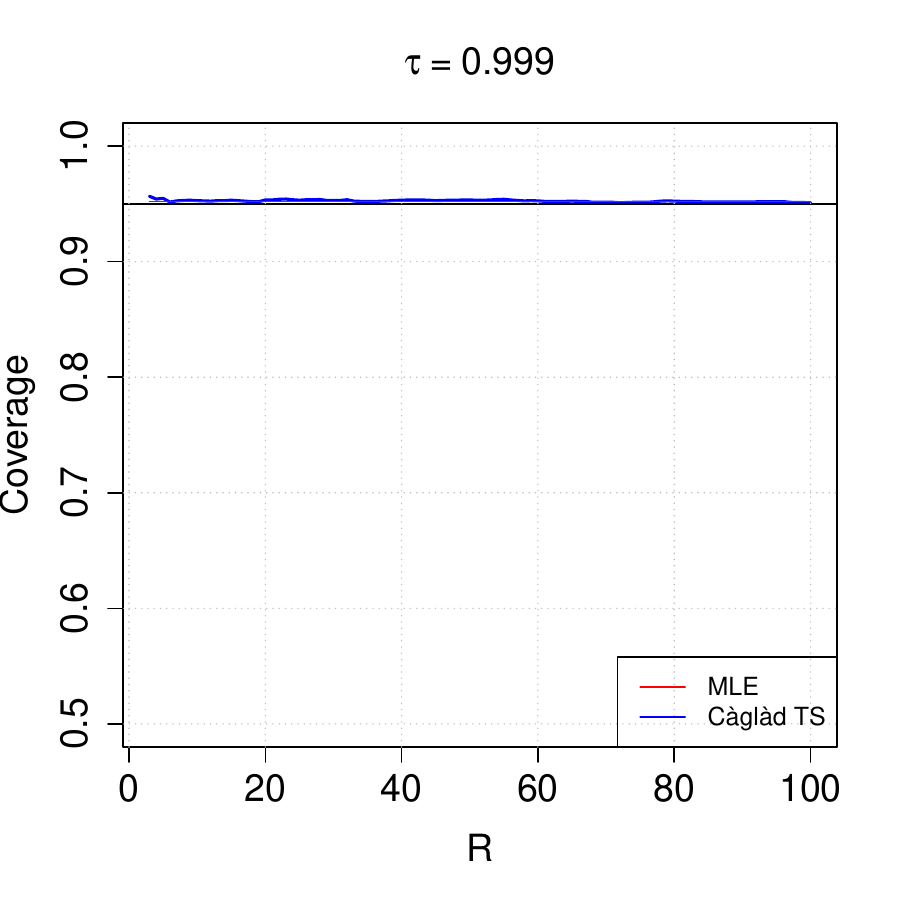}
		\caption{$T=500$}
	\end{subfigure}
	
	\caption{GEV: coverage of confidence intervals based on the true sampling variance.}
	\label{fig:gev_coverage_true}
\end{figure}

\begin{figure}[H]
	\centering

	\begin{subfigure}[H]{\textwidth}

		\centering
		\includegraphics[width=0.24\textwidth]{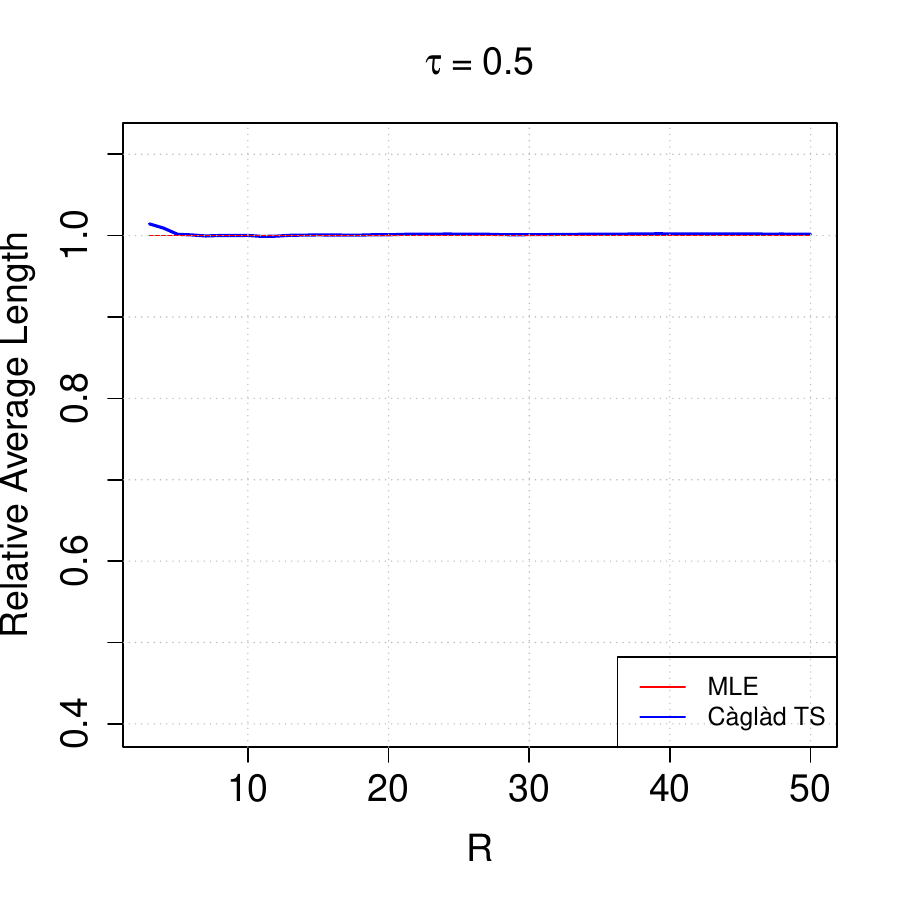}   \includegraphics[width=0.24\textwidth]{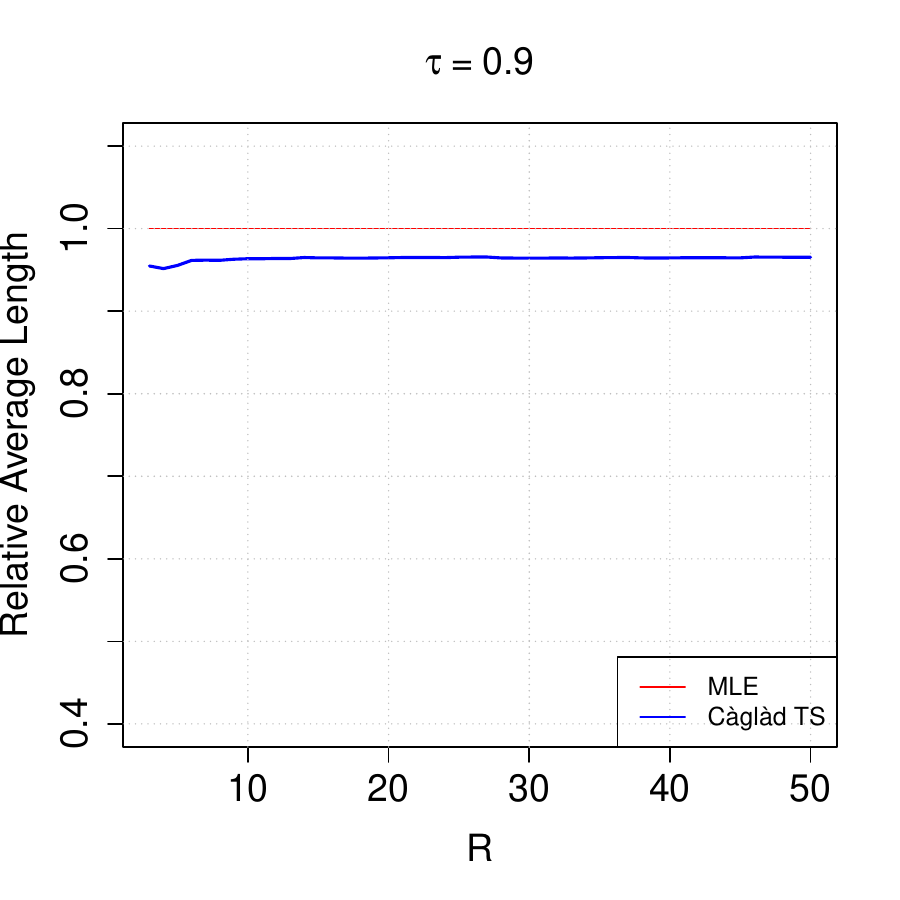}
		\includegraphics[width=0.24\textwidth]{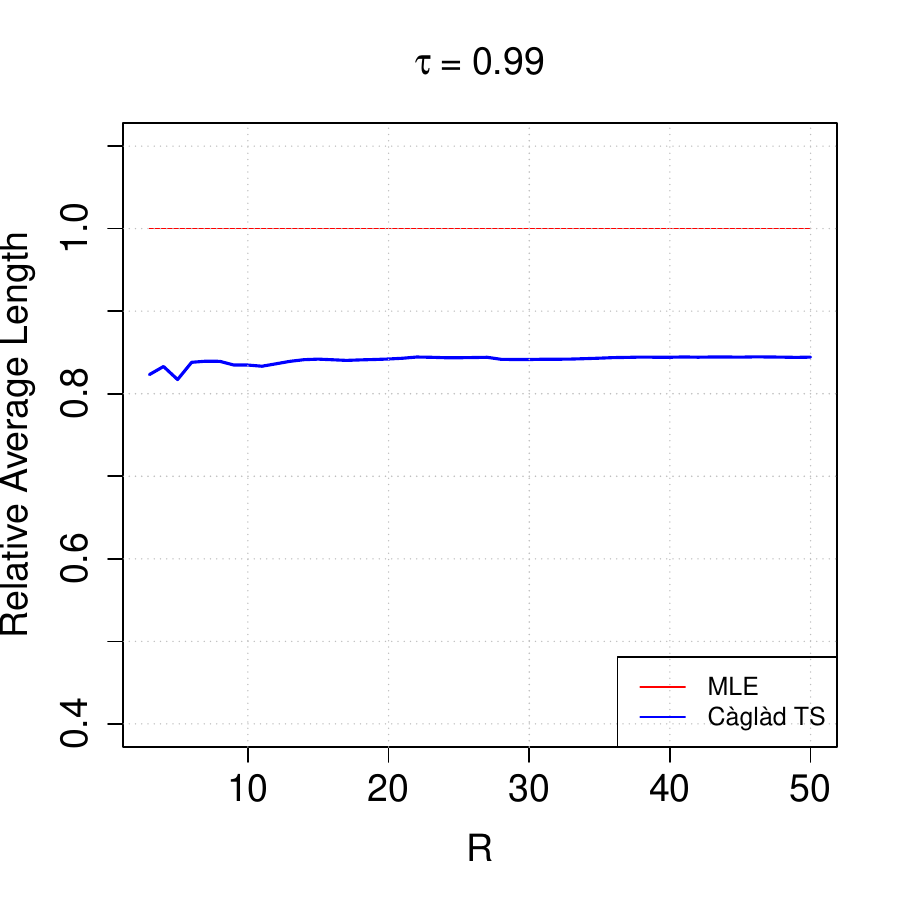}   \includegraphics[width=0.24\textwidth]{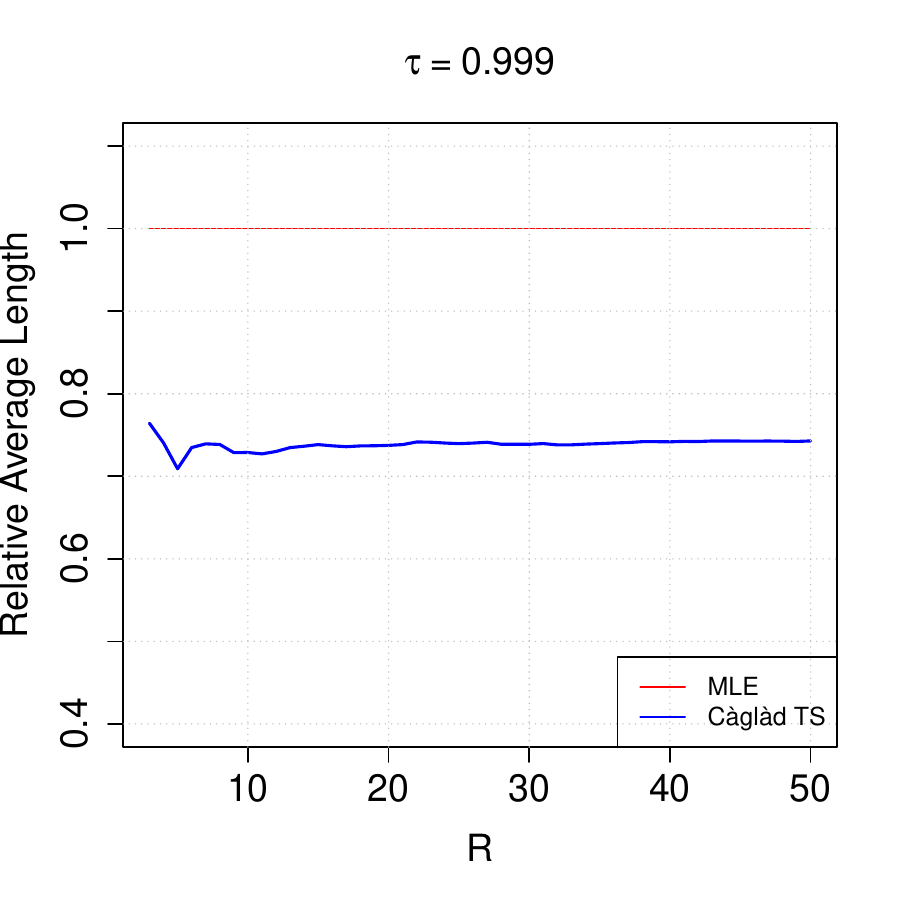}
		\caption{$T=50$}
	\end{subfigure}

	\begin{subfigure}[H]{\textwidth}

		\centering
		\includegraphics[width=0.24\textwidth]{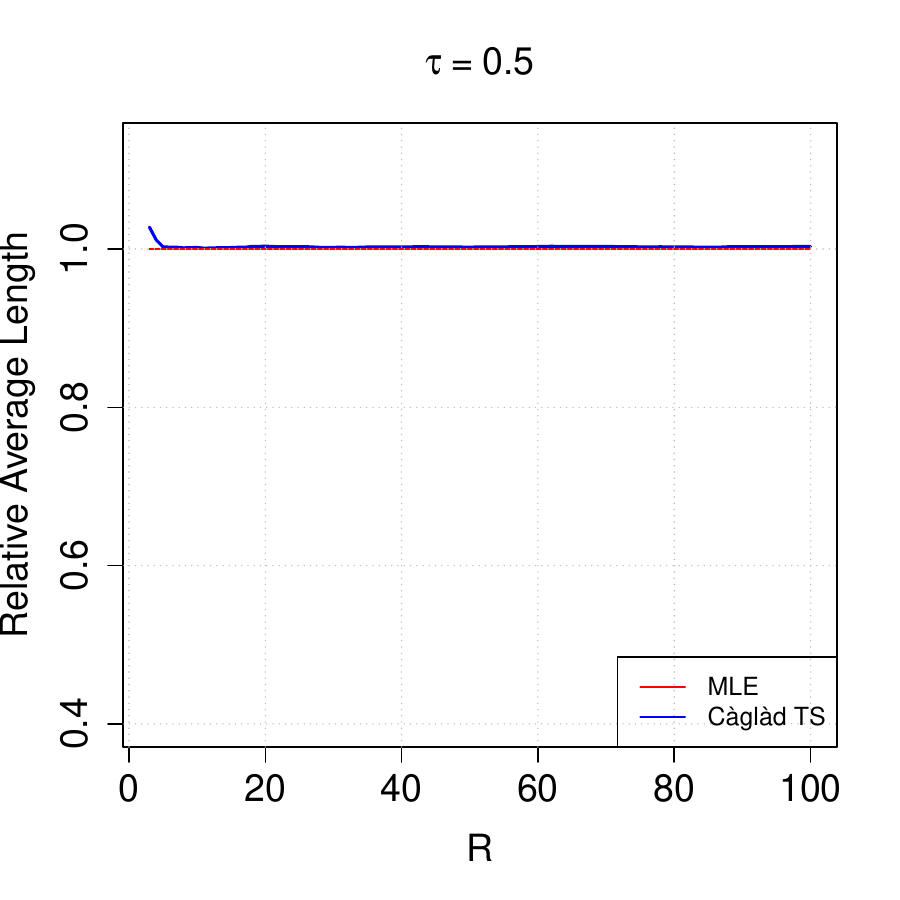}   \includegraphics[width=0.24\textwidth]{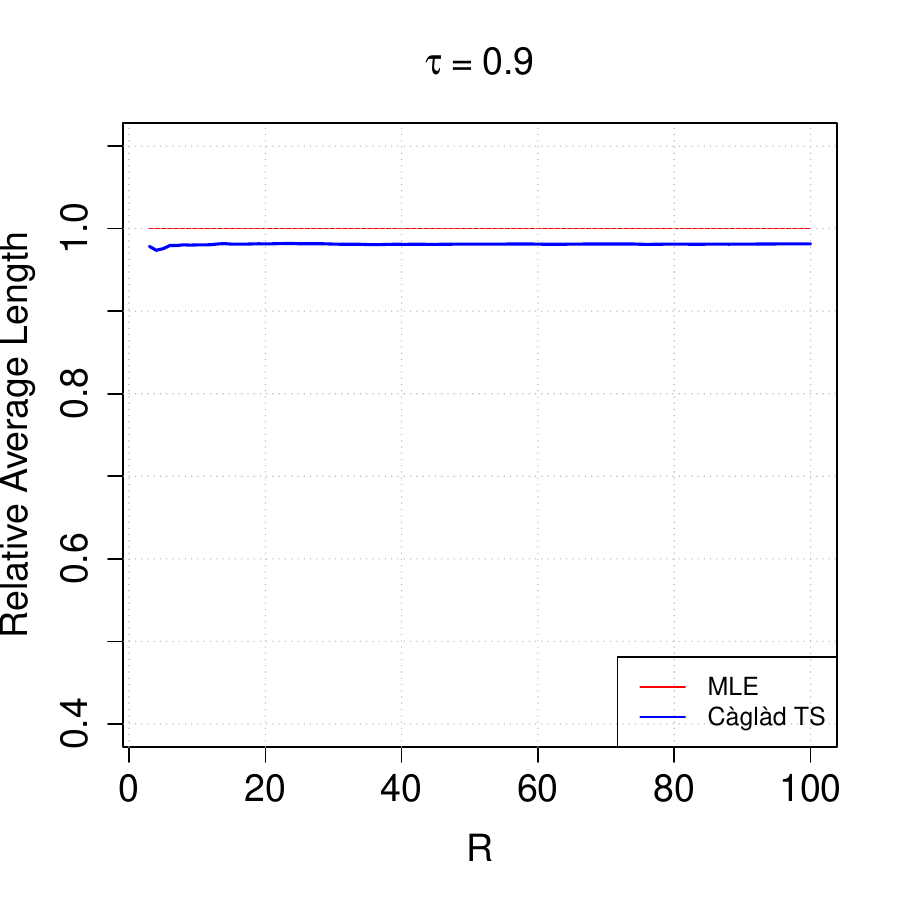}
		\includegraphics[width=0.24\textwidth]{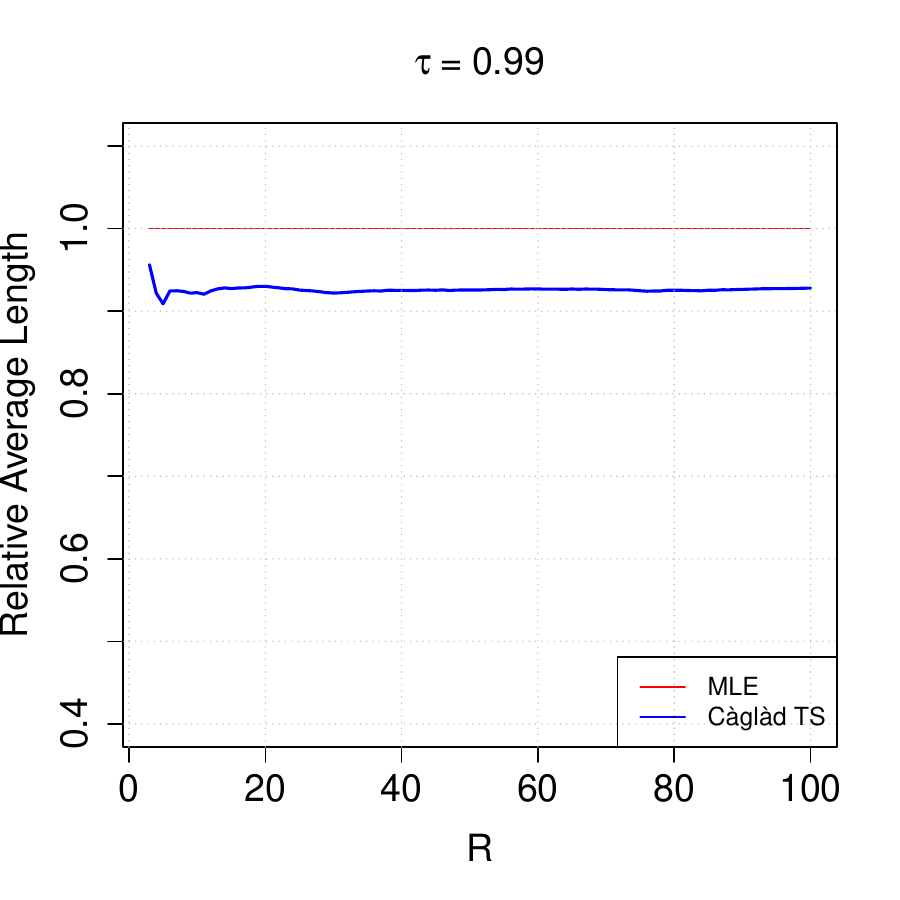}   \includegraphics[width=0.24\textwidth]{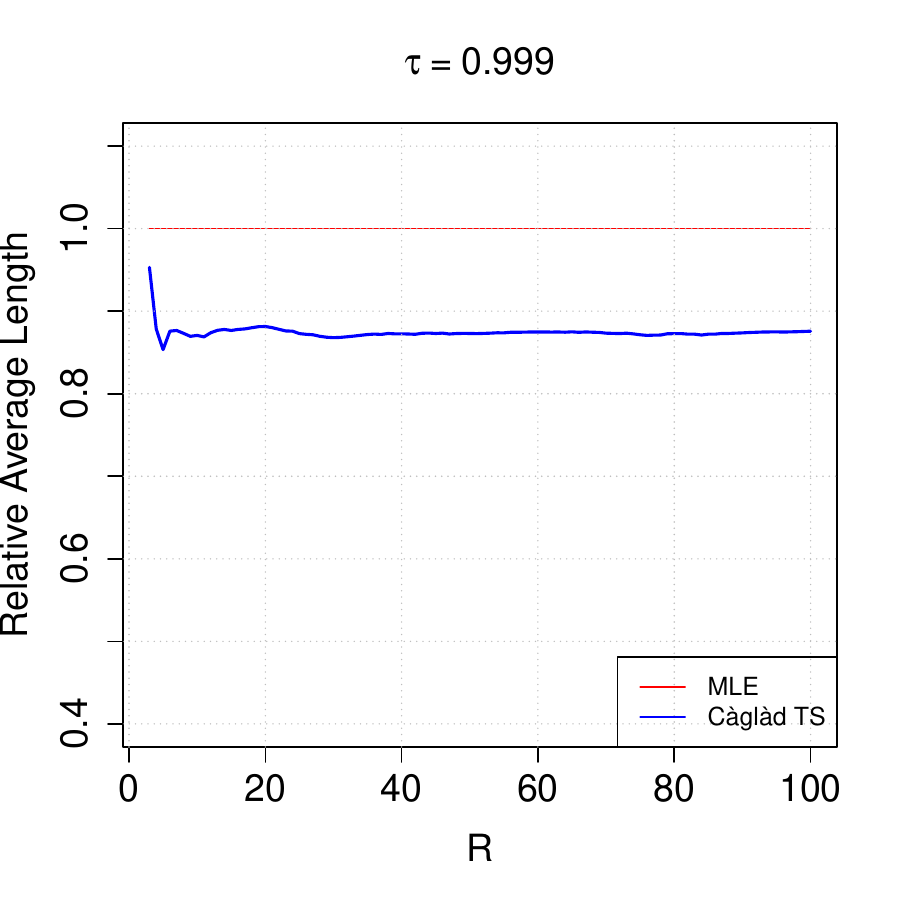}
		\caption{$T=100$}
	\end{subfigure}

	\begin{subfigure}[H]{\textwidth}

		\centering
		\includegraphics[width=0.24\textwidth]{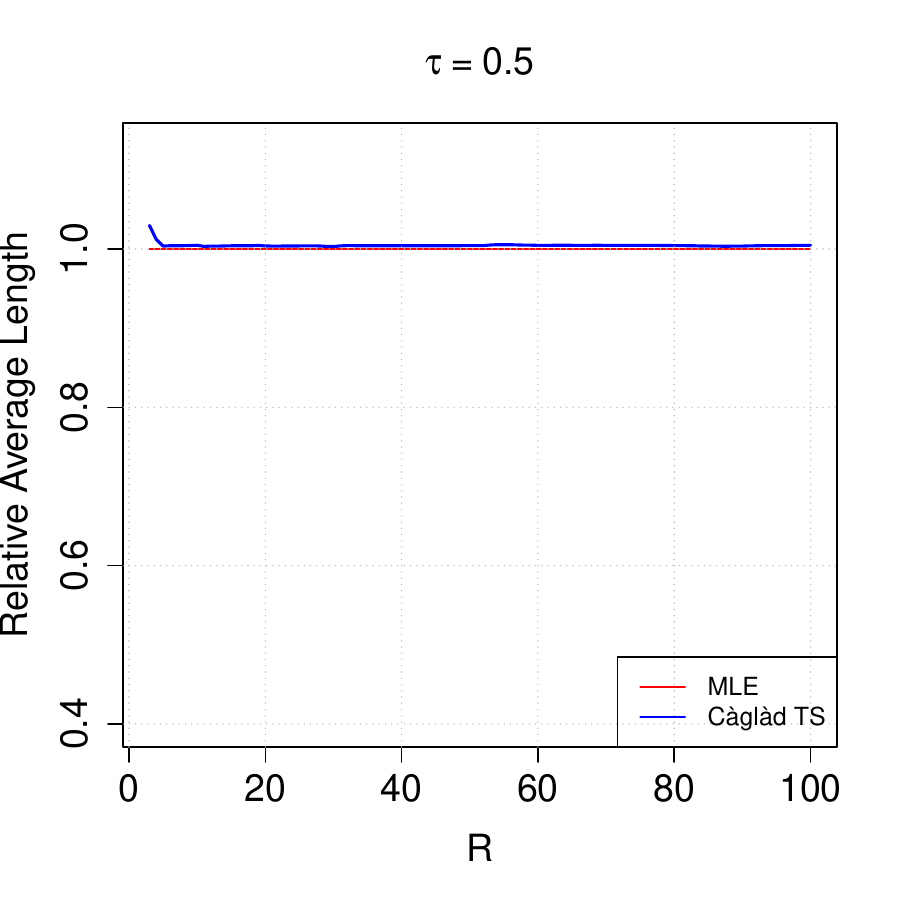}   \includegraphics[width=0.24\textwidth]{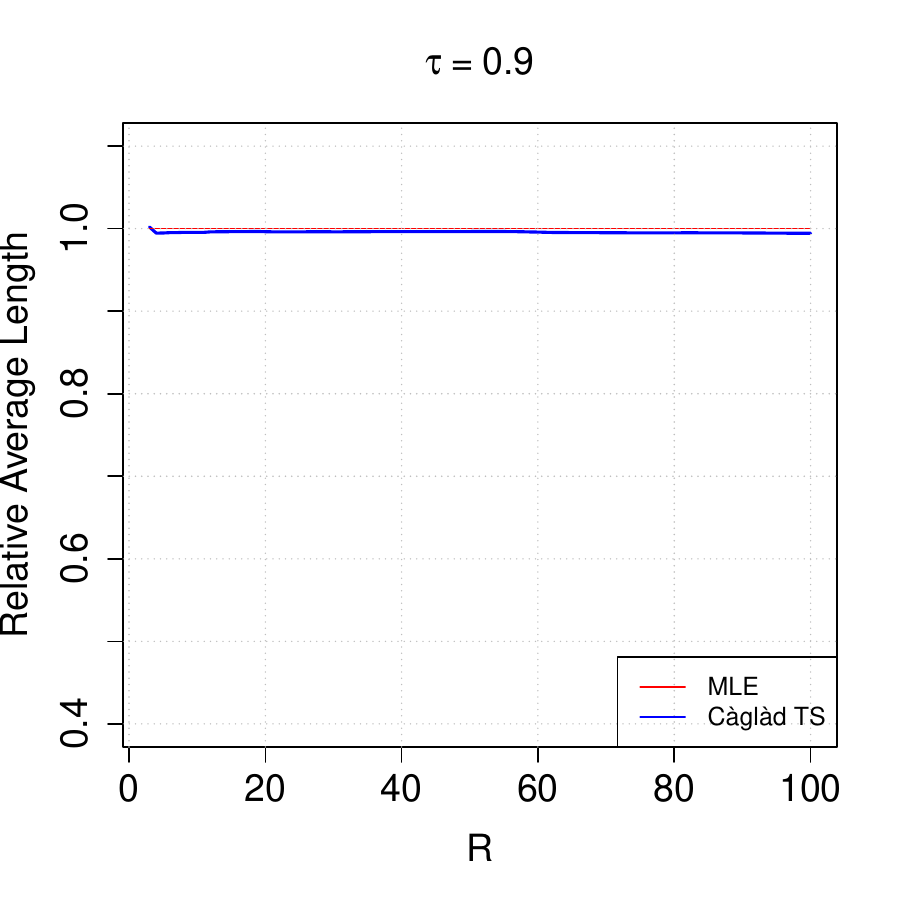}
		\includegraphics[width=0.24\textwidth]{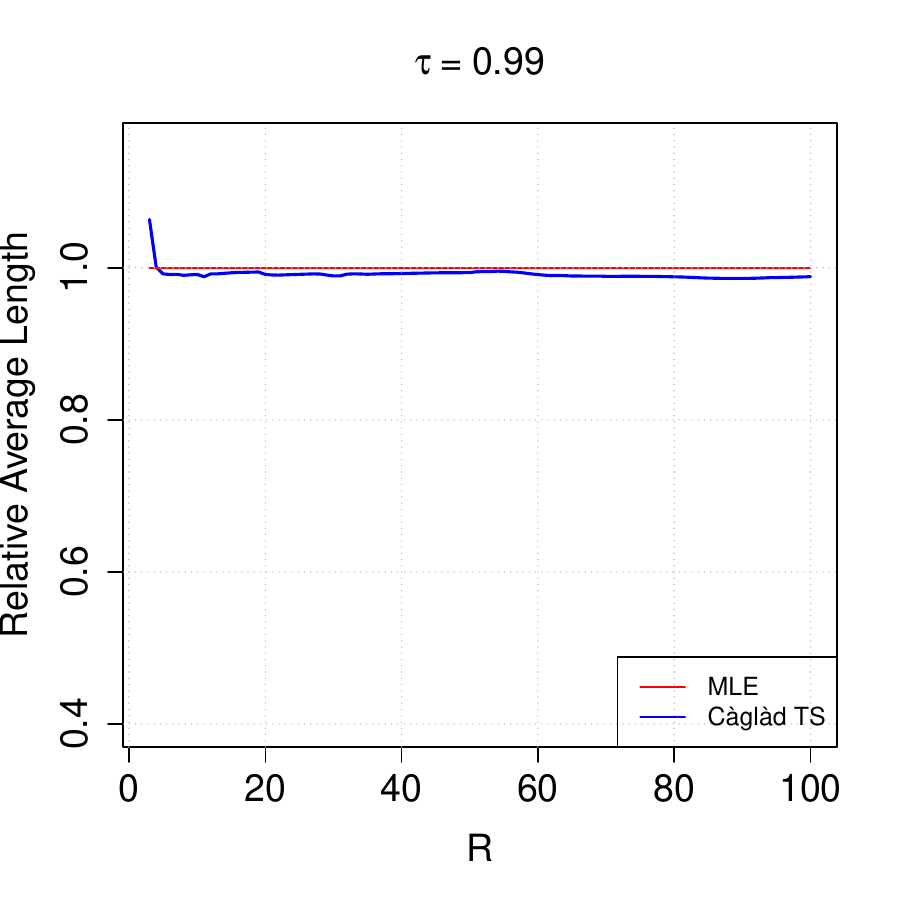}   \includegraphics[width=0.24\textwidth]{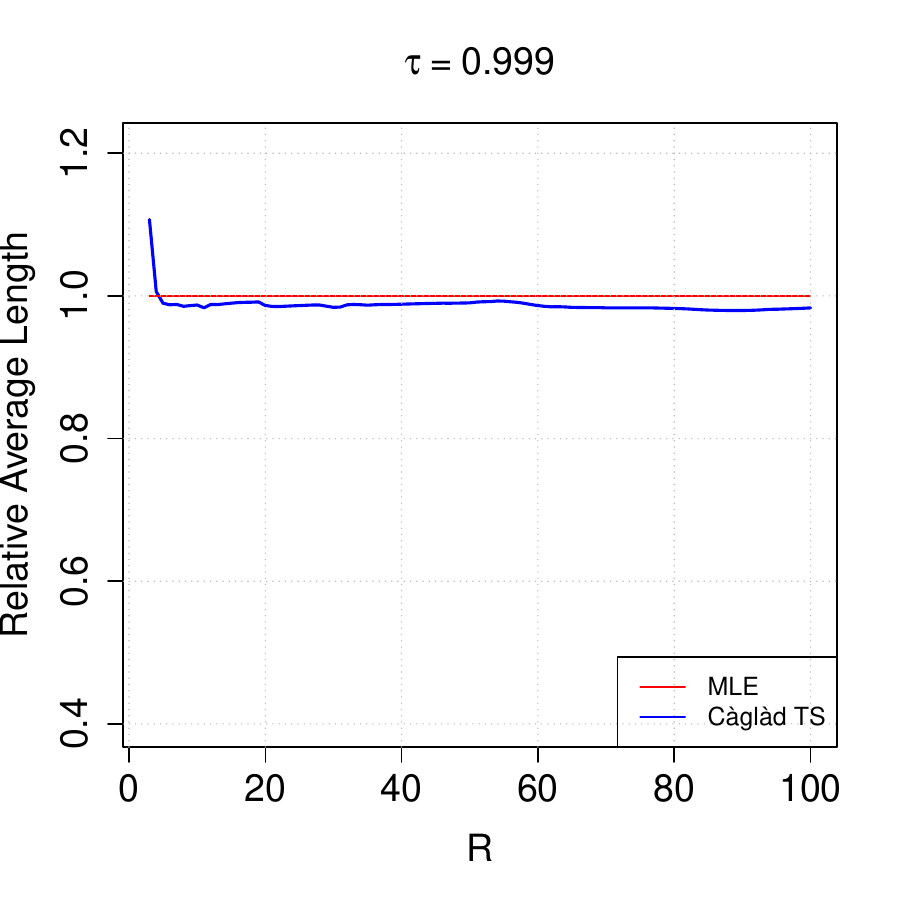}
		\caption{$T=500$}
	\end{subfigure}
	
	\caption{GEV: relative length of confidence (vis-à-vis the unfeasible MLE-based CI) intervals based on the true sampling variance.}
	\label{fig:gev_length_true}
\end{figure}

\begin{figure}[H]
	\centering

	\begin{subfigure}[H]{\textwidth}

		\centering
		\includegraphics[width=0.24\textwidth]{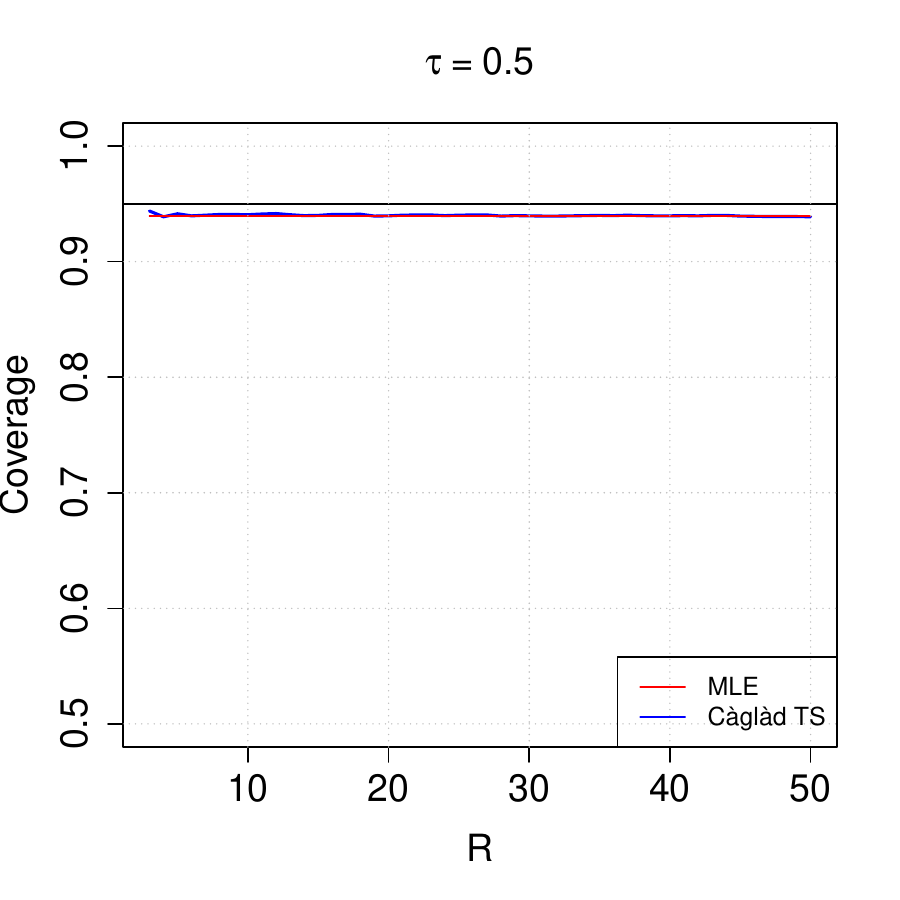}   \includegraphics[width=0.24\textwidth]{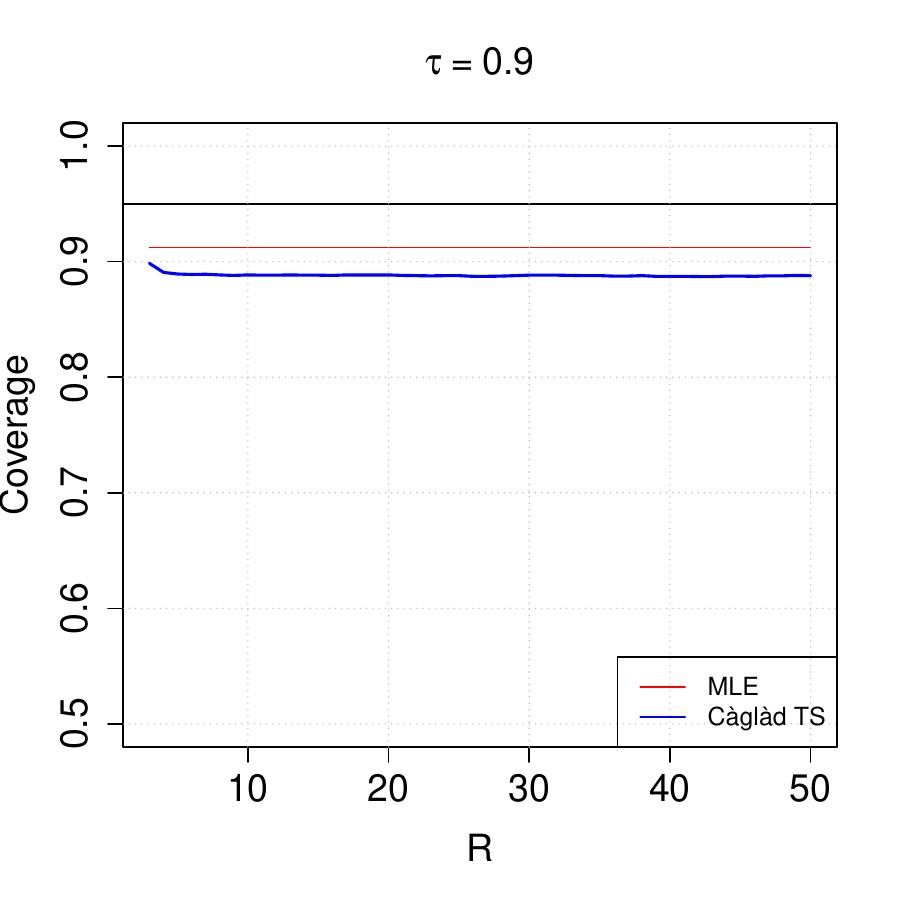}
		\includegraphics[width=0.24\textwidth]{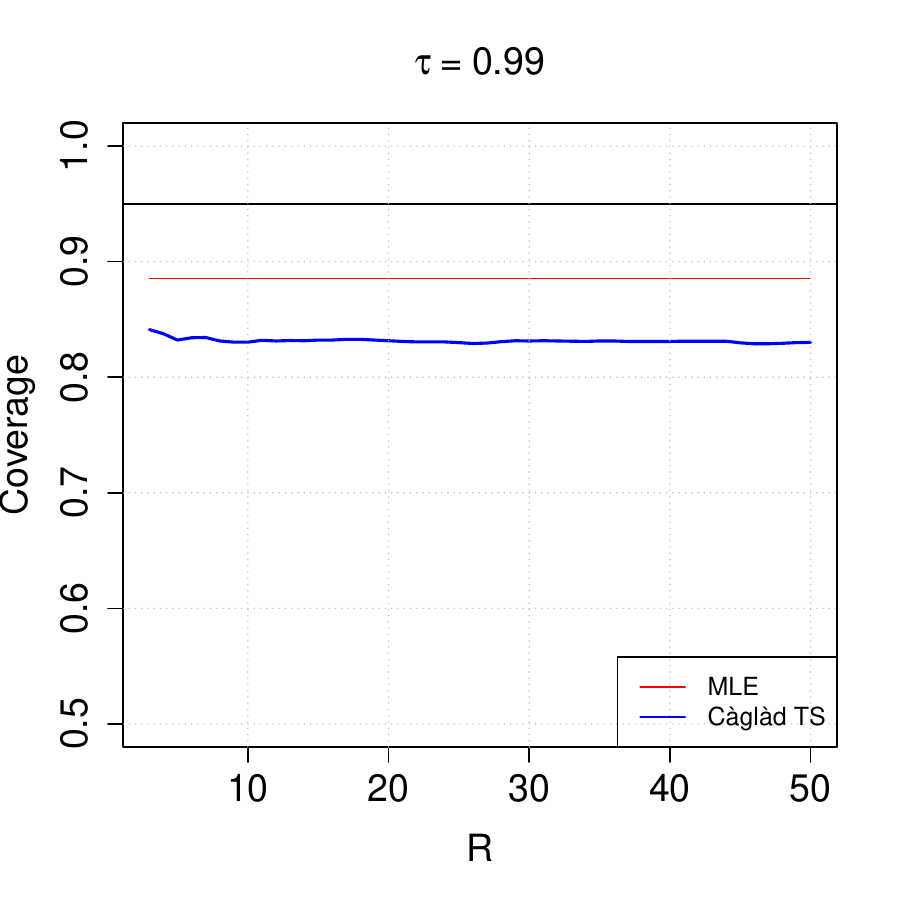}   \includegraphics[width=0.24\textwidth]{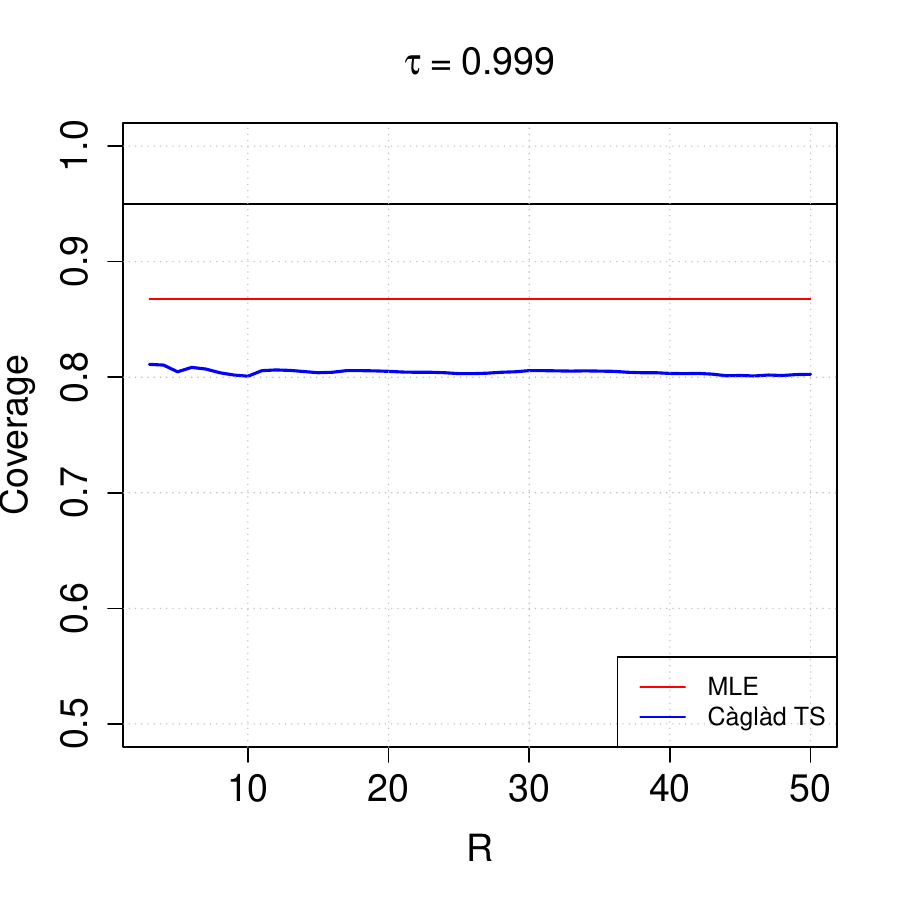}
		\caption{$T=50$}
	\end{subfigure}

	\begin{subfigure}[H]{\textwidth}

		\centering
		\includegraphics[width=0.24\textwidth]{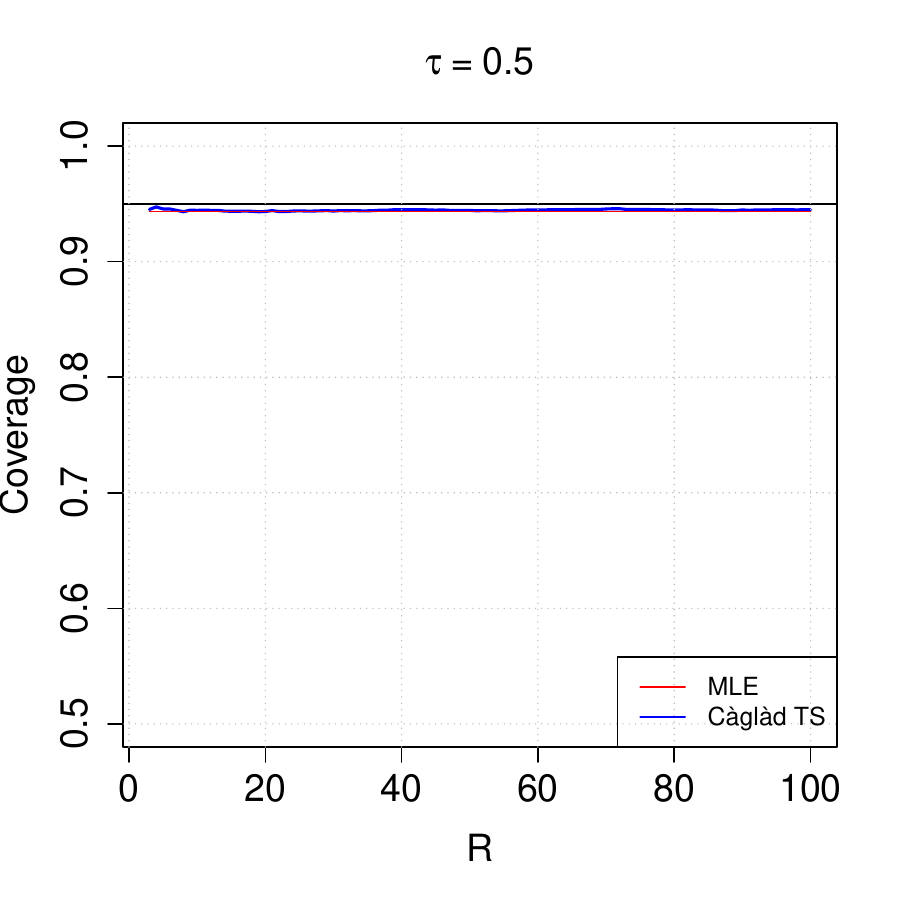}   \includegraphics[width=0.24\textwidth]{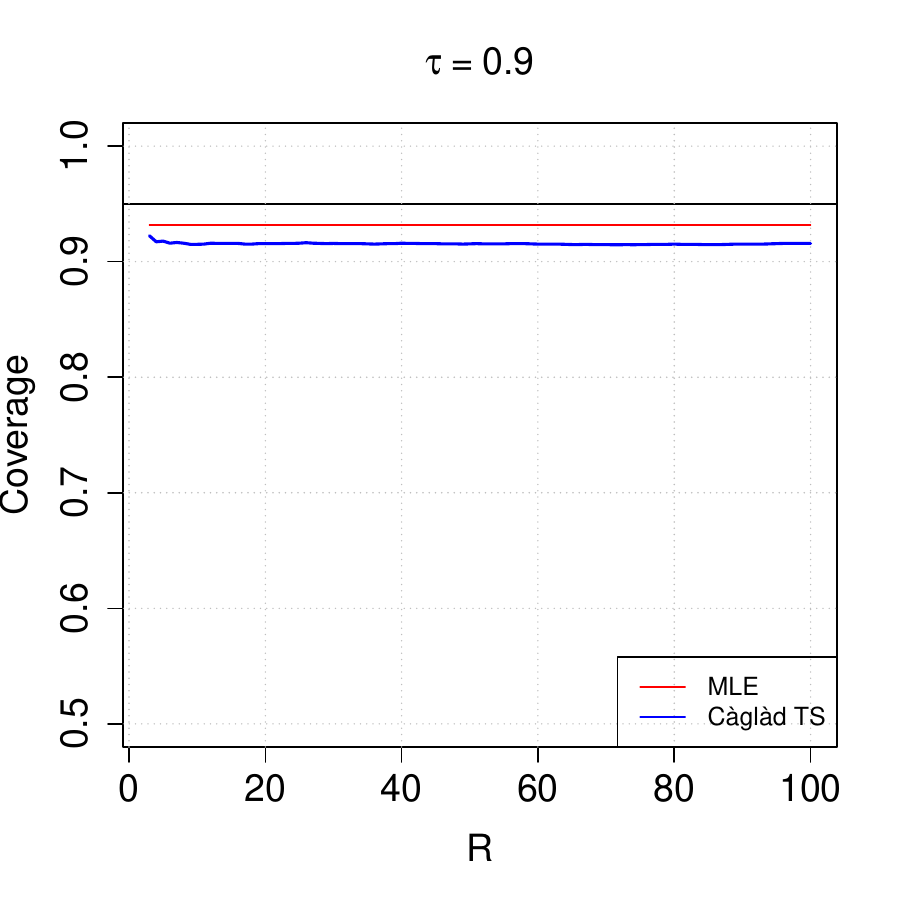}
		\includegraphics[width=0.24\textwidth]{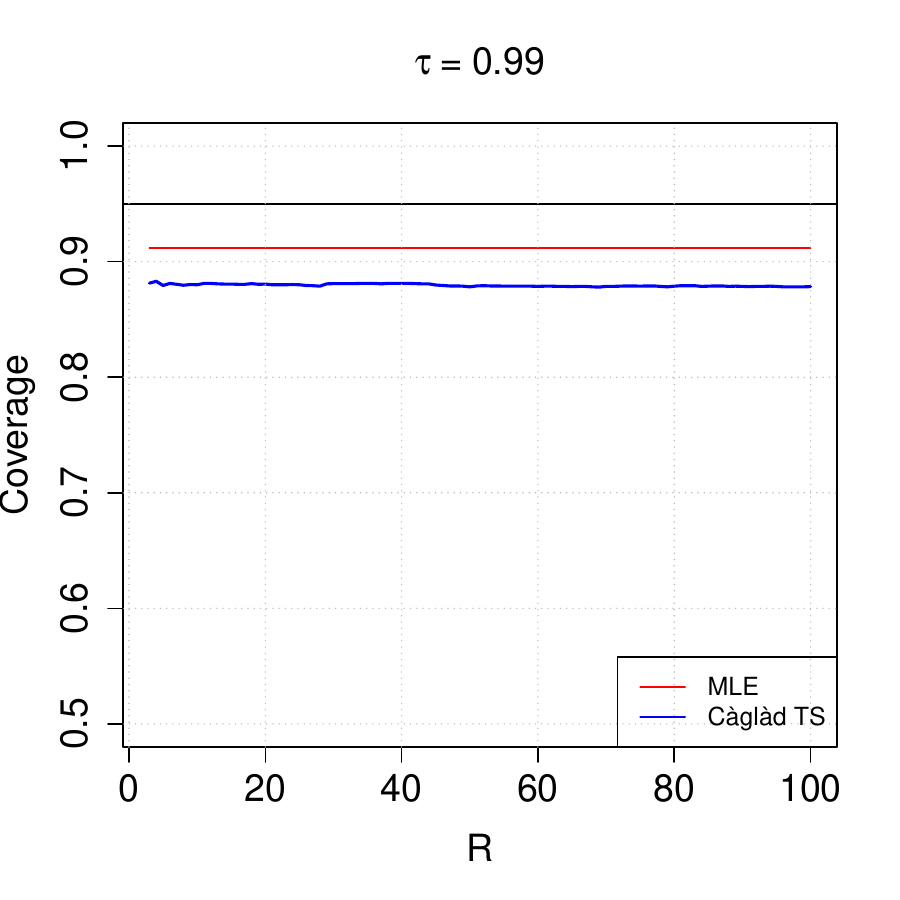}   \includegraphics[width=0.24\textwidth]{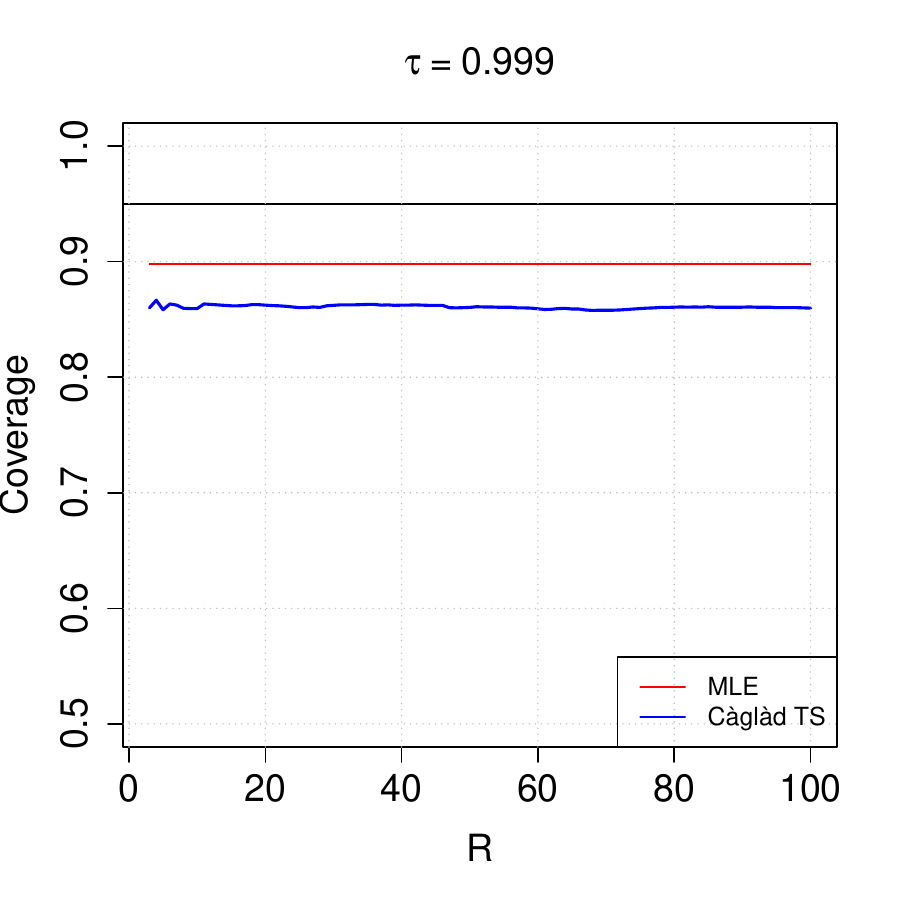}
		\caption{$T=100$}
	\end{subfigure}

	\begin{subfigure}[H]{\textwidth}

		\centering
		\includegraphics[width=0.24\textwidth]{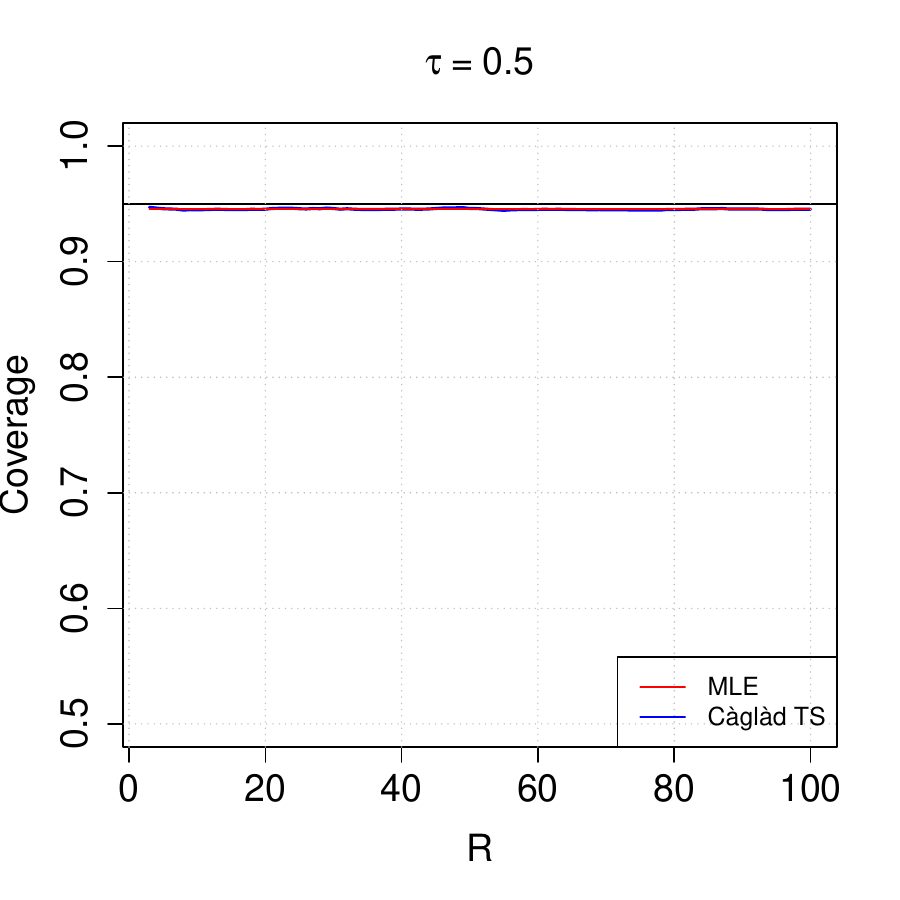}   \includegraphics[width=0.24\textwidth]{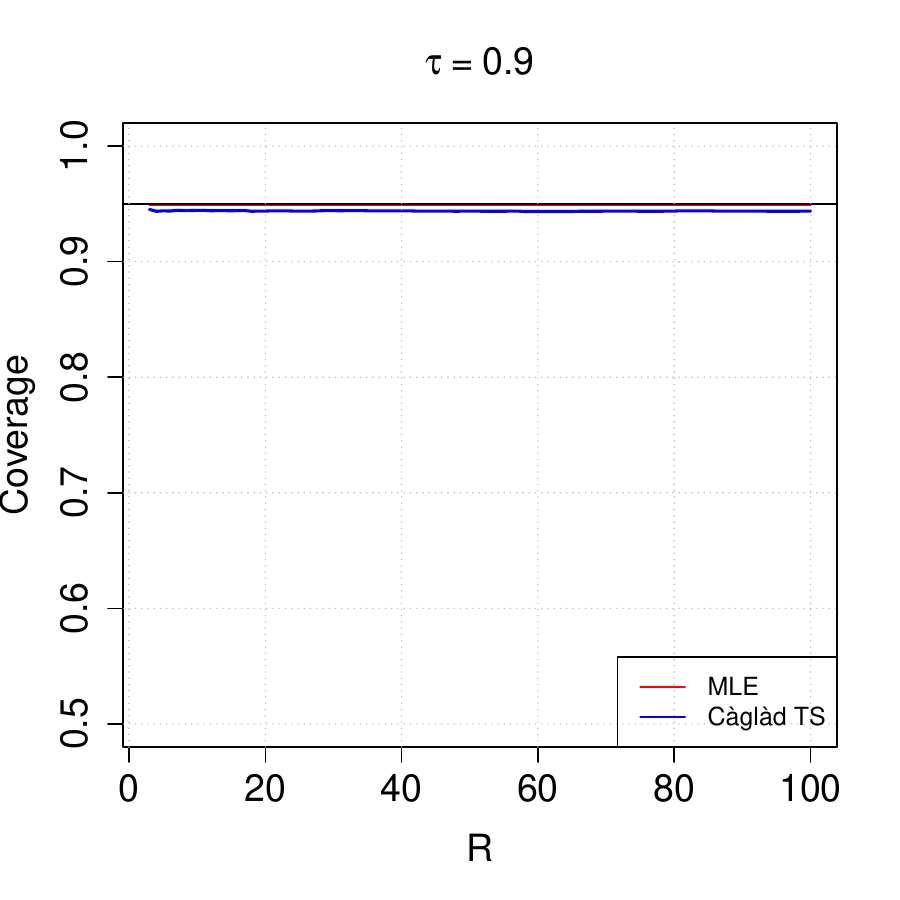}
		\includegraphics[width=0.24\textwidth]{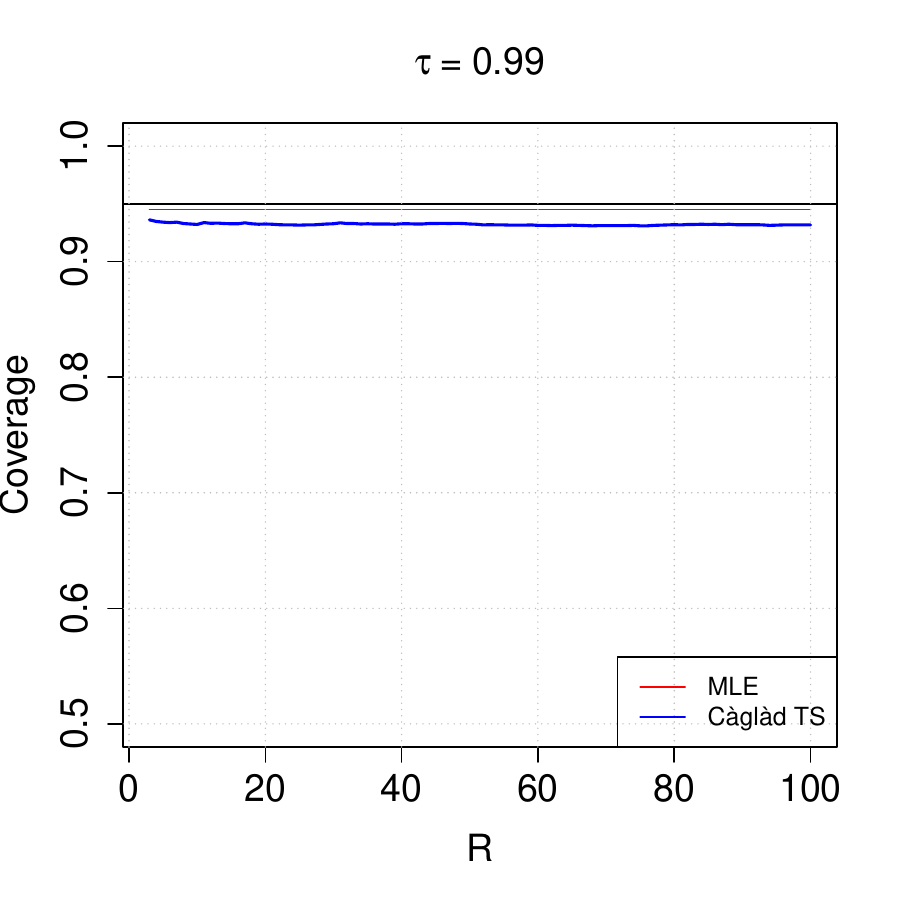}   \includegraphics[width=0.24\textwidth]{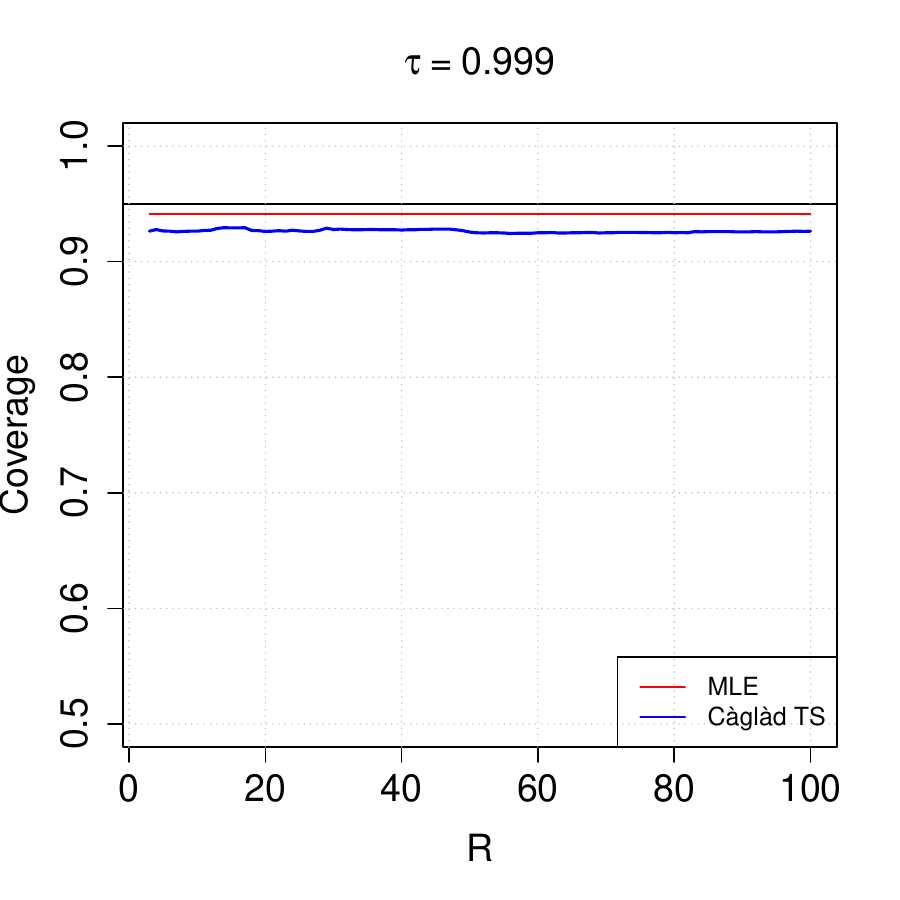}
		\caption{$T=500$}
	\end{subfigure}
	
	\caption{GEV: coverage of confidence intervals that rely on an estimator of the asymptotic variance.}
	\label{fig:gev_coverage_est}
\end{figure}

\begin{figure}[H]
	\centering

	\begin{subfigure}[H]{\textwidth}

		\centering
		\includegraphics[width=0.24\textwidth]{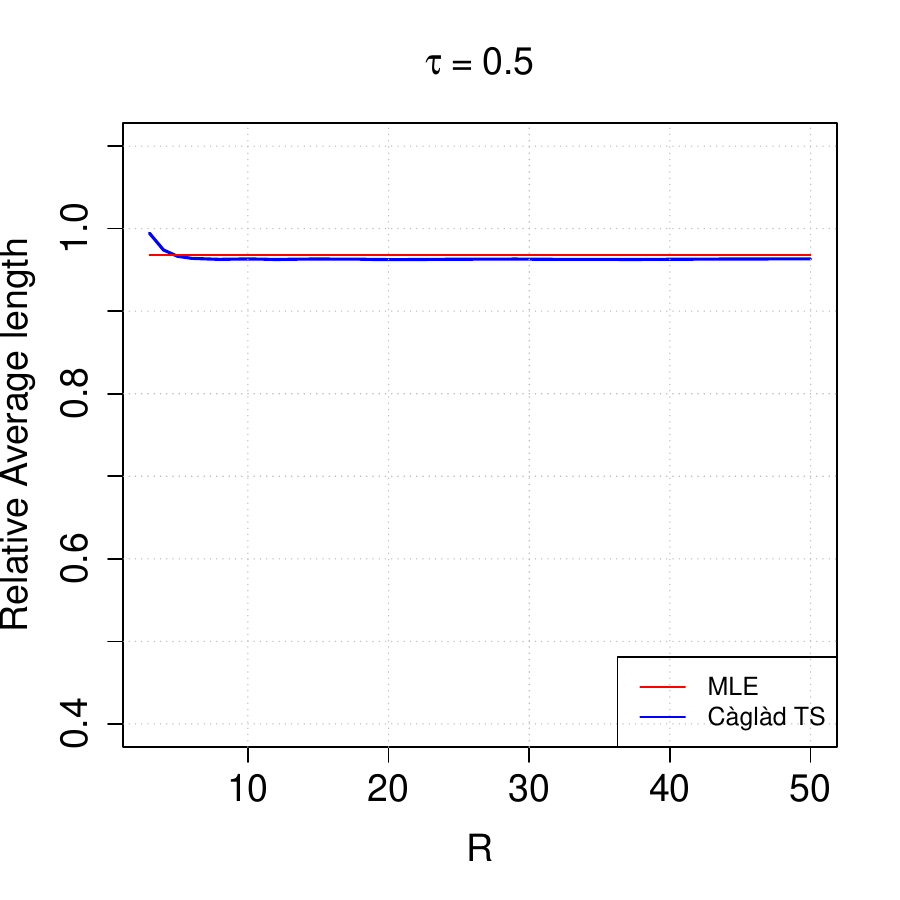}   \includegraphics[width=0.24\textwidth]{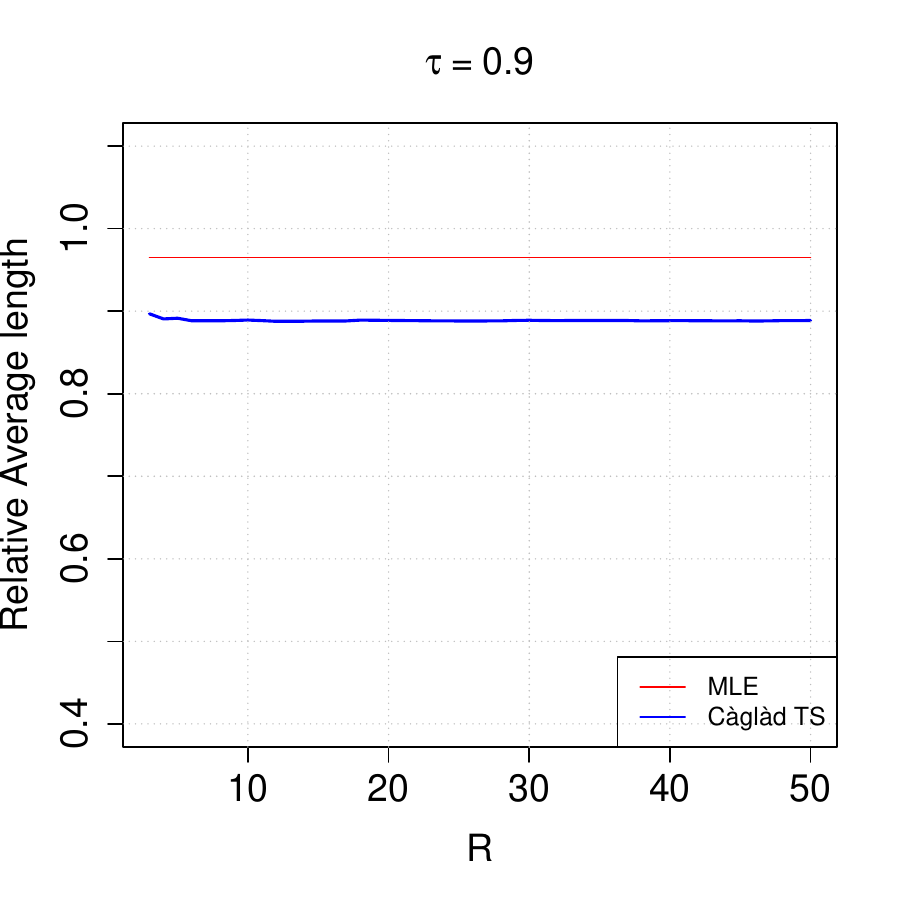}
		\includegraphics[width=0.24\textwidth]{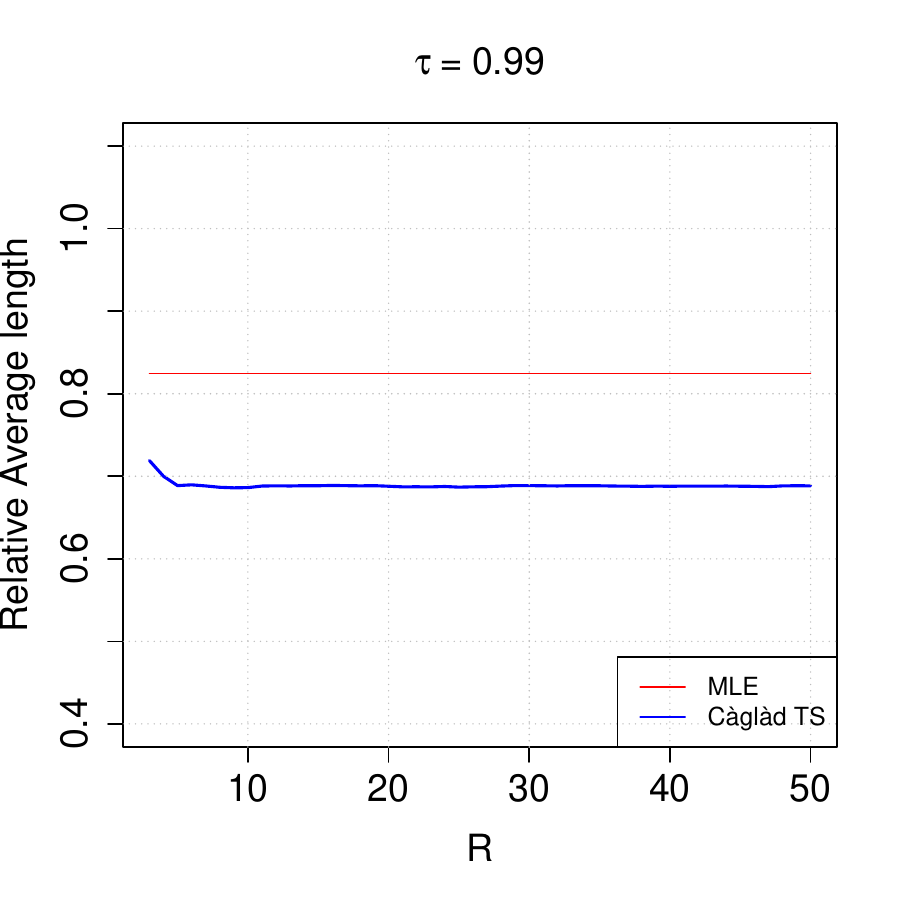}   \includegraphics[width=0.24\textwidth]{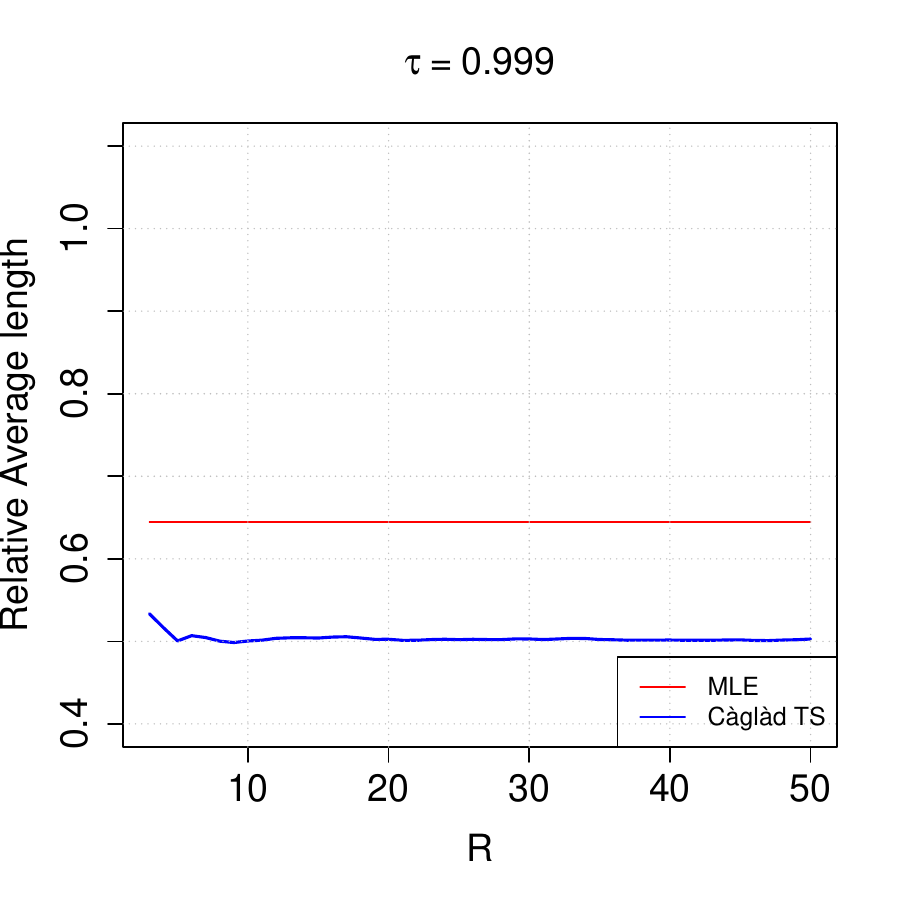}
		\caption{$T=50$}
	\end{subfigure}

	\begin{subfigure}[H]{\textwidth}

		\centering
		\includegraphics[width=0.24\textwidth]{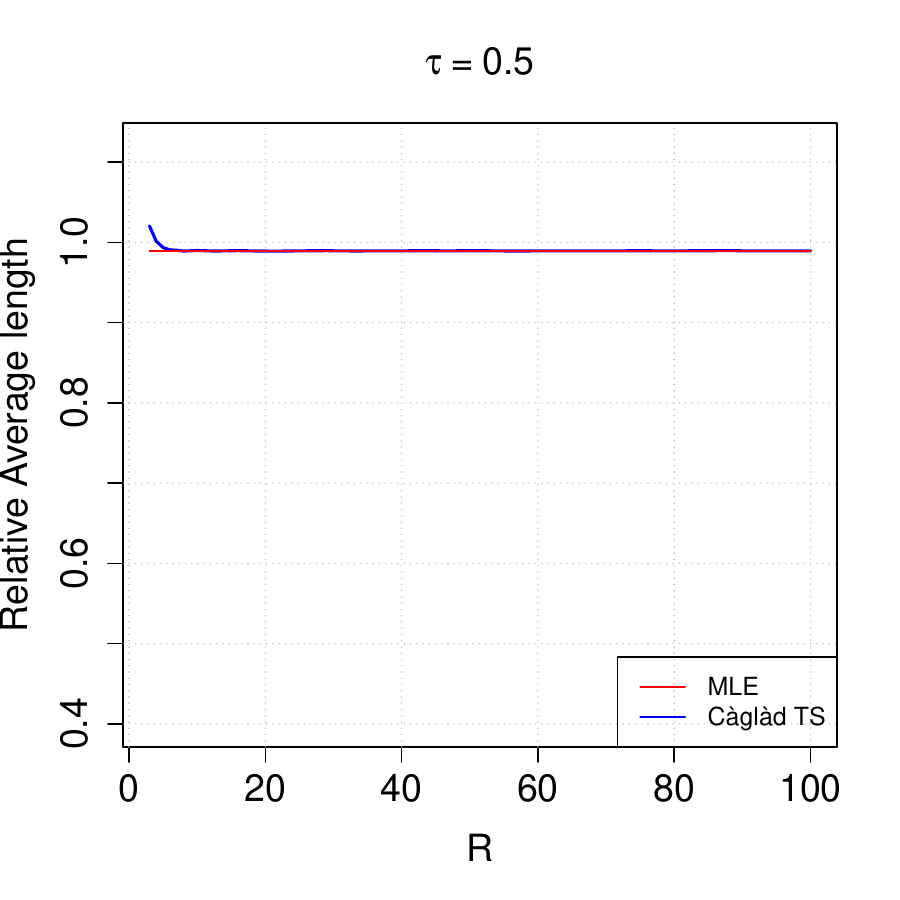}   \includegraphics[width=0.24\textwidth]{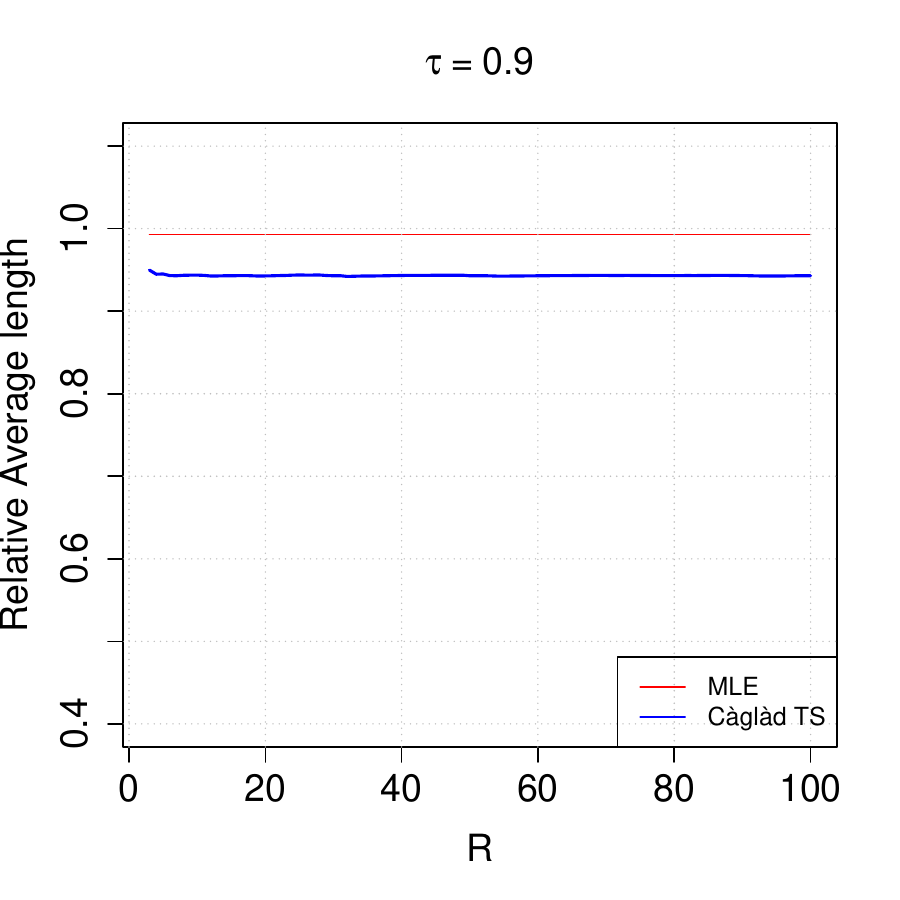}
		\includegraphics[width=0.24\textwidth]{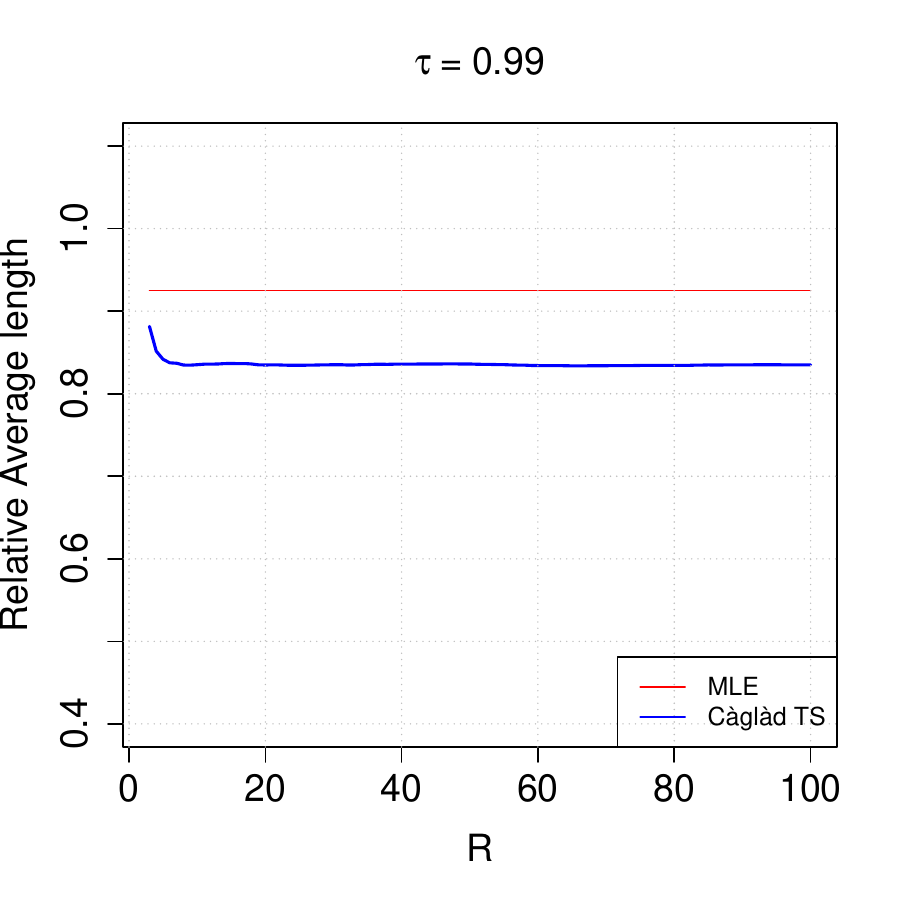}   \includegraphics[width=0.24\textwidth]{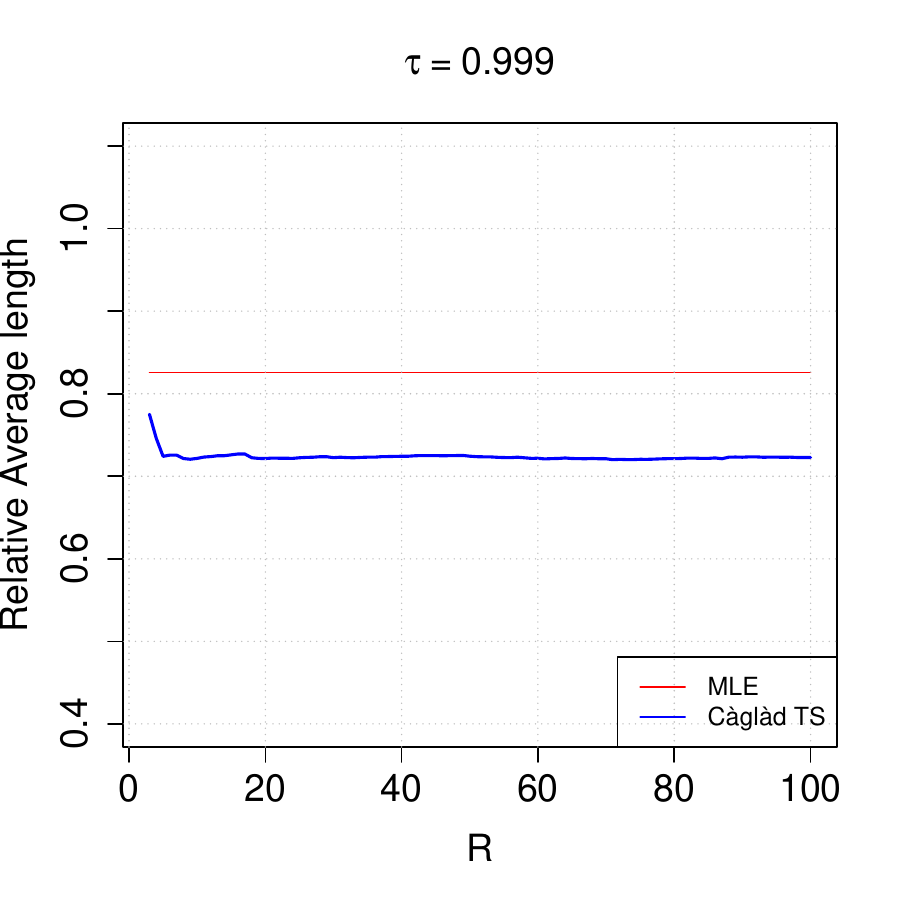}
		\caption{$T=100$}
	\end{subfigure}

	\begin{subfigure}[H]{\textwidth}

		\centering
		\includegraphics[width=0.24\textwidth]{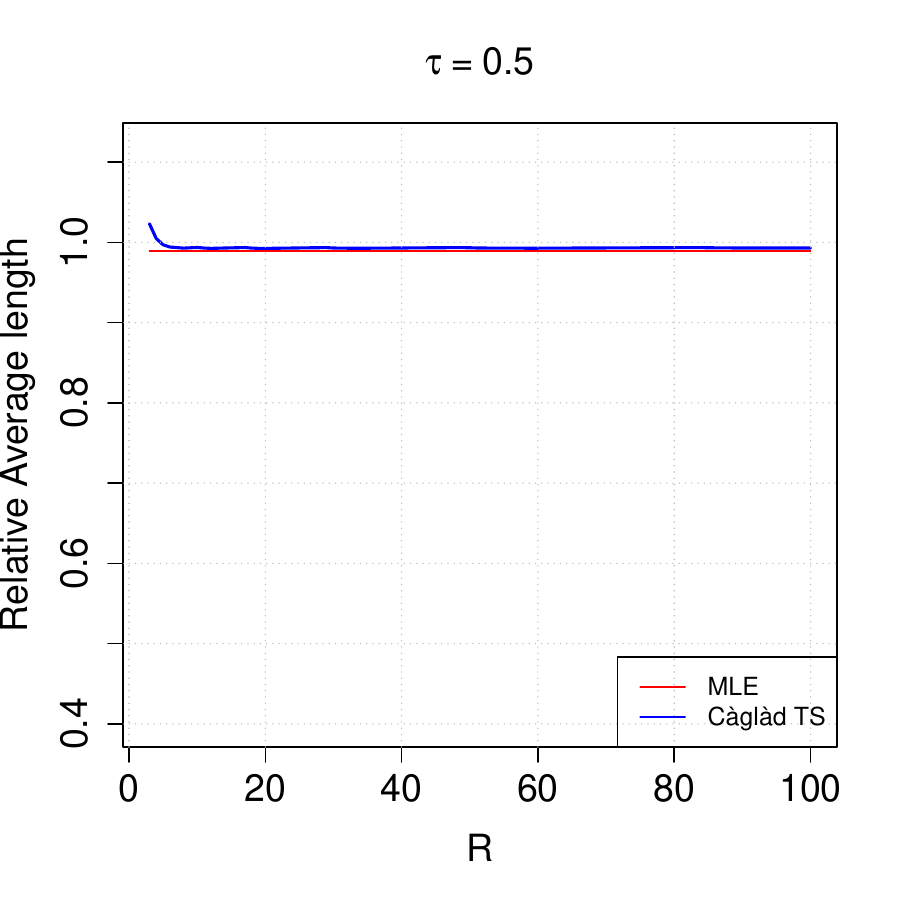}   \includegraphics[width=0.24\textwidth]{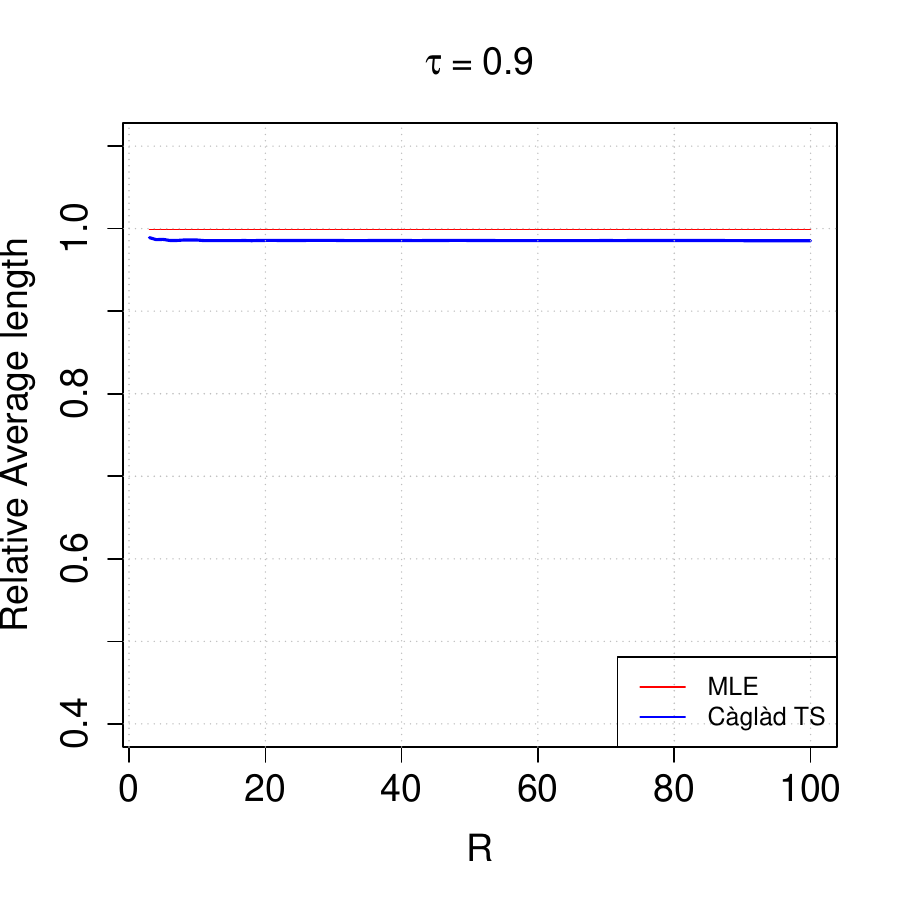}
		\includegraphics[width=0.24\textwidth]{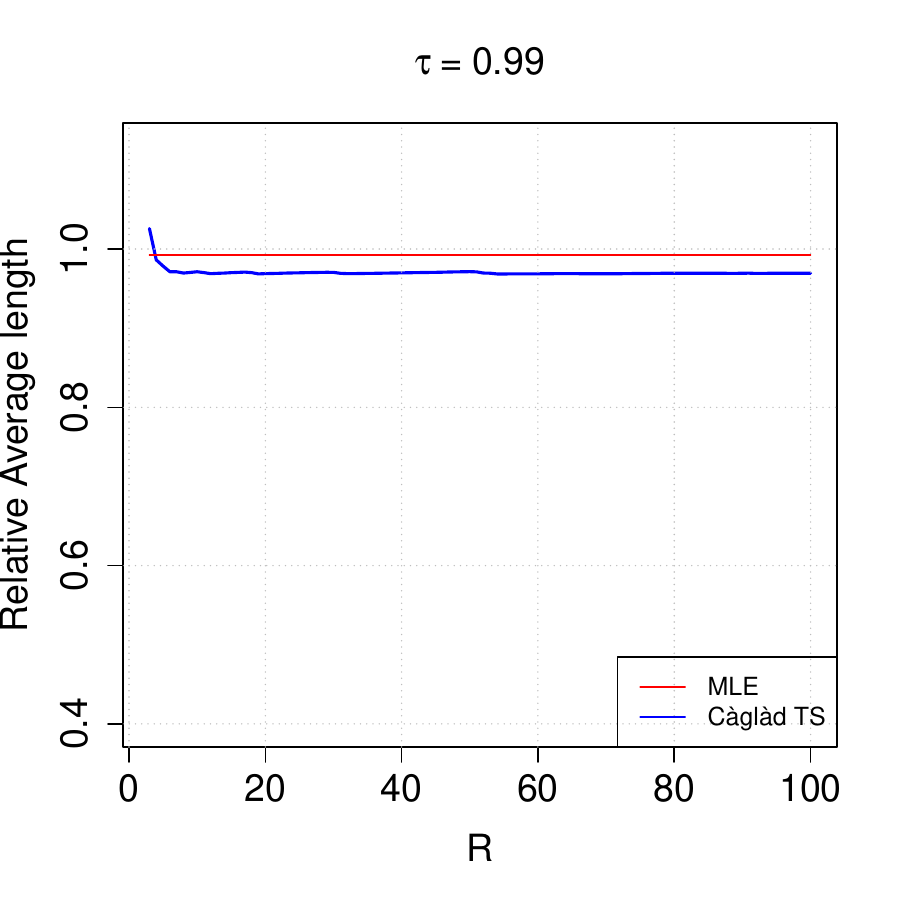}   \includegraphics[width=0.24\textwidth]{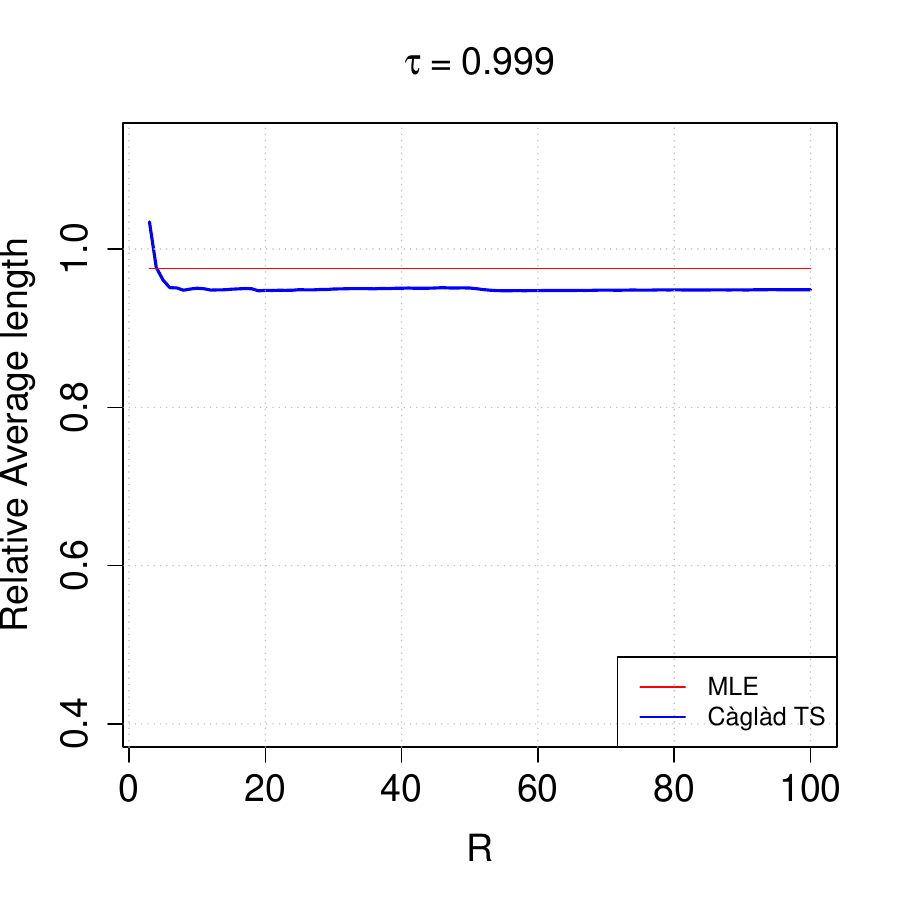}
		\caption{$T=500$}
	\end{subfigure}
	
	\caption{GEV: relative length (vis-à-vis the unfeasible MLE-based CI) of confidence intervals that rely on an estimator of the asymptotic variance.}
	\label{fig:gev_length_est}
\end{figure}

\begin{figure}[H]
	\centering

	\begin{subfigure}[H]{\textwidth}

		\centering
		\includegraphics[width=0.32\textwidth]{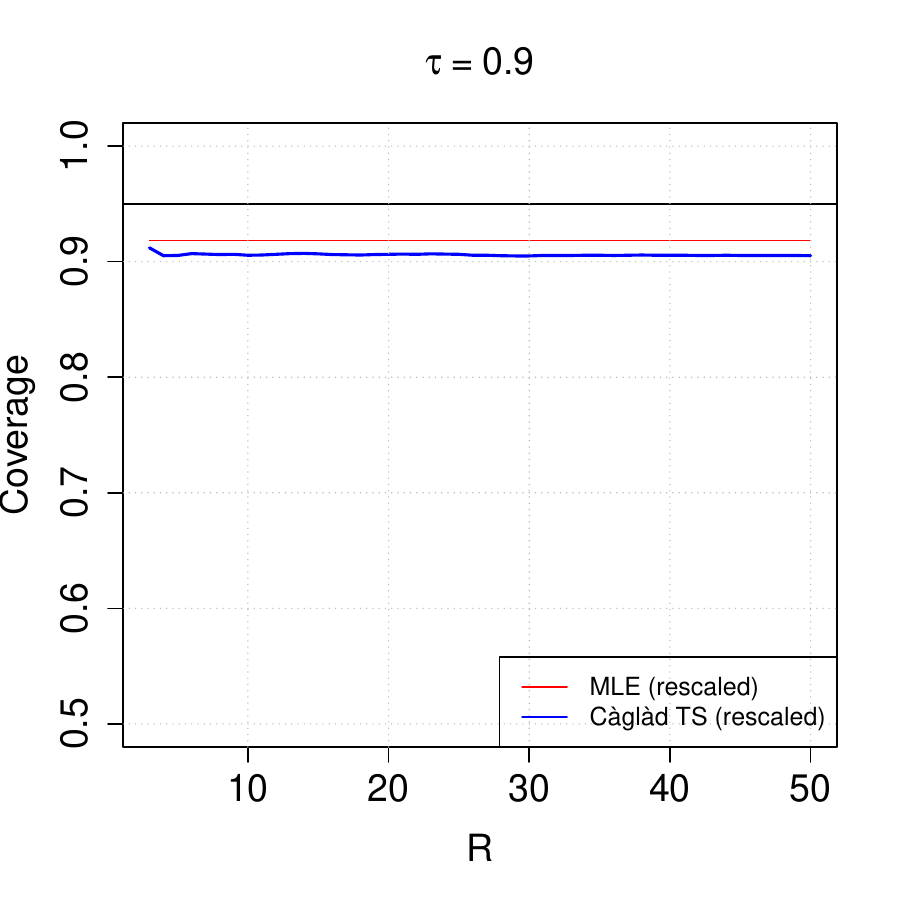}
		\includegraphics[width=0.32\textwidth]{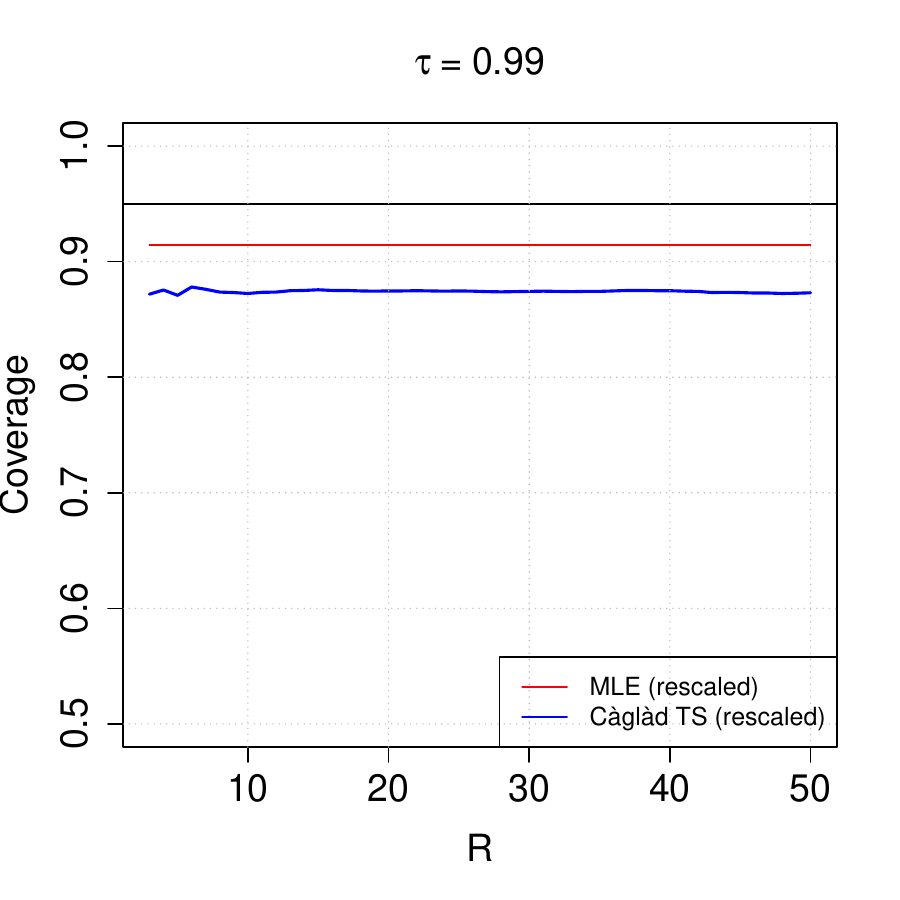}   \includegraphics[width=0.32\textwidth]{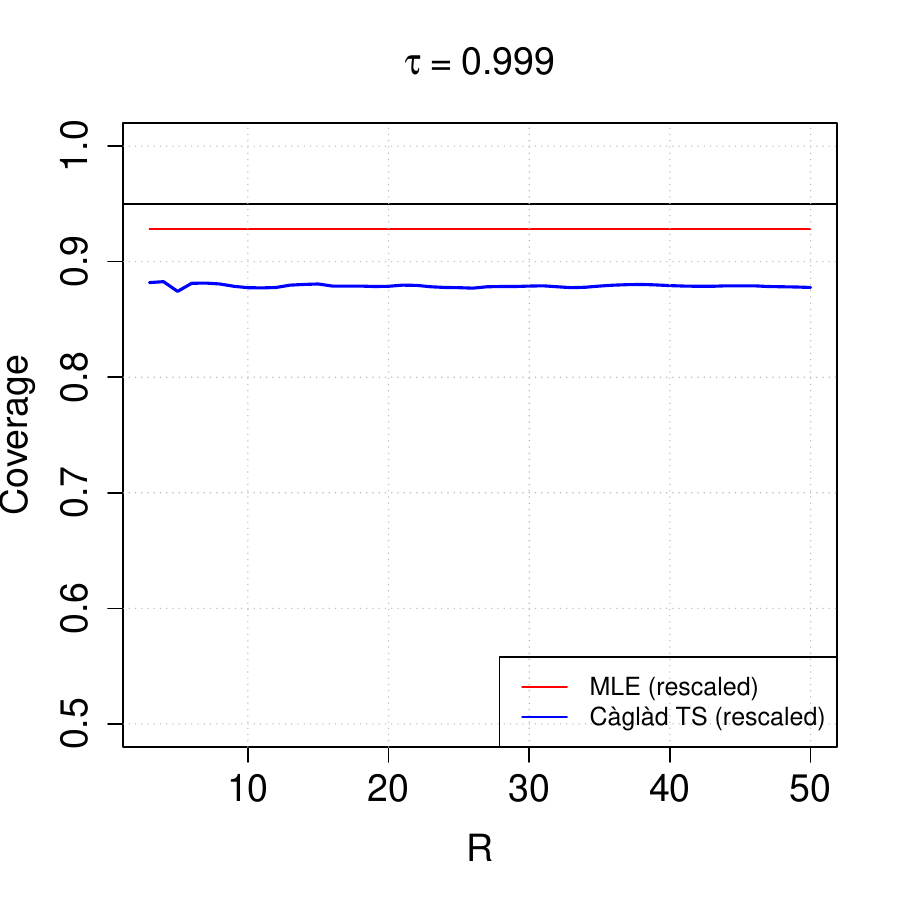}
		\caption{$T=50$}
	\end{subfigure}

	\begin{subfigure}[H]{\textwidth}

		\centering
		\includegraphics[width=0.32\textwidth]{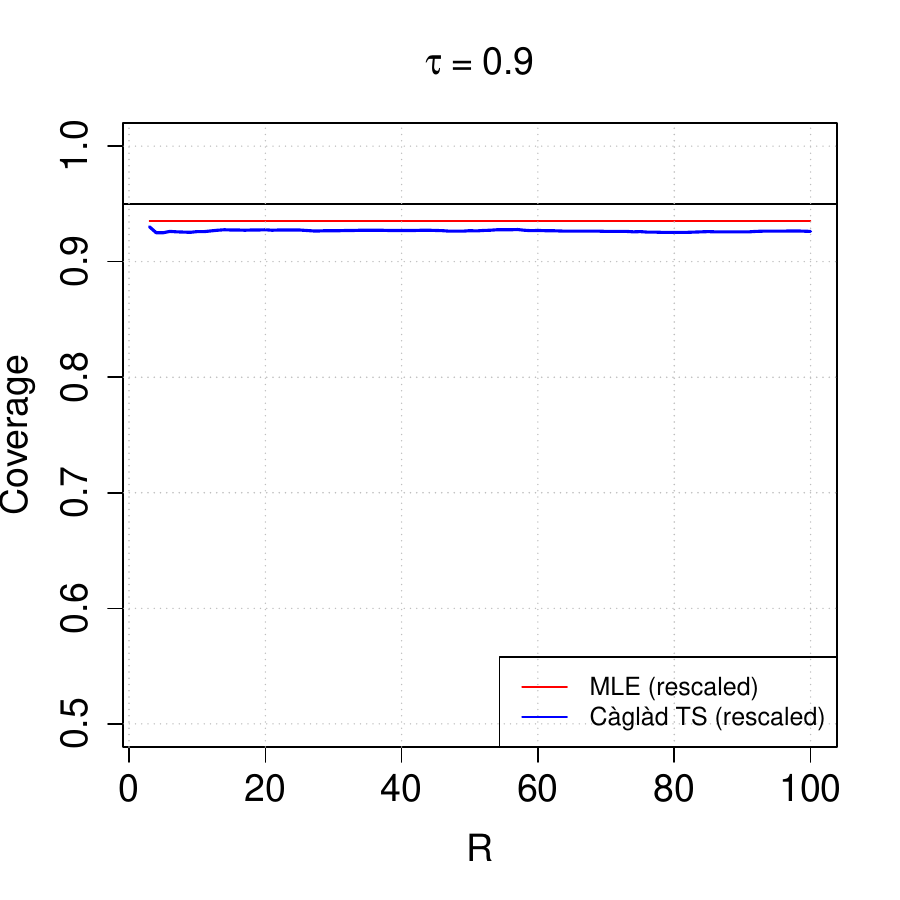}
		\includegraphics[width=0.32\textwidth]{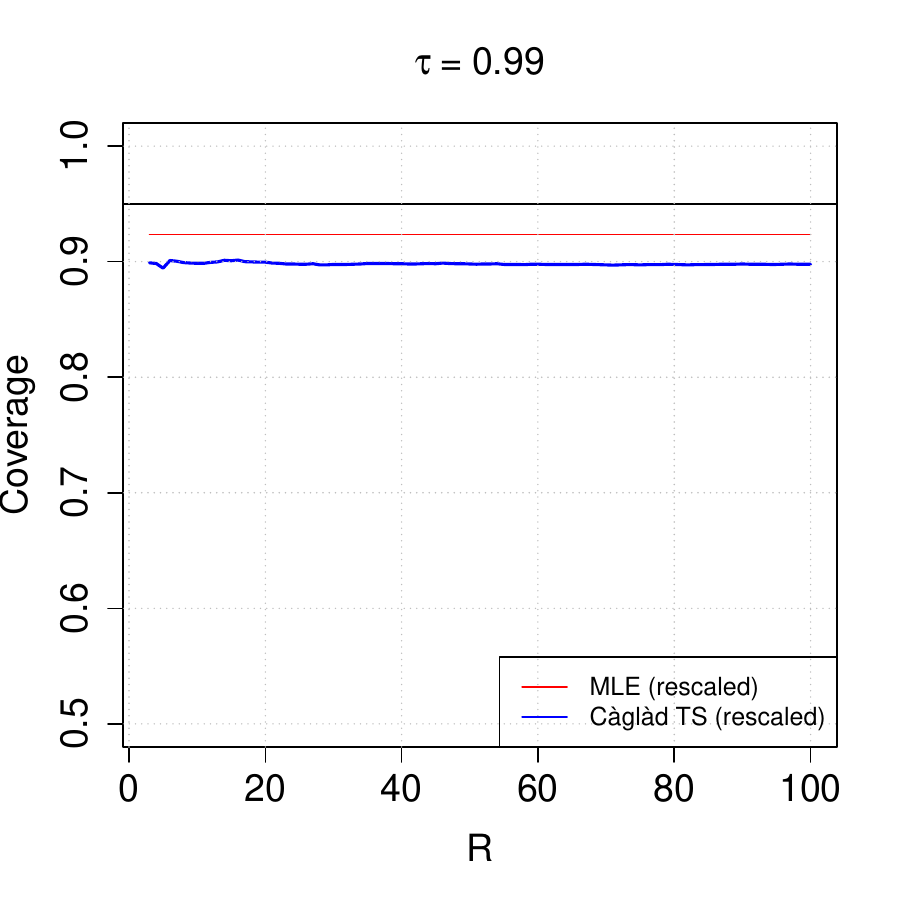}   \includegraphics[width=0.32\textwidth]{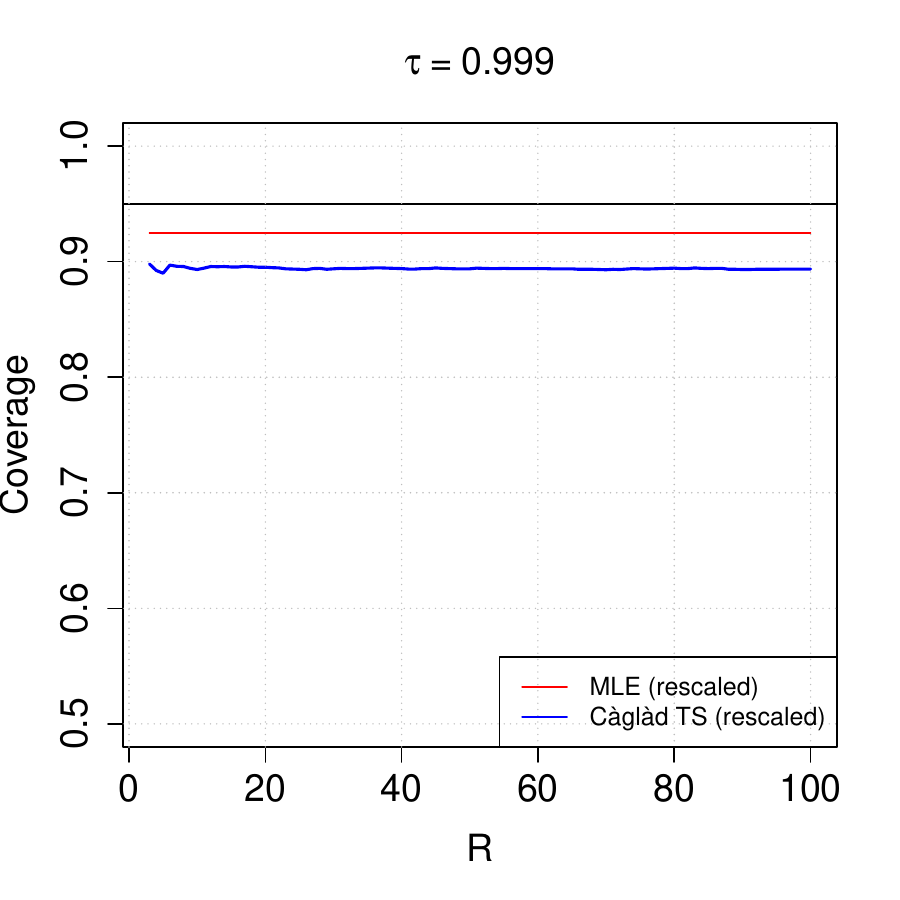}
		\caption{$T=100$}
	\end{subfigure}
	
	\caption{GEV: coverage of unfeasible confidence intervals that rely on rescaled estimators of the asymptotic variance.}
	\label{fig:gev_coverage_rescale}
\end{figure}

\begin{figure}[H]
	\centering

	\begin{subfigure}[H]{\textwidth}

		\centering
		\includegraphics[width=0.32\textwidth]{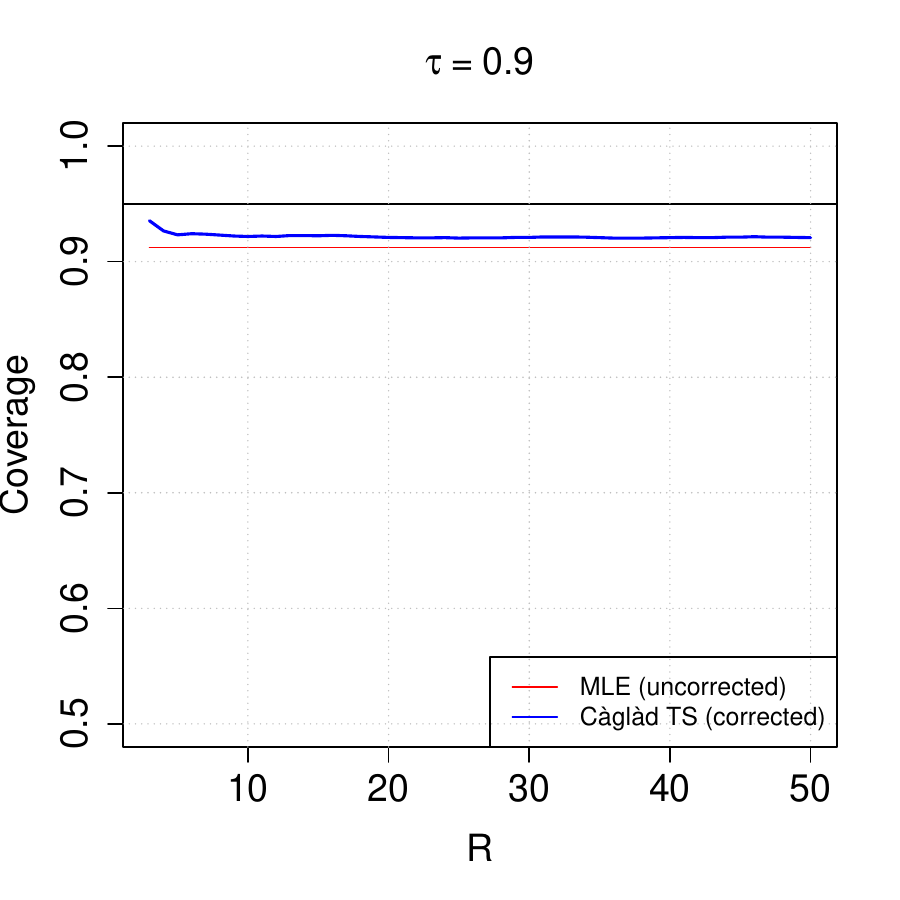}
		\includegraphics[width=0.32\textwidth]{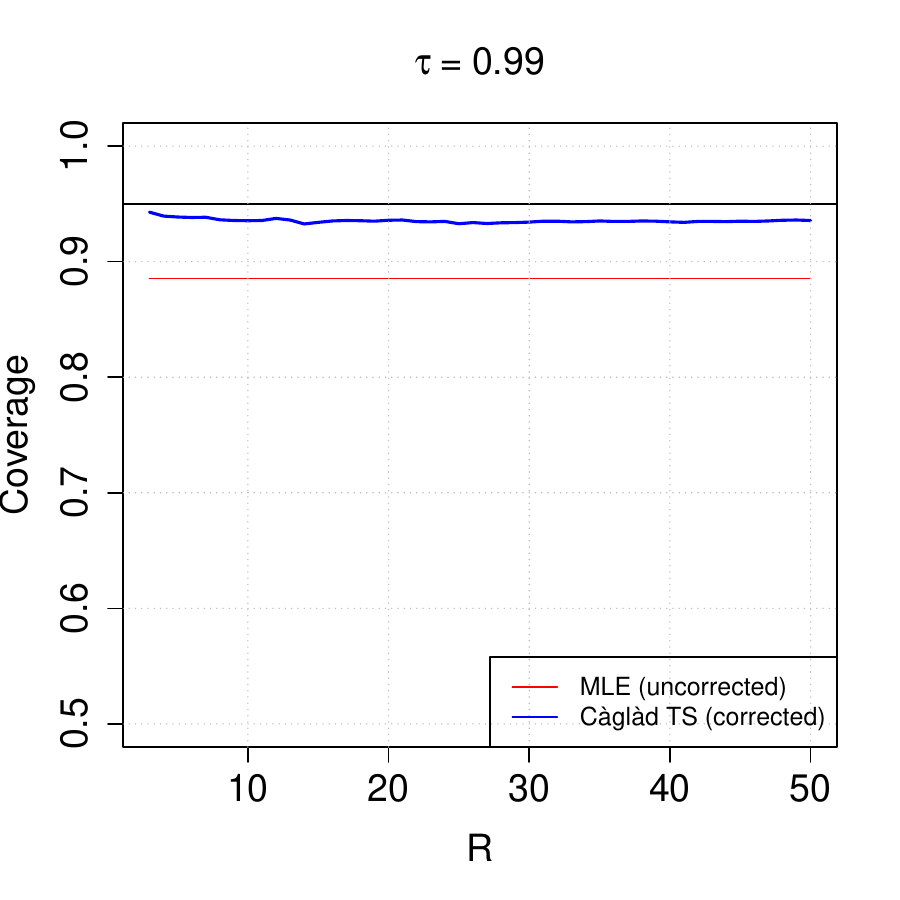}   \includegraphics[width=0.32\textwidth]{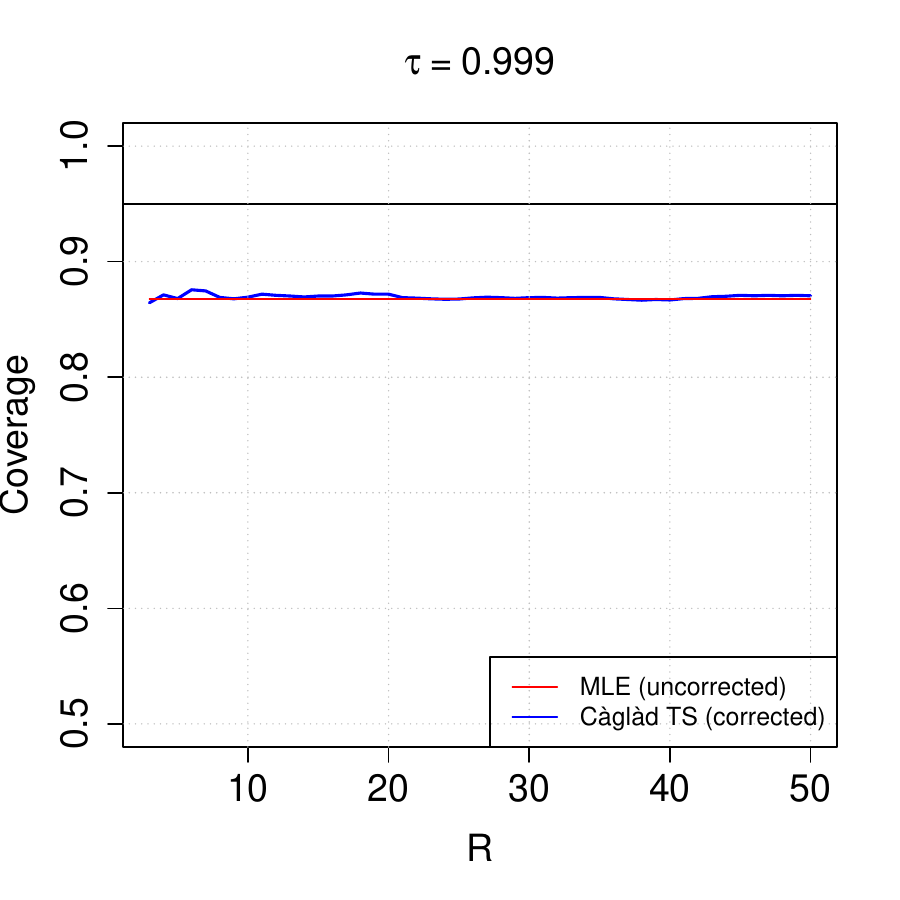}
		\caption{$T=50$}
	\end{subfigure}

	\begin{subfigure}[H]{\textwidth}

		\centering
		\includegraphics[width=0.32\textwidth]{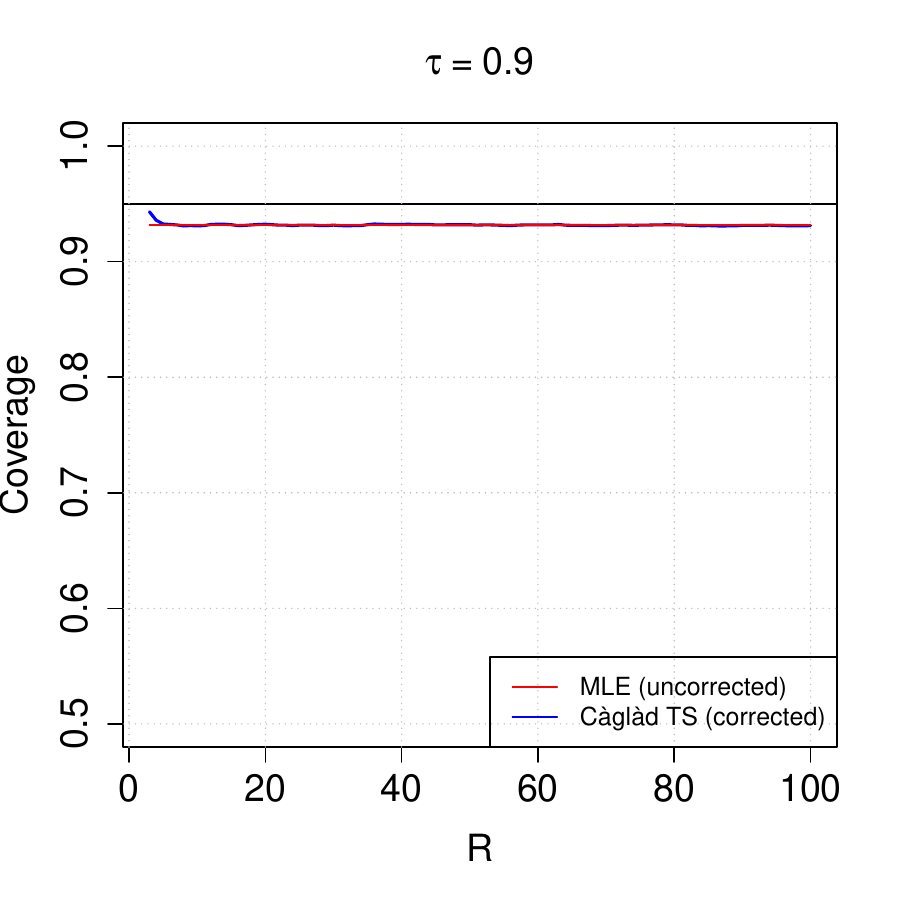}
		\includegraphics[width=0.32\textwidth]{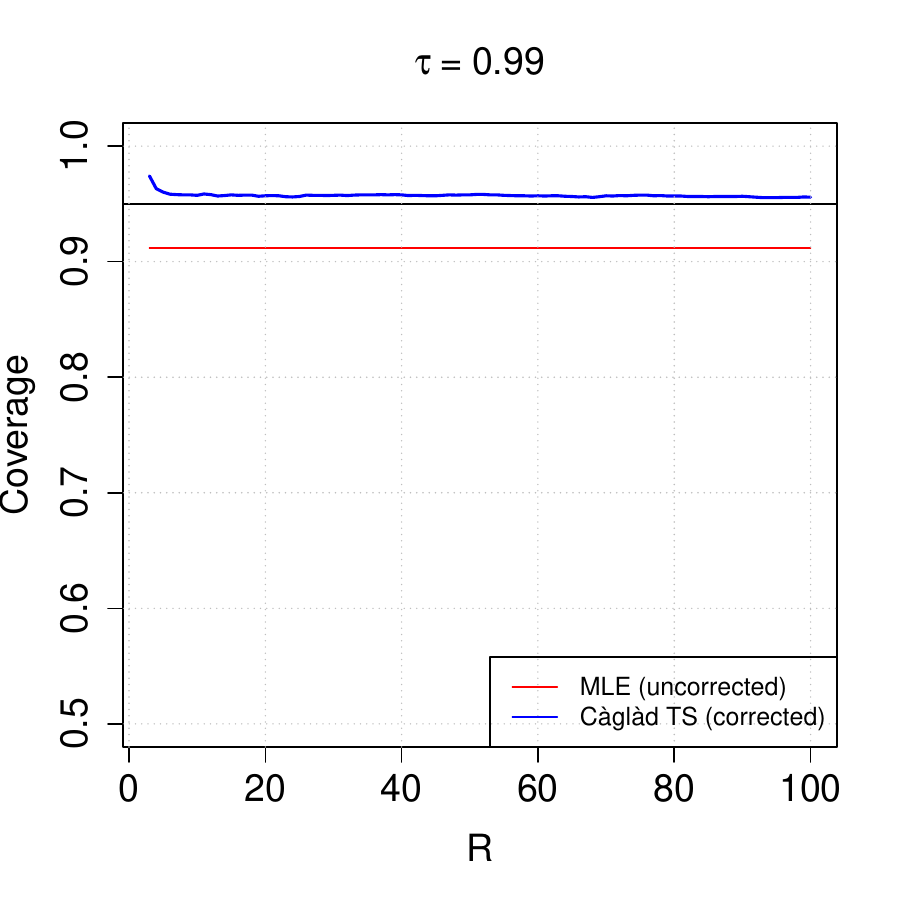}   \includegraphics[width=0.32\textwidth]{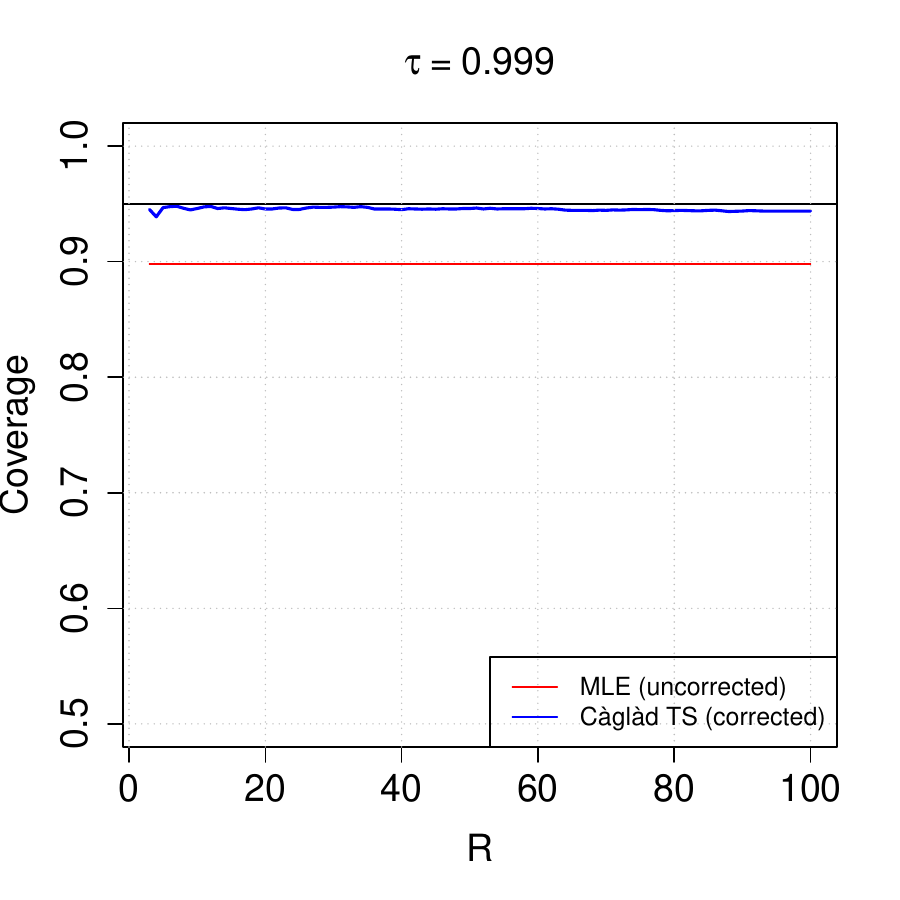}
		\caption{$T=100$}
	\end{subfigure}

	\caption{GEV: coverage of feasible confidence intervals that rely on corrected quantiles.}
	\label{fig:gev_coverage_correct}
\end{figure}

\begin{figure}[H]
	\centering

	\begin{subfigure}[H]{\textwidth}

		\centering
		\includegraphics[width=0.32\textwidth]{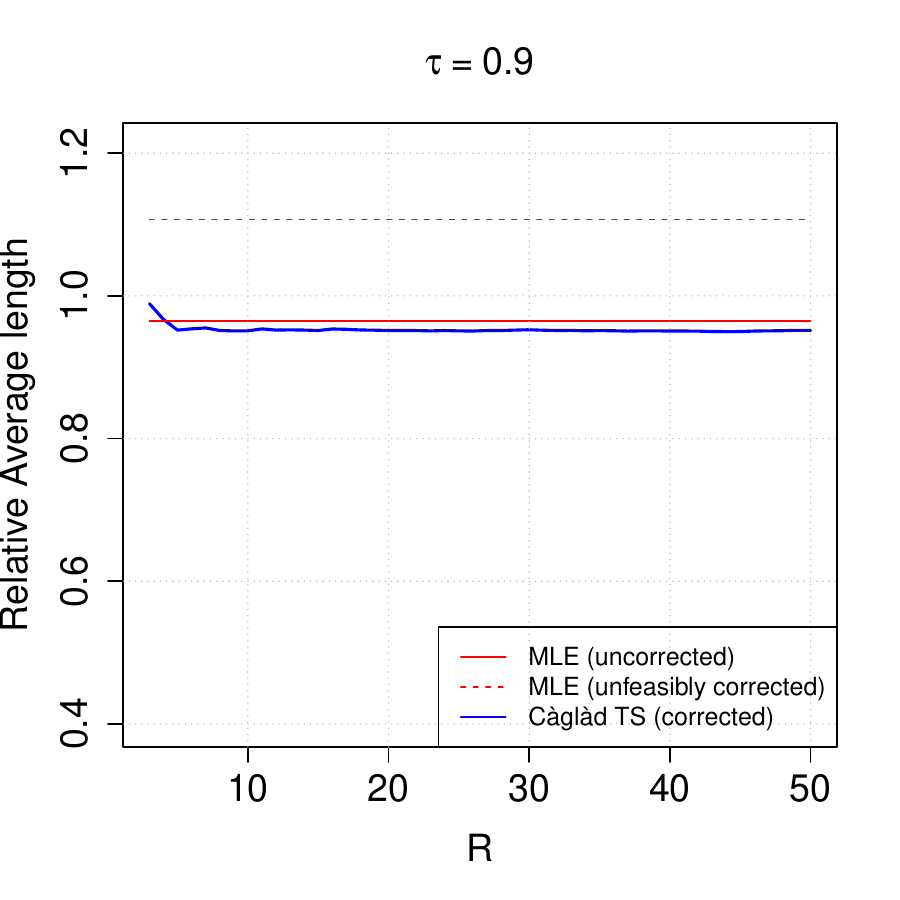}
		\includegraphics[width=0.32\textwidth]{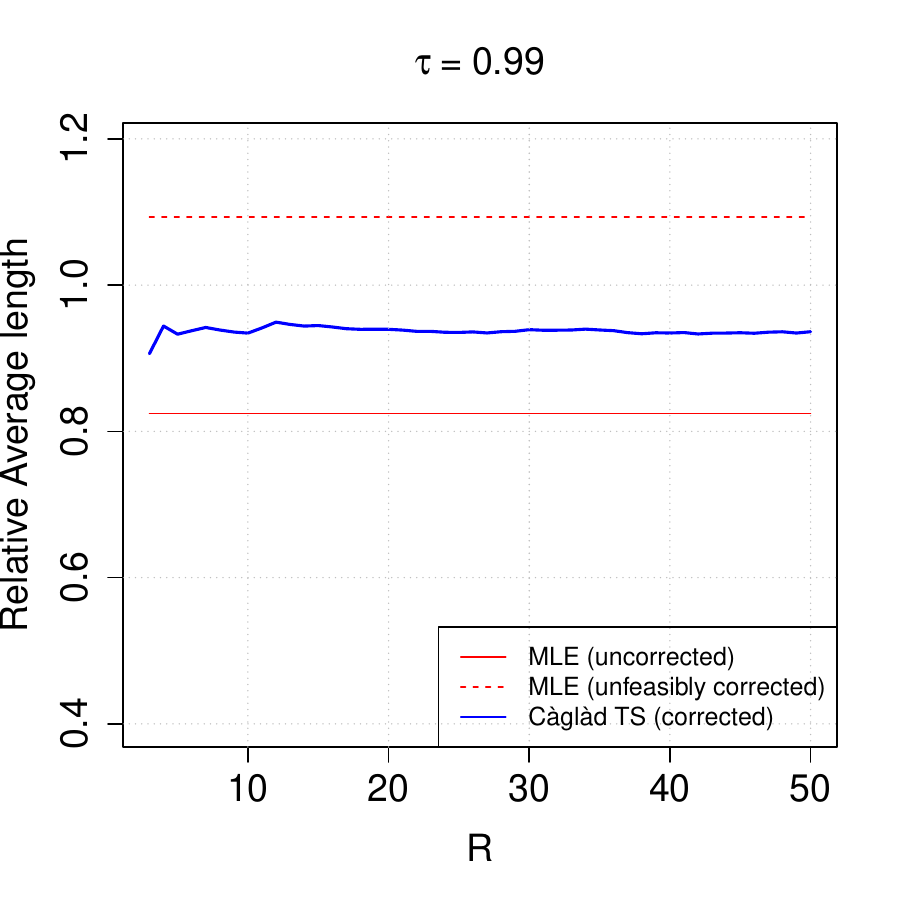}   \includegraphics[width=0.32\textwidth]{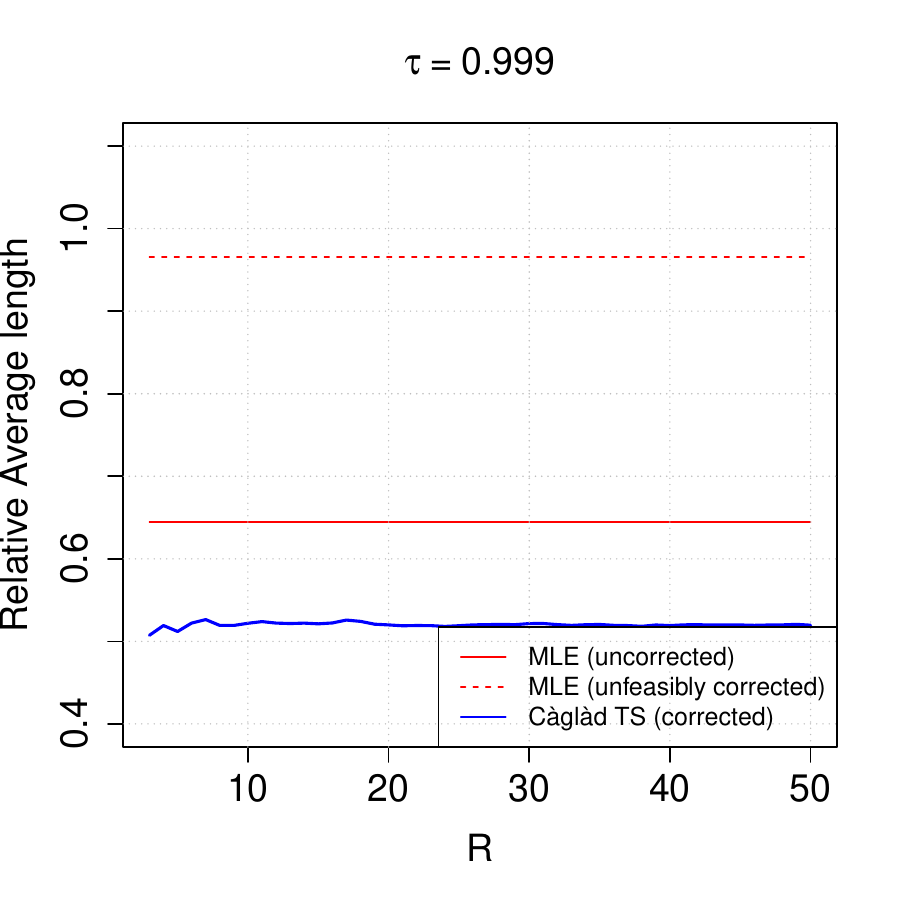}
		\caption{$T=50$}
	\end{subfigure}

	\begin{subfigure}[H]{\textwidth}

		\centering
		\includegraphics[width=0.32\textwidth]{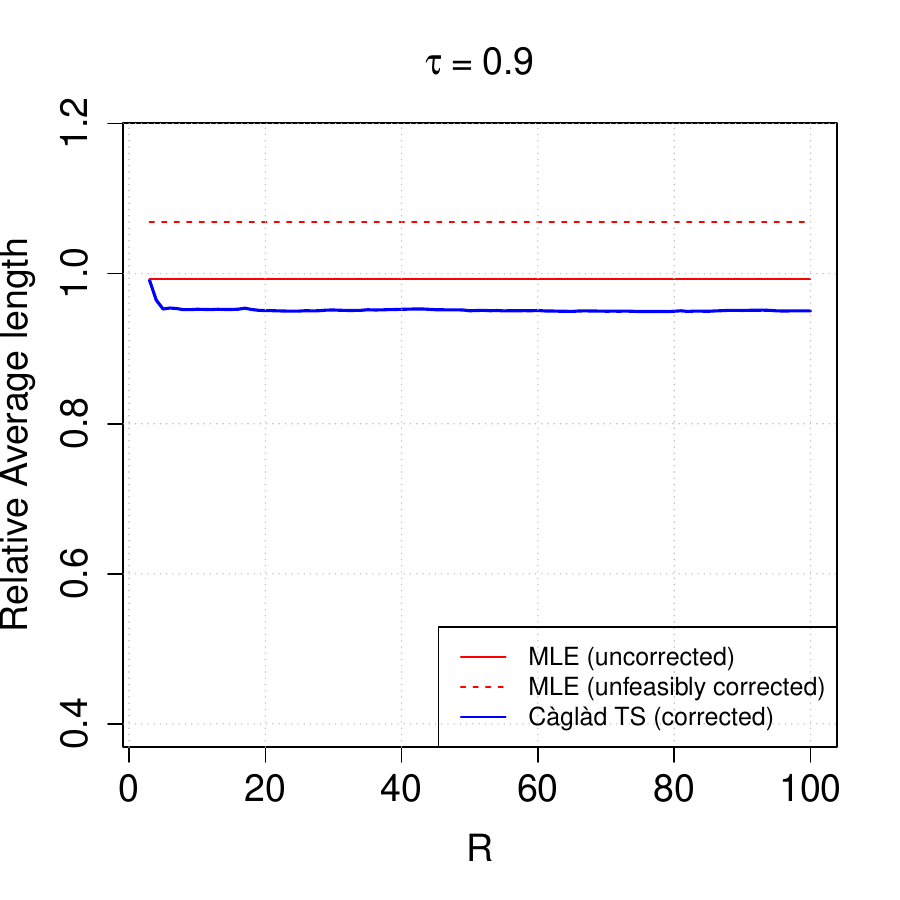}
		\includegraphics[width=0.32\textwidth]{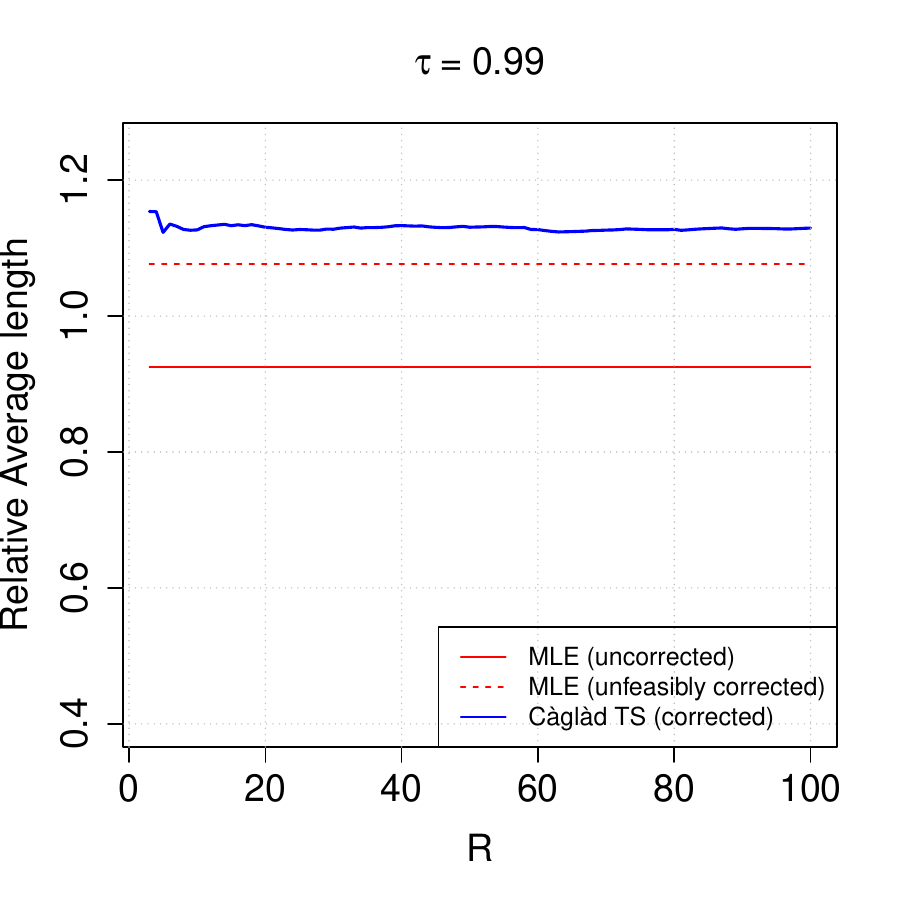}   \includegraphics[width=0.32\textwidth]{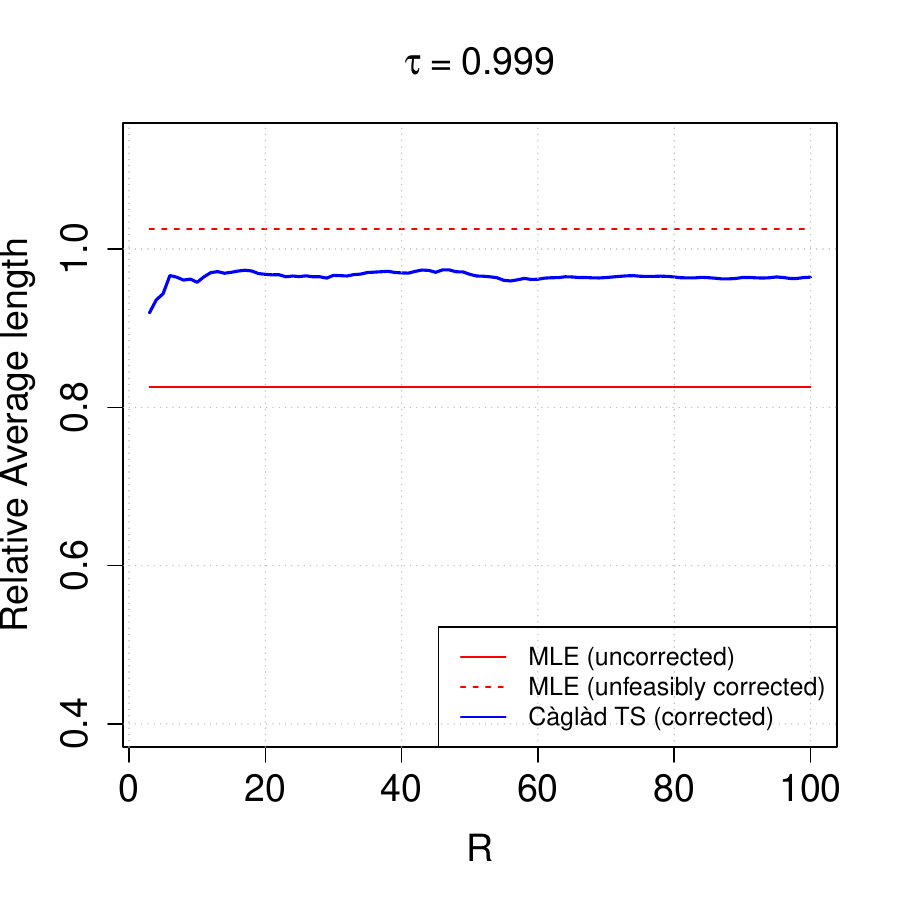}
		\caption{$T=100$}
	\end{subfigure}

	\caption{GEV: relative length (vis-à-vis the unfeasible MLE-based CI that relies on the true sampling variance) of feasible confidence intervals that rely on corrected quantiles.}
	\label{fig:gev_length_correct}
\end{figure}

%% file: plots/gpd/coverage_plots.tex
\begin{figure}[H]
	\centering

	\begin{subfigure}[H]{\textwidth}

		\centering
		\includegraphics[width=0.24\textwidth]{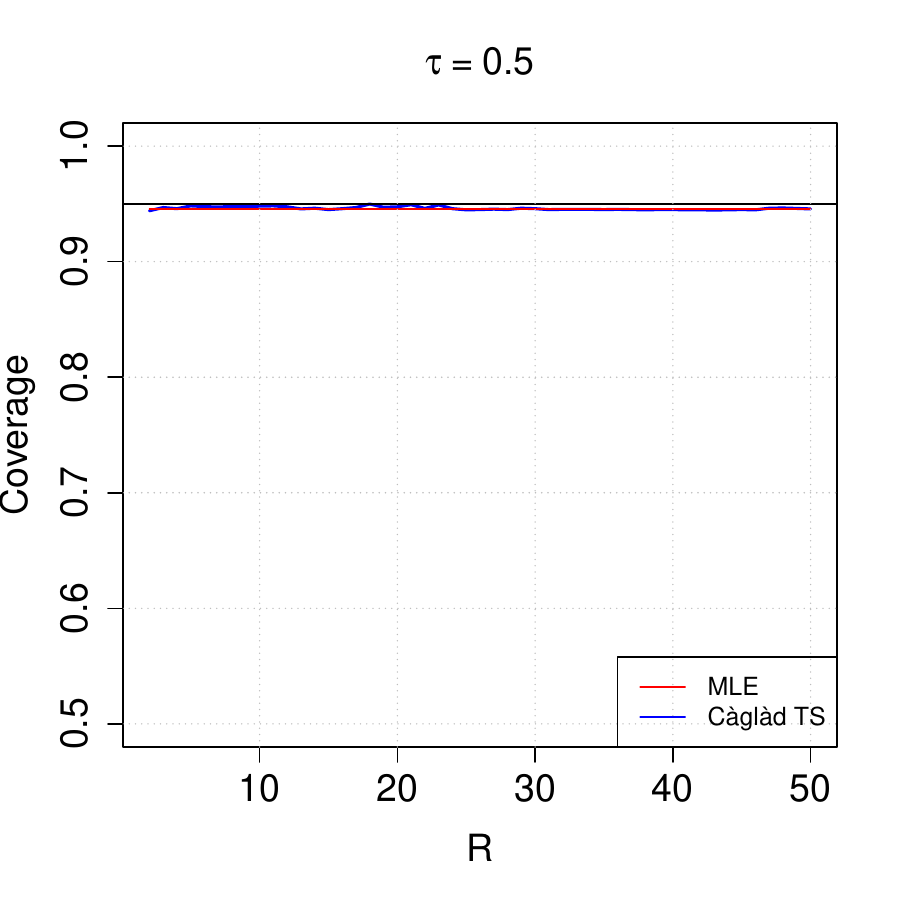}   \includegraphics[width=0.24\textwidth]{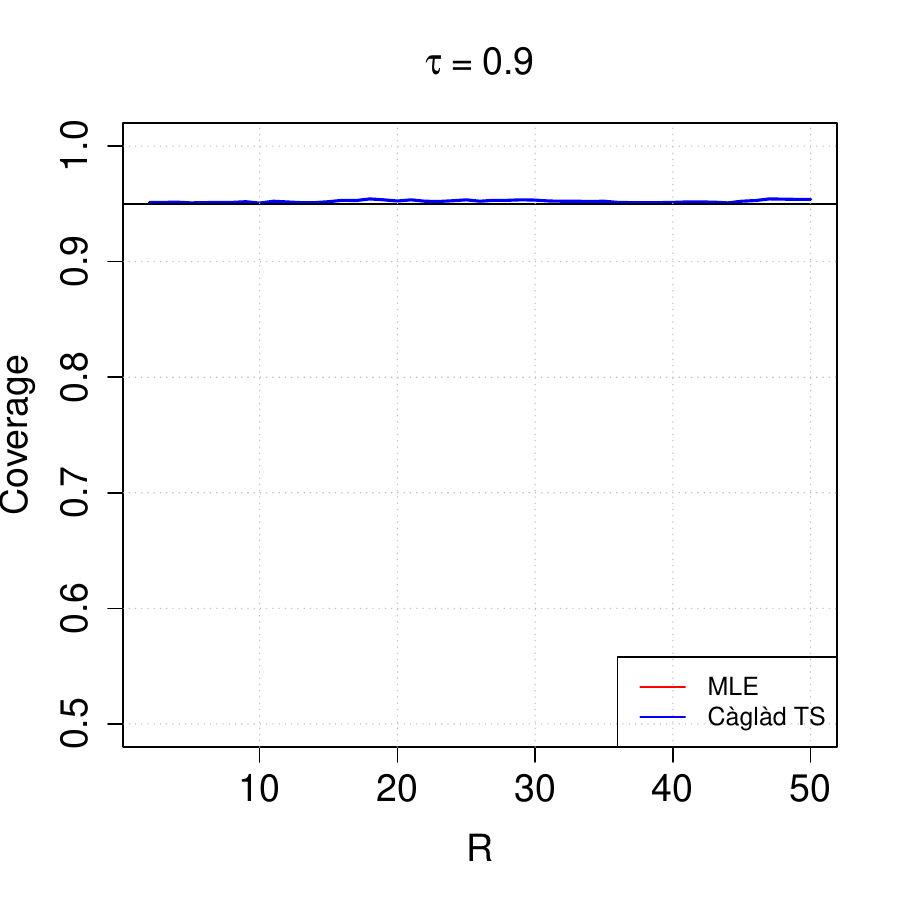}
		\includegraphics[width=0.24\textwidth]{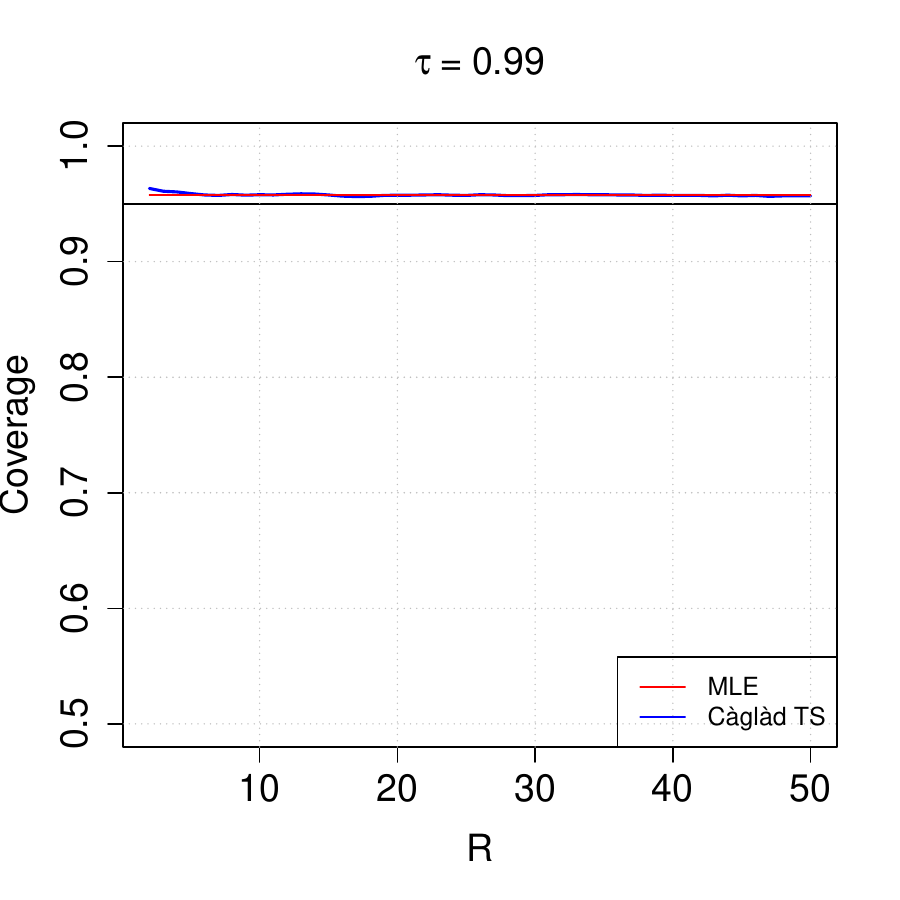}   \includegraphics[width=0.24\textwidth]{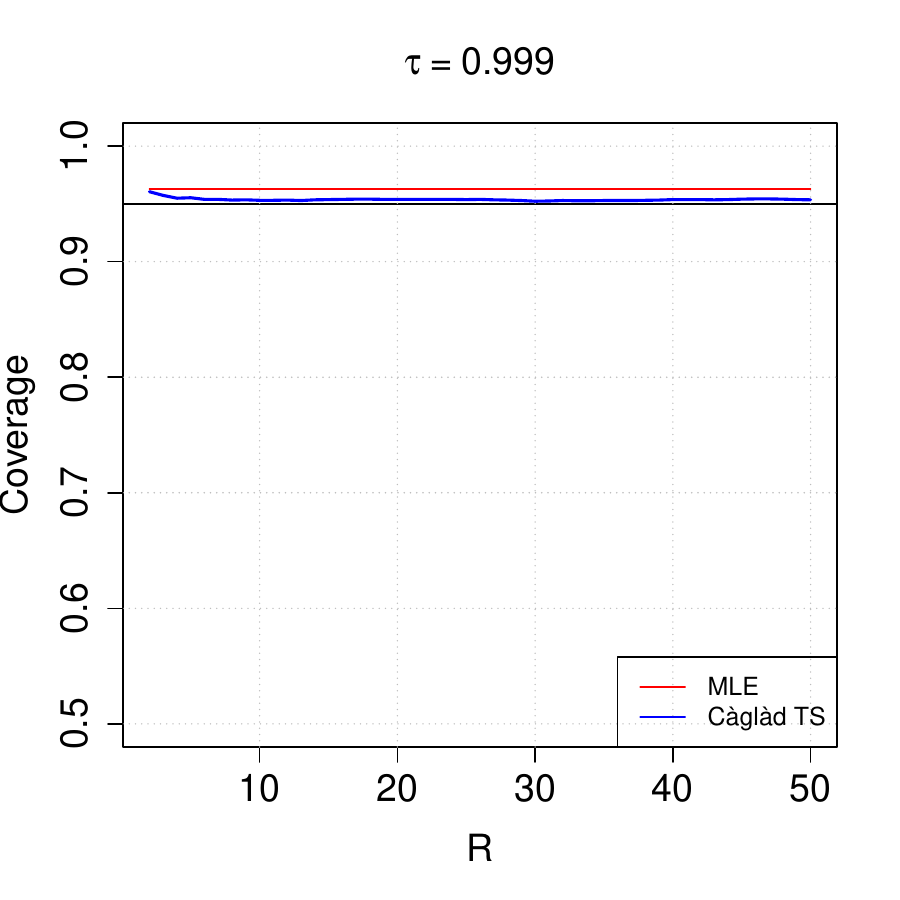}
		\caption{$T=50$}
	\end{subfigure}

	\begin{subfigure}[H]{\textwidth}

		\centering
		\includegraphics[width=0.24\textwidth]{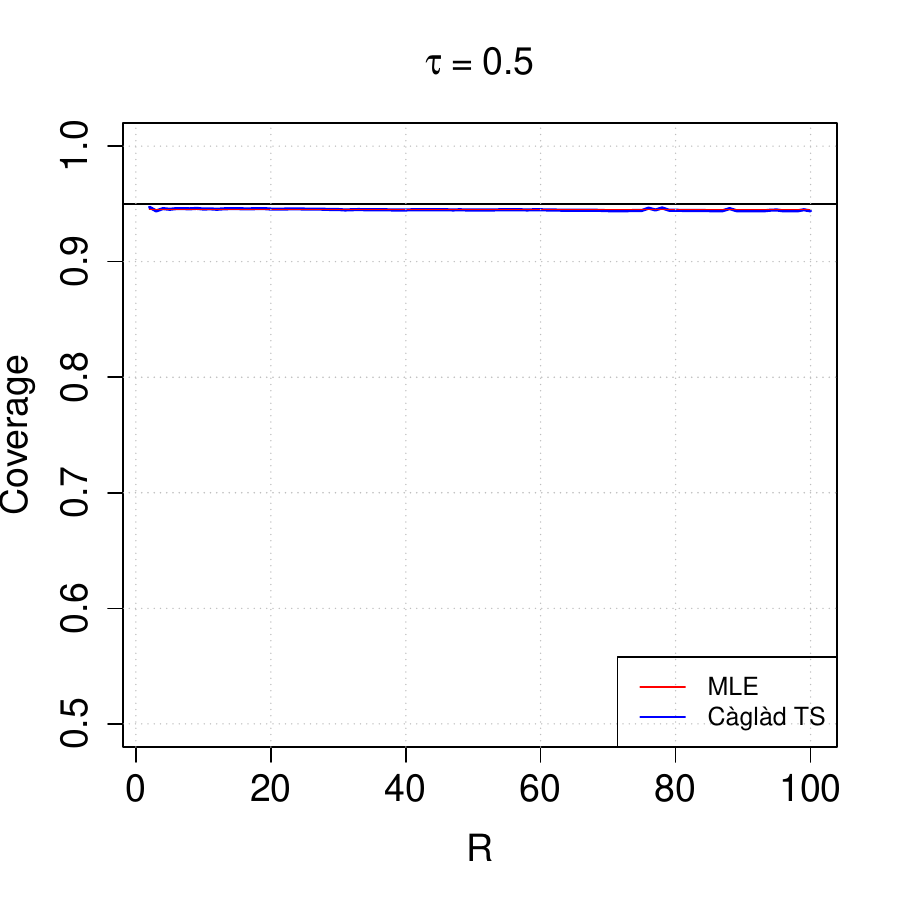}   \includegraphics[width=0.24\textwidth]{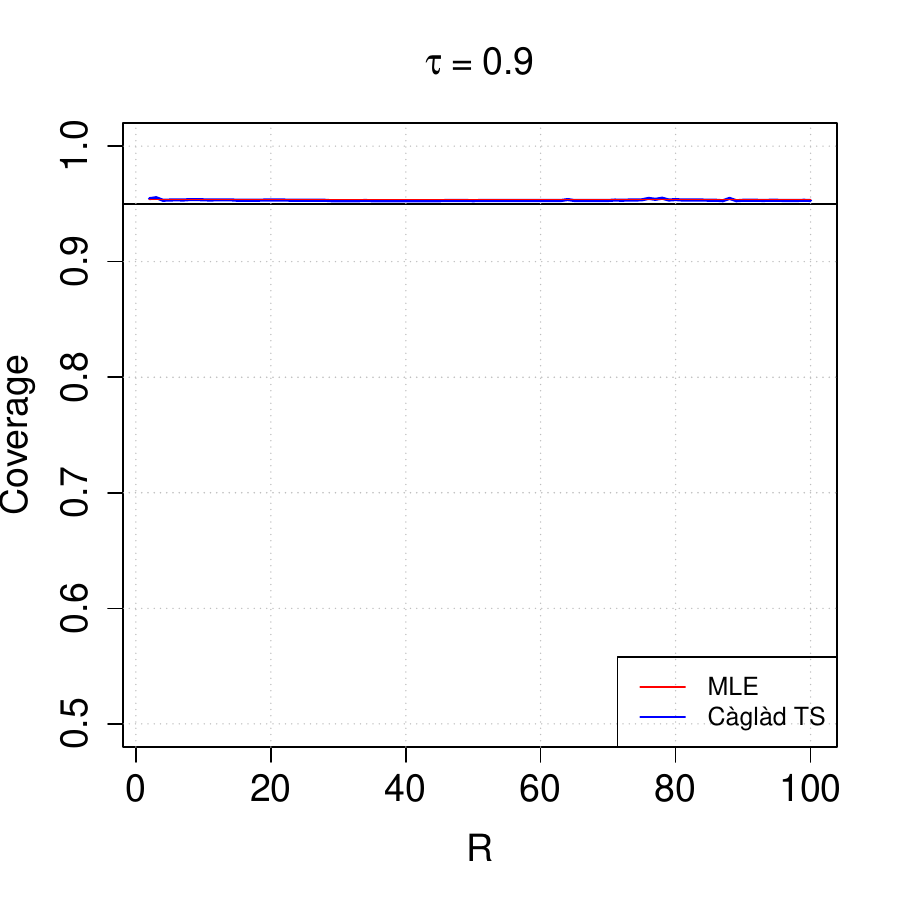}
		\includegraphics[width=0.24\textwidth]{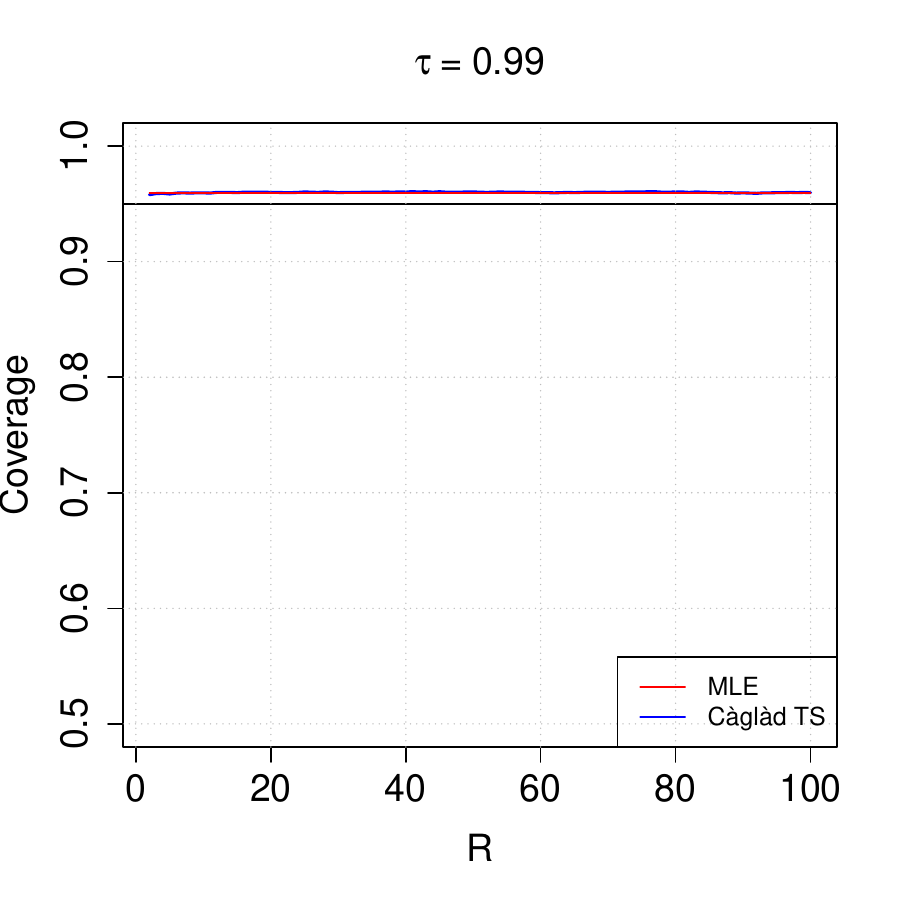}   \includegraphics[width=0.24\textwidth]{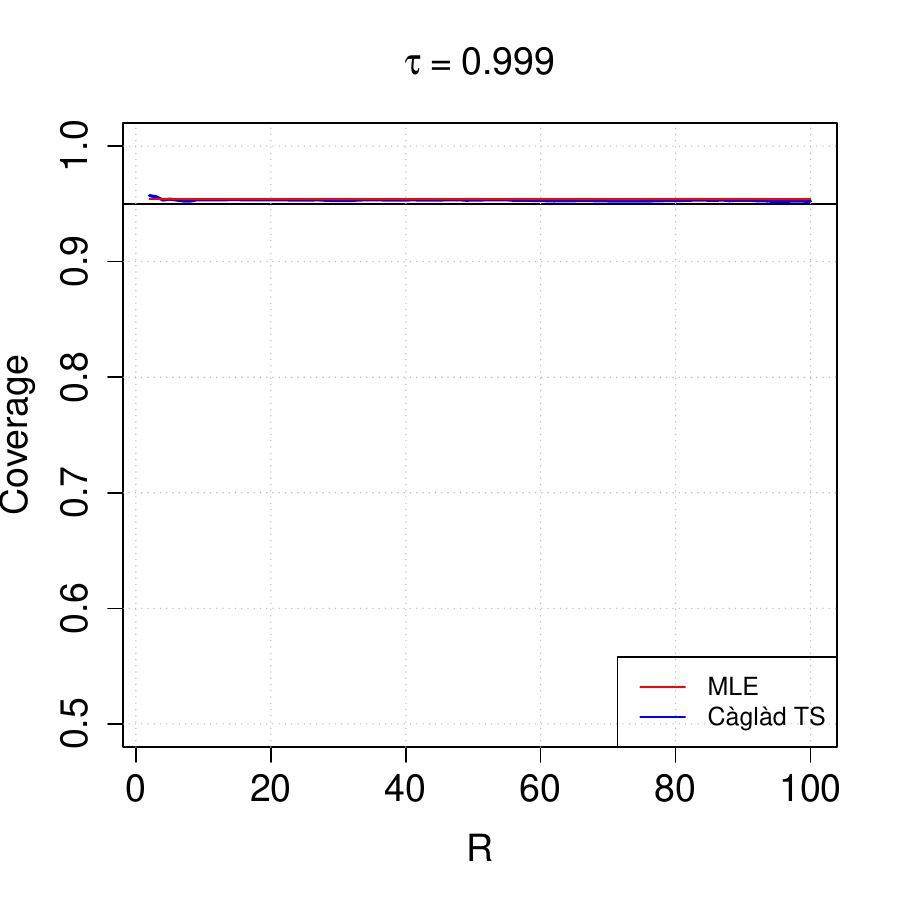}
		\caption{$T=100$}
	\end{subfigure}

	\begin{subfigure}[H]{\textwidth}

		\centering
		\includegraphics[width=0.24\textwidth]{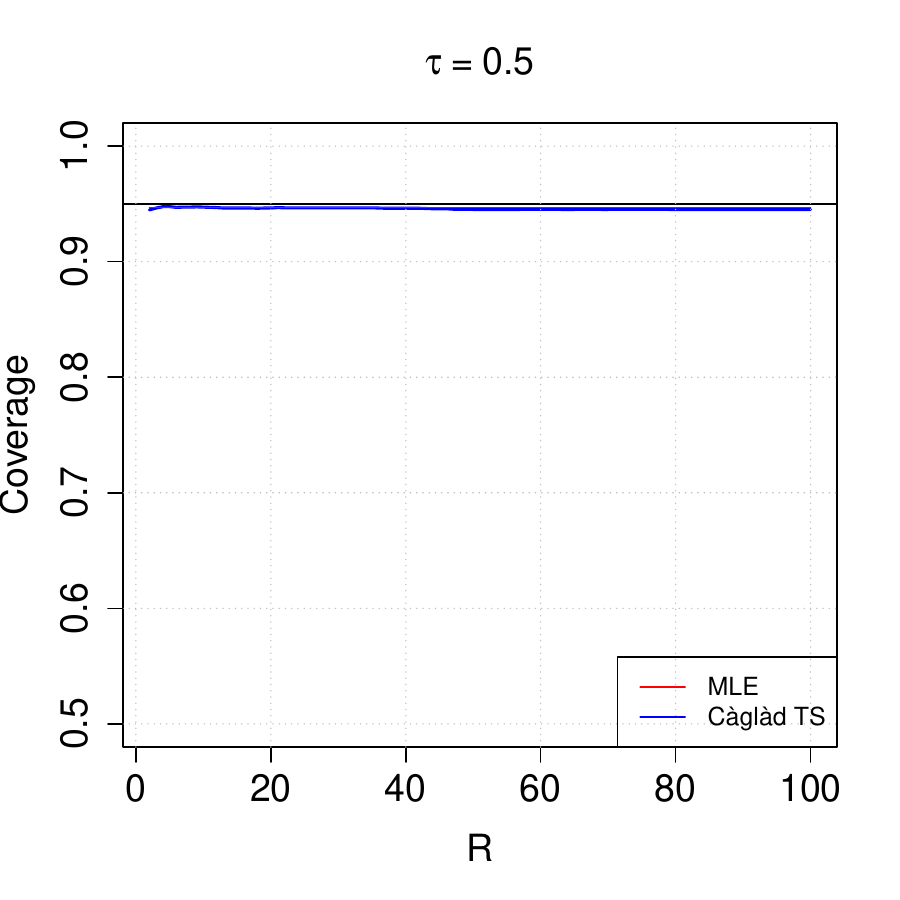}   \includegraphics[width=0.24\textwidth]{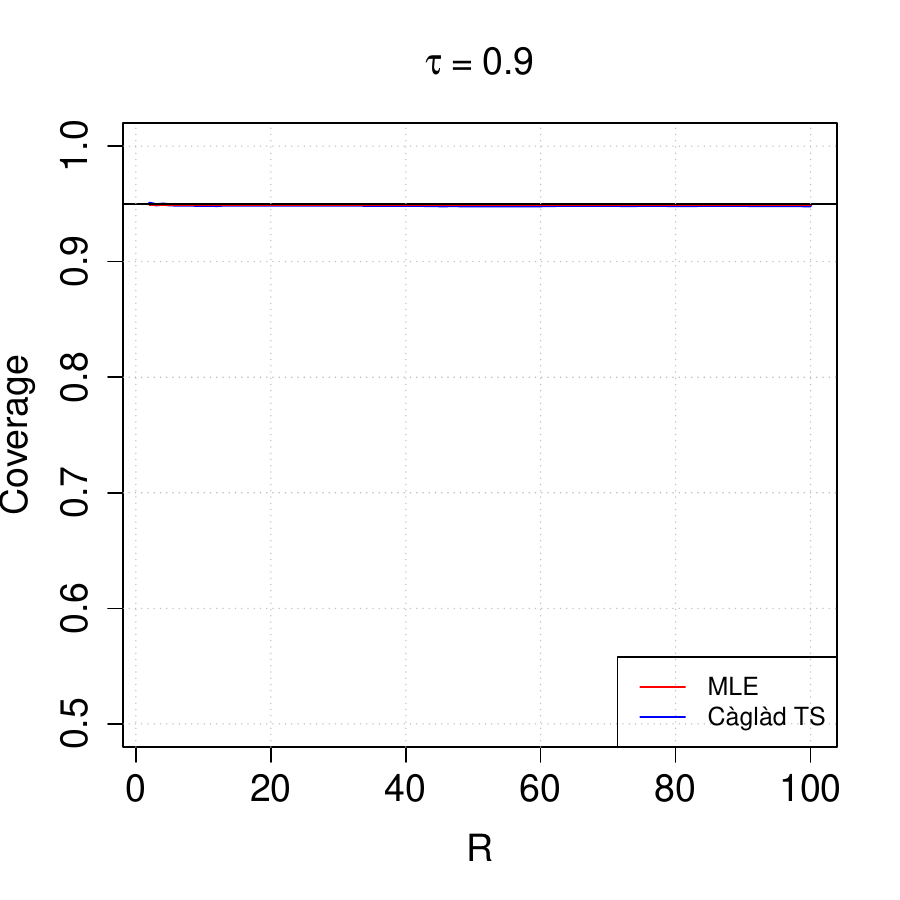}
		\includegraphics[width=0.24\textwidth]{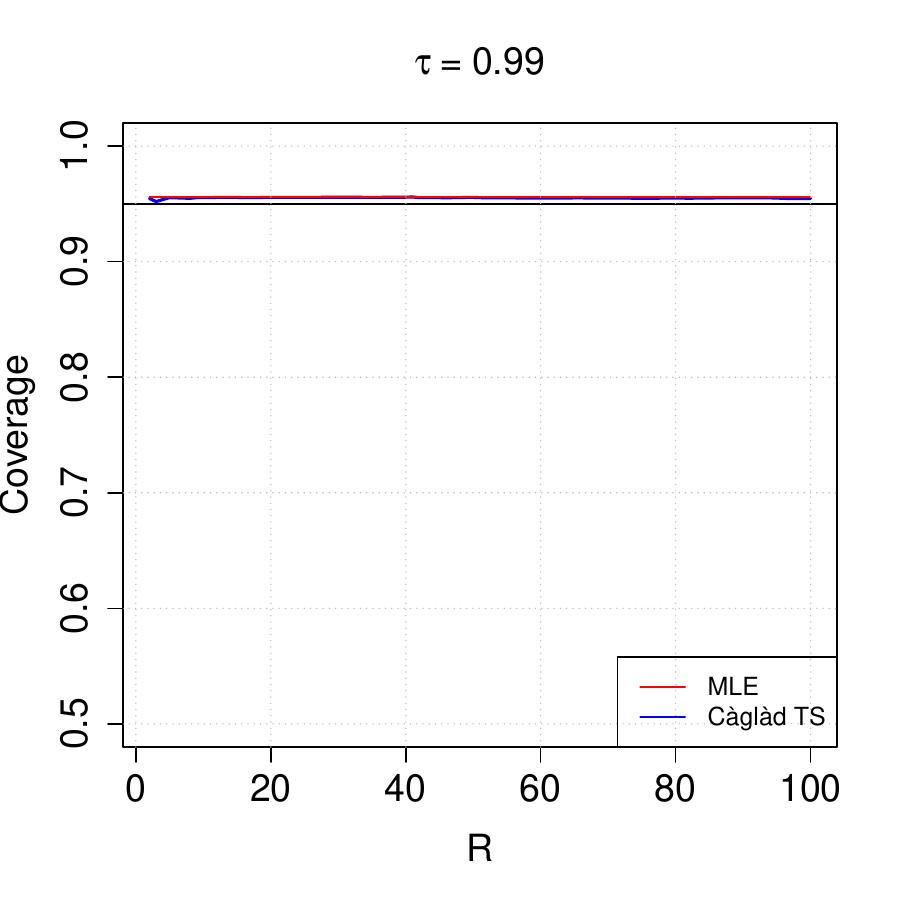}   \includegraphics[width=0.24\textwidth]{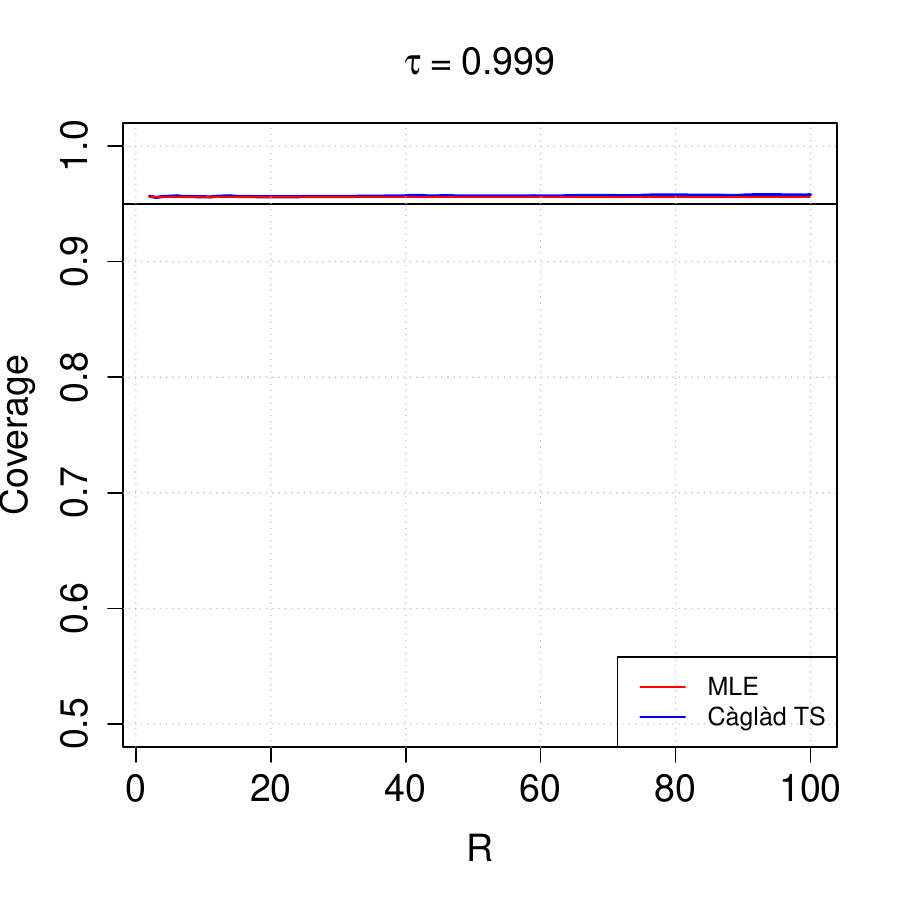}
		\caption{$T=500$}
	\end{subfigure}
	
	\caption{GPD: coverage of confidence intervals based on the true sampling variance.}
	\label{fig:gpd_coverage_true}
\end{figure}

\begin{figure}[H]
	\centering

	\begin{subfigure}[H]{\textwidth}

		\centering
		\includegraphics[width=0.24\textwidth]{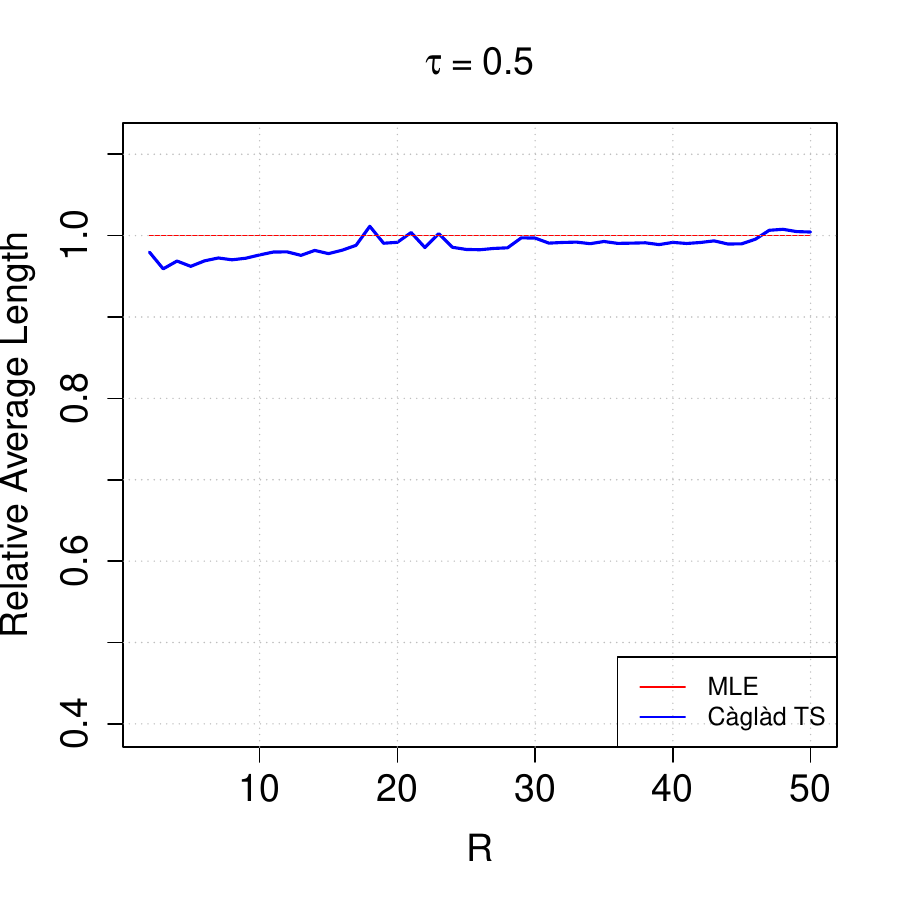}   \includegraphics[width=0.24\textwidth]{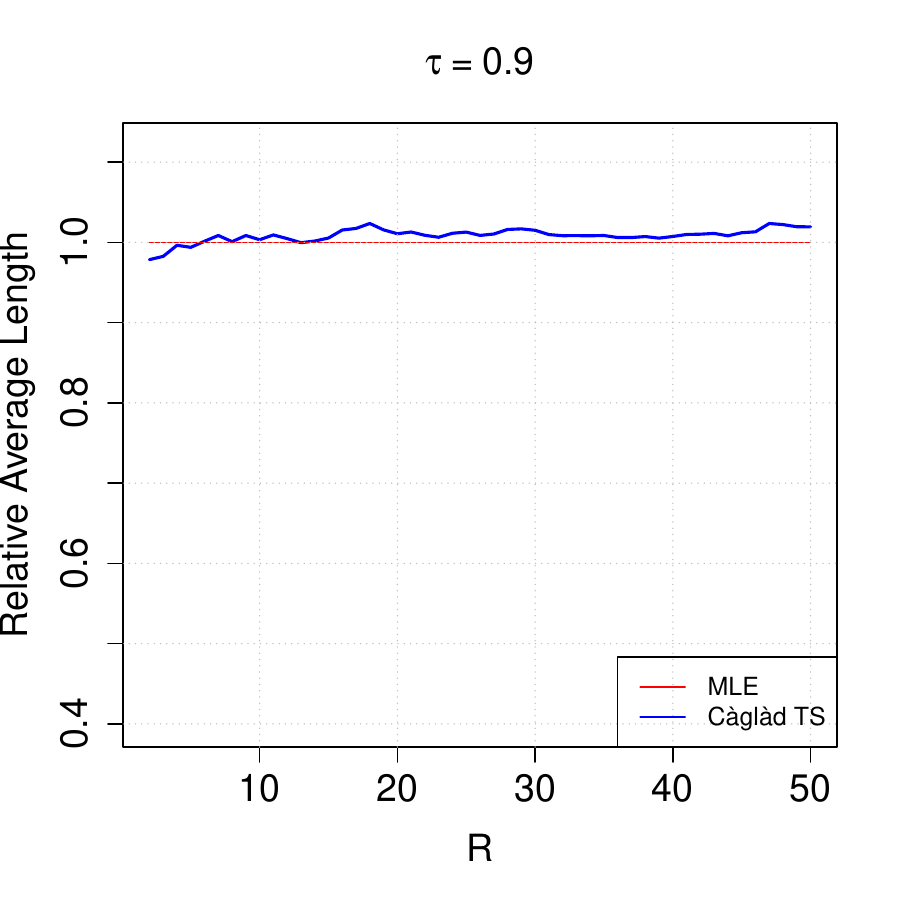}
		\includegraphics[width=0.24\textwidth]{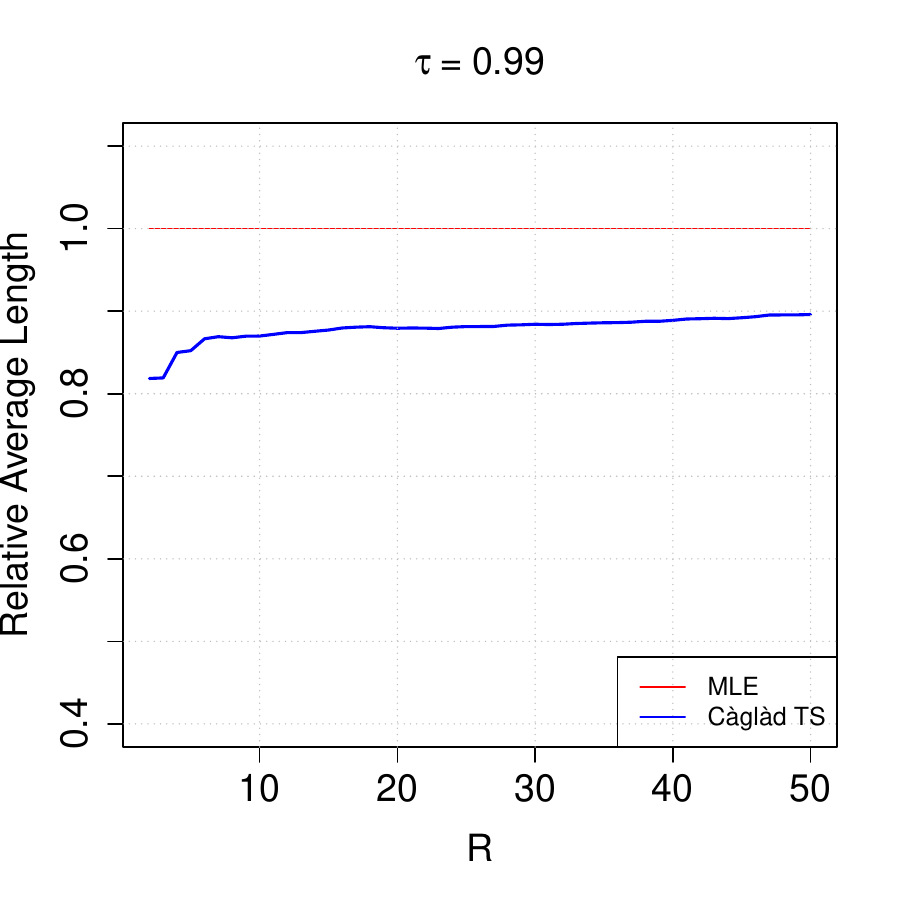}   \includegraphics[width=0.24\textwidth]{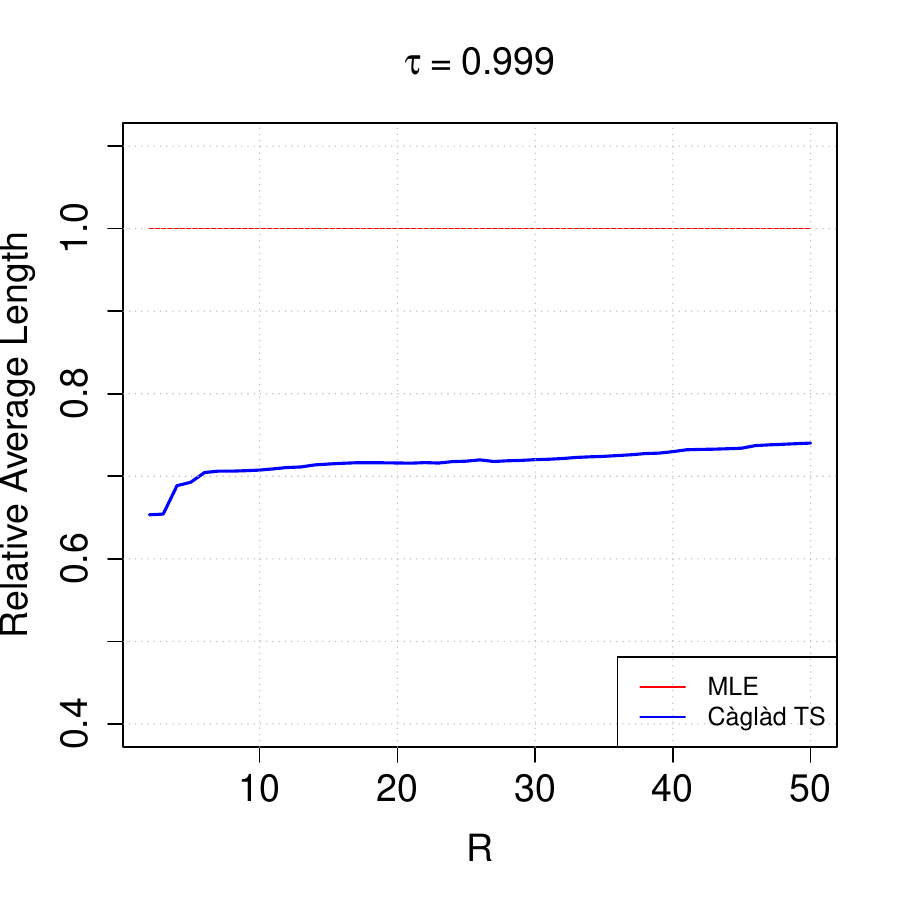}
		\caption{$T=50$}
	\end{subfigure}

	\begin{subfigure}[H]{\textwidth}

		\centering
		\includegraphics[width=0.24\textwidth]{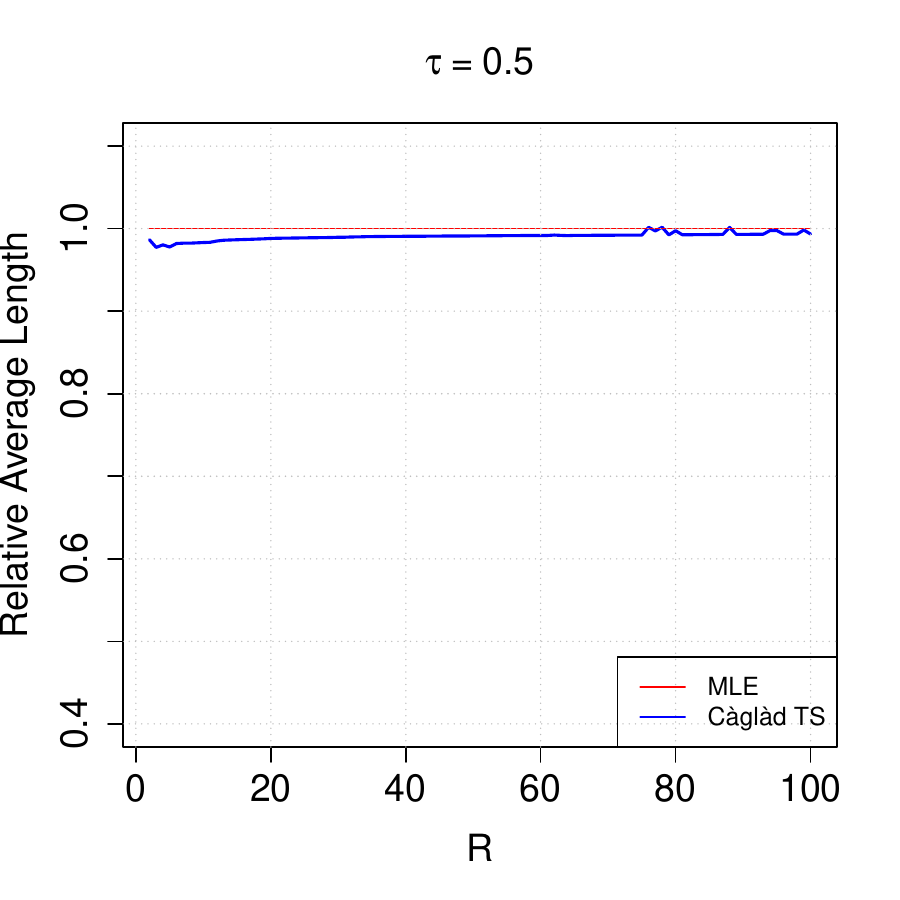}   \includegraphics[width=0.24\textwidth]{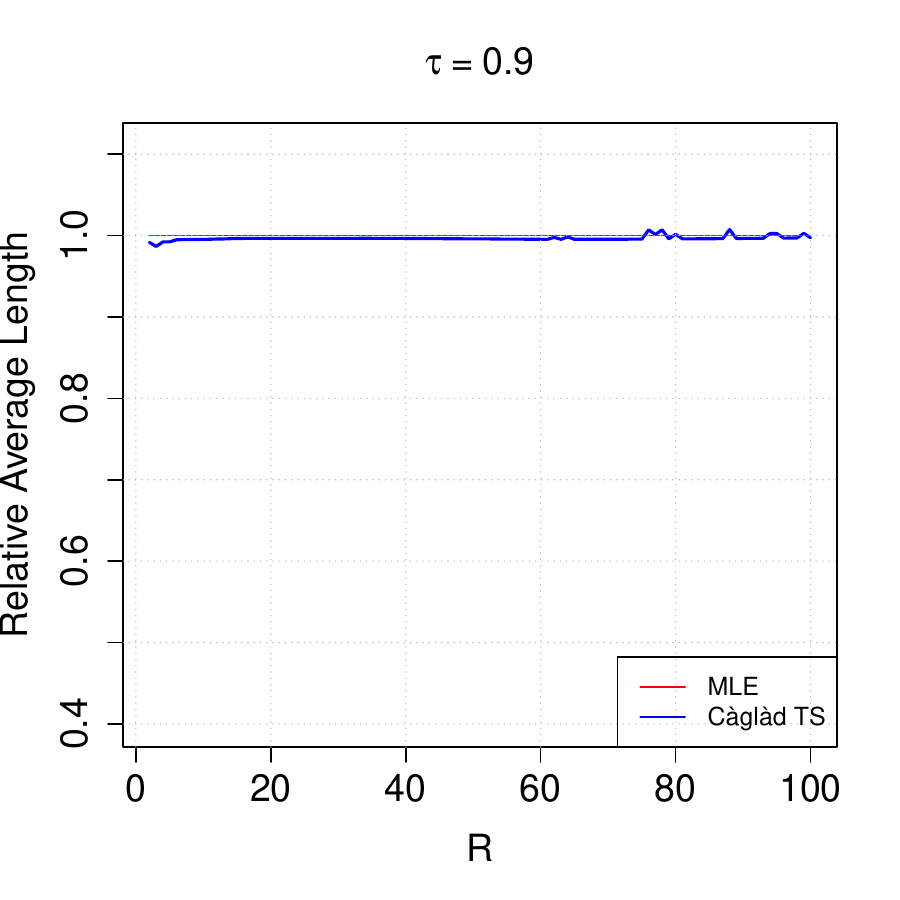}
		\includegraphics[width=0.24\textwidth]{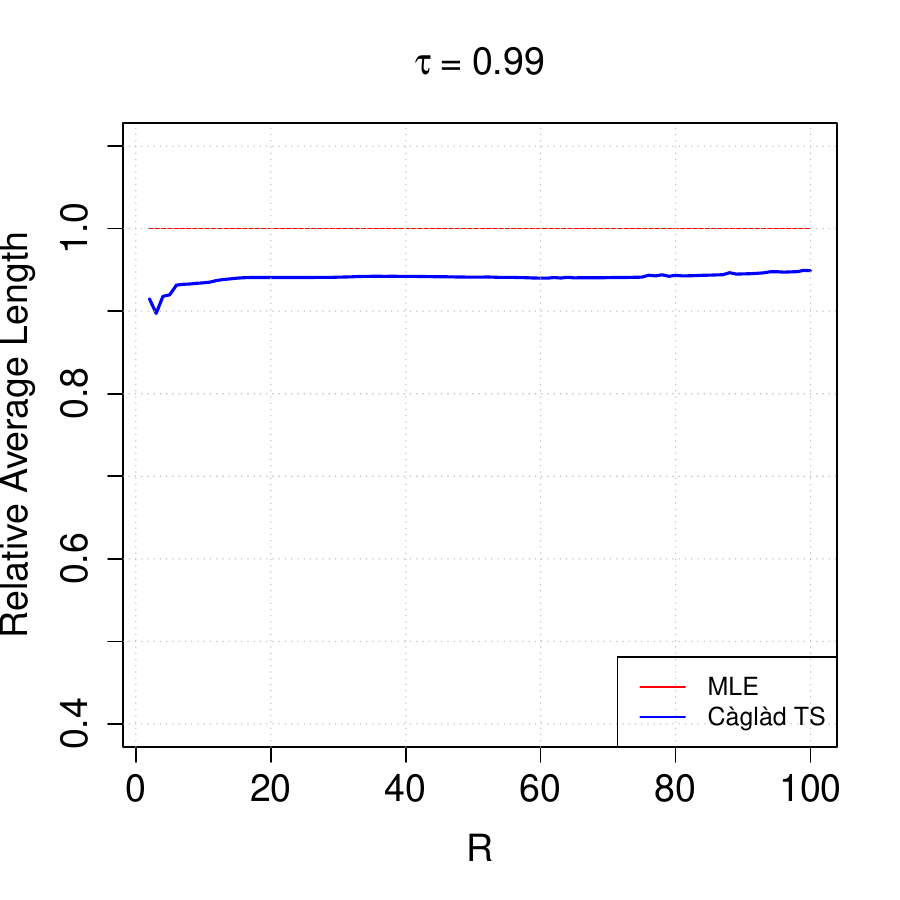}   \includegraphics[width=0.24\textwidth]{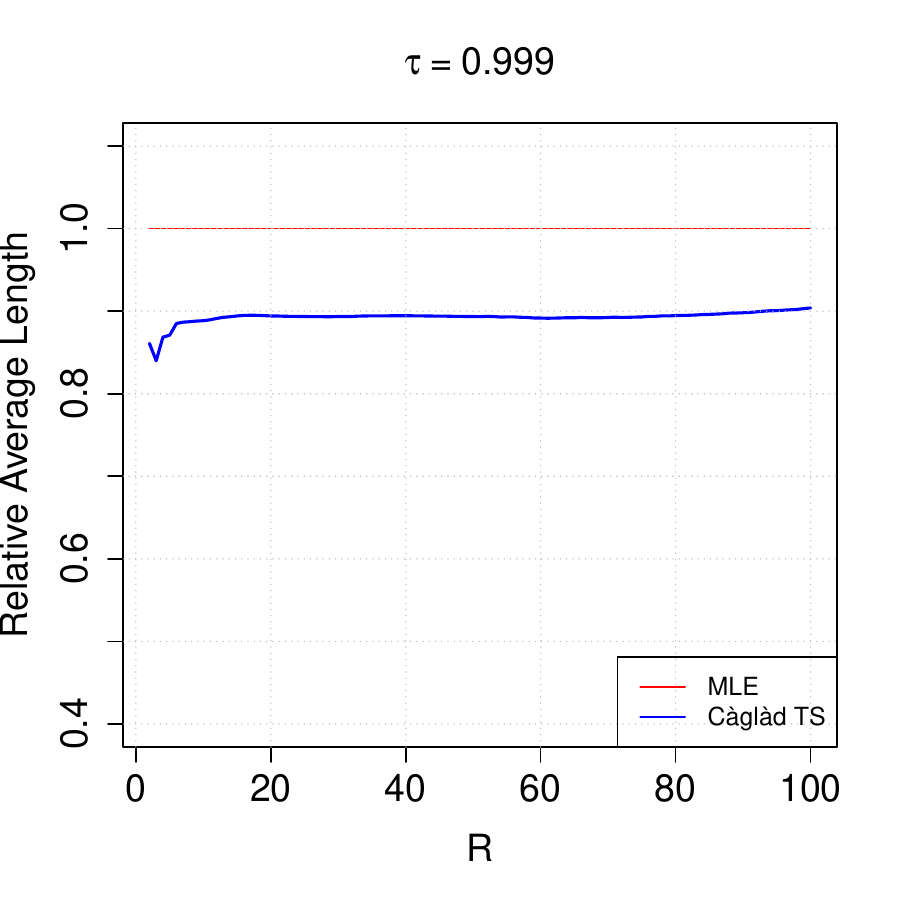}
		\caption{$T=100$}
	\end{subfigure}

	\begin{subfigure}[H]{\textwidth}

		\centering
		\includegraphics[width=0.24\textwidth]{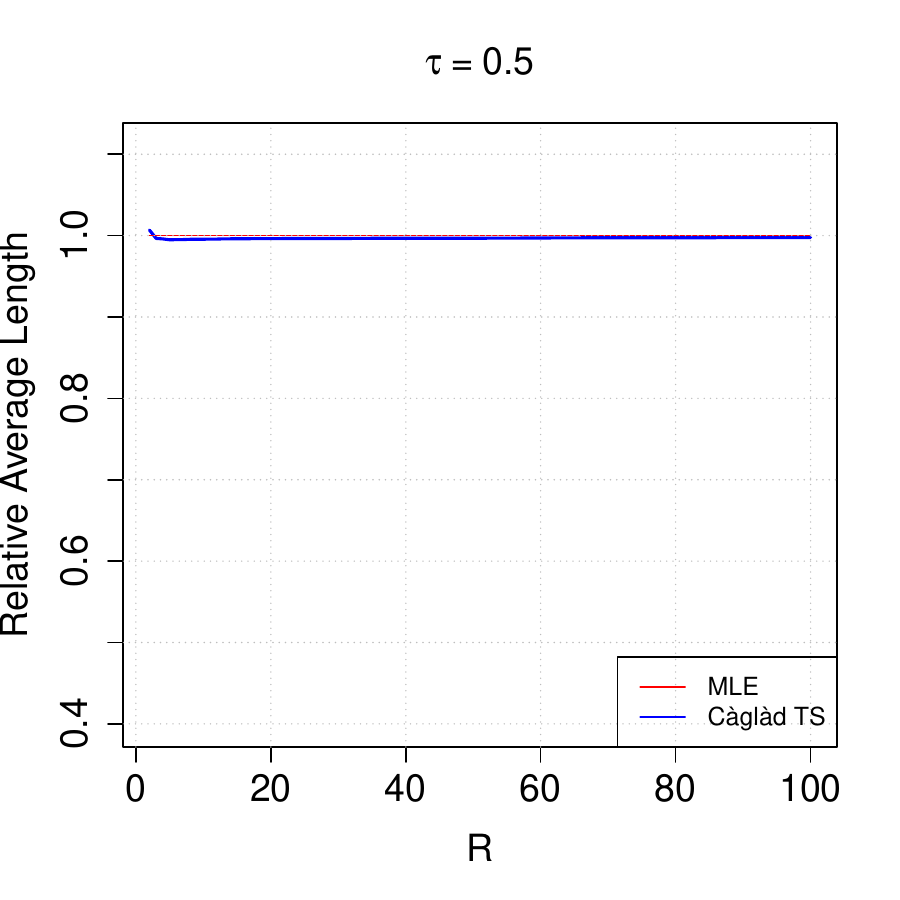}   \includegraphics[width=0.24\textwidth]{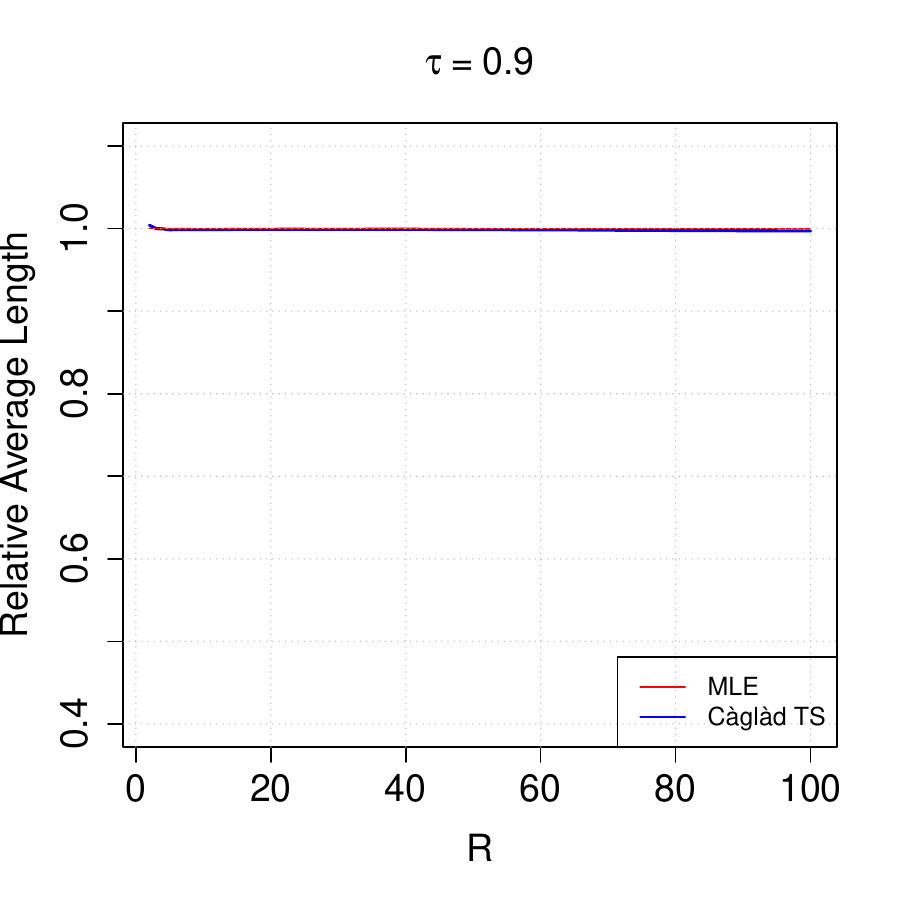}
		\includegraphics[width=0.24\textwidth]{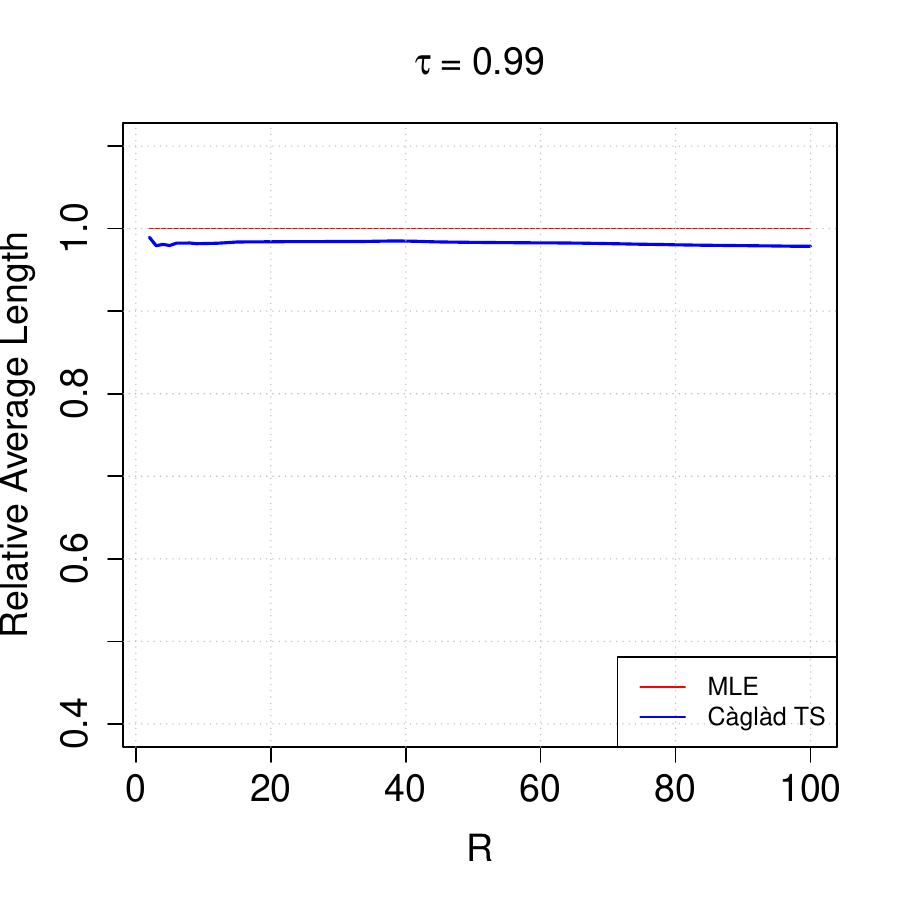}   \includegraphics[width=0.24\textwidth]{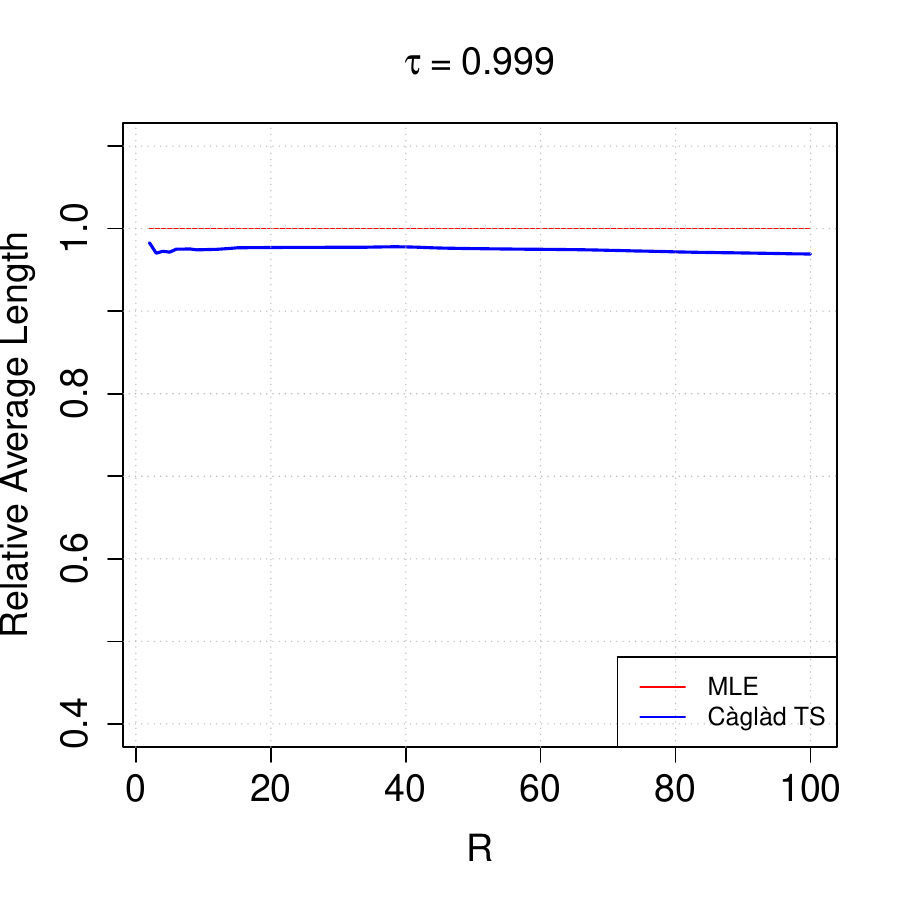}
		\caption{$T=500$}
	\end{subfigure}
	
	\caption{GPD: relative length of confidence (vis-à-vis the unfeasible MLE-based CI) intervals based on the true sampling variance.}
	\label{fig:gpd_length_true}
\end{figure}

\begin{figure}[H]
	\centering

	\begin{subfigure}[H]{\textwidth}

		\centering
		\includegraphics[width=0.24\textwidth]{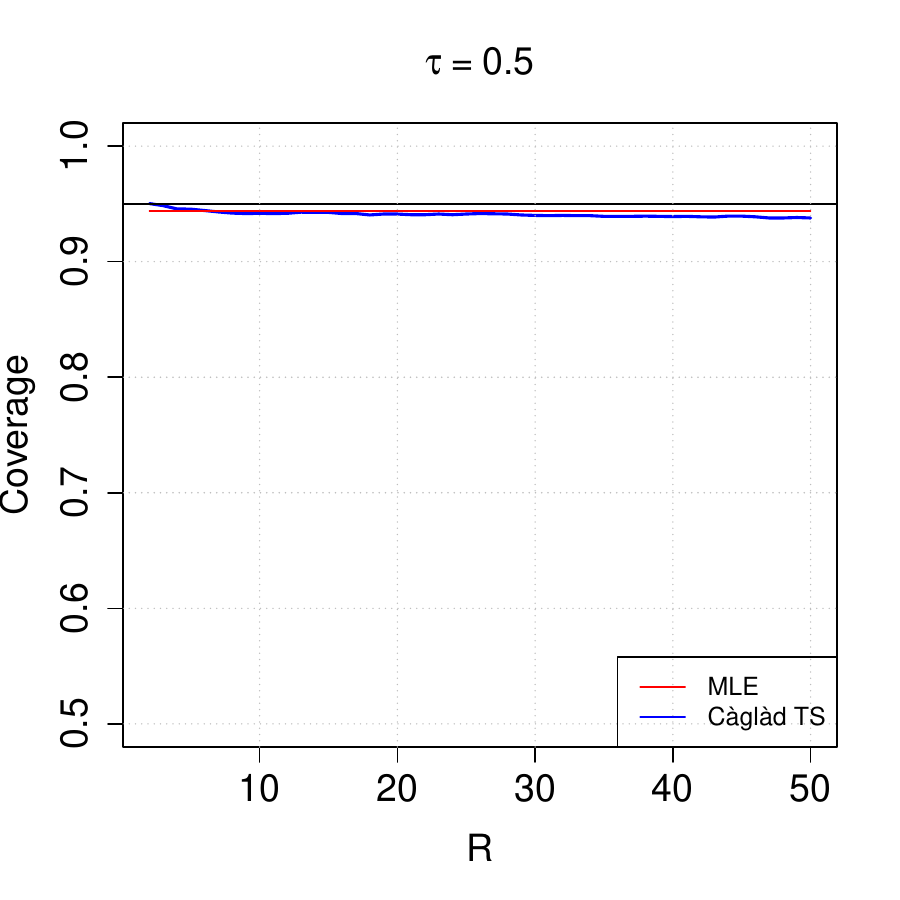}   \includegraphics[width=0.24\textwidth]{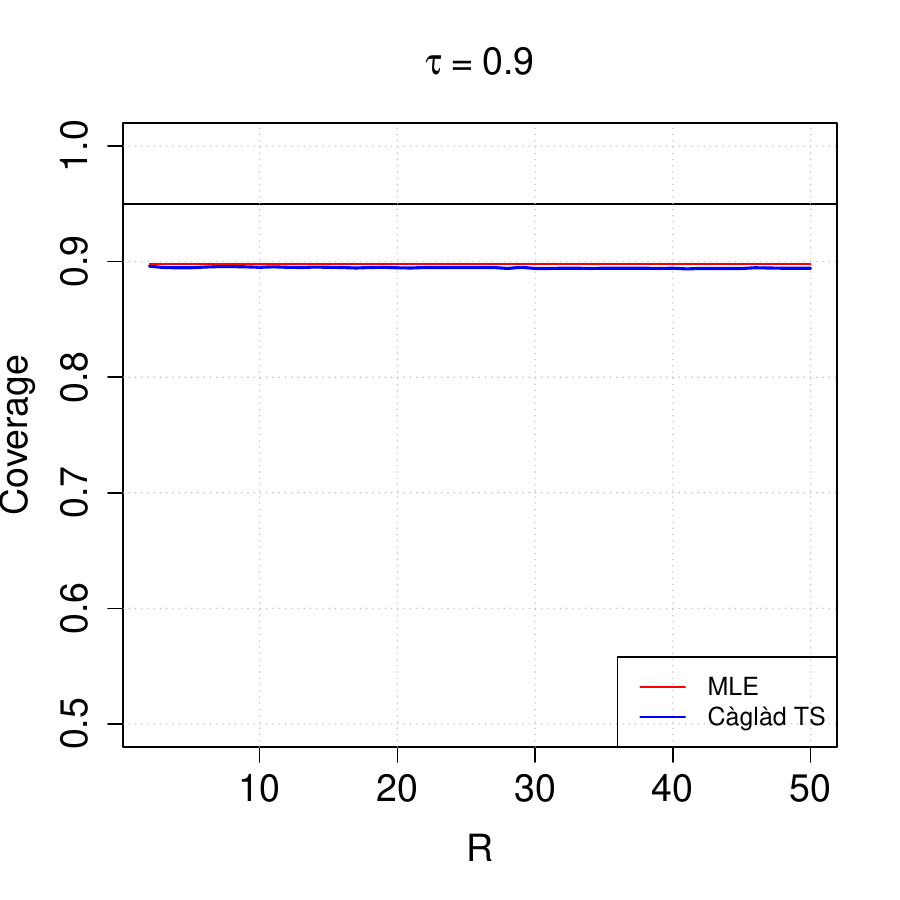}
		\includegraphics[width=0.24\textwidth]{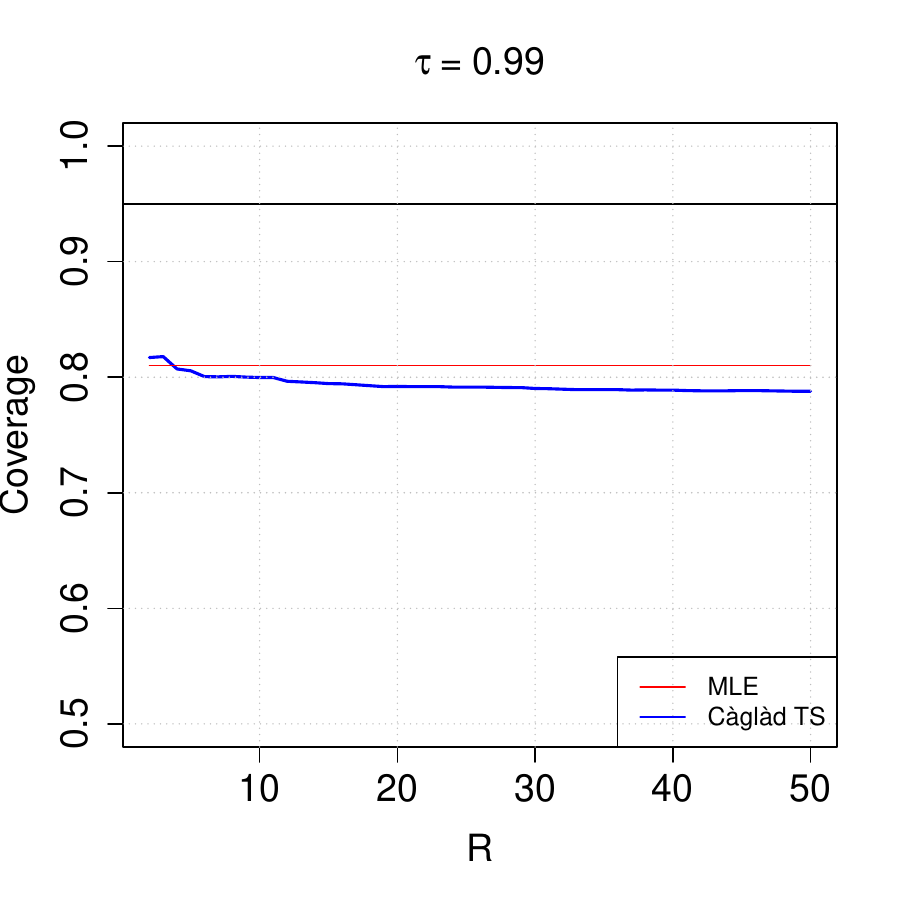}   \includegraphics[width=0.24\textwidth]{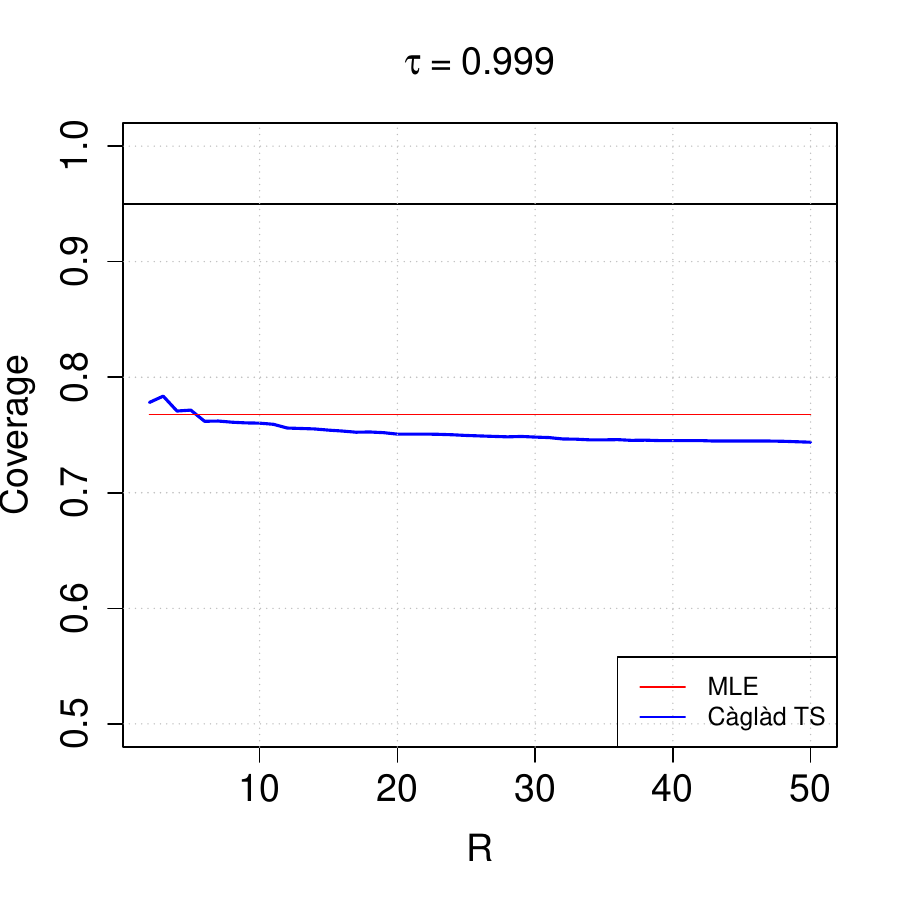}
		\caption{$T=50$}
	\end{subfigure}

	\begin{subfigure}[H]{\textwidth}

		\centering
		\includegraphics[width=0.24\textwidth]{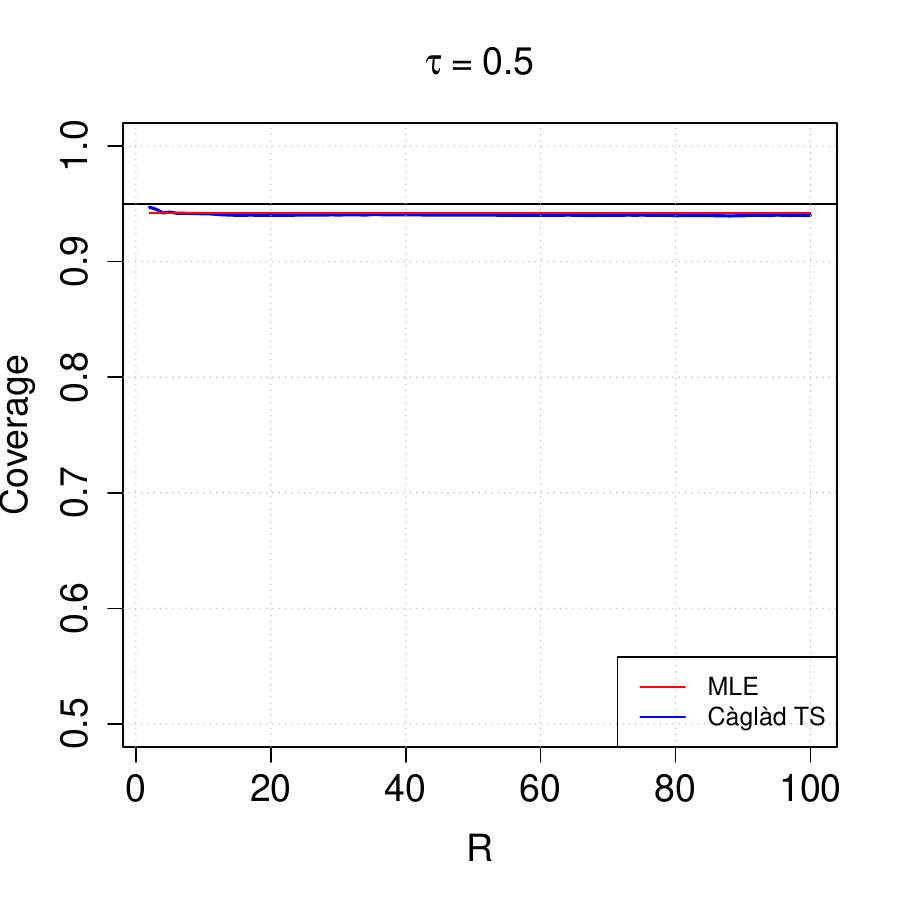}   \includegraphics[width=0.24\textwidth]{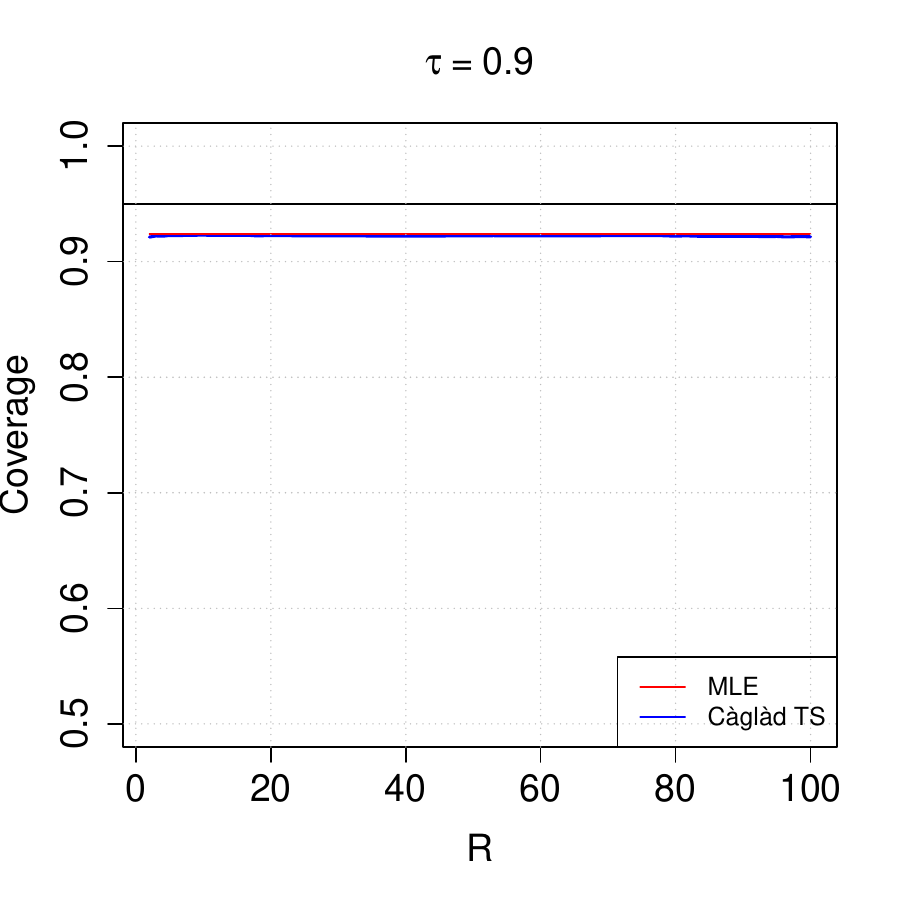}
		\includegraphics[width=0.24\textwidth]{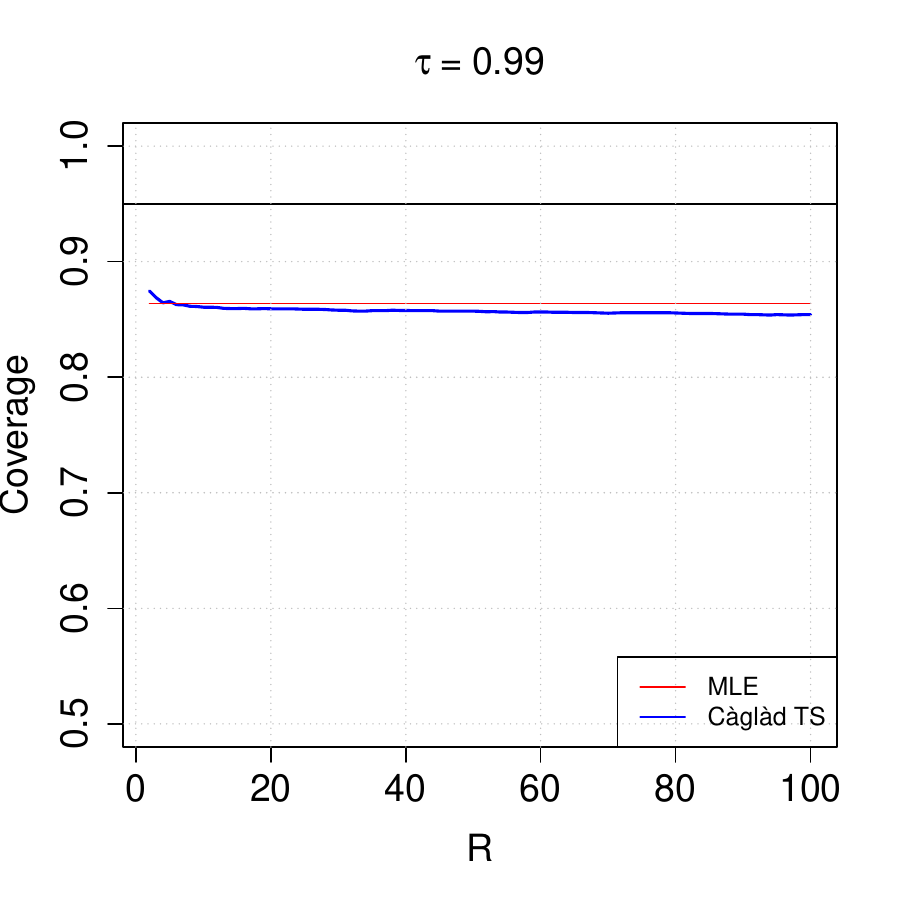}   \includegraphics[width=0.24\textwidth]{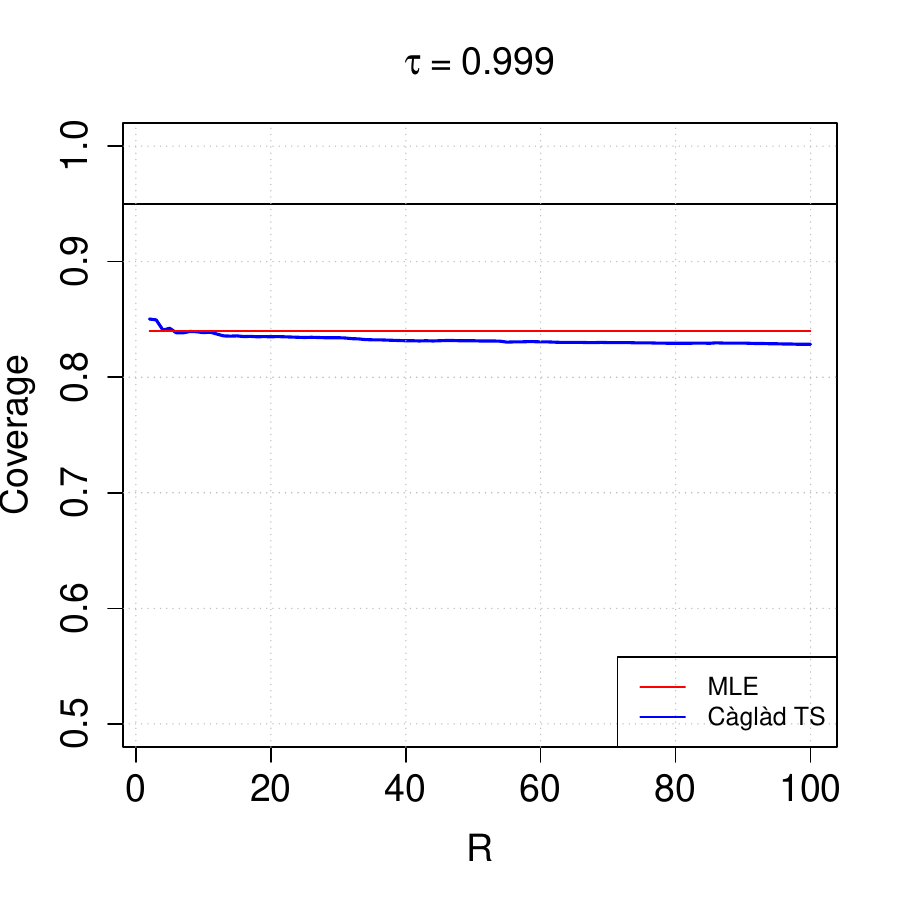}
		\caption{$T=100$}
	\end{subfigure}

	\begin{subfigure}[H]{\textwidth}

		\centering
		\includegraphics[width=0.24\textwidth]{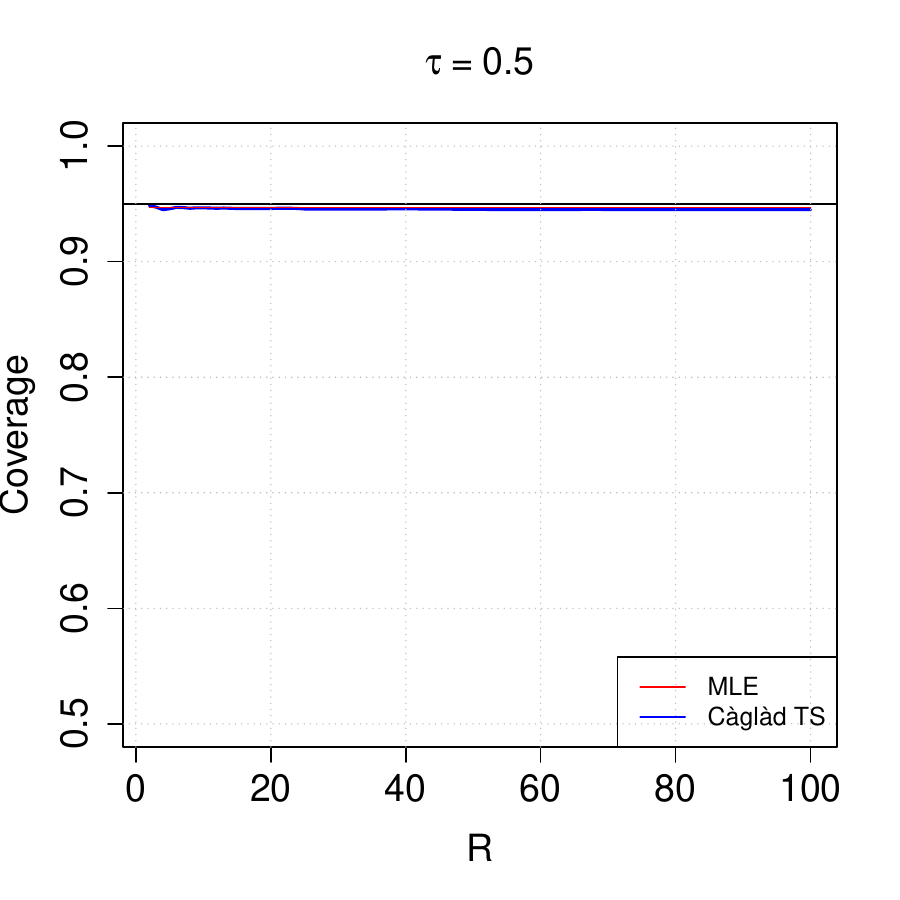}   \includegraphics[width=0.24\textwidth]{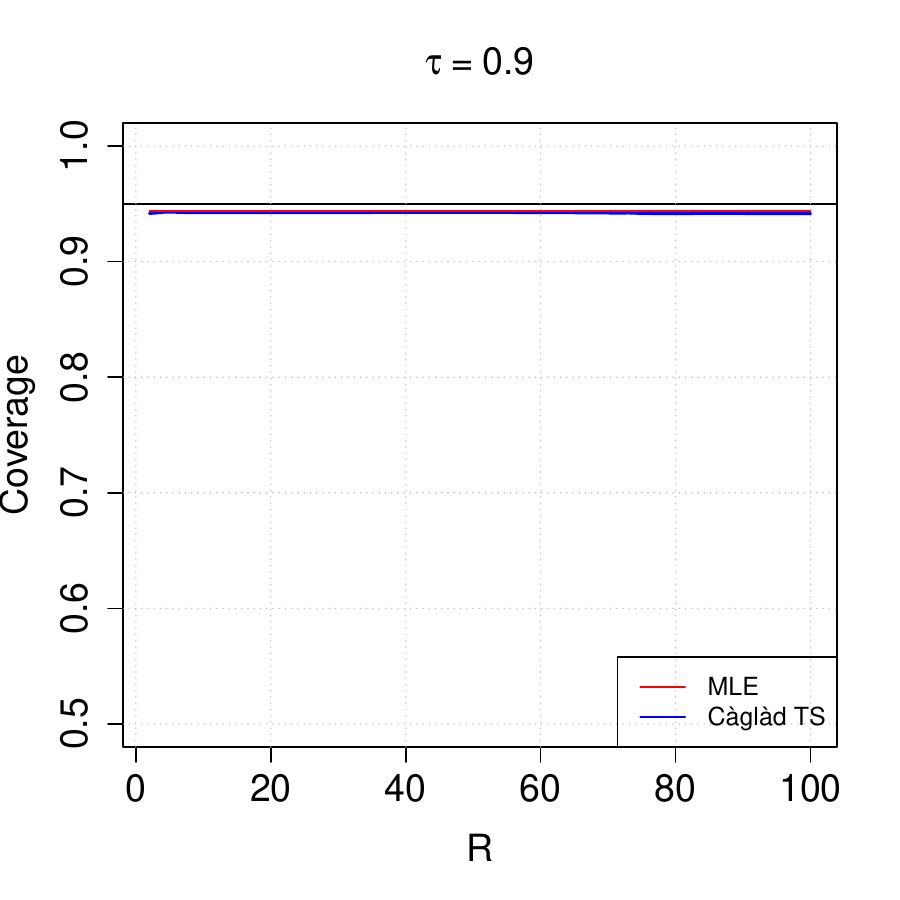}
		\includegraphics[width=0.24\textwidth]{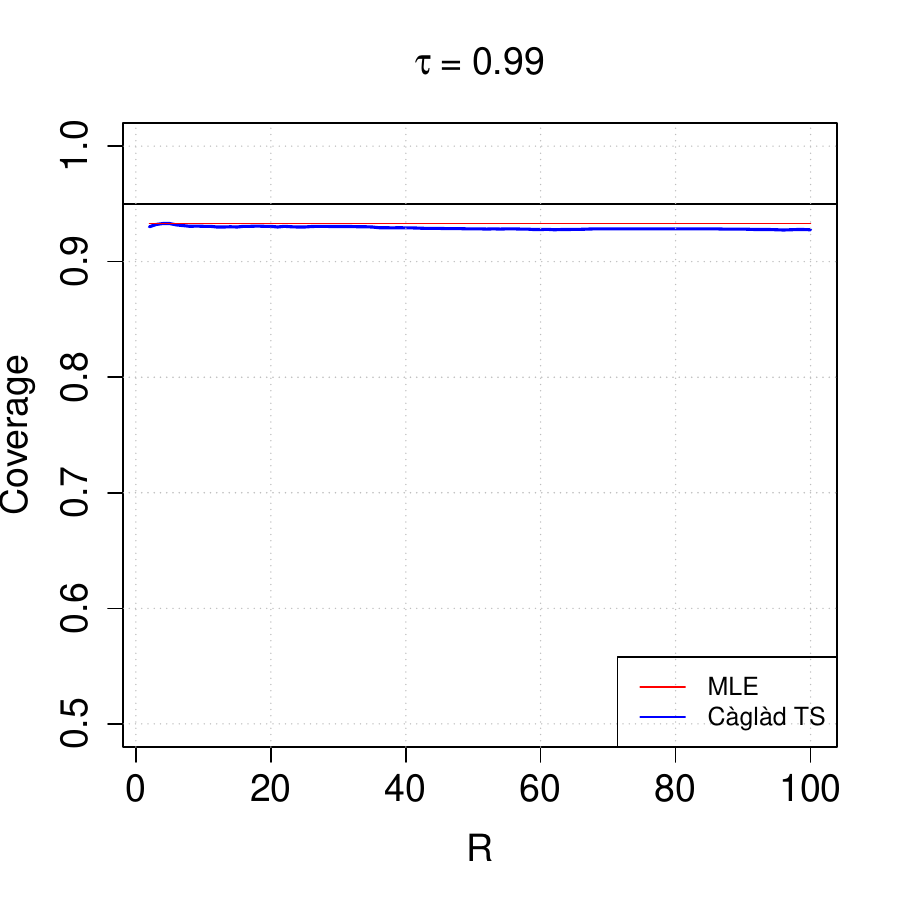}   \includegraphics[width=0.24\textwidth]{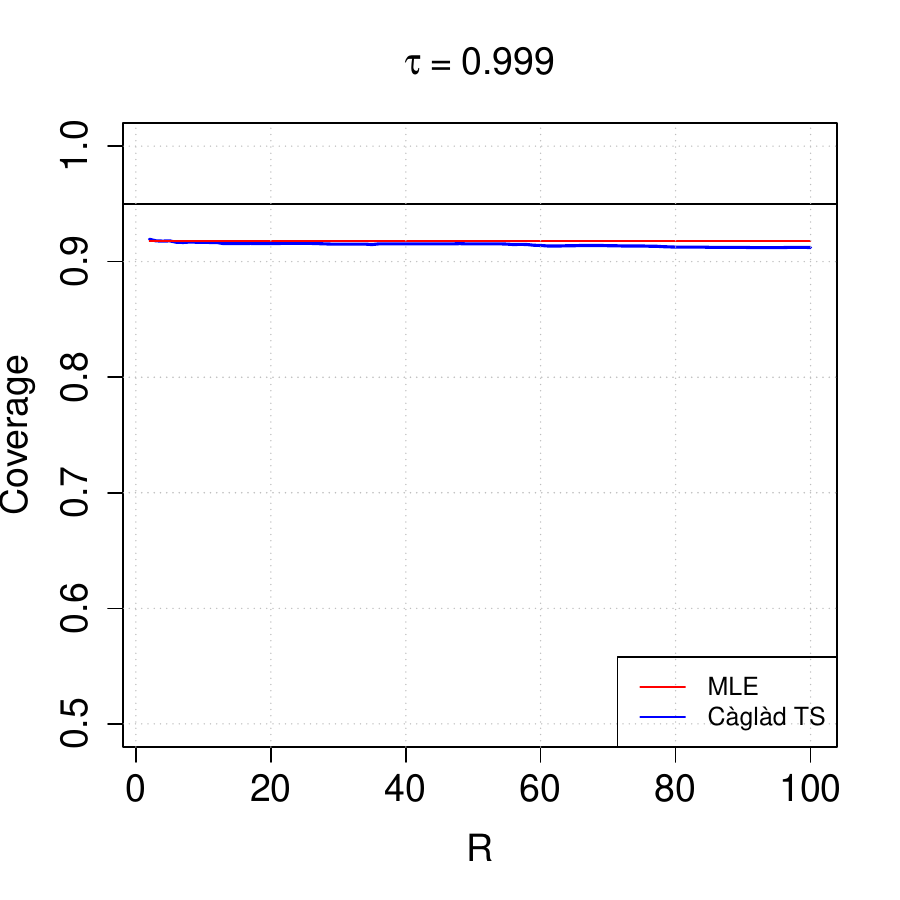}
		\caption{$T=500$}
	\end{subfigure}
	
	\caption{GPD: coverage of confidence intervals that rely on an estimator of the asymptotic variance.}
	\label{fig:gpd_coverage_est}
\end{figure}

\begin{figure}[H]
	\centering

	\begin{subfigure}[H]{\textwidth}

		\centering
		\includegraphics[width=0.24\textwidth]{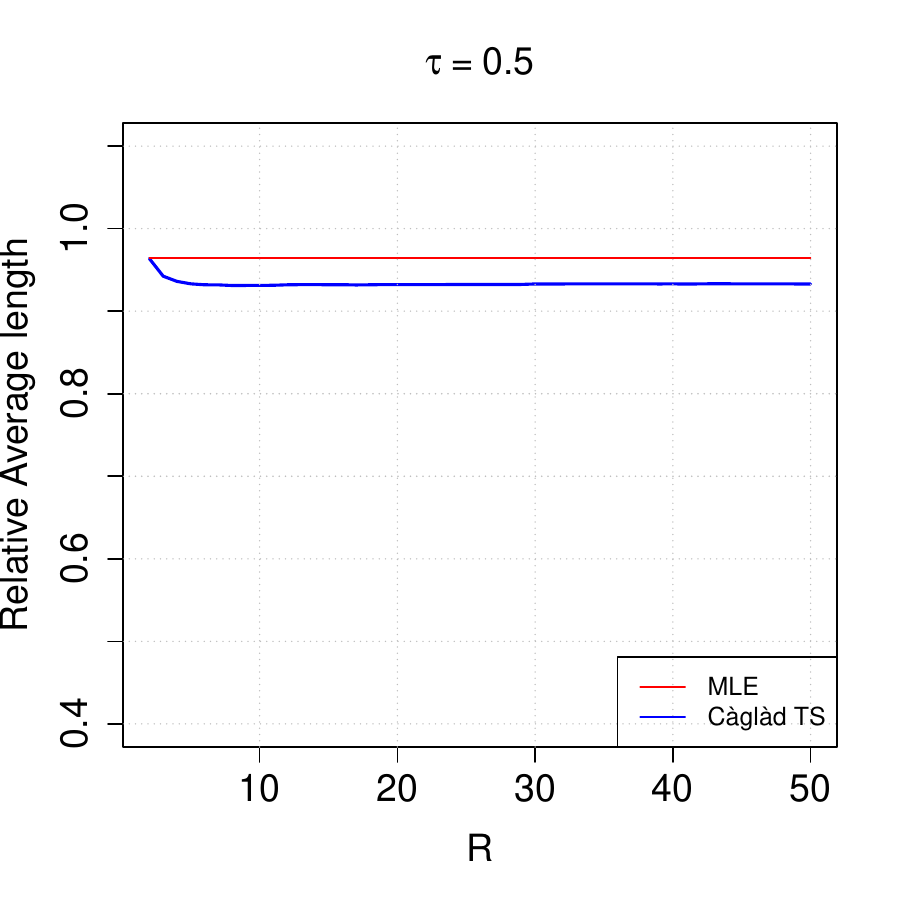}   \includegraphics[width=0.24\textwidth]{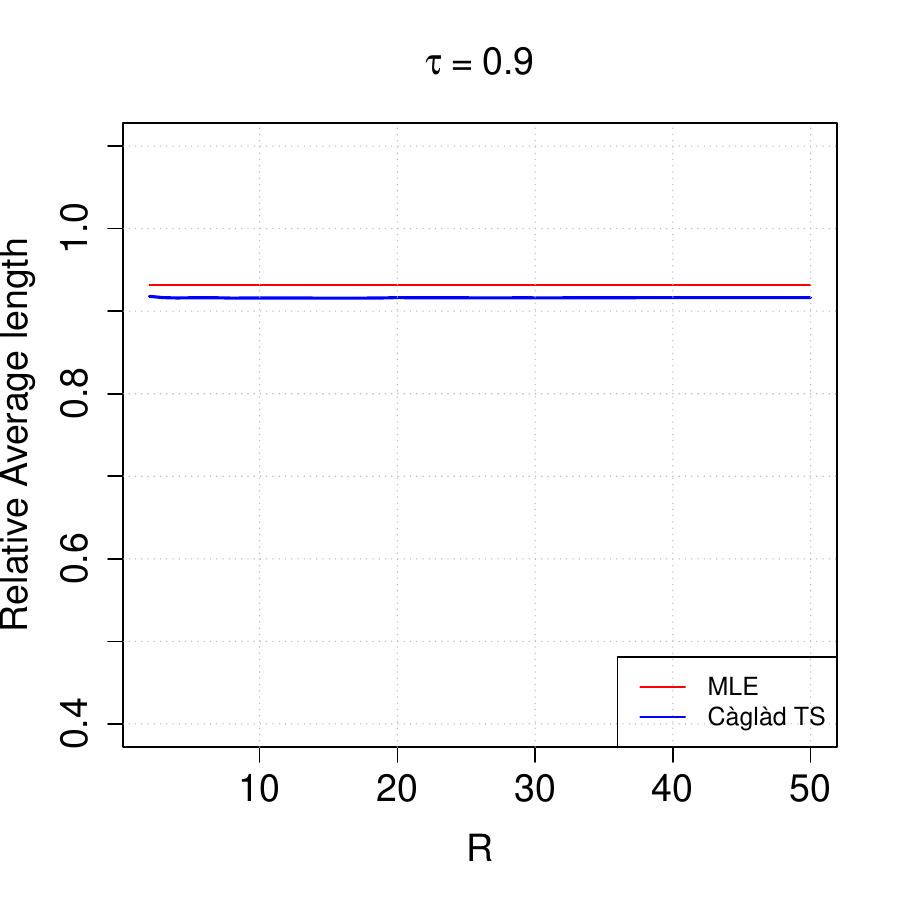}
		\includegraphics[width=0.24\textwidth]{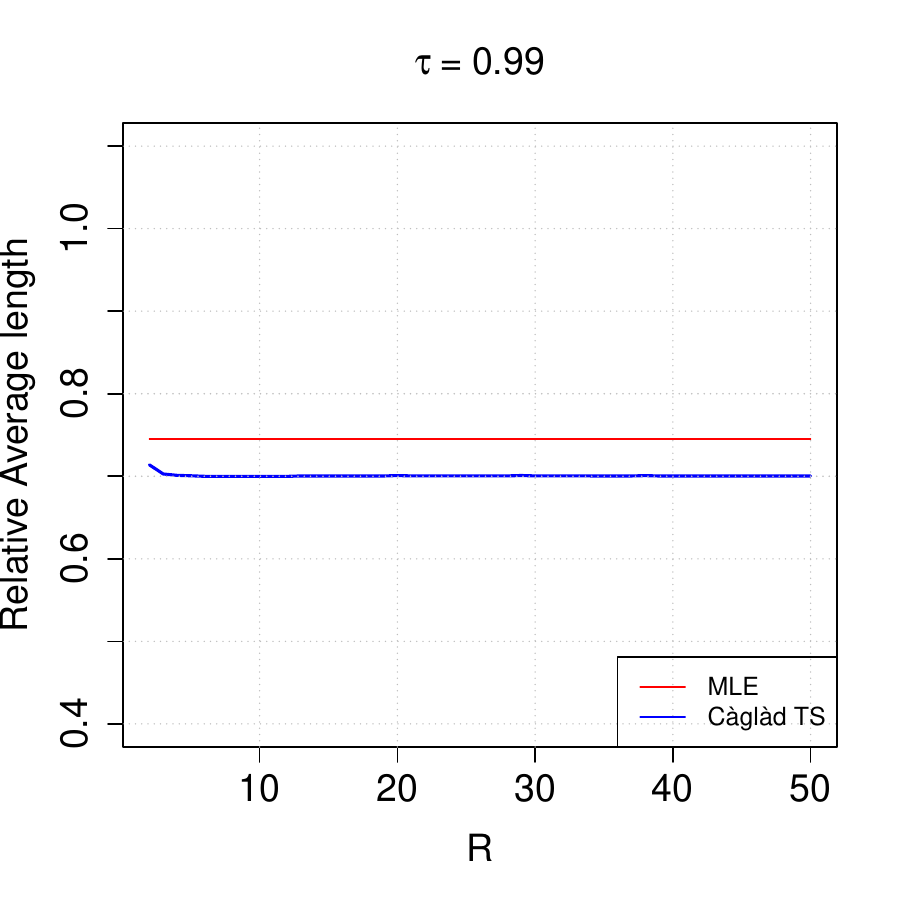}   \includegraphics[width=0.24\textwidth]{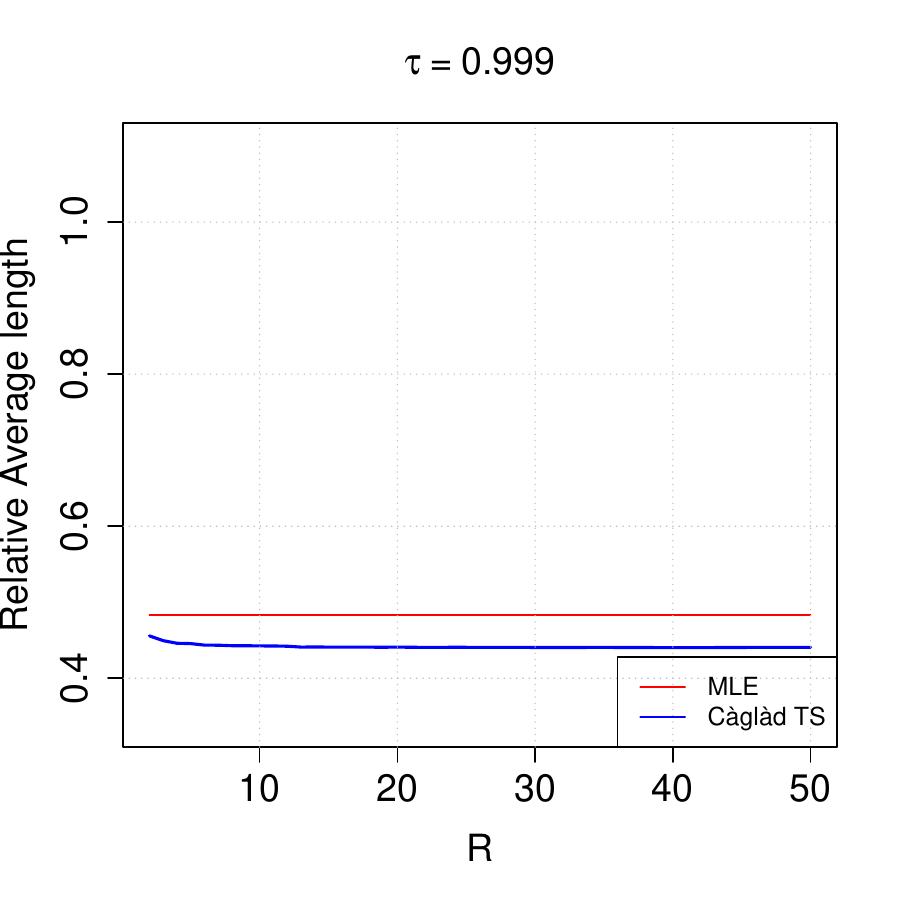}
		\caption{$T=50$}
	\end{subfigure}

	\begin{subfigure}[H]{\textwidth}

		\centering
		\includegraphics[width=0.24\textwidth]{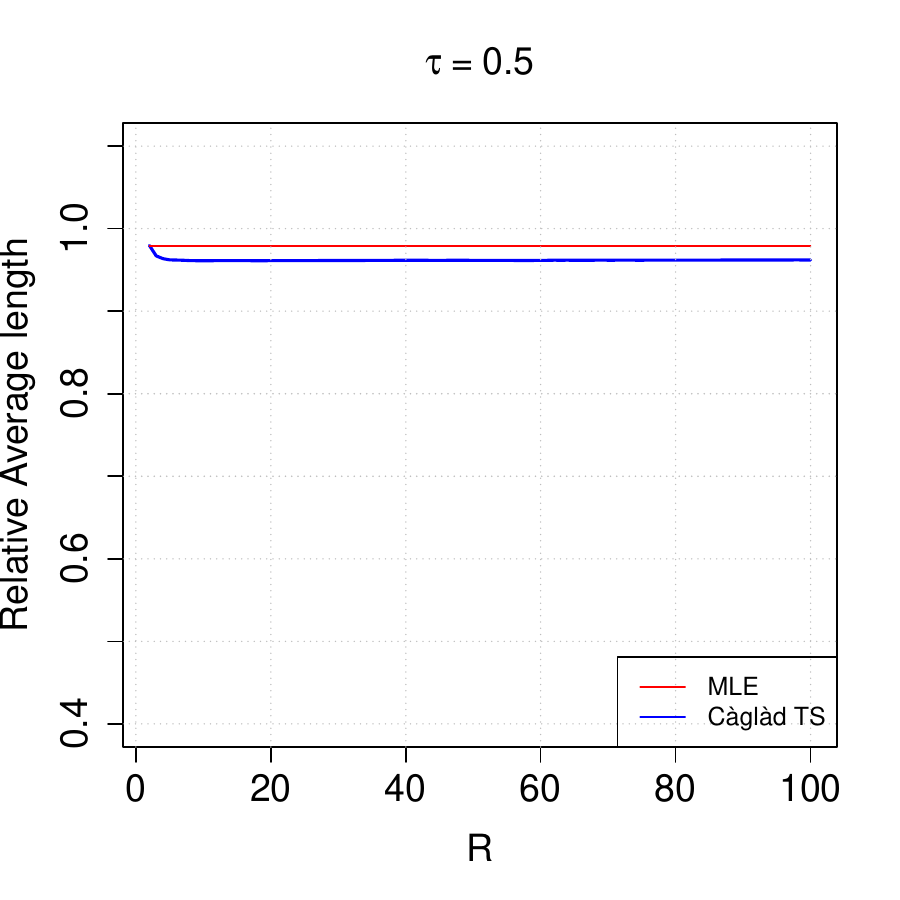}   \includegraphics[width=0.24\textwidth]{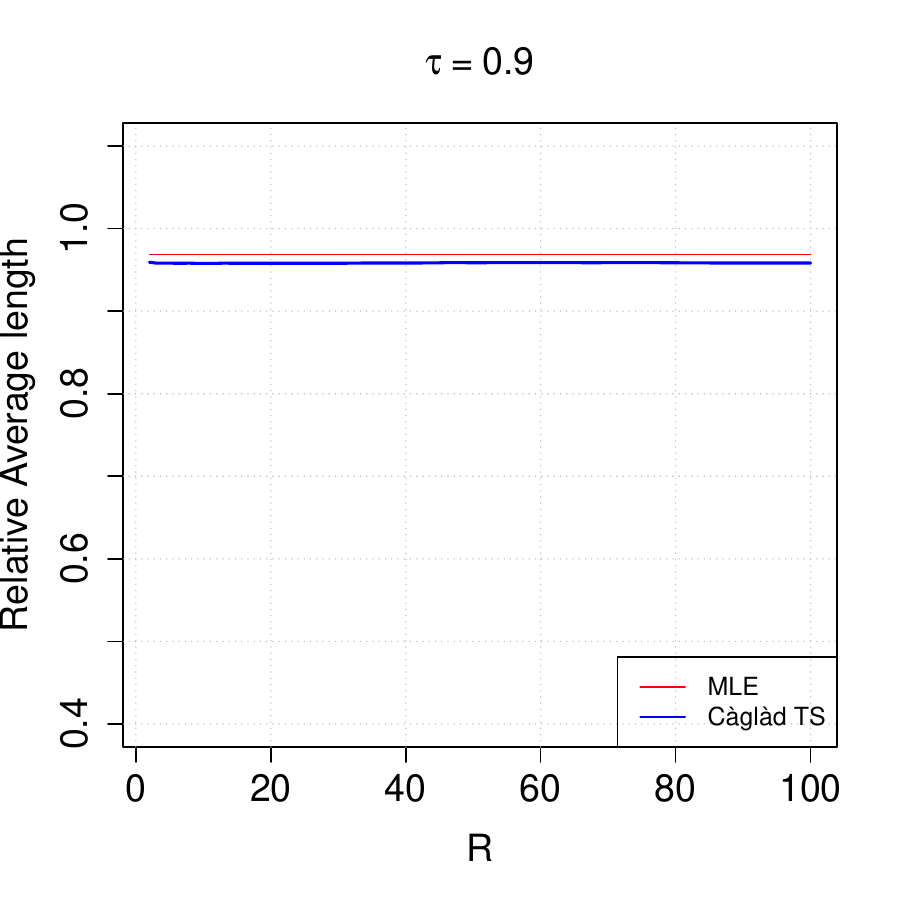}
		\includegraphics[width=0.24\textwidth]{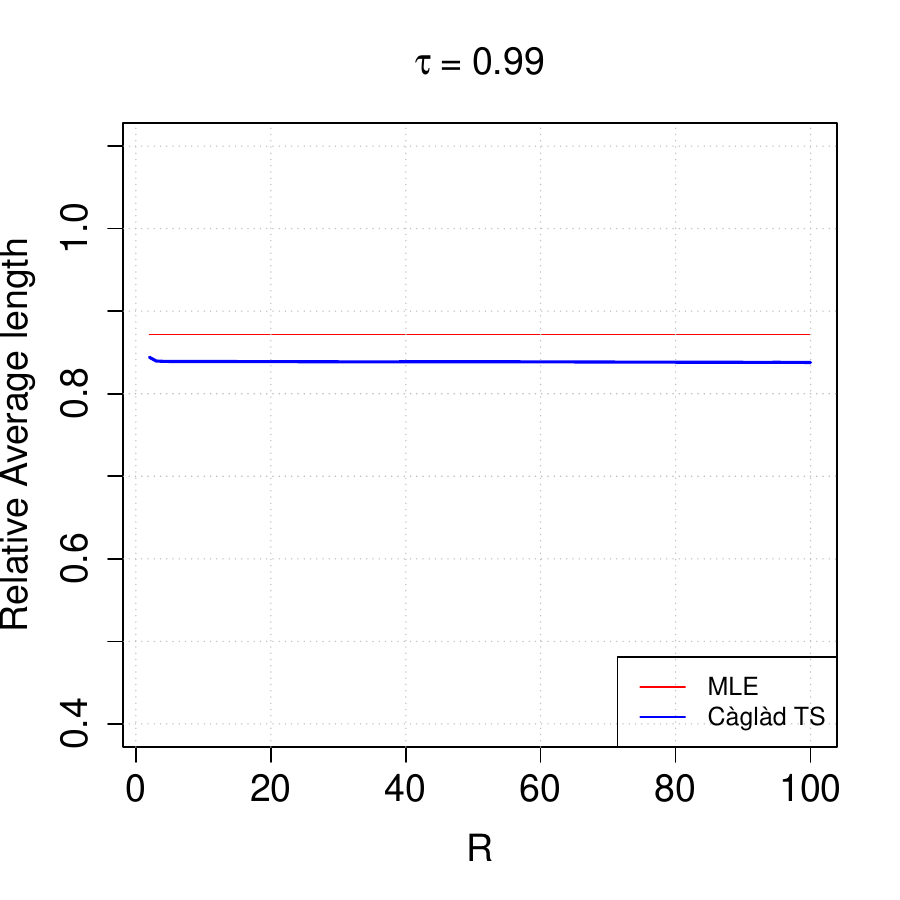}   \includegraphics[width=0.24\textwidth]{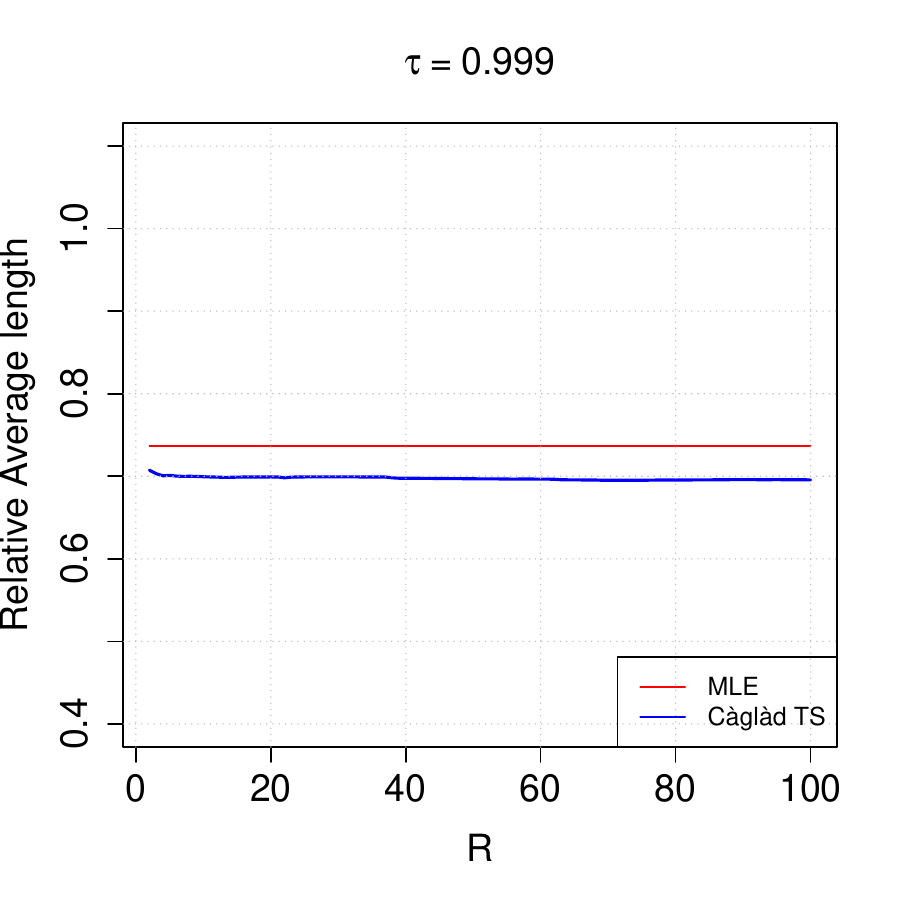}
		\caption{$T=100$}
	\end{subfigure}

	\begin{subfigure}[H]{\textwidth}

		\centering
		\includegraphics[width=0.24\textwidth]{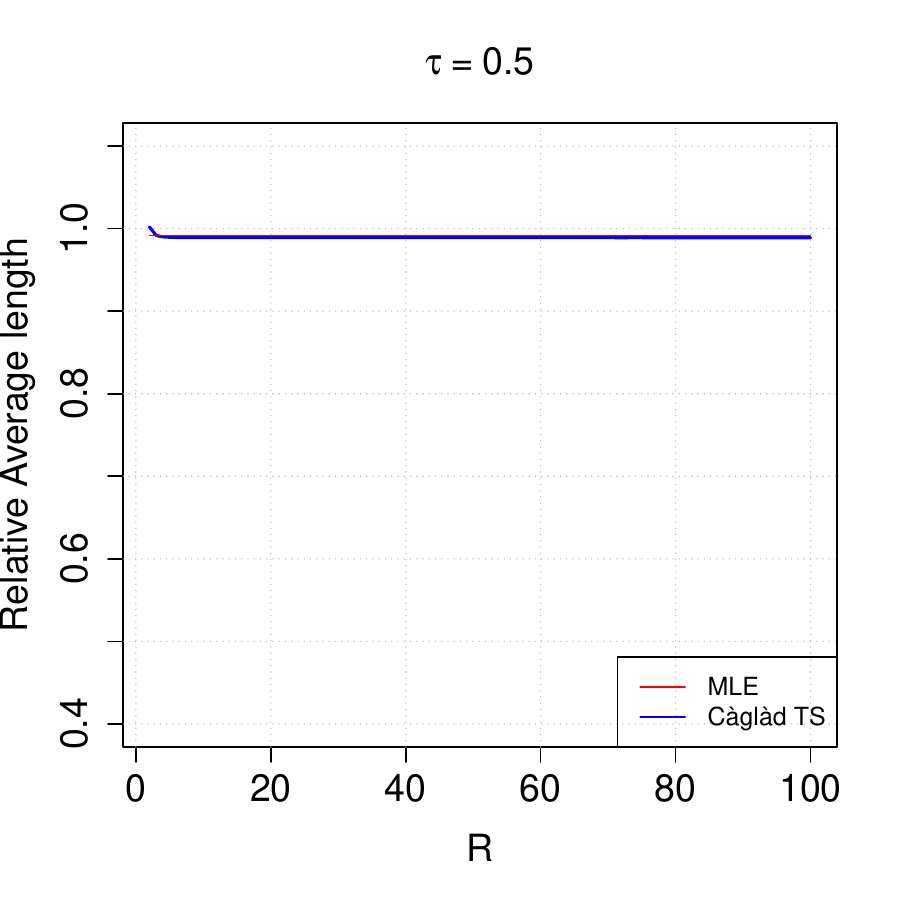}   \includegraphics[width=0.24\textwidth]{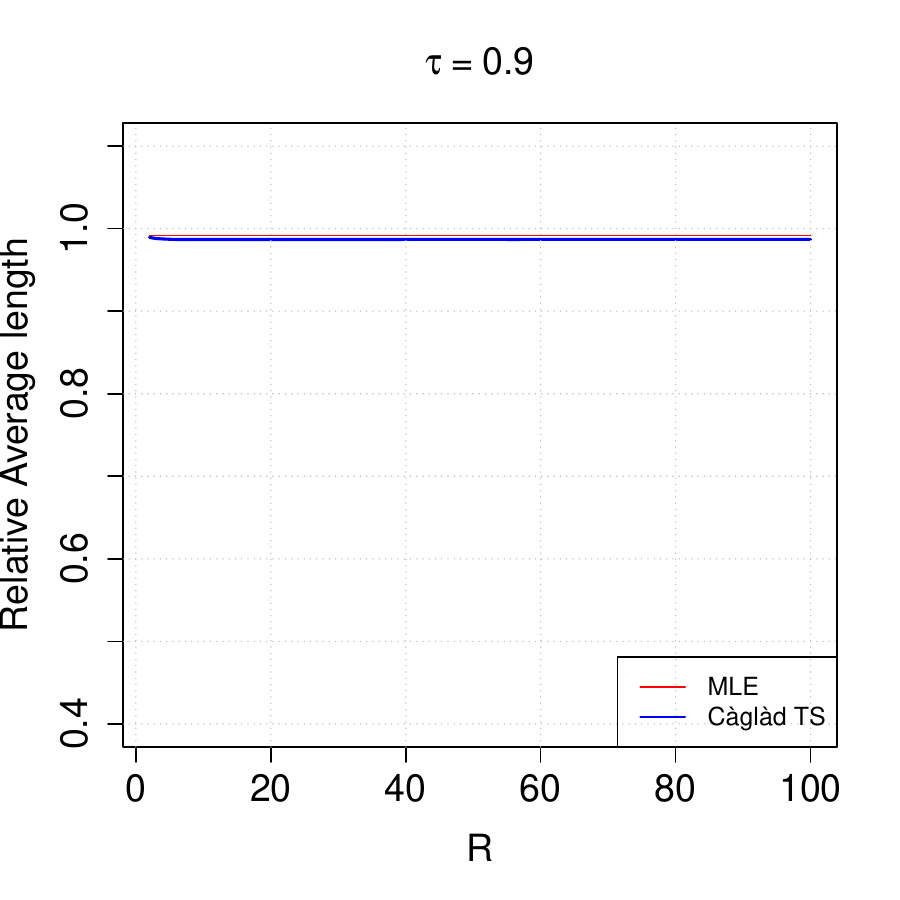}
		\includegraphics[width=0.24\textwidth]{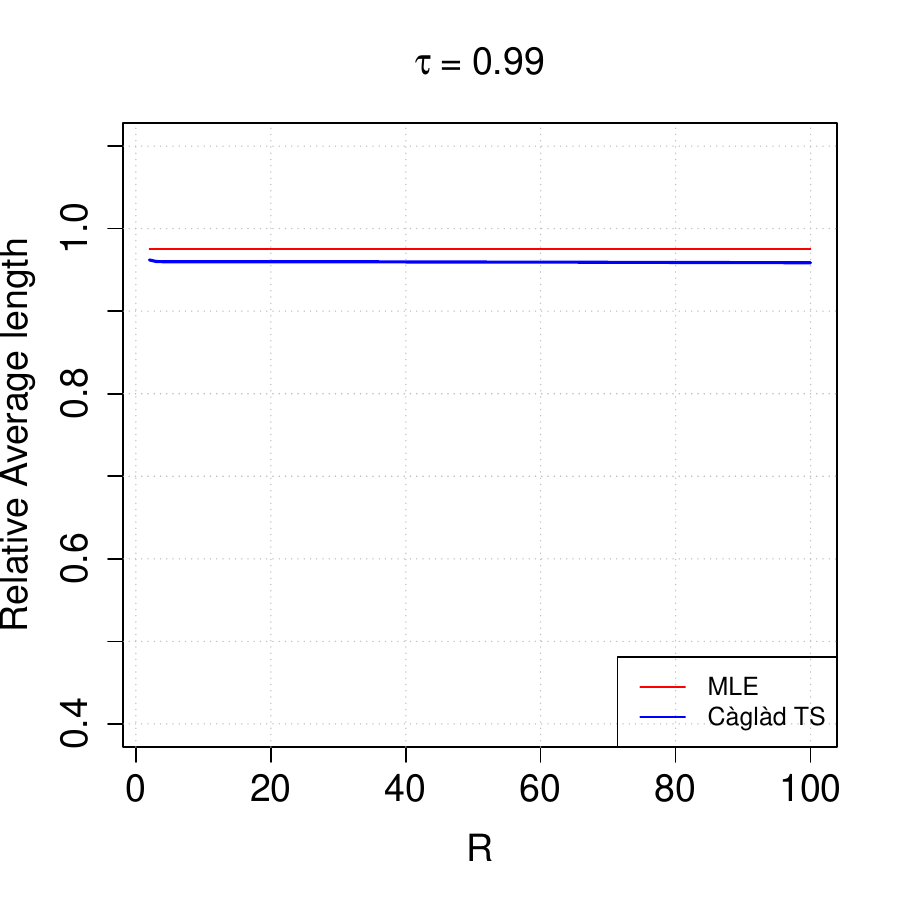}   \includegraphics[width=0.24\textwidth]{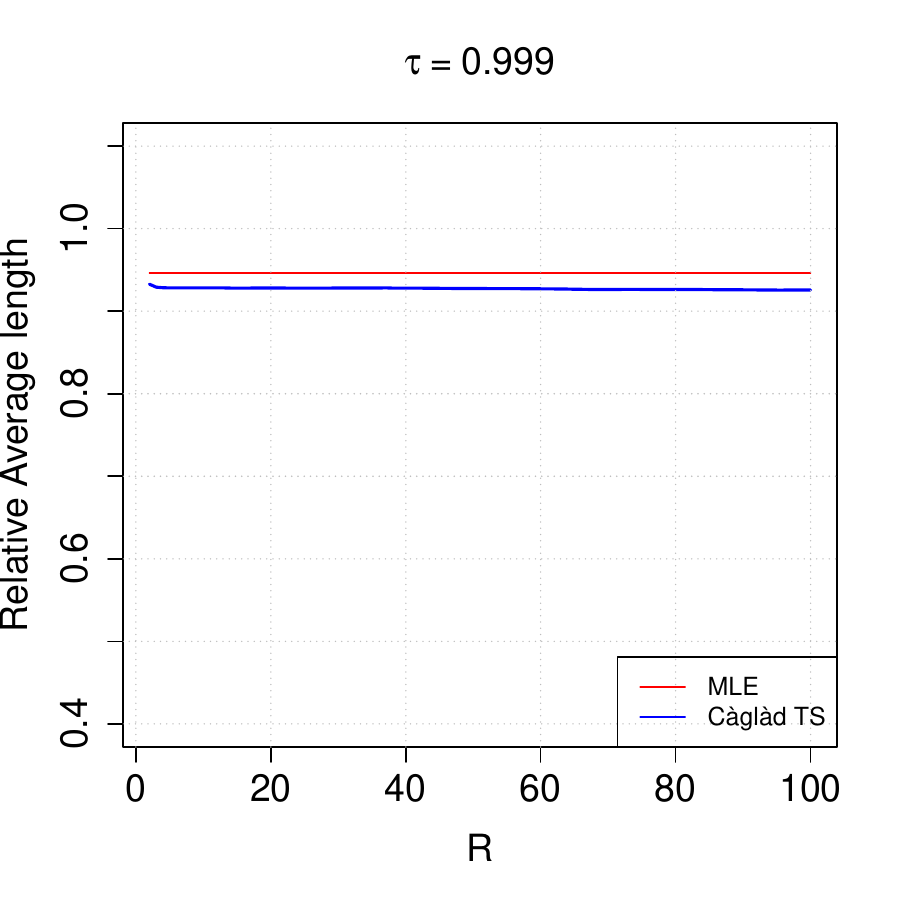}
		\caption{$T=500$}
	\end{subfigure}
	
	\caption{GPD: relative length (vis-à-vis the unfeasible MLE-based CI) of confidence intervals that rely on an estimator of the asymptotic variance.}
	\label{fig:gpd_length_est}
\end{figure}

\begin{figure}[H]
	\centering

	\begin{subfigure}[H]{\textwidth}

		\centering
		\includegraphics[width=0.32\textwidth]{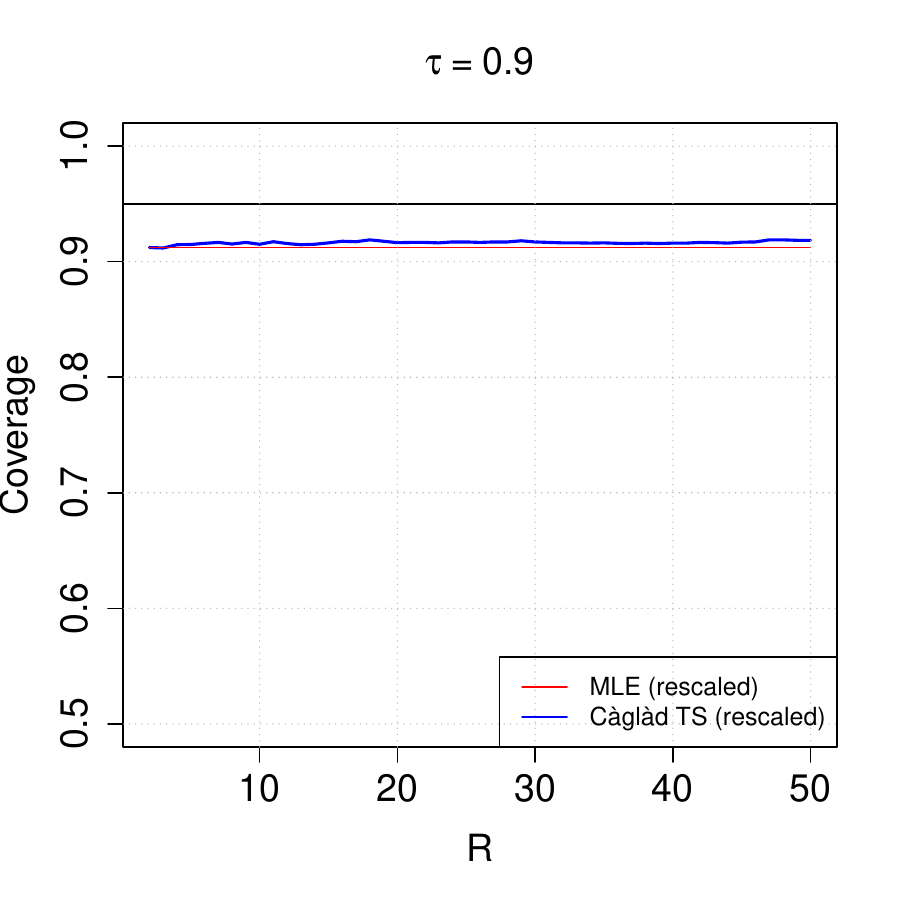}
		\includegraphics[width=0.32\textwidth]{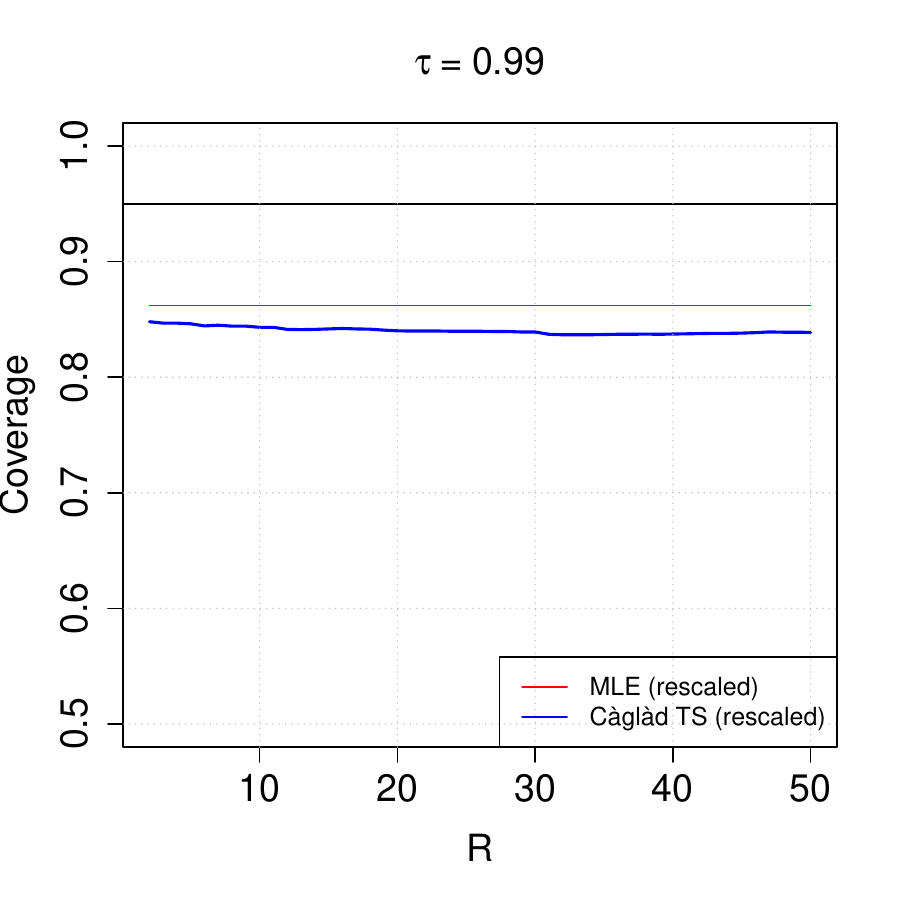}   \includegraphics[width=0.32\textwidth]{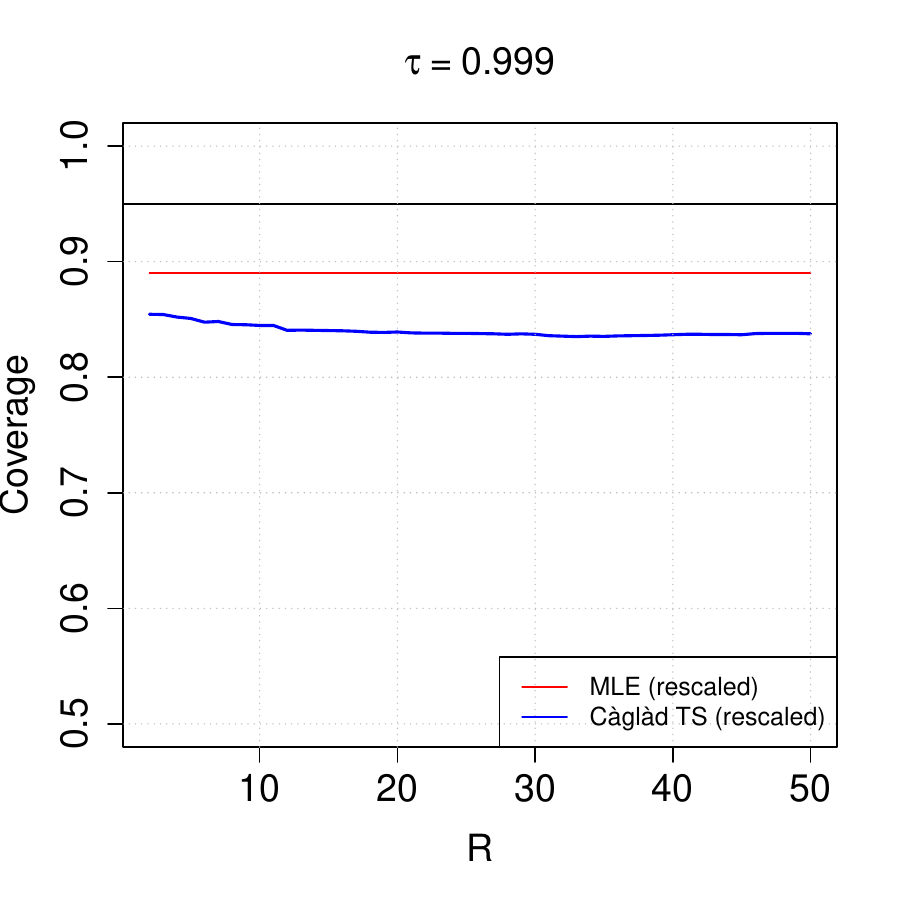}
		\caption{$T=50$}
	\end{subfigure}

	\begin{subfigure}[H]{\textwidth}

		\centering
		\includegraphics[width=0.32\textwidth]{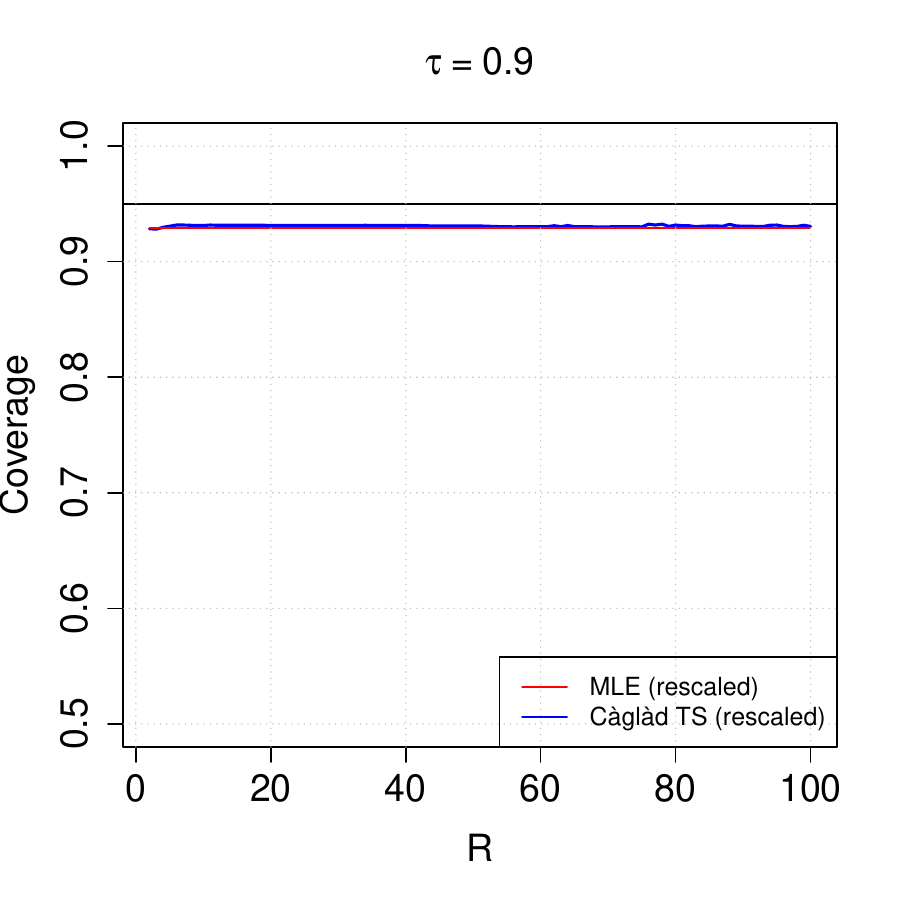}
		\includegraphics[width=0.32\textwidth]{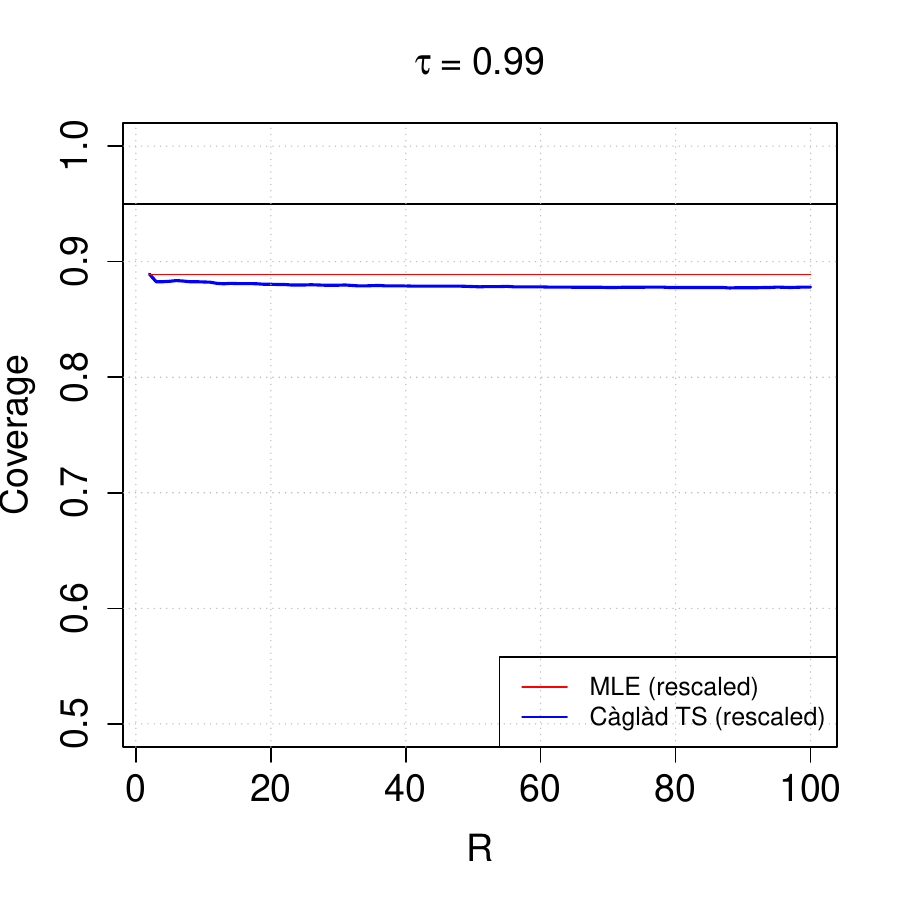}   \includegraphics[width=0.32\textwidth]{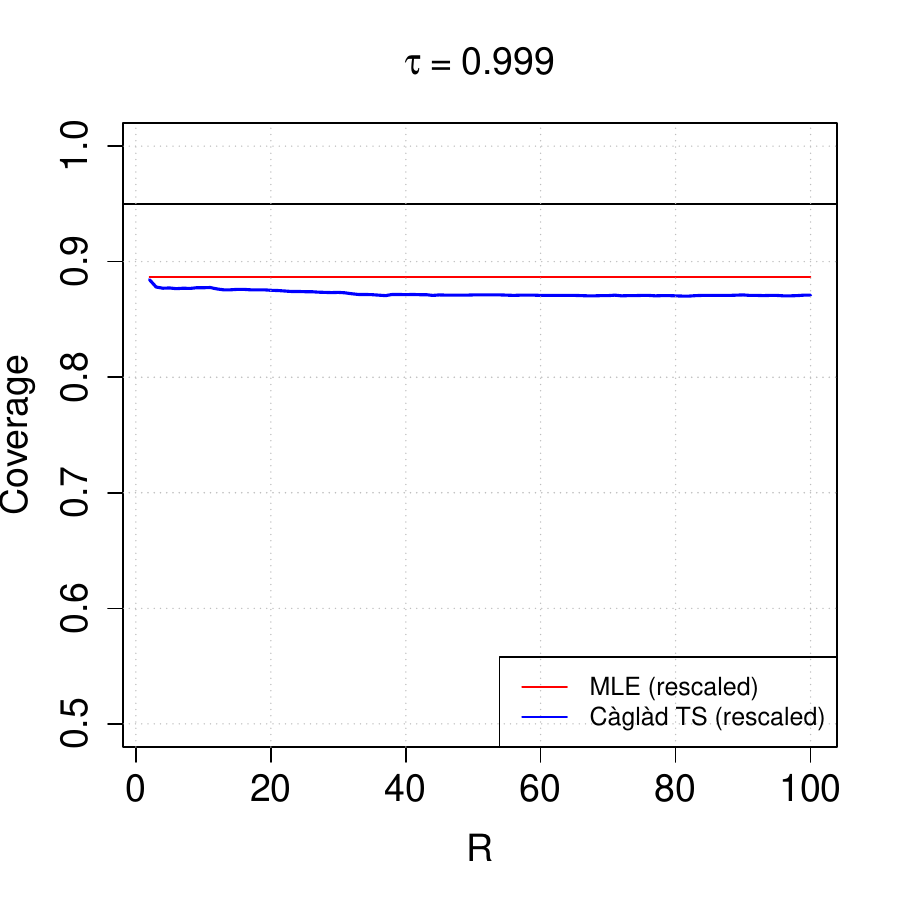}
		\caption{$T=100$}
	\end{subfigure}
	
	\caption{GPD: coverage of unfeasible confidence intervals that rely on rescaled estimators of the asymptotic variance.}
	\label{fig:gpd_coverage_rescale}
\end{figure}

\begin{figure}[H]
	\centering

	\begin{subfigure}[H]{\textwidth}

		\centering
		\includegraphics[width=0.32\textwidth]{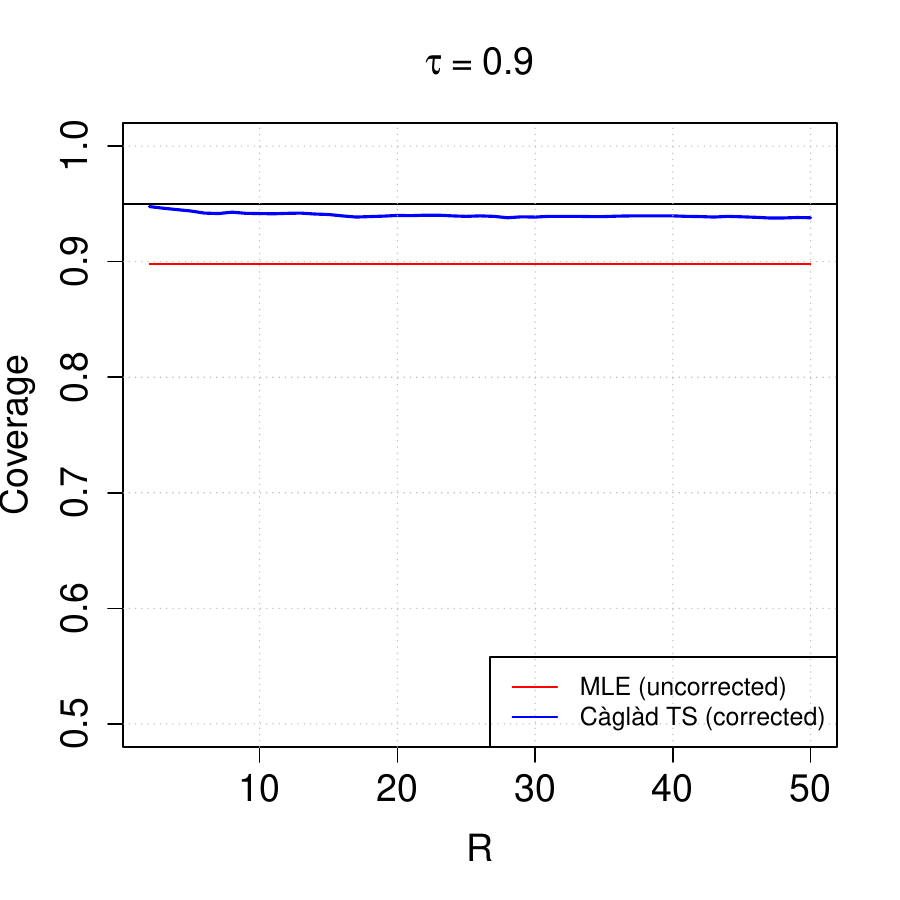}
		\includegraphics[width=0.32\textwidth]{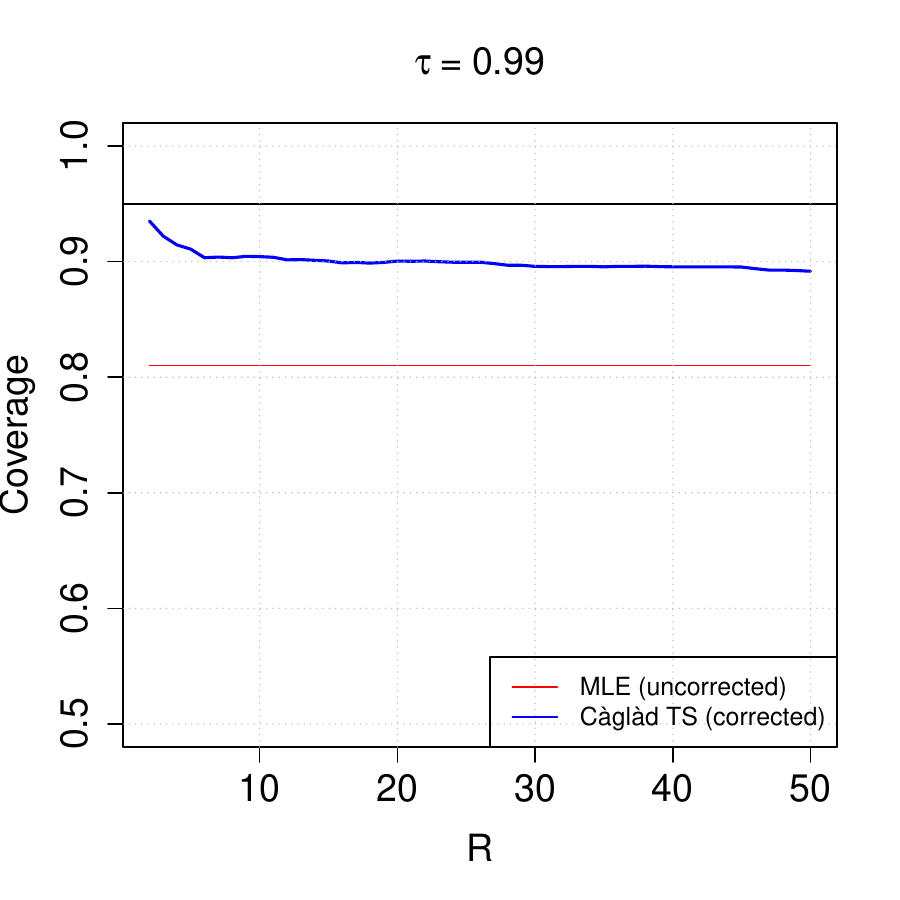}   \includegraphics[width=0.32\textwidth]{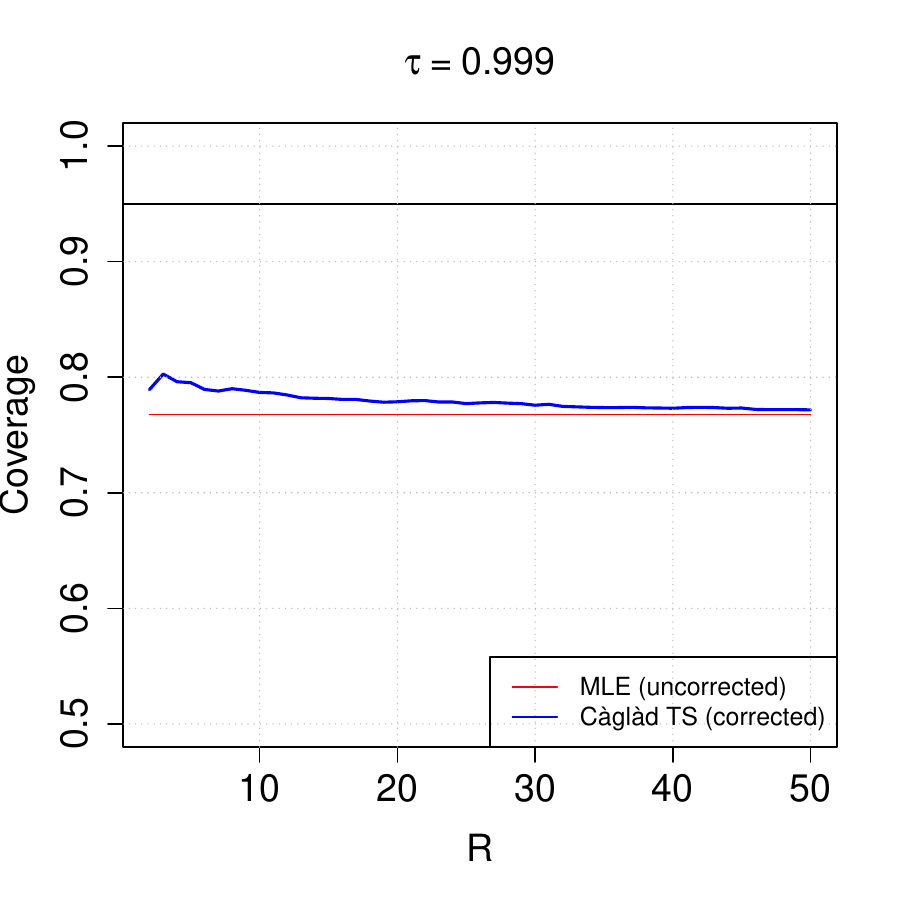}
		\caption{$T=50$}
	\end{subfigure}

	\begin{subfigure}[H]{\textwidth}

		\centering
		\includegraphics[width=0.32\textwidth]{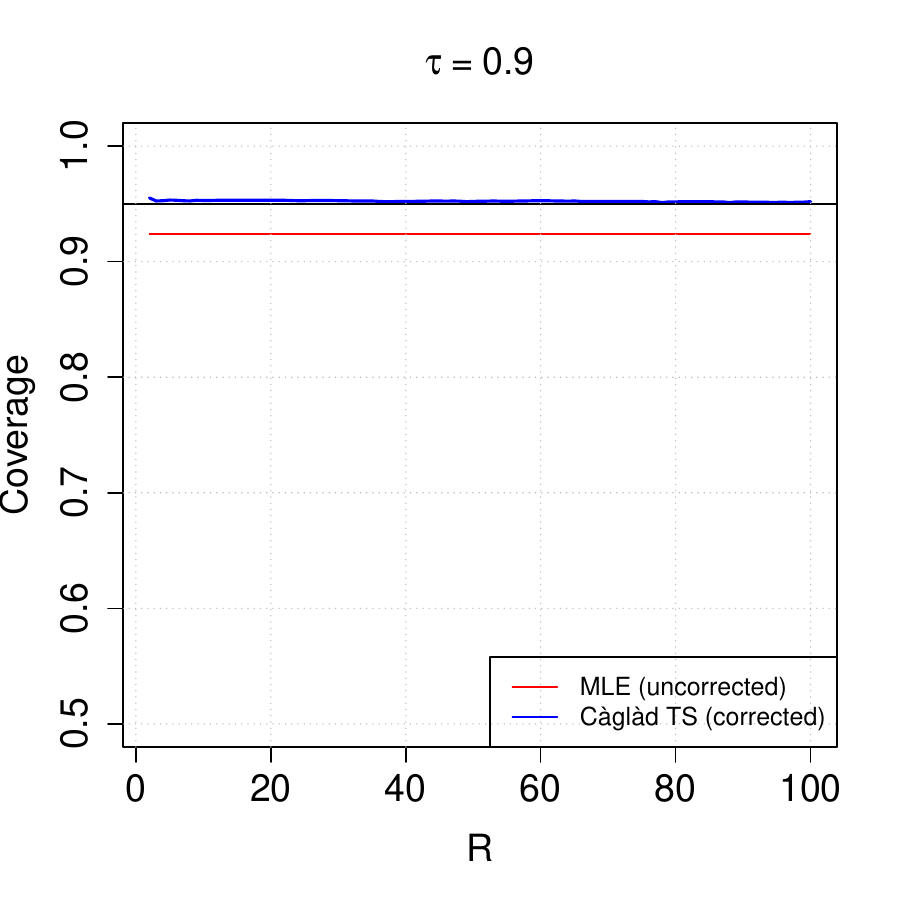}
		\includegraphics[width=0.32\textwidth]{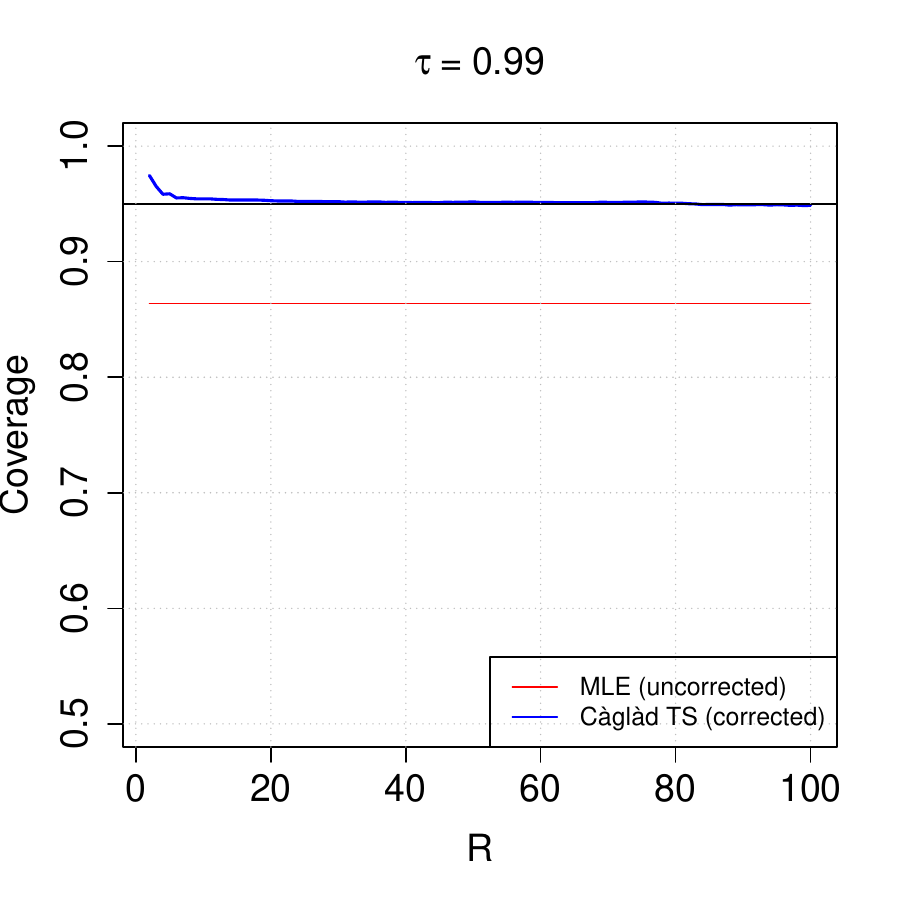}   \includegraphics[width=0.32\textwidth]{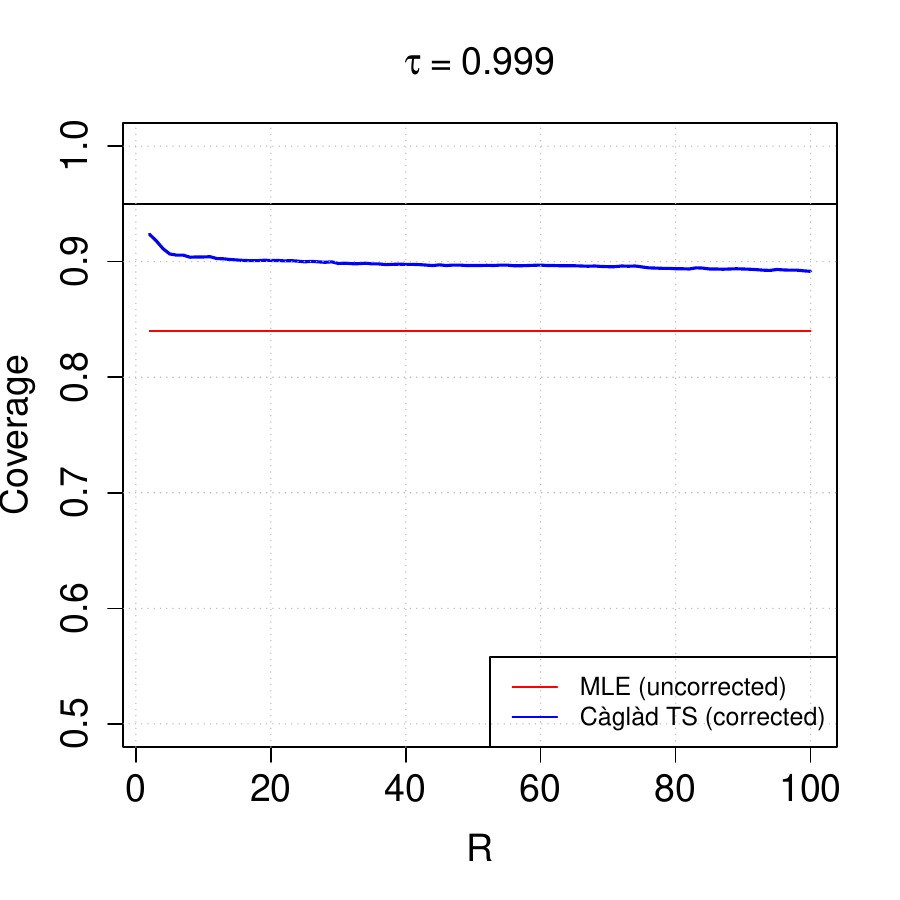}
		\caption{$T=100$}
	\end{subfigure}

	\caption{GPD: coverage of feasible confidence intervals that rely on corrected quantiles.}
	\label{fig:gpd_coverage_correct}
\end{figure}

\begin{figure}[H]
	\centering

	\begin{subfigure}[H]{\textwidth}

		\centering
		\includegraphics[width=0.32\textwidth]{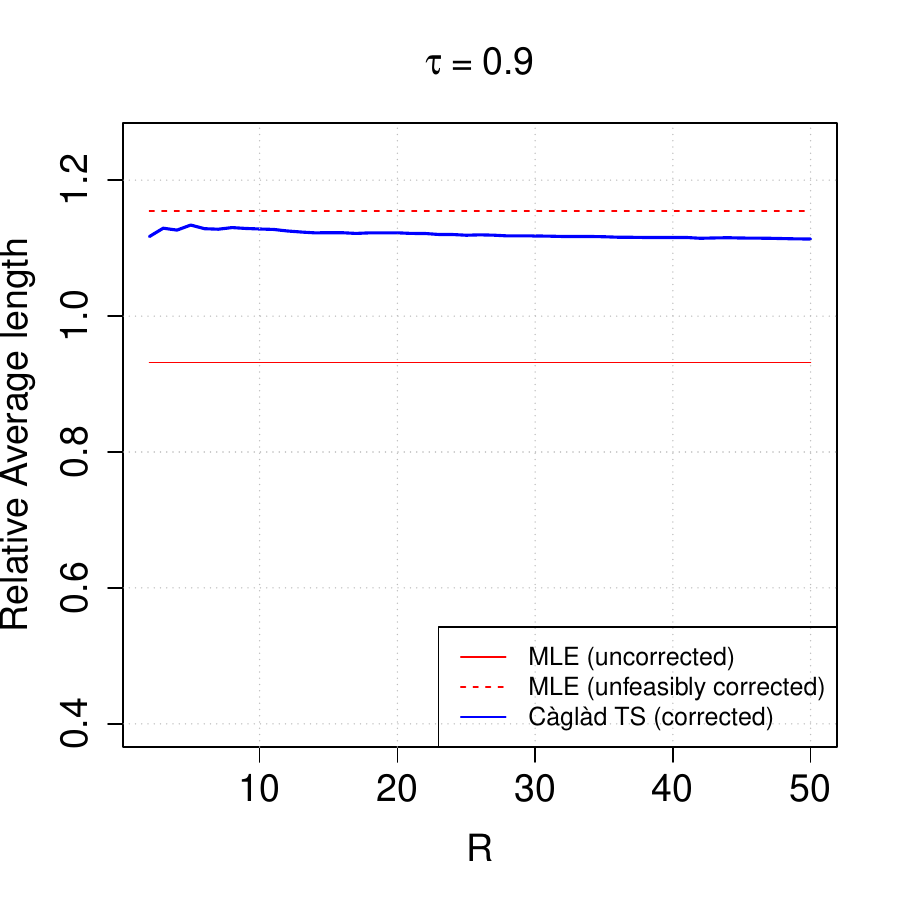}
		\includegraphics[width=0.32\textwidth]{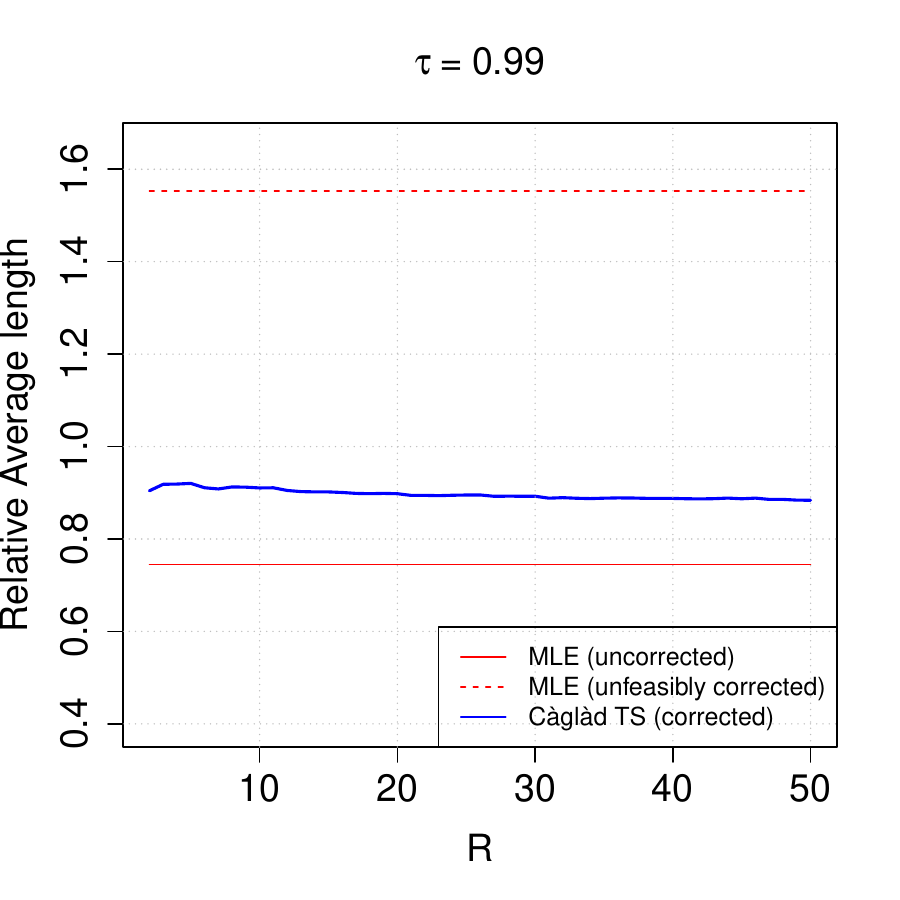}   \includegraphics[width=0.32\textwidth]{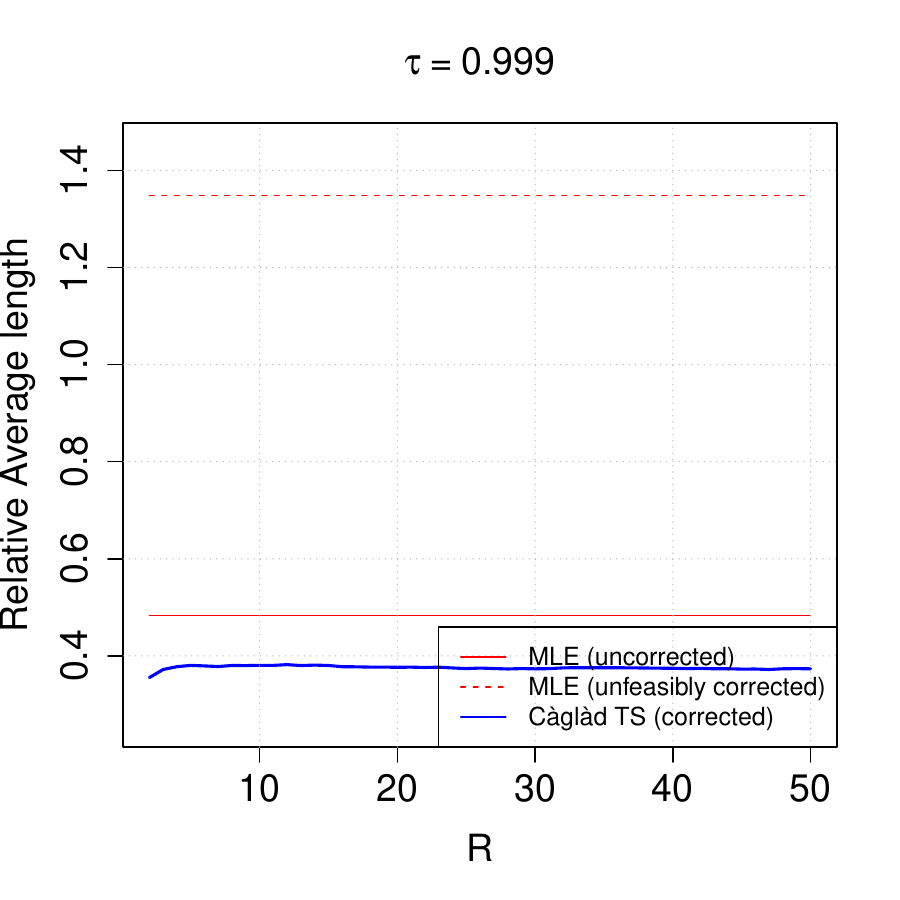}
		\caption{$T=50$}
	\end{subfigure}

	\begin{subfigure}[H]{\textwidth}

		\centering
		\includegraphics[width=0.32\textwidth]{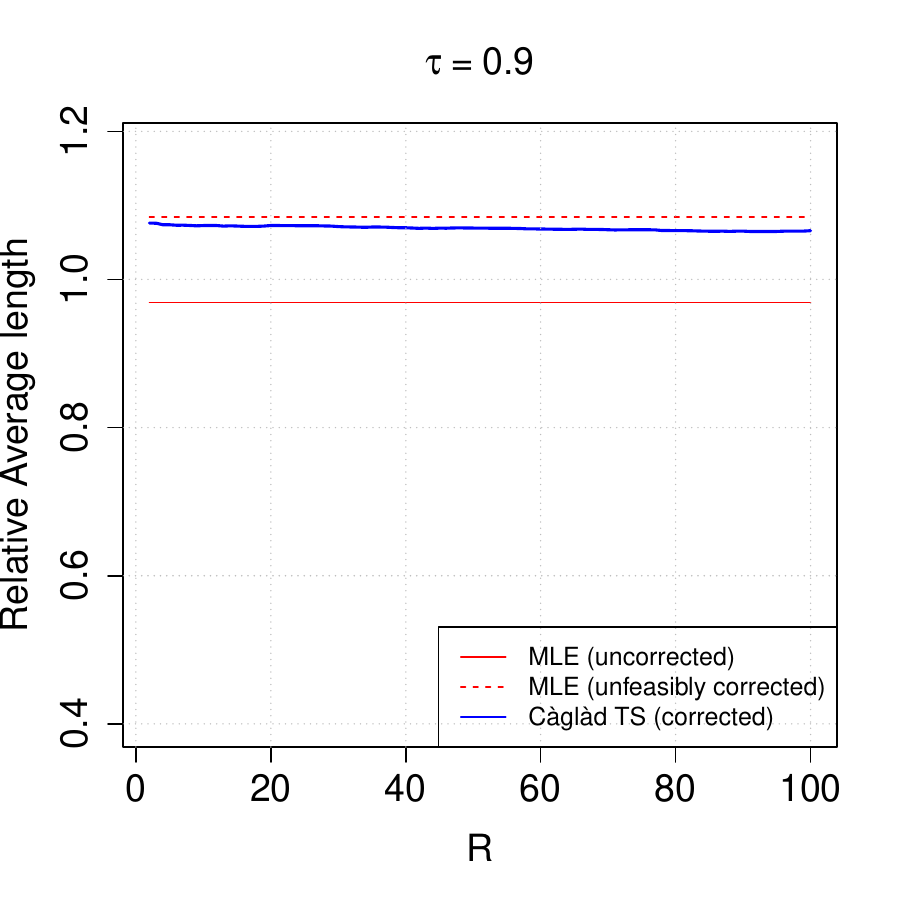}
		\includegraphics[width=0.32\textwidth]{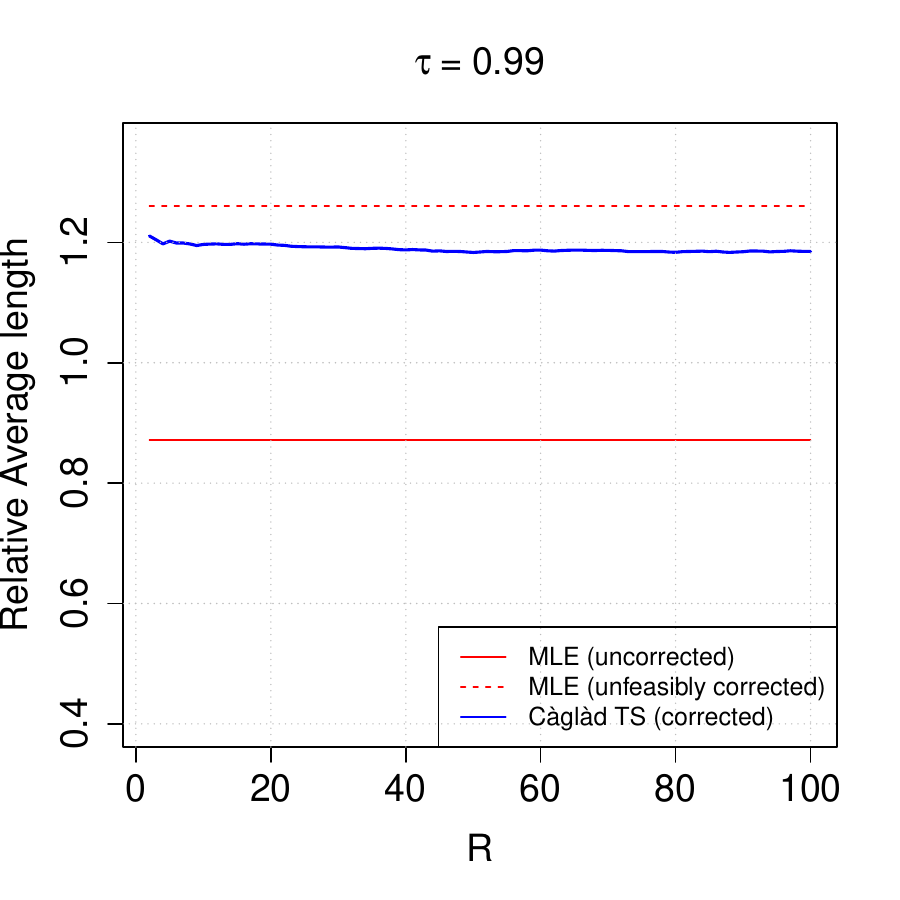}   \includegraphics[width=0.32\textwidth]{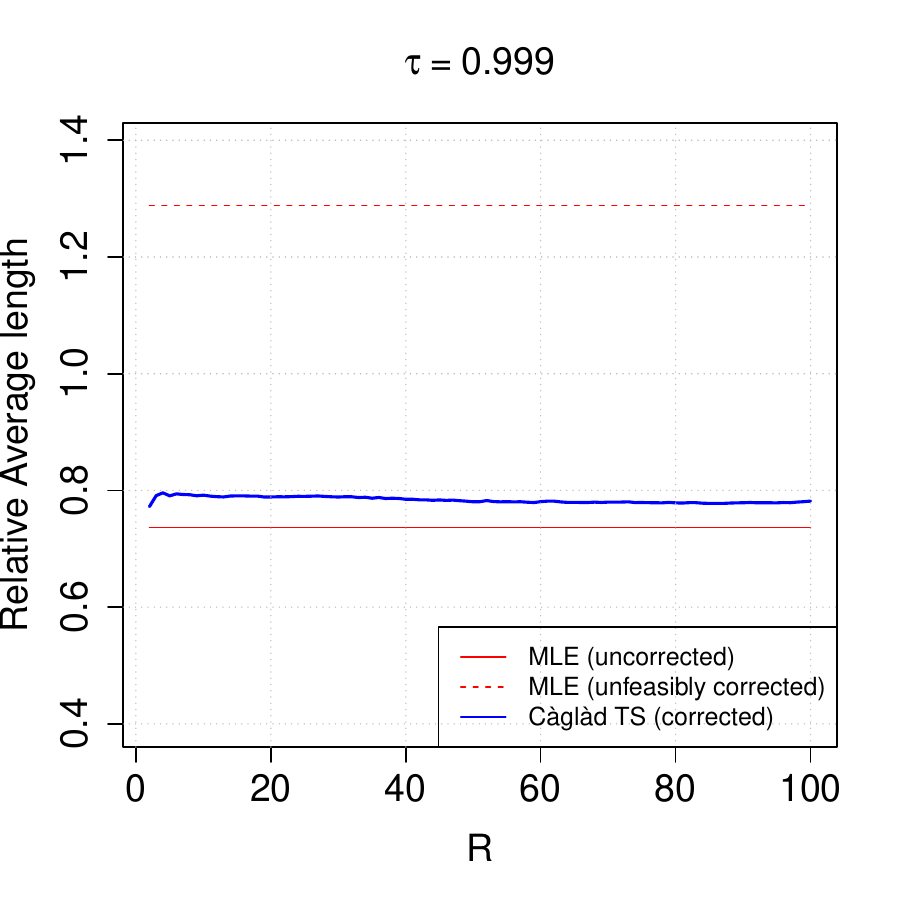}
		\caption{$T=100$}
	\end{subfigure}

	\caption{GPD: relative length (vis-à-vis the unfeasible MLE-based CI that relies on the true sampling variance) of feasible confidence intervals that rely on corrected quantiles.}
	\label{fig:gpd_length_correct}
\end{figure}

%% file: tables/rmse_algorithm.tex
    \begin{algorithm}[H]
	\caption{Pseudo-code for higher-order mean-squared error calculation}\label{alg:rmse}
	\begin{algorithmic}[1]
		\small
		
		\Require Maximum number of $L$-moments $\overline{R}$, number of bootstrap simulations $B$. Target quantile $\tau$.
		\Ensure Higher-order MSE estimates of quantile estimators based on the Càglád two-step estimators $\hat{\theta}_R$, $R\in \{1,\ldots \bar{R}\}$, and a MSE-minimizing choice $R^*$.
		\State Estimate $\theta_0$ using the Càglád estimator with $R=d$, denoted by $\tilde{\theta}$.
		\State Compute and store the derivatives pertaining to expansion of the target quantile: $\nabla_{\theta}Q(\tau|\tilde{\theta})$, $\nabla_{\theta\theta'}Q(\tau|\tilde{\theta})$, 
		
		\State Compute and store the derivatives pertaining to estimation of the Jacobian of the objective function: $\{\int_0^1 \nabla_\theta Q(u|\tilde{\theta}) P_r(u)du, \int_0^1 \nabla_{\theta\theta'} Q(u|\tilde{\theta}) P_r(u)du \}$, $r=1,\ldots, \overline{R}$. 
		\State Compute and store the evaluation of the following functions and its partial derivatives, with respect to $\theta$ at $\tilde{\theta}$:	 $\int_0^1 \int_0^1 \frac{(U\land V - UV)}{f_{\theta}(Q(U|\theta))f_{\theta}(Q(V|\theta))} U^r V^s dU dV$, for $(r,s)\in \{1,\ldots,R\}^2$. These quantities pertain to estimation of the optimal weights.
		\State Compute and store the vector of L-moments $\hat{L} \gets Q(u|\tilde{\theta})\boldsymbol{P}_{\overline{R}}(u)du$.
		\For{b=1 to B} 
		\State Generate a sample of size $T$ from $F_{\tilde{\theta}}$. Denote it by $\mathcal{Z}_b$.
			\State Calculate the empirical quantiles of $\mathcal{Z}_b$, $\hat{Q}_b$, and compute and store $\tilde{h}_{b} \gets \sqrt{T}(\int_0^1\hat{Q}_b(u)\boldsymbol{P}_{\overline{R}}(u)du-\hat{L})$.
			\EndFor
		\For{R=1 to $\overline{R}$}
					\For{b=1 to $\overline{B}$}
					\State Compute the quantities $\Tilde{\Theta}^1_{T}$ and  $\Tilde{\Theta}^2_{T}$ of the higher-order expansion \eqref{eq_higher_taylor} using the first $R$ entries of $\tilde{h}_{b}$ and the derivatives and estimated optimal weights of the first $R$ L-moments. Store these quantities as  $\Tilde{\Theta}^1_{R,b}$ and $\Tilde{\Theta}^2_{R,b}$.
					\EndFor
					\State Estimate and store the MSE of $Q(\tau|\hat{\theta}_R)$ as: $$\widehat{\operatorname{MSE}(Q(\tau|\hat{\theta}_R))} \gets \frac{1}{B}\sum_{b=1}^B\Big(\nabla_{\theta'}Q(\tau|\tilde{\theta}) ( \Tilde{\Theta}^1_{R,b} +T^{-\frac{1}{2}}\tilde{\Theta}^2_{R,b})+ ( \Tilde{\Theta}^1_{R,b} +T^{-\frac{1}{2}}\tilde{\Theta}^2_{R,b})' \nabla_{\theta \theta'} Q(\tau|\tilde{\theta})( \tilde{\Theta}^1_{R,b} +T^{-\frac{1}{2}}\tilde{\Theta}^2_{R,b})\Big)^2 $$
					\EndFor
					\State Set $R^* \gets \operatorname{argmin}_{R \in \{1,\ldots, \overline{R}\}}\widehat{\operatorname{MSE}(Q(\tau|\hat{\theta}_R))}$.
	\end{algorithmic}
\end{algorithm}

%% file: tables/gev/gev_table_select.tex
\begin{table}[H]
\centering
\caption{GEV : relative RMSE under different selection procedures} 
\label{gev_table_select}
\adjustbox{width=\textwidth}{
\begin{tabular}{|l|cccc|cccc|cccc|}
  \hline
  & \multicolumn{ 4 } {c|}{$T = 50 $}&\multicolumn{ 4 } {c|}{$T = 100 $}&\multicolumn{ 4 } {c|}{$T = 500 $} \\ & $\tau = 0.5 $ & $\tau = 0.9 $ & $\tau = 0.99 $ & $\tau = 0.999 $ & $\tau = 0.5 $ & $\tau = 0.9 $ & $\tau = 0.99 $ & $\tau = 0.999 $ & $\tau = 0.5 $ & $\tau = 0.9 $ & $\tau = 0.99 $ & $\tau = 0.999 $ \\ 
   \hline
FS &  1.026 &  0.962 &  0.821 &  0.737 &  1.031 &  0.983 &  0.950 &  0.928 &  1.028 &  1.004 &  1.061 &  1.095 \\ 
   &  (3) &  (3) &  (3) &  (3) &  (3) &  (3) &  (3) &  (3) &  (3) &  (3) &  (3) &  (3) \\ 
   \hline
TS RMSE &  1.008 &  0.964 &  0.794 &  0.674 &  1.004 &  0.987 &  0.923 &  0.865 &  1.005 &  0.999 &  1.006 &  1.009 \\ 
   &  (16.81) &  (3.66) &  (3.3) &  (3.41) &  (33.02) &  (4.17) &  (4.32) &  (4.57) &  (20.29) &  (35.51) &  (40.91) &  (43.17) \\ 
   \hline
TS Lasso &  4.983 &  1.897 &  1.803 & $>10$ &  8.188 &  3.059 &  2.924 & $>10$ & $>10$ &  4.352 &  2.481 &  3.630 \\ 
   &  (7.99) &  (7.99) &  (7.99) &  (7.99) &  (9.24) &  (9.24) &  (9.24) &  (9.24) &  (9.95) &  (9.95) &  (9.95) &  (9.95) \\ 
   \hline
TS Post-Lasso &  1.017 &  0.975 &  0.857 &  0.781 &  1.010 &  0.988 &  0.928 &  0.866 &  1.006 &  0.999 &  0.999 &  0.993 \\ 
   &  (7.99) &  (7.99) &  (7.99) &  (7.99) &  (9.24) &  (9.24) &  (9.24) &  (9.24) &  (9.95) &  (9.95) &  (9.95) &  (9.95) \\ 
   \hline
\end{tabular}
}
\end{table}

%% file: tables/gpd/gpd_table_select.tex
\begin{table}[H]
\centering
\caption{GPD : relative RMSE under different selection procedures} 
\label{gpd_table_select}
\adjustbox{width=\textwidth}{
\begin{tabular}{|l|cccc|cccc|cccc|}
  \hline
  & \multicolumn{ 4 } {c|}{$T = 50 $}&\multicolumn{ 4 } {c|}{$T = 100 $}&\multicolumn{ 4 } {c|}{$T = 500 $} \\ & $\tau = 0.5 $ & $\tau = 0.9 $ & $\tau = 0.99 $ & $\tau = 0.999 $ & $\tau = 0.5 $ & $\tau = 0.9 $ & $\tau = 0.99 $ & $\tau = 0.999 $ & $\tau = 0.5 $ & $\tau = 0.9 $ & $\tau = 0.99 $ & $\tau = 0.999 $ \\ 
   \hline
FS &                                                   0.984 &                                                   0.981 &                                                   0.824 &                                                   0.648 &                                                    0.988 &                                                    0.993 &                                                    0.917 &                                                    0.856 &                                                  1.007 &                                                  1.005 &                                                  0.990 &                                                  0.982 \\ 
   &  (2) &  (2) &  (2) &  (2) &  (2) &  (2) &  (2) &  (2) &  (2) &  (2) &  (2) &  (2) \\ 
   \hline
TS RMSE &                                                   0.964 &                                                   0.994 &                                                   0.817 &                                                   0.640 &                                                    0.980 &                                                    0.990 &                                                    0.896 &                                                    0.828 &                                                  0.997 &                                                  0.999 &                                                  0.978 &                                                  0.970 \\ 
   &  (2.86) &  (4.43) &  (2.61) &  (2.96) &  (3.59) &  (4.31) &  (3.02) &  (3.09) &  (5.34) &  (48.42) &  (31.11) &  (29.71) \\ 
   \hline
TS Lasso & $>10$ & $>10$ & $>10$ & $>10$ & $>10$ & $>10$ & $>10$ & $>10$ & $>10$ & $>10$ & $>10$ & $>10$ \\ 
   &  (3.52) &  (3.52) &  (3.52) &  (3.52) &  (3.7) &  (3.7) &  (3.7) &  (3.7) &  (3.78) &  (3.78) &  (3.78) &  (3.78) \\ 
   \hline
TS Post-Lasso &                                                   0.995 &                                                   0.999 &                                                   0.891 &                                                   0.741 &                                                    0.992 &                                                    0.999 &                                                    0.950 &                                                    0.905 &                                                  0.998 &                                                  1.000 &                                                  0.985 &                                                  0.977 \\ 
   &  (3.52) &  (3.52) &  (3.52) &  (3.52) &  (3.7) &  (3.7) &  (3.7) &  (3.7) &  (3.78) &  (3.78) &  (3.78) &  (3.78) \\ 
   \hline
\end{tabular}
}
\end{table}

%% file: paper.bbl
\begin{thebibliography}{}

\bibitem[\protect\citeauthoryear{Abadie, Gu, and Shen}{Abadie
  et~al.}{2023}]{Abadie2019}
Abadie, A., J.~Gu, and S.~Shen (2023).
\newblock Instrumental variable estimation with first-stage heterogeneity.
\newblock {\em Journal of Econometrics\/}.

\bibitem[\protect\citeauthoryear{Ahidar-Coutrix and Berthet}{Ahidar-Coutrix and
  Berthet}{2016}]{Coutrix2016}
Ahidar-Coutrix, A. and P.~Berthet (2016).
\newblock Convergence of multivariate quantile surfaces.
\newblock \textit{arXiv:1607.02604}.

\bibitem[\protect\citeauthoryear{Ai and Chen}{Ai and Chen}{2003}]{Ai2003}
Ai, C. and X.~Chen (2003).
\newblock Efficient estimation of models with conditional moment restrictions
  containing unknown functions.
\newblock {\em Econometrica\/}~{\em 71\/}(6), 1795--1843.

\bibitem[\protect\citeauthoryear{Alvarez and Biderman}{Alvarez and
  Biderman}{2024}]{alvarez2024learning}
Alvarez, L. and C.~Biderman (2024).
\newblock The learning effects of subsidies to bundled goods: a semiparametric
  approach.
\newblock \textit{arXiv:2311.01217}.

\bibitem[\protect\citeauthoryear{Alvarez and Orestes}{Alvarez and
  Orestes}{2023}]{alvarez2023quantile}
Alvarez, L. and V.~Orestes (2023).
\newblock Quantile mixture models: Estimation and inference.
\newblock Working paper.

\bibitem[\protect\citeauthoryear{Alvarez and Ferman}{Alvarez and
  Ferman}{2024}]{AlvarezFerman2024}
Alvarez, L. A.~F. and B.~Ferman (2024).
\newblock On “imputation of counterfactual outcomes when the errors are
  predictable”: Discussions on misspecification and suggestions of
  sensitivity analyses.
\newblock {\em Journal of Business \& Economic Statistics\/}~{\em 42\/}(4),
  1123--1127.

\bibitem[\protect\citeauthoryear{Anderson and Hsiao}{Anderson and
  Hsiao}{1982}]{Anderson1982}
Anderson, T. and C.~Hsiao (1982).
\newblock Formulation and estimation of dynamic models using panel data.
\newblock {\em Journal of Econometrics\/}~{\em 18\/}(1), 47--82.

\bibitem[\protect\citeauthoryear{Arellano and Bond}{Arellano and
  Bond}{1991}]{Arellano1991}
Arellano, M. and S.~Bond (1991, 04).
\newblock {Some Tests of Specification for Panel Data: Monte Carlo Evidence and
  an Application to Employment Equations}.
\newblock {\em The Review of Economic Studies\/}~{\em 58\/}(2), 277--297.

\bibitem[\protect\citeauthoryear{Athey, Bickel, Chen, Imbens, and
  Pollmann}{Athey et~al.}{2023}]{Athey2021}
Athey, S., P.~J. Bickel, A.~Chen, G.~W. Imbens, and M.~Pollmann (2023, 07).
\newblock {Semi-parametric estimation of treatment effects in randomised
  experiments}.
\newblock {\em Journal of the Royal Statistical Society Series B: Statistical
  Methodology\/}~{\em 85\/}(5), 1615--1638.

\bibitem[\protect\citeauthoryear{Athey, Tibshirani, and Wager}{Athey
  et~al.}{2019}]{Athey2019}
Athey, S., J.~Tibshirani, and S.~Wager (2019).
\newblock {Generalized random forests}.
\newblock {\em The Annals of Statistics\/}~{\em 47\/}(2), 1148 -- 1178.

\bibitem[\protect\citeauthoryear{Barrio, Giné, and Utzet}{Barrio
  et~al.}{2005}]{Barrio2005}
Barrio, E.~D., E.~Giné, and F.~Utzet (2005).
\newblock {Asymptotics for L2 functionals of the empirical quantile process,
  with applications to tests of fit based on weighted Wasserstein distances}.
\newblock {\em Bernoulli\/}~{\em 11\/}(1), 131 -- 189.

\bibitem[\protect\citeauthoryear{Belloni, Chen, Chernozhukov, and
  Hansen}{Belloni et~al.}{2012}]{Belloni2012}
Belloni, A., D.~Chen, V.~Chernozhukov, and C.~Hansen (2012).
\newblock Sparse models and methods for optimal instruments with an application
  to eminent domain.
\newblock {\em Econometrica\/}~{\em 80\/}(6), 2369--2429.

\bibitem[\protect\citeauthoryear{Belloni, Chernozhukov, Chetverikov, and
  Fernández-Val}{Belloni et~al.}{2019}]{Belloni2019}
Belloni, A., V.~Chernozhukov, D.~Chetverikov, and I.~Fernández-Val (2019).
\newblock Conditional quantile processes based on series or many regressors.
\newblock {\em Journal of Econometrics\/}~{\em 213\/}(1), 4 -- 29.
\newblock Annals: In Honor of Roger Koenker.

\bibitem[\protect\citeauthoryear{Biderman}{Biderman}{2018}]{Biderman2018}
Biderman, C. (2018, October).
\newblock E-hailing and last mile connectivity in sao paulo.
\newblock {\em American Economic Association Registry for Randomized Controlled
  Trials\/}.
\newblock {AEARCTR-0003518}.

\bibitem[\protect\citeauthoryear{Billingsley}{Billingsley}{2012}]{Billingsley2012}
Billingsley, P. (2012).
\newblock {\em Probability and Measure}.
\newblock Wiley.

\bibitem[\protect\citeauthoryear{Boulange, Hanasaki, Yamazaki, and
  Pokhrel}{Boulange et~al.}{2021}]{Boulange2021}
Boulange, J., N.~Hanasaki, D.~Yamazaki, and Y.~Pokhrel (2021).
\newblock Role of dams in reducing global flood exposure under climate change.
\newblock {\em Nature communications\/}~{\em 12\/}(1), 1--7.

\bibitem[\protect\citeauthoryear{Broniatowski and Decurninge}{Broniatowski and
  Decurninge}{2016}]{Broniatowski2016}
Broniatowski, M. and A.~Decurninge (2016).
\newblock {Estimation for models defined by conditions on their L-moments}.
\newblock {\em IEEE Transactions on Information Theory\/}~{\em 62\/}(9),
  5181--5198.

\bibitem[\protect\citeauthoryear{Carrasco and Florens}{Carrasco and
  Florens}{2014}]{Carrasco2014}
Carrasco, M. and J.-P. Florens (2014).
\newblock On the asymptotic efficiency of gmm.
\newblock {\em Econometric Theory\/}~{\em 30\/}(2), 372–406.

\bibitem[\protect\citeauthoryear{Castellacci}{Castellacci}{2012}]{castellacci2012formula}
Castellacci, G. (2012).
\newblock A formula for the quantiles of mixtures of distributions with
  disjoint supports.
\newblock {\em Available at SSRN 2055022\/}.

\bibitem[\protect\citeauthoryear{Cattaneo, Feng, and and}{Cattaneo
  et~al.}{2021}]{Cattaneo2021}
Cattaneo, M.~D., Y.~Feng, and R.~T. and (2021).
\newblock Prediction intervals for synthetic control methods.
\newblock {\em Journal of the American Statistical Association\/}~{\em
  116\/}(536), 1865--1880.
\newblock PMID: 35756161.

\bibitem[\protect\citeauthoryear{Chernozhukov, Chetverikov, Demirer, Duflo,
  Hansen, Newey, and Robins}{Chernozhukov et~al.}{2018}]{Chernozhukov2018}
Chernozhukov, V., D.~Chetverikov, M.~Demirer, E.~Duflo, C.~Hansen, W.~Newey,
  and J.~Robins (2018, 01).
\newblock {Double/debiased machine learning for treatment and structural
  parameters}.
\newblock {\em The Econometrics Journal\/}~{\em 21\/}(1), C1--C68.

\bibitem[\protect\citeauthoryear{Chernozhukov, Escanciano, Ichimura, Newey, and
  Robins}{Chernozhukov et~al.}{2022}]{Chernozhukov2022}
Chernozhukov, V., J.~C. Escanciano, H.~Ichimura, W.~K. Newey, and J.~M. Robins
  (2022).
\newblock Locally robust semiparametric estimation.
\newblock {\em Econometrica\/}~{\em 90\/}(4), 1501--1535.

\bibitem[\protect\citeauthoryear{Chernozhukov and
  Fern{\'a}ndez-Val}{Chernozhukov and
  Fern{\'a}ndez-Val}{2011}]{chernozhukov2011inference}
Chernozhukov, V. and I.~Fern{\'a}ndez-Val (2011).
\newblock Inference for extremal conditional quantile models, with an
  application to market and birthweight risks.
\newblock {\em The Review of Economic Studies\/}~{\em 78\/}(2), 559--589.

\bibitem[\protect\citeauthoryear{Chernozhukov, Wüthrich, and Zhu}{Chernozhukov
  et~al.}{2021a}]{CWZ2021}
Chernozhukov, V., K.~Wüthrich, and Y.~Zhu (2021a).
\newblock Distributional conformal prediction.
\newblock {\em Proceedings of the National Academy of Sciences\/}~{\em
  118\/}(48), e2107794118.

\bibitem[\protect\citeauthoryear{Chernozhukov, Wüthrich, and Zhu}{Chernozhukov
  et~al.}{2021b}]{CWZ2021b}
Chernozhukov, V., K.~Wüthrich, and Y.~Zhu (2021b).
\newblock An exact and robust conformal inference method for counterfactual and
  synthetic controls.
\newblock {\em Journal of the American Statistical Association\/}~{\em
  116\/}(536), 1849--1864.

\bibitem[\protect\citeauthoryear{Crump, Hotz, Imbens, and Mitnik}{Crump
  et~al.}{2009}]{Crump2009}
Crump, R.~K., V.~J. Hotz, G.~W. Imbens, and O.~A. Mitnik (2009, 01).
\newblock {Dealing with limited overlap in estimation of average treatment
  effects}.
\newblock {\em Biometrika\/}~{\em 96\/}(1), 187--199.

\bibitem[\protect\citeauthoryear{Csorgo and Revesz}{Csorgo and
  Revesz}{1978}]{Csorgo1978}
Csorgo, M. and P.~Revesz (1978, 07).
\newblock Strong approximations of the quantile process.
\newblock {\em Annals of Statistics\/}~{\em 6\/}(4), 882--894.

\bibitem[\protect\citeauthoryear{Das}{Das}{2021}]{Das2021}
Das, S. (2021).
\newblock Extreme rainfall estimation at ungauged locations: information that
  needs to be included in low-lying monsoon climate regions like bangladesh.
\newblock {\em Journal of Hydrology\/}~{\em 601}, 126616.

\bibitem[\protect\citeauthoryear{Dey, Mazucheli, and Nadarajah}{Dey
  et~al.}{2018}]{Dey2018}
Dey, S., J.~Mazucheli, and S.~Nadarajah (2018, may).
\newblock {Kumaraswamy distribution: different methods of estimation}.
\newblock {\em Computational and Applied Mathematics\/}~{\em 37\/}(2),
  2094--2111.

\bibitem[\protect\citeauthoryear{Donald, Imbens, and Newey}{Donald
  et~al.}{2003}]{Donald2003}
Donald, S.~G., G.~W. Imbens, and W.~K. Newey (2003).
\newblock Empirical likelihood estimation and consistent tests with conditional
  moment restrictions.
\newblock {\em Journal of Econometrics\/}~{\em 117\/}(1), 55--93.

\bibitem[\protect\citeauthoryear{Donald, Imbens, and Newey}{Donald
  et~al.}{2009}]{Donald2009}
Donald, S.~G., G.~W. Imbens, and W.~K. Newey (2009).
\newblock Choosing instrumental variables in conditional moment restriction
  models.
\newblock {\em Journal of Econometrics\/}~{\em 152\/}(1), 28--36.

\bibitem[\protect\citeauthoryear{Donald and Newey}{Donald and
  Newey}{2001}]{Donald2001}
Donald, S.~G. and W.~K. Newey (2001).
\newblock Choosing the number of instruments.
\newblock {\em Econometrica\/}~{\em 69\/}(5), 1161--1191.

\bibitem[\protect\citeauthoryear{Ferman and Pinto}{Ferman and
  Pinto}{2021}]{Ferman2021}
Ferman, B. and C.~Pinto (2021).
\newblock Synthetic controls with imperfect pretreatment fit.
\newblock {\em Quantitative Economics\/}~{\em 12\/}(4), 1197--1221.

\bibitem[\protect\citeauthoryear{Ferrari and Yang}{Ferrari and
  Yang}{2010}]{Ferrari2010}
Ferrari, D. and Y.~Yang (2010, 04).
\newblock Maximum l q -likelihood estimation.
\newblock {\em Annals of Statistics\/}~{\em 38\/}(2), 753--783.

\bibitem[\protect\citeauthoryear{Foss, Korshunov, Zachary, et~al.}{Foss
  et~al.}{2011}]{foss2011introduction}
Foss, S., D.~Korshunov, S.~Zachary, et~al. (2011).
\newblock {\em An introduction to heavy-tailed and subexponential
  distributions}, Volume~6.
\newblock Springer.

\bibitem[\protect\citeauthoryear{Fotopoulos and Ahn}{Fotopoulos and
  Ahn}{1994}]{Fotopoulos1994}
Fotopoulos, S. and S.~Ahn (1994).
\newblock Strong approximation of the quantile processes and its applications
  under strong mixing properties.
\newblock {\em Journal of Multivariate Analysis\/}~{\em 51\/}(1), 17--45.

\bibitem[\protect\citeauthoryear{Freyberger and Rai}{Freyberger and
  Rai}{2018}]{Freyberger2018}
Freyberger, J. and Y.~Rai (2018).
\newblock Uniform confidence bands: Characterization and optimality.
\newblock {\em Journal of Econometrics\/}~{\em 204\/}(1), 119--130.

\bibitem[\protect\citeauthoryear{G{\"o}tze, Naumov, Spokoiny, and
  Ulyanov}{G{\"o}tze et~al.}{2019}]{Gotze2019}
G{\"o}tze, F., A.~Naumov, V.~Spokoiny, and V.~Ulyanov (2019).
\newblock {Large ball probabilities, Gaussian comparison and
  anti-concentration}.
\newblock {\em Bernoulli\/}~{\em 25\/}(4A), 2538 -- 2563.

\bibitem[\protect\citeauthoryear{Gourieroux and Jasiak}{Gourieroux and
  Jasiak}{2008}]{Gourieroux2008}
Gourieroux, C. and J.~Jasiak (2008).
\newblock {Dynamic quantile models}.
\newblock {\em Journal of Econometrics\/}~{\em 147\/}(1), 198--205.

\bibitem[\protect\citeauthoryear{Guerre and Sabbah}{Guerre and
  Sabbah}{2012}]{Guerre2012}
Guerre, E. and C.~Sabbah (2012).
\newblock Uniform bias study and bahadur representation for local polynomial
  estimators of the conditional quantile function.
\newblock {\em Econometric Theory\/}~{\em 28\/}(1), 87--129.

\bibitem[\protect\citeauthoryear{Gupta and Kundu}{Gupta and
  Kundu}{2001}]{Gupta2001}
Gupta, R.~D. and D.~Kundu (2001).
\newblock {Generalized exponential distribution: Different method of
  estimations}.
\newblock {\em Journal of Statistical Computation and Simulation\/}~{\em
  69\/}(4), 315--337.

\bibitem[\protect\citeauthoryear{Hahn, Kuersteiner, and Newey}{Hahn
  et~al.}{2002}]{hahn2002higher}
Hahn, J., G.~Kuersteiner, and W.~Newey (2002).
\newblock Higher order properties of bootstrap and jackknife bias corrected
  maximum likelihood estimators.
\newblock {\em Unpublished manuscript\/}.

\bibitem[\protect\citeauthoryear{Han and Phillips}{Han and
  Phillips}{2006}]{Han2006}
Han, C. and P.~C.~B. Phillips (2006).
\newblock Gmm with many moment conditions.
\newblock {\em Econometrica\/}~{\em 74\/}(1), 147--192.

\bibitem[\protect\citeauthoryear{Hansen}{Hansen}{1982}]{Hansen1982}
Hansen, L.~P. (1982).
\newblock Large sample properties of generalized method of moments estimators.
\newblock {\em Econometrica\/}~{\em 50\/}(4), 1029--1054.

\bibitem[\protect\citeauthoryear{Hosking}{Hosking}{1986}]{hosking1986theory}
Hosking, J.~R. (1986).
\newblock {\em The theory of probability weighted moments}.
\newblock IBM Research Division, TJ Watson Research Center New York, USA.

\bibitem[\protect\citeauthoryear{Hosking}{Hosking}{2007}]{Hosking2007}
Hosking, J.~R. (2007).
\newblock {Some theory and practical uses of trimmed L-moments}.
\newblock {\em Journal of Statistical Planning and Inference\/}~{\em 137\/}(9),
  3024--3039.

\bibitem[\protect\citeauthoryear{Hosking}{Hosking}{1990}]{Hosking1990}
Hosking, J. R.~M. (1990, sep).
\newblock {L-Moments: Analysis and Estimation of Distributions Using Linear
  Combinations of Order Statistics}.
\newblock {\em Journal of the Royal Statistical Society: Series B
  (Methodological)\/}~{\em 52\/}(1), 105--124.

\bibitem[\protect\citeauthoryear{Hosking}{Hosking}{2024}]{Hosking2024}
Hosking, J. R.~M. (2024).
\newblock {\em L-Moments}.
\newblock R package, version 3.2.

\bibitem[\protect\citeauthoryear{Hosking and Wallis}{Hosking and
  Wallis}{1987}]{Hosking1987}
Hosking, J. R.~M. and J.~R. Wallis (1987).
\newblock Parameter and quantile estimation for the generalized pareto
  distribution.
\newblock {\em Technometrics\/}~{\em 29\/}(3), 339--349.

\bibitem[\protect\citeauthoryear{Hosking, Wallis, and Wood}{Hosking
  et~al.}{1985}]{Hosking1985}
Hosking, J. R.~M., J.~R. Wallis, and E.~F. Wood (1985).
\newblock Estimation of the generalized extreme-value distribution by the
  method of probability-weighted moments.
\newblock {\em Technometrics\/}~{\em 27\/}(3), 251--261.

\bibitem[\protect\citeauthoryear{Ichimura and Newey}{Ichimura and
  Newey}{2022}]{Ichimura2021}
Ichimura, H. and W.~K. Newey (2022).
\newblock The influence function of semiparametric estimators.
\newblock {\em Quantitative Economics\/}~{\em 13\/}(1), 29--61.

\bibitem[\protect\citeauthoryear{Karvanen}{Karvanen}{2006}]{karvanen2006estimation}
Karvanen, J. (2006).
\newblock Estimation of quantile mixtures via l-moments and trimmed l-moments.
\newblock {\em Computational Statistics \& Data Analysis\/}~{\em 51\/}(2),
  947--959.

\bibitem[\protect\citeauthoryear{Kennedy}{Kennedy}{2023}]{kennedy2023semiparametric}
Kennedy, E.~H. (2023).
\newblock Semiparametric doubly robust targeted double machine learning: a
  review.

\bibitem[\protect\citeauthoryear{Kerstens, Mounir, and {De Woestyne}}{Kerstens
  et~al.}{2011}]{Kerstens2011}
Kerstens, K., A.~Mounir, and I.~V. {De Woestyne} (2011).
\newblock {Non-parametric frontier estimates of mutual fund performance using
  C- and L-moments: Some specification tests}.
\newblock {\em Journal of Banking and Finance\/}~{\em 35\/}(5), 1190--1201.

\bibitem[\protect\citeauthoryear{Kivaranovic, Ristl, Posch, and
  Leeb}{Kivaranovic et~al.}{2020}]{Kivaranovic2020}
Kivaranovic, D., R.~Ristl, M.~Posch, and H.~Leeb (2020).
\newblock Conformal prediction intervals for the individual treatment effect.

\bibitem[\protect\citeauthoryear{Koenker and Machado}{Koenker and
  Machado}{1999}]{Koenker1999}
Koenker, R. and J.~A. Machado (1999).
\newblock Gmm inference when the number of moment conditions is large.
\newblock {\em Journal of Econometrics\/}~{\em 93\/}(2), 327--344.

\bibitem[\protect\citeauthoryear{Komunjer and Vuong}{Komunjer and
  Vuong}{2010}]{Komunjer2010}
Komunjer, I. and Q.~Vuong (2010).
\newblock Semiparametric efficiency bound in time-series models for conditional
  quantiles.
\newblock {\em Econometric Theory\/}~{\em 26\/}(2), 383--405.

\bibitem[\protect\citeauthoryear{Kreyszig}{Kreyszig}{1989}]{Kreyszig1989}
Kreyszig, E. (1989).
\newblock {\em Introductory Functional Analysis with Applications}.
\newblock Wiley.

\bibitem[\protect\citeauthoryear{Landwehr, Matalas, and Wallis}{Landwehr
  et~al.}{1979}]{Landwehr1979}
Landwehr, J.~M., N.~C. Matalas, and J.~R. Wallis (1979).
\newblock Probability weighted moments compared with some traditional
  techniques in estimating gumbel parameters and quantiles.
\newblock {\em Water Resources Research\/}~{\em 15\/}(5), 1055--1064.

\bibitem[\protect\citeauthoryear{Lei and Candès}{Lei and
  Candès}{2021}]{Lei2021}
Lei, L. and E.~J. Candès (2021, 10).
\newblock Conformal inference of counterfactuals and individual treatment
  effects.
\newblock {\em Journal of the Royal Statistical Society Series B: Statistical
  Methodology\/}~{\em 83\/}(5), 911--938.

\bibitem[\protect\citeauthoryear{Li, Zwiers, Zhang, Li, Sun, and Wehner}{Li
  et~al.}{2021}]{Li2021}
Li, C., F.~Zwiers, X.~Zhang, G.~Li, Y.~Sun, and M.~Wehner (2021).
\newblock Changes in annual extremes of daily temperature and precipitation in
  cmip6 models.
\newblock {\em Journal of Climate\/}~{\em 34\/}(9), 3441--3460.

\bibitem[\protect\citeauthoryear{Li, Li, Qin, Lin, and Yang}{Li
  et~al.}{2021}]{Yang2021b}
Li, Y., R.~Li, Y.~Qin, C.~Lin, and Y.~Yang (2021).
\newblock Robust group variable screening based on maximum lq-likelihood
  estimation.
\newblock {\em Statistics in Medicine\/}~{\em 40\/}(30), 6818--6834.

\bibitem[\protect\citeauthoryear{Luo et~al.}{Luo et~al.}{2015}]{Luo2016}
Luo, Y. et~al. (2015).
\newblock {\em High-dimensional econometrics and model selection}.
\newblock Ph.\ D. thesis, Massachusetts Institute of Technology.

\bibitem[\protect\citeauthoryear{Mason}{Mason}{1984}]{Mason1984}
Mason, D.~M. (1984, 02).
\newblock Weak convergence of the weighted empirical quantile process in
  $l^2(0, 1)$.
\newblock {\em Annals of Probability\/}~{\em 12\/}(1), 243--255.

\bibitem[\protect\citeauthoryear{Meinshausen}{Meinshausen}{2006}]{Meinshausen2006}
Meinshausen, N. (2006).
\newblock Quantile regression forests.
\newblock {\em Journal of Machine Learning Research\/}~{\em 7\/}(35), 983--999.

\bibitem[\protect\citeauthoryear{Newey}{Newey}{1990}]{Newey1990}
Newey, W.~K. (1990).
\newblock Semiparametric efficiency bounds.
\newblock {\em Journal of Applied Econometrics\/}~{\em 5\/}(2), 99--135.

\bibitem[\protect\citeauthoryear{Newey and McFadden}{Newey and
  McFadden}{1994}]{Newey1994}
Newey, W.~K. and D.~McFadden (1994).
\newblock Chapter 36 large sample estimation and hypothesis testing.
\newblock In {\em Handbook of Econometrics}, Volume~4, pp.\  2111 -- 2245.
  Elsevier.

\bibitem[\protect\citeauthoryear{Newey and Smith}{Newey and
  Smith}{2004}]{Newey2004}
Newey, W.~K. and R.~J. Smith (2004).
\newblock Higher order properties of gmm and generalized empirical likelihood
  estimators.
\newblock {\em Econometrica\/}~{\em 72\/}(1), 219--255.

\bibitem[\protect\citeauthoryear{Newey and West}{Newey and
  West}{1987}]{Newey1987}
Newey, W.~K. and K.~D. West (1987).
\newblock A simple, positive semi-definite, heteroskedasticity and
  autocorrelation consistent covariance matrix.
\newblock {\em Econometrica\/}~{\em 55\/}(3), 703--708.

\bibitem[\protect\citeauthoryear{Okui}{Okui}{2009}]{Okui2009}
Okui, R. (2009).
\newblock The optimal choice of moments in dynamic panel data models.
\newblock {\em Journal of Econometrics\/}~{\em 151\/}(1), 1--16.

\bibitem[\protect\citeauthoryear{Parzen}{Parzen}{1960}]{Parzen1960}
Parzen, E. (1960).
\newblock {\em Modern probability theory and its applications}.
\newblock Wiley.

\bibitem[\protect\citeauthoryear{Pfanzagl and Wefelmeyer}{Pfanzagl and
  Wefelmeyer}{1978}]{Pfanzagl1978}
Pfanzagl, J. and W.~Wefelmeyer (1978).
\newblock A third-order optimum property of the maximum likelihood estimator.
\newblock {\em Journal of Multivariate Analysis\/}~{\em 8\/}(1), 1--29.

\bibitem[\protect\citeauthoryear{Portnoy}{Portnoy}{1991}]{Portnoy1991}
Portnoy, S. (1991).
\newblock Asymptotic behavior of regression quantiles in non-stationary,
  dependent cases.
\newblock {\em Journal of Multivariate Analysis\/}~{\em 38\/}(1), 100 -- 113.

\bibitem[\protect\citeauthoryear{Rothenberg}{Rothenberg}{1971}]{Rothenberg1971}
Rothenberg, T.~J. (1971).
\newblock Identification in parametric models.
\newblock {\em Econometrica\/}~{\em 39\/}(3), 577--91.

\bibitem[\protect\citeauthoryear{Sankarasubramanian and
  Srinivasan}{Sankarasubramanian and Srinivasan}{1999}]{Sankarasubramanian1999}
Sankarasubramanian, A. and K.~Srinivasan (1999).
\newblock {Investigation and comparison of sampling properties of L-moments and
  conventional moments}.
\newblock {\em Journal of Hydrology\/}~{\em 218\/}(1-2), 13--34.

\bibitem[\protect\citeauthoryear{{\v{S}}imkov{\'{a}}}{{\v{S}}imkov{\'{a}}}{2017}]{Simkova2017}
{\v{S}}imkov{\'{a}}, T. (2017).
\newblock {Statistical inference based on L-moments}.
\newblock {\em Statistika\/}~{\em 97\/}(1), 44--58.

\bibitem[\protect\citeauthoryear{van~der Vaart}{van~der
  Vaart}{1998}]{Vaart1998}
van~der Vaart, A.~W. (1998).
\newblock {\em Asymptotic Statistics}.
\newblock Cambridge Series in Statistical and Probabilistic Mathematics.
  Cambridge University Press.

\bibitem[\protect\citeauthoryear{van~der Vaart and Wellner}{van~der Vaart and
  Wellner}{1996}]{Vaart1996}
van~der Vaart, A.~W. and J.~A. Wellner (1996).
\newblock {\em Weak convergence and empirical processes}.
\newblock Springer.

\bibitem[\protect\citeauthoryear{Wang and Hutson}{Wang and
  Hutson}{2013}]{Wang2013}
Wang, D. and A.~D. Hutson (2013).
\newblock {Joint confidence region estimation of L-moment ratios with an
  extension to right censored data}.
\newblock {\em Journal of Applied Statistics\/}~{\em 40\/}(2), 368--379.

\bibitem[\protect\citeauthoryear{Wang}{Wang}{1997}]{Wang1997}
Wang, Q.~J. (1997).
\newblock {LH moments for statistical analysis of extreme events}.
\newblock {\em Water Resources Research\/}~{\em 33\/}(12), 2841--2848.

\bibitem[\protect\citeauthoryear{Wooldridge}{Wooldridge}{2010}]{Wooldridge2010}
Wooldridge, J. (2010).
\newblock {\em Econometric analysis of cross section and panel data}.
\newblock Cambridge, Mass: MIT Press.

\bibitem[\protect\citeauthoryear{Yang, Bian, Liu, Jiang, Lu, Gao, and
  Song}{Yang et~al.}{2021}]{Yang2021a}
Yang, J., Z.~Bian, J.~Liu, B.~Jiang, W.~Lu, X.~Gao, and H.~Song (2021).
\newblock No-reference quality assessment for screen content images using
  visual edge model and adaboosting neural network.
\newblock {\em IEEE Transactions on Image Processing\/}~{\em 30}, 6801--6814.

\bibitem[\protect\citeauthoryear{Yoshihara}{Yoshihara}{1995}]{Yoshihara1995}
Yoshihara, K. (1995).
\newblock The bahadur representation of sample quantiles for sequences of
  strongly mixing random variables.
\newblock {\em Statistics \& Probability Letters\/}~{\em 24\/}(4), 299 -- 304.

\bibitem[\protect\citeauthoryear{Yu}{Yu}{1996}]{Yu1996}
Yu, H. (1996).
\newblock A note on strong approximation for quantile processes of strong
  mixing sequences.
\newblock {\em Statistics \& Probability Letters\/}~{\em 30\/}(1), 1 -- 7.

\bibitem[\protect\citeauthoryear{Yu and Jones}{Yu and Jones}{1998}]{Yu1998}
Yu, K. and M.~C. Jones (1998).
\newblock Local linear quantile regression.
\newblock {\em Journal of the American Statistical Association\/}~{\em
  93\/}(441), 228--237.

\end{thebibliography}
